%% file: main.tex
\documentclass[aps, prx,amsmath,amssymb,longbibliography,showpacs,letterpaper,twocolumn,nofootinbib,floatfix,superscriptaddress]{revtex4-1}

\include{packages}

\definecolor{lime}{HTML}{A6CE39}
\DeclareRobustCommand{\orcidicon}{
	\begin{tikzpicture}
	\draw[lime, fill=lime] (0,0) 
	circle [radius=0.16] 
	node[white] {{\fontfamily{qag}\selectfont \tiny ID}};
	\draw[white, fill=white] (-0.0665,0.095) 
	circle [radius=0.005];
	\end{tikzpicture}
	\hspace{-2mm}
}

\foreach \x in {A, ..., Z}{\expandafter\xdef\csname orcid\x\endcsname{\noexpand\href{https://orcid.org/\csname orcidauthor\x\endcsname}
			{\noexpand\orcidicon}}
   }

\begin{document}

\title{From the Dawn of Neutrino Astronomy to a New View of the Extreme Universe}
\begin{abstract}
Over the past decade, neutrino astronomy has emerged as a new window into the extreme and hidden universe.
Current-generation experiments have detected high-energy neutrinos of astrophysical origin and identified the first sources, opening the field to discovery.
Looking ahead, the authors of this \textit{Perspective} identify seven major open questions in neutrino astrophysics and particle physics that could lead to transformative discoveries over the next 20 years.
These multidisciplinary questions range from understanding the vicinity of a black hole to unveiling the nature of neutrino mass, among other topics.
Additionally, we critically review the current experimental capabilities and their limitations and, from there, discuss the interplay between different proposed neutrino telescope technologies and analysis techniques.
The authors firmly believe that achieving the immense discovery potential over the next two decades demands a model of global partnership and specialized, complementary detectors.
This collaborative neutrino telescope network will pave the way for a thriving multimessenger era, transforming our understanding of neutrino physics, astrophysics, and the extreme universe.
\end{abstract}
\date{\today}

\author{C.~A.~Arg{\"u}elles\orcidA{}}
\email{carguelles@fas.harvard.edu}
\affiliation{Department of Physics \& Laboratory for Particle Physics and Cosmology, Harvard University, Cambridge, MA 02138, USA}

\author{F.~Halzen\orcidB{}}
\email{halzen@icecube.wisc.edu}
\affiliation{Department of Physics \& Wisconsin IceCube Particle Astrophysics Center, University of Wisconsin, Madison, WI 53706, USA}

\author{N.~Kurahashi\orcidC{}}
\email{naoko@icecube.wisc.edu}
\affiliation{Department of Physics, Drexel University, 3141 Chestnut Street, Philadelphia, PA 19104, USA}

\maketitle


\section{Introduction}\label{sec:intro}

Astronomy views the Universe at wavelengths of light, reaching from radio waves to gamma rays.
The spectacularly successful method of multiwavelength astronomy~\cite{Hill:2018trh} eventually fails at the smallest wavelengths or the highest energies where the Universe turns dark to light~\cite{Gould:1966pza}.
Above the threshold for pair production $\gamma+\gamma_{\rm cmb} \rightarrow e^+ + e^-$, the gamma rays interact with microwave photons, $\SI{411}\cm^{-3}$~\cite{ParticleDataGroup:2014cgo}, to produce electron-positron pairs.
These shower down in the cosmic microwave background (CMB) and in other background electromagnetic fields permeating the Universe to reach our telescopes as multiple photons of lower energy.
At some point, we barely see the center of our own galaxy. 
In this extreme Universe known to harbor powerful cosmic particle accelerators, weakly interacting neutrinos are the only astronomical messengers.  

Imagined more than half a century ago~\cite{Spiering:2012xe}, high-energy neutrino astronomy became a reality when 
a cubic kilometer of transparent natural Antarctic ice one mile below the geographic South Pole was transformed into the largest particle detector ever built~\cite{IceCube:2016zyt}.
It discovered neutrinos in the $\si\TeV \sim \si\PeV$ energy range originating beyond our galaxy, opening a new window for astronomy~\cite{Aartsen:2013jdh,IceCube:2014stg}.
The observed energy density of neutrinos in the extreme universe exceeds the energy in gamma rays~\cite{Fang:2022trf}; it even outshines the neutrino flux from the nearby sources in our galaxy~\cite{IceCube:2023ame}.
The Galactic plane only appears as a faint glow at the ten percent level of the extragalactic flux, in sharp contrast with any other wavelength of light where the Milky Way is the dominant feature in the sky.
This observation implies the existence of sources of high-energy neutrinos in other galaxies that are not present in our own~\cite{Fang:2023azx}.

After accumulating a decade of data with a detector with gradually improved sensitivity, the first high-energy neutrino sources emerged in the neutrino sky: the active galaxies NGC~1068, NGC~4151, PKS~1424+240~\cite{IceCube:2022der,IceCube:2023jds}, and TXS~0506+056~\cite{IceCube:2018dnn,IceCube:2018cha}.
The observations point to the acceleration of protons and the production of neutrinos in the obscured dense cores surrounding the supermassive black holes at their centers, typically within a distance of only $\sim10$ Schwarzschild radii~\cite{IceCube:2022der,Murase:2022dog}.
TXS~0506+056 had previously been identified as a neutrino source in a multimessenger campaign triggered by a neutrino of $\SI{290}\TeV$ energy, IC170922, and by the independent observation of a neutrino burst from the same source in archival IceCube data in 2014~\cite{IceCube:2018dnn,IceCube:2018cha}; see also Refs.~\cite{Baikal-GVD:2023qib,ANTARES:2018osx} for hints from water-based neutrino telescopes.

With indications that neutrinos originate in the vicinity of the central black hole of active galaxies, which can only be reached by radio telescopes, neutrino astronomy represents an extraordinary opportunity for discovery.
This includes resolving the century-old problem of where cosmic rays originate and how they reach their phenomenal energies.
Since neutrinos originate from the decay of charged pions, only sources accelerating cosmic rays that interact with radiation fields or gas surrounding the accelerator to produce pions can be neutrino sources.
With energies exceeding by a factor of one million those produced with earthbound accelerator beams, cosmic neutrinos also present a remarkable opportunity for studying neutrinos themselves.

In contrast with its exceptional discovery potential, neutrino astronomy faces the evident lack of statistics for future progress, not only in neutrinos but also in identified sources, with only a handful in a decade.
A weakness in the current status represents a future challenge and opportunity.
The sensitivity of IceCube has recently been significantly enhanced by progress in the characterization of the optics of ice and by exploiting the rapid progress in machine learning to collect larger data samples that are reconstructed in neutrinos with improved energy and angular resolution.
The observations of NGC~1068 and the Galactic plane directly resulted from these improvements.

Efforts to upgrade the performance of IceCube continue~\cite{Ishihara:2019aao}.
The IceCube project, which built on the experience of its predecessor AMANDA~\cite{Andres:1999hm},  has instrumented a cubic kilometer of natural Antarctic ice with 5,160~light sensors to configure a Cherenkov detector~\cite{IceCube:2016zyt}.
IceCube is comprised of 10-inch photomultiplier tubes deployed like beads on 86 one-kilometer-long electric cables, separated by $\SI{125}\m$.
This instrumentation maps the secondary muon tracks and showers initiated by neutrinos interacting inside or near the detector to reconstruct their direction, energy, and flavor.
Further progress requires a next-generation instrument with a sensitivity improved by an order of magnitude.
The design for such an instrument exists~\cite{IceCube-Gen2:2020qha}, and its construction has been strongly endorsed by both the astronomy~\cite{NAP26141} and particle physics communities~\cite{P5:2023wyd}.

In parallel, instruments the size of IceCube are under construction in Lake Baikal~\cite{Shoibonov:2019gfj}, the Mediterranean Sea~\cite{Sanguineti:2023qfa}, and the Pacific Ocean~\cite{Bailly:2021dxn} off the coast of Canada.
Pioneered by the DUMAND~\cite{Roberts:1992re} and ANTARES~\cite{ANTARES:2011hfw} projects, these new initiatives introduce multiple advantages: as complementary instruments using deep water as the Cherenkov medium, by collectively adding to the growth of statistics of IceCube, and by improving our global field of view of the neutrino sky because of their different geographic locations; examples of such analyses can be found in Refs.~\cite{ANTARES:2020srt,ANTARES:2018nyb}.
Projects in China aim for larger detectors: TRIDENT matches the next-generation IceCube detector~\cite{Ye:2023dch} and HUNT envisions a volume of $\SI{30}\km^3$~\cite{Huang:2023mzt}.
Additionally, specialized detectors that aim to find astrophysical tau neutrinos with high purity in the energy range of IceCube observations are also currently being developed~\cite{ACKERMANN202255}---notably, TRINITY~\cite{Otte:2023osf} using Cherenkov telescopes and TAMBO~\cite{TAMBO:2023plw} using particle detectors at the end-point of current astrophysical neutrino observations.
As we will discuss in this \textit{Perspective}, these small, relatively low-cost experiments will serve as an essential complement to the larger optical neutrino telescopes.

In the future, we anticipate rapid progress with the collection of larger statistical samples of neutrinos with superior directional pointing covering a larger energy range.
Further progress will be made possible by multiple instruments able to cross-check individual results, an advantage that has been demonstrated over time in high-energy physics laboratories by the deployment of several complementary detectors at a single accelerator.
Eventually, we anticipate that data from these multiple instruments will be combined in real time, as is already routinely done for the observed strains of the gravitational wave detectors---LIGO~\cite{LIGOScientific:2014pky}, Virgo~\cite{VIRGO:2014yos}, and KAGRA~\cite{KAGRA:2020tym}---yielding improved sensitivity~\cite{KAGRA:2013rdx,IceCube:2022mma}.

In parallel, many diverse efforts are underway to develop novel techniques to extend the IceCube spectrum to EeV energies~\cite{GRAND:2018iaj,POEMMA:2020ykm,PUEO:2020bnn,RNO-G:2020rmc,Prohira:2019glh,Allison:2018ynt,Adams:2017fjh}, driven mainly by radio neutrino detection techniques~\cite{Huege:2017khw}.
Unfortunately, the steeply falling flux and the slow growth of the neutrino cross section at these energies require extremely large arrays to obtain a handful of events for the most pessimistic flux predictions.
Given the lack of overlap in energy with the current observations---and thus the discovery opportunities they represent---we will not comment further on these efforts.  

In this \textit{Perspective}, we attempt to anticipate a broad-brush description of the future of the field, building on the present results and realizing that the most important impact of the rapid expansion of neutrino astronomy will result from unanticipated discoveries.
We will start by reviewing the astronomical discoveries made by the first generation of instruments and the important measurements they contributed to neutrino physics.
We will discuss how novel observations raise new scientific questions that are harder to resolve and how we anticipate that the field will move from discovery into astronomy in the future by deploying a new generation of instruments.
For the readers that are not familiar with the experimental details of neutrino telescopes and the different energy ranges involved, we encourage you to review~\cref{sec:astro-fluxes} and~\cref{sec:opt-det}.

\section{Questions that can lead to transformational discoveries}\label{sec:questinos}

In this section, we pose seven questions that we believe can be answered over the following decades and can yield discoveries that will significantly impact physics and astronomy.
At the end of each subsection, for each question, we have written a short take-home message in italics preceded by the symbol {\smaller \NibSolidRight}.
For a reader who wants a first glance at our vision of the future, we encourage you to read these short answers and our parting words in~\cref{sec:conclusion}.

\subsection{Where are gamma rays produced together with neutrinos?}\label{subsec:gamma}

\begin{figure}[ht]
\centering
\includegraphics[width=0.6\linewidth, trim=0 150 0 150, clip=true]{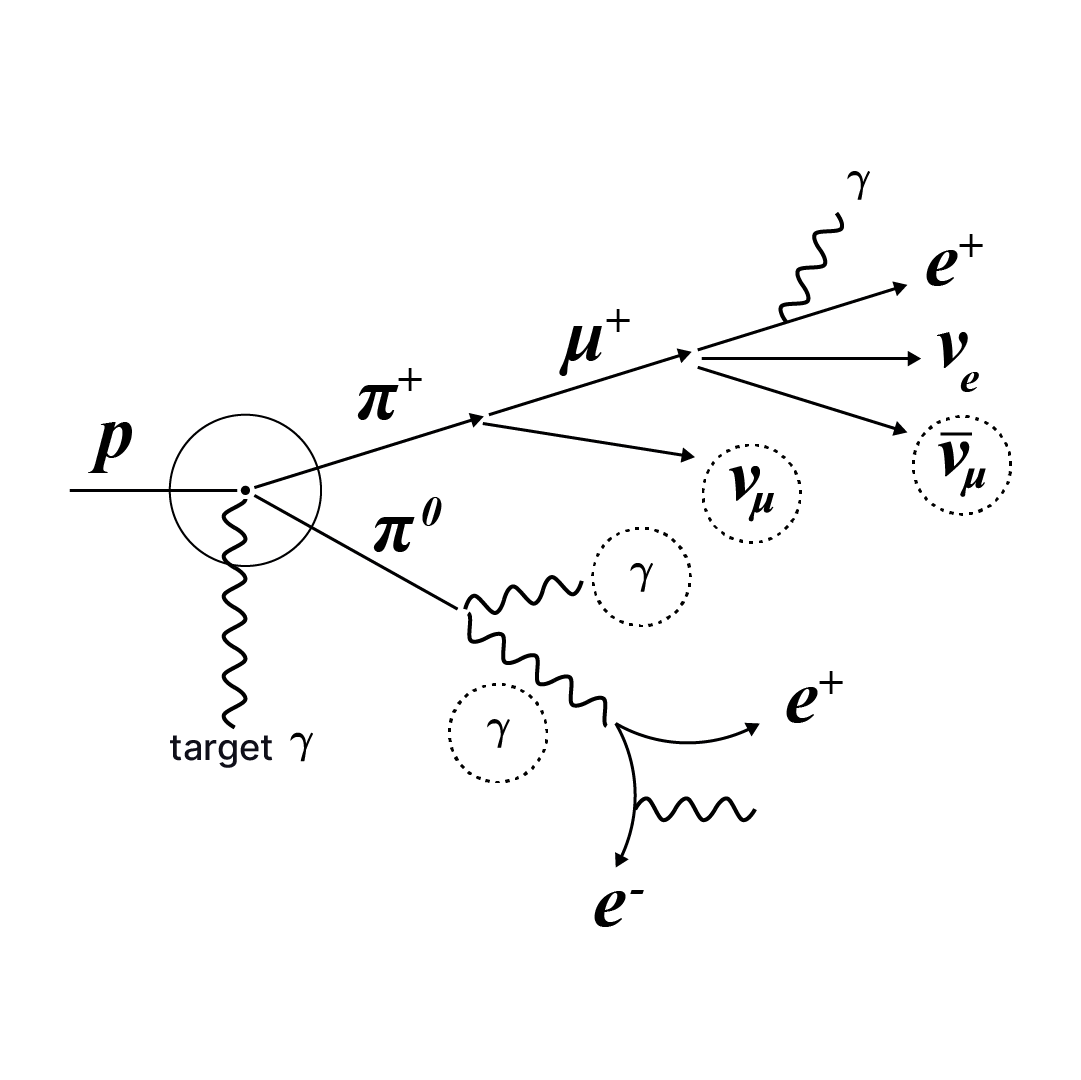}
\caption{\textbf{\textit{Flow diagram showing the production of charged and neutral pions in $p\gamma$ interactions.}}
The circles indicate equal energy going into pairs of gamma rays and muon neutrinos and antineutrinos.
Because the charged pion energy is shared roughly equally among four particles and the neutral pion energy among two photons, the photons have twice the energy of the neutrinos.
Unlike the weakly interacting neutrinos, the gamma rays and electrons may lose energy in the target and will lose additional energy in the background light (EBL) before reaching Earth.}
\label{fig:flow}
\end{figure}

The rationale for searching for cosmic-ray sources by observing neutrinos is straightforward: in relativistic particle flows, for instance, onto black holes, some of the gravitational energy released in the accretion of matter is transformed into the acceleration of protons or heavier nuclei.
These subsequently interact with radiation and/or ambient matter in the vicinity of the black hole to produce pions and other secondary particles that decay into neutrinos.
Both neutrinos and gamma rays are produced with roughly equal rates; while neutral pions decay into two gamma rays, $\pi^0\to\gamma+\gamma$, the charged pions decay into three high-energy neutrinos ($\nu$) and antineutrinos ($\bar\nu$) via the decay chain $\pi^+\to\mu^++\nu_\mu$ followed by $\mu^+\to e^++\bar\nu_\mu +\nu_e$; see~\cref{fig:flow}.
Based on this simplified flow diagram, we expect equal fluxes of gamma rays and muon neutrinos.
The flow diagram of~\cref{fig:flow} implies a multimessenger correspondence between the pionic gamma-ray and three-flavor neutrino flux~\cite{Ahlers:2015lln}:
\begin{equation}
    E_\gamma^2 \frac{dN_{\gamma}}{dE_\gamma} \approx \frac{4}{K_\pi }\, \frac{1}{3} E_\nu^2 \frac{dN}{dE_\nu}  \Big| _{E_\nu = {E_\gamma}/2},
\label{eq:nutogamma}
\end{equation}
where $K_\pi$ is the ratio of charged and neutral pions produced, with $K_\pi \approx 2 (1)$ for $pp (p\gamma$) interactions.

This powerful relation connects neutrinos and pionic gamma rays with no reference to the cosmic-ray beam producing the neutrinos; it simply reflects the fact that a $\pi^0$ decays into two gamma rays for every charged pion, producing a $\nu_\mu +\bar\nu_\mu$ pair. 
Also, from the fact that in the photoproduction process 20\% of the initial proton energy is transferred to the pion\footnote{This is referred to as the inelasticity $\kappa_{p\gamma} \simeq 0.2$.}, we obtain that the gamma ray carries one-tenth of the proton energy and the neutrino approximately half of that.

Because high-energy photons inevitably accompany cosmic neutrinos, neutrino astronomy is a multimessenger astronomy.
Unlike neutrinos, gamma rays interact with microwave photons and other diffuse sources of extragalactic background light (EBL) while propagating to Earth.
The gamma rays lose energy by $e^+e^-$ pair production, and the resulting electromagnetic shower subdivides the initial photon energy into multiple photons with reduced energy reaching our telescopes~\cite{Gould:1966pza,Berezinsky:2016feh}.

Observing neutrinos as a signature for the acceleration of cosmic-ray protons or nuclei is powerful because the alternative possibility of identifying pionic photons has turned out to be challenging because they must be isolated from photons radiated by high-energy electrons or upscattered to high energy by inverse Compton scattering.

\begin{figure}[t]
\centering
\includegraphics[width=\linewidth, trim=0 50 0 80, clip=true]{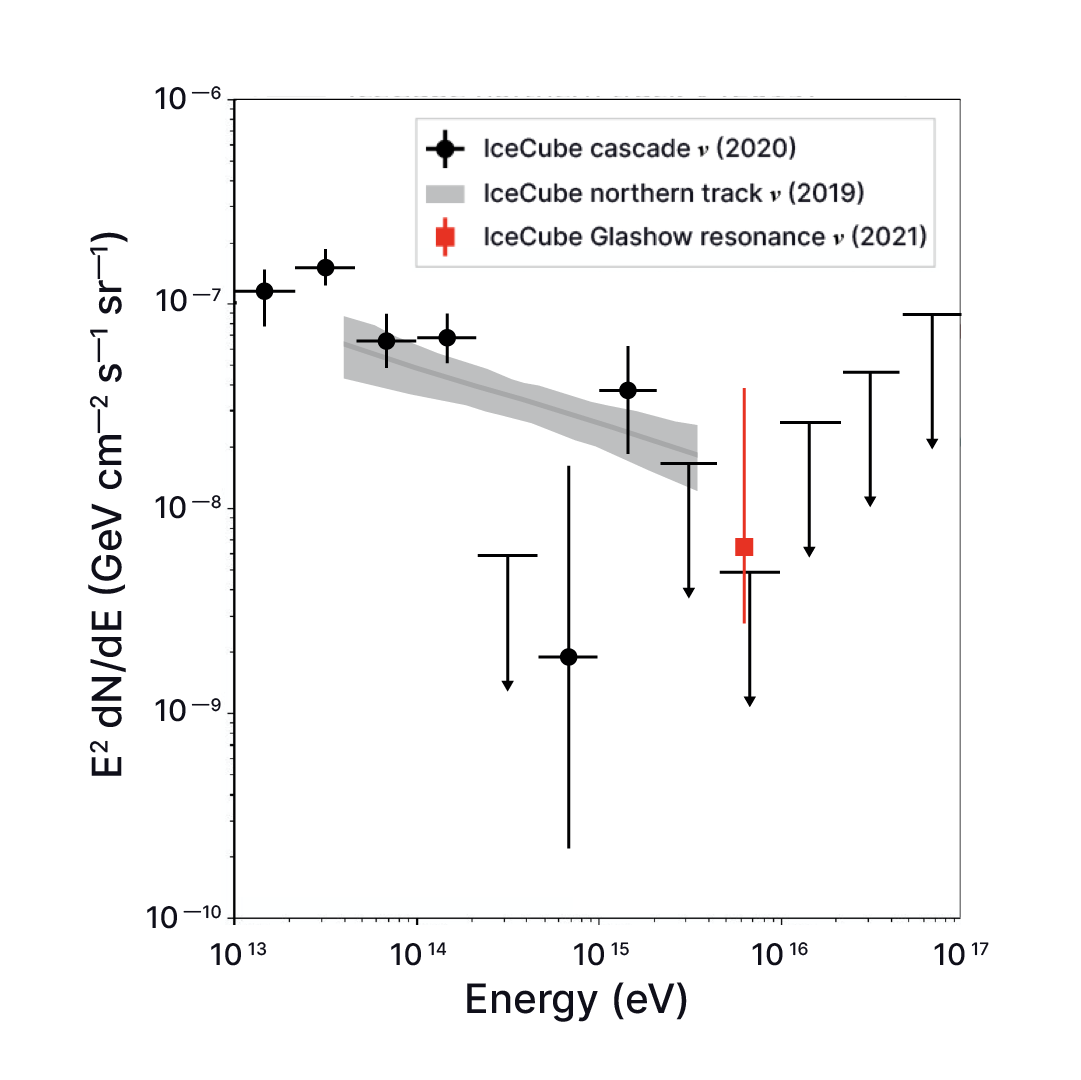}
\caption{\textbf{\textit{Current all-sky measurements of the high-energy astrophysical neutrino emission.}} 
The flux of cosmic muon neutrinos~\cite{Aartsen:2017mau} inferred from the $9.5$-year upgoing-muon track analysis (solid line) with $1\sigma$ uncertainty range (shaded) is compared with the flux of showers initiated by electron and tau neutrinos~\cite{IceCube:2020acn}, when assuming standard oscillations.
The measurements are consistent with the assumption that each neutrino flavor contributes an identical flux to the diffuse spectrum.
Both are consistent with the flux derived from the observation of a single Glashow-resonance event~\cite{IceCube:2021rpz} shown in red.}
\label{fig:showerstracks-2}
\end{figure}

After accumulating a decade of data, the observed flux of neutrinos reaching us from the cosmos is shown in~\cref{fig:showerstracks-2}.
It has been measured using two different methods to separate the high-energy cosmic neutrinos from the large backgrounds of cosmic-ray muons, 3,000 per second, and neutrinos, one every few minutes, produced by cosmic rays in the atmosphere.
The first method identifies muon neutrinos of cosmic origin in samples of muon tracks produced by upgoing muon neutrinos reaching the South Pole from the Northern Hemisphere~\cite{IceCube:2015qii,Abbasi:2021qfz}; see also similar analyses with ANTARES~\cite{ANTARES:2024ihw} and Baikal-GVD~\cite{Baikal-GVD:2022fis} as well as a forecast for KM3NeT~\cite{KM3NeT:2024uhg}.
In this case, Earth is used as a passive shield for the large background of cosmic-ray muons.
By separating the flux of high-energy cosmic neutrinos from the lower energy flux of neutrinos of atmospheric origin, a cosmic high-energy component $\rm dN/dE \sim E^{-\gamma}$ with $\gamma = 2.37\pm0.09$ is isolated above an energy of $\sim \SI{100}\TeV$~\cite{Aartsen:2017mau}; see~\cref{fig:showerstracks-2}.
Also shown are the results of a second search that exclusively identifies showers, often known as ``cascades,'' initiated by electron and tau neutrinos that interact inside the detector and can be isolated from the atmospheric background to energies as low as $\SI{10}\TeV$~\cite{IceCube:2020acn,IceCube:2020fpi}.
Reaching us over long baselines, the cosmic neutrino flux has oscillated to a ratio $\nu_e:\nu_\tau:\nu_\mu$ of approximately 1:1:1.

Another important method to search for astrophysical neutrinos is to select events whose interaction vertex is contained within the detector's fiducial volume, called ``starting events.''
Starting events have the advantage of a reduced background from Southern Hemisphere downgoing muons~\cite{Schonert:2008is,Gaisser:2014bja,Arguelles:2018awr}; in particular, the cascade channel is especially sensitive due to the smaller atmospheric background~\cite{Beacom:2004jb}.
This methodology led to the first observation of astrophysical neutrinos~\cite{IceCube:2013low,IceCube:2014stg} using high-energy starting events (``HESE'')~\cite{IceCube:2020wum} and has enabled the study of inelasticity of the muon neutrino interaction~\cite{IceCube:2018pgc}.

Additionally, IceCube has independently confirmed the existence of neutrinos of cosmic origin by the observation of high-energy tau neutrinos~\cite{Abbasi:2020zmr} and by the identification of a Glashow resonant event, where a weak intermediate $W^-$ boson is produced in the resonant interaction of an electron antineutrino with an atomic electron: $\bar{\nu}_e + e^- \rightarrow W^- \rightarrow q + \bar{q}$~\cite{IceCube:2021rpz}.
Found in a dedicated search for partially contained showers of very high energy~\cite{Lu:2017nti}, the reconstructed energy of the Glashow event is $\SI{6.3}\PeV$, matching the laboratory energy to produce a $W^-$ of mass $\SI{80.37}\GeV$.
Given its high energy, the initial neutrino is cosmic in origin and provides an independent discovery of cosmic neutrinos at the level of $5.2\sigma$.

\begin{figure*}[t]
\centering
\includegraphics[width=0.9\textwidth]{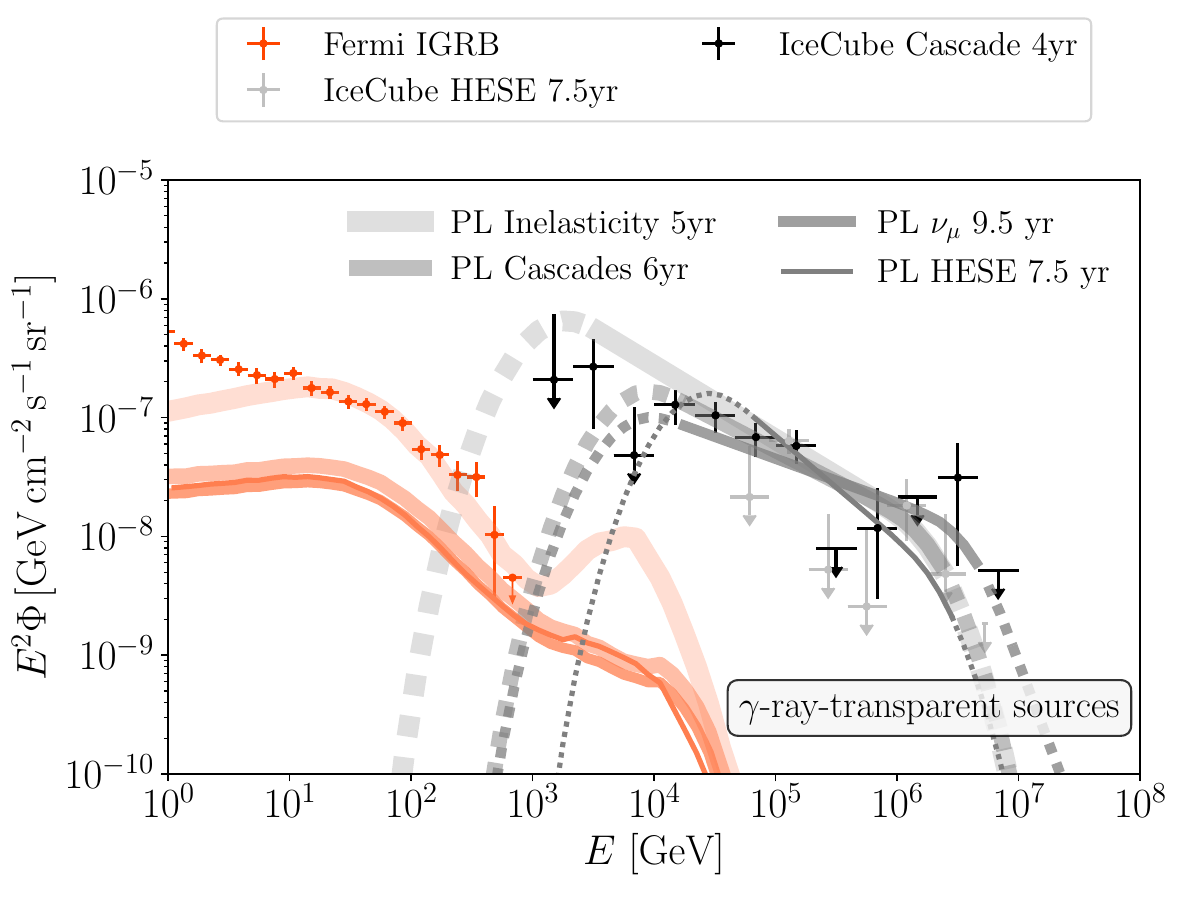}
\caption{\textbf{\textit{Gamma-ray-transparent sources?}} Cosmic neutrino spectra (gray curves) and $\gamma$-ray flux resulting from the accompanying gamma rays (orange curves).
The cosmic neutrino spectra are chosen to be compatible with different power-law (PL) fits to distinct IceCube data sets.
The data points are measurements of the diffuse cosmic neutrino flux~\citep{IceCube:2017zho, IceCube:2020wum}, the extragalactic gamma-ray background (EGB), the isotropic gamma-ray background (IGRB)~\citep{Fermi-LAT:2014ryh} from $\SI{0.1}\GeV$ to $\SI{1}\TeV$, and the diffuse MeV gamma-ray background~\citep{1999PhDT.......284W,Watanabe:1998ds,Strong:2004ry}.
Neutrino spectra correspond to the best-fit single power-law models for the different IceCube analyses.
Fluxes below and above the sensitivity range for the IceCube analyses are not constrained by the data and are shown as dashed curves.
Pionic gamma rays from hadronic interactions are assumed to leave the sources without attenuation, propagate in the EBL, and cascade down to GeV-TeV energies.} 
\label{fig:seda} 
\end{figure*}

As already discussed, unlike neutrinos, gamma rays interact with photons associated with the extragalactic background light while propagating to Earth.
The gamma-rays lose energy by $e^+e^-$ pair production, and the resulting electromagnetic shower subdivides the initial photon energy into multiple photons of reduced energy reaching Earth.
This significantly modifies the implications of the gamma-neutrino correspondence, which we illustrate~\cite{Fang:2022trf} by first parametrizing the neutrino spectrum as a power law,
\begin{equation}
  \frac{dN}{dE_\nu}  \propto E_\nu^{-\gamma_{\rm astro}},\, E_{\nu, \rm min} \leq E_\nu \leq E_{\nu, \rm max},\,
\label{eq:powerlaw}
\end{equation}
for four recent measurements taken from Refs.~\cite{IceCube:2020wum, IceCube:2020acn, IceCube:2021uhz, IceCube:2018pgc}.
Conservatively, the minimum and maximum energies are limited to the range of neutrino energies the particular analysis is sensitive to.
For each neutrino flux measurement, we derive the accompanying gamma-ray flux from the fit to the neutrino spectrum using the interface of~\cref{eq:nutogamma}.
This pionic gamma-ray flux is subsequently injected into the EBL, assuming that the sources follow the star-formation (SFR) history in a flat $\Lambda$CDM universe~\citep{Hopkins:2006bw}.
\Cref{fig:seda} shows the diffuse neutrino fluxes and those of their accompanying gamma-ray counterparts.
As the values of the neutrino flux spectral indices are comparable for the four measurements, the flux of the showers produced by $\nu_e$ and $\nu_\tau$  is essentially determined by the value of $E_{\nu,\rm min}$ where the energy peaks.
We conclude that when $E_{\nu,\rm min}\lesssim \SI{10}\TeV$, the showers initiated by the pionic photons contribute up to $\sim 30-50\%$ of the diffuse extragalactic gamma-ray flux between $\SI{30}\GeV$ and $\SI{300}\GeV$ and nearly 100\% above $\SI{500}\GeV$.
The cascade flux exceeds the observed gamma-ray flux above $\sim \SI{10}\GeV$ when $E_{\nu,\rm min} \leq \SI{10}\TeV$, assuming that the measured power-law distribution continues.

In the end, the room left to accommodate secondary photons from TeV-PeV gamma rays and electrons is small, and it is difficult to avoid the conclusion that the extragalactic gamma-ray flux accompanying cosmic neutrinos exceeds the diffuse gamma-ray flux observed by Fermi-LAT~\cite{Fermi-LAT:2009ihh}, especially because blazars produce $\sim 80\%$ of the latter at the highest gamma-ray energies that are not the main sources of neutrinos~\cite{IceCube:2016qvd}. 
There is no contradiction here; we infer from the calculations that the pionic gamma rays already lose energy in the target producing the neutrinos prior to propagating in the EBL.
As a result, pionic gamma rays emerge below Fermi threshold, at MeV energies or below.
The observed diffuse neutrino flux originates from gamma-ray-obscured sources. 
The analysis presented here reinforces the conclusions of previous analyses; see Refs.~\cite{Murase:2013rfa,Murase:2015xka, Capanema:2020rjj, Capanema:2020oet}.

We should emphasize that the fact that powerful neutrino sources are gamma-ray-obscured should not come as a surprise~\cite{Ambrosone:2024zrf}.
The photon and proton opacities in a neutrino-producing target are related by their cross sections (up to a kinematic factor associated with the different thresholds of the two interactions; see Ref.~\cite{Svensson:1987nlx}) 
\begin{equation}
\tau_{\gamma\gamma} \simeq \frac{\sigma_{\gamma\gamma}}{\kappa_{p\gamma} \sigma_{p\gamma}} \, \tau_{p\gamma} \simeq 10^3\,  \tau_{p\gamma}\,,
\label{opacitygamma}
\end{equation}
where $\kappa_{p\gamma} \sim 0.2$ is the inelasticity in $p\gamma$ interactions. For instance, we should not expect neutrinos to be significantly produced in blazar jets that are transparent to very high-energy gamma rays.
In contrast, the highly obscured dense cores close to supermassive black holes in active galaxies represent an excellent opportunity to produce neutrinos, besides providing opportunities for accelerating protons.

\vspace{0.2cm} \NibSolidRight \textit{
Understanding the production regions of neutrinos in astrophysical sources requires the use of multimessenger data.
This does not come as a surprise since neutrino astrophysics is multimessenger astrophysics, as the production of neutrinos is associated with the production of gamma rays.
However, the current data indicates that the intense neutrino sources are actually gamma-ray-opaque.
We predict that upcoming sub-GeV gamma-ray instruments will find correlations with high-energy astrophysical neutrino sources. 
}

\subsection{What does an improved observation of Galactic neutrinos tell us?\label{subsec:galactic-neutrinos}}
The Milky Way is an ideal observational target to answer many of the questions raised in the previous decade of neutrino astronomy.
Our own galaxy is expected to host high-energy sources of cosmic rays and, thus, neutrinos.
Cosmic-ray propagation, a challenge to model~\cite{GALPROP} and of long-standing interest, will enter a new era once diffuse Galactic neutrinos are measured accurately.
Neutrino-gamma-ray connections can be studied at much closer distances than by studying extragalactic sources. 

High-energy neutrinos are produced when Galactic cosmic rays interact at their acceleration sites and during propagation through the interstellar medium.
The Galactic plane has, therefore, long been hypothesized as a neutrino source, leading to various searches performed over the last decade~\cite{Halzen:2016seh,ANTARES:2017nlh,IceCube:2017trr}.
In 2023, IceCube announced the first detection of the Milky Way in neutrinos at a statistical significance of 4.5$\sigma$~\cite{IceCube:2023ame}, thus starting the field of Galactic neutrino astronomy.
IceCube's observation of the diffuse Galactic plane represents the first non-electromagnetic image of our galaxy as a whole and the first high-energy neutrino observations from our galaxy (MeV-scale neutrinos and, more recently, lower-energy neutrinos have been observed for decades from the Sun).

The high-energy neutrino observations from our galaxy open a new door to search for Galactic PeVatrons, sources capable of cosmic-ray acceleration to PeV energies~\cite{Bustamante_2023}.
While gamma rays have led the search for PeVatrons, there is a strong chance that their interactions in the Galaxy obscure sources.
Using future neutrino observations, together with cosmic-ray and gamma-ray observations, we will be able to identify Galactic PeVatrons. 

With the Galactic plane observed, the Galactic Center (GC) emission also becomes interesting.
The GC is a region that promises high activity based on the presence of a supermassive black hole (SMBH) at the position of Sgr~A*.
SMBHs can be sources of cosmic-ray emissions and their secondaries, neutrinos and gamma rays.
Because of the density of matter in the GC region, future neutrino observations may be able to probe deeper than any other observations in cosmic rays, gamma rays, or any other electromagnetic wavelengths, thanks to their large interaction length (i.e., ``neutrino depth").

The pressing question regarding the observed diffuse flux is how much individual Galactic point sources, below the current detection threshold, contribute to it.
Identifying many types of individual Galactic sources remains a challenge.
Future detectors with better capabilities will provide a more complete answer~\cite{Ambrosone_2024}.
This is a great opportunity for the upcoming KM3NeT neutrino observatory in the Mediterranean~\cite{KM3NeT:2018wnd}, which is currently under construction.
Future detectors planned in the Pacific Ocean are also promising as they are water-based detectors, as discussed in Section~\ref{sec:technologies}.

The 2023 observation of the diffuse Galactic plane was based on testing templates.
The templates of the spatial and energy distribution of Galactic neutrinos were generated by models~\cite{fermi-diffusegp, kra} that consider the matter distribution in the Galaxy, Galactic cosmic-ray production, and their propagation as well as a model template of neutral pion distribution.
Currently, the observational limit in neutrinos means we are constrained to testing existing models based on gamma-ray and cosmic-ray observations.
Once neutrino observations become more sensitive, they then inform the modeling of gamma-ray production and cosmic-ray propagation.

Beyond model testing, a model-free observation of the spatial distribution of the neutrino flux in the Galaxy provides a comparison to the gamma-ray data to shed light on the gamma-ray-neutrino connection, as described in~\ref{subsec:gamma}.
The model-independent observation of the energy spectrum is also a pressing question for the field.
This can only be answered by collecting larger statistical and more sensitive observations of the Galactic plane. 
It remains to be seen whether there is any energy region where the Galactic diffuse flux outshines the extragalactic diffuse flux in neutrinos. 

\vspace{0.2cm} \NibSolidRight \textit{Galactic neutrino astronomy provides an opportunity to study sources of the closest distance, the closest black hole, and discover Galactic PeVatrons.
Detailed observations of the Galactic neutrino diffuse flux, and comparison to the counterpart gamma-ray flux, will lead to understanding the gamma-ray-neutrino connection, cosmic-ray production and propagation, and matter distribution in the Galaxy.}

\subsection{What will we learn from probing the black hole environment with neutrinos?\label{subsec:black-holes}}

We discussed in Section~\ref{subsec:gamma} that the gamma rays accompanying the diffuse cosmic neutrino flux emerge from their sources at energies below the Fermi-LAT threshold, at MeV energies, or below.
The identification of the first cosmic-neutrino sources confirms indeed that cosmic neutrinos originate in sites that are opaque to gamma rays.

\begin{figure}[ht]
\centering
\includegraphics[width=0.9\columnwidth]{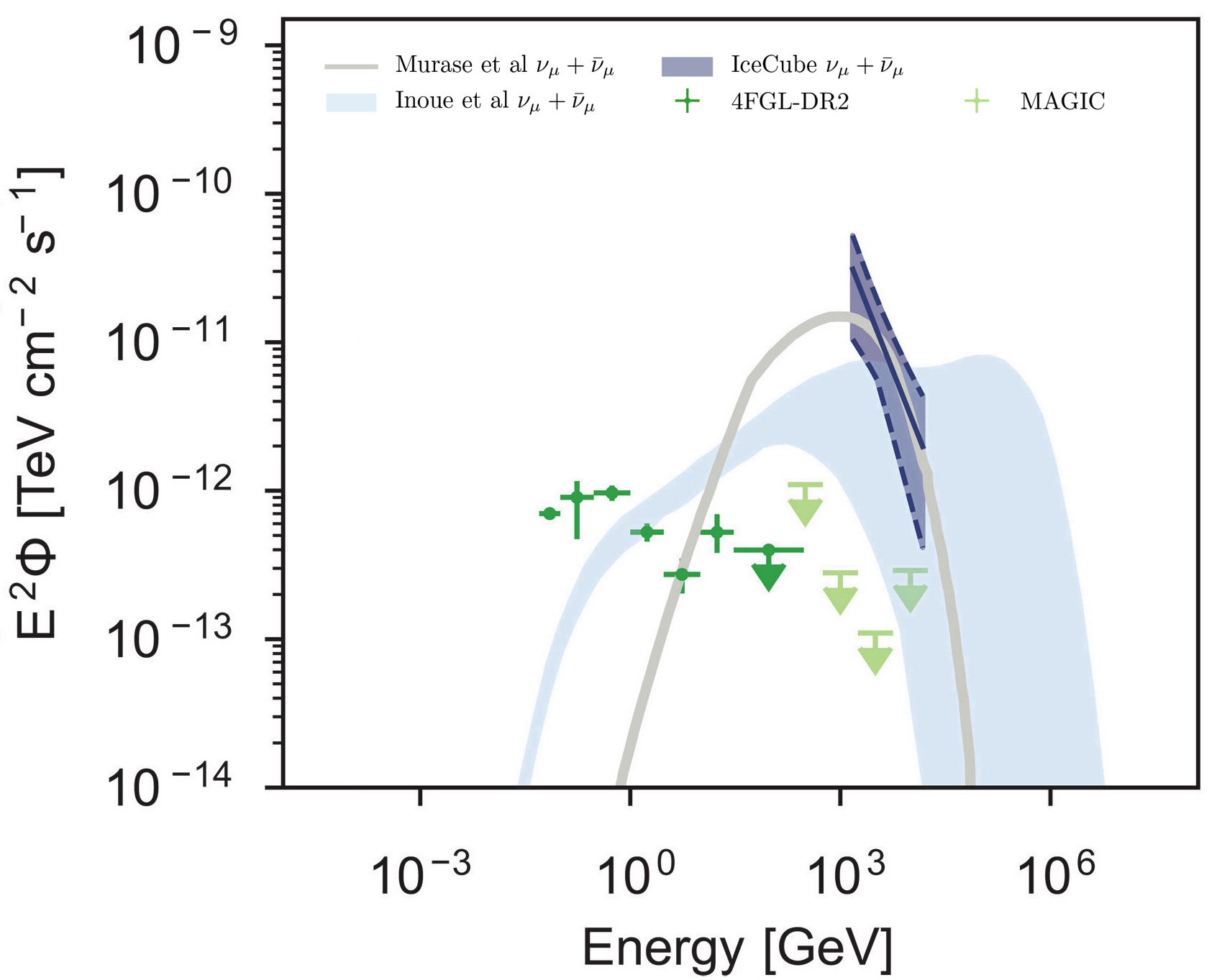}
\caption{\textbf{\textit{Multimessenger spectral energy distribution of NGC 1068.}}
Dark and light green points indicate gamma-ray observations at 0.1 to $\SI{100}\GeV$~\cite{Fermi-LAT:2019pir,Fermi-LAT:2019yla} and $>\SI{200}\GeV$~\cite{Acciari:2019raw}, respectively.
Arrows indicate upper limits, and error bars are $1\sigma$ confidence intervals.
The solid, dark blue line shows our best-fitting neutrino spectrum, with the dark blue shaded region indicating the 95\% confidence region.
We restrict this spectrum to the range between 1.5 and $\SI{15}\TeV$, where the flux measurement is well constrained.
Two theoretical predictions are shown for comparison: The light blue shaded region and the gray line show the NGC~1068 neutrino emission models from Refs.~\cite{Inoue:2019fil}, \cite{Inoue:2019yfs}, and~\cite{Murase:2019vdl}, respectively.
All fluxes are multiplied by the energy squared $\rm E^2$.}
\label{fig:ngcdatatheory}
\end{figure}

The multiwavelength observations of the active galaxy NGC~1068 represent an interesting case study.
The neutrino flux measured by IceCube is shown in~\cref{fig:ngcdatatheory} along with the gamma-ray data.
Note that the MAGIC telescope's upper limits exclude the presence of the accompanying pionic gamma-ray flux by more than one order of magnitude in the TeV energy range; the source is indeed gamma-ray obscured~\cite{Fang:2023vdg}.

\begin{figure}[ht]
\centering
\includegraphics[width=0.9\columnwidth]{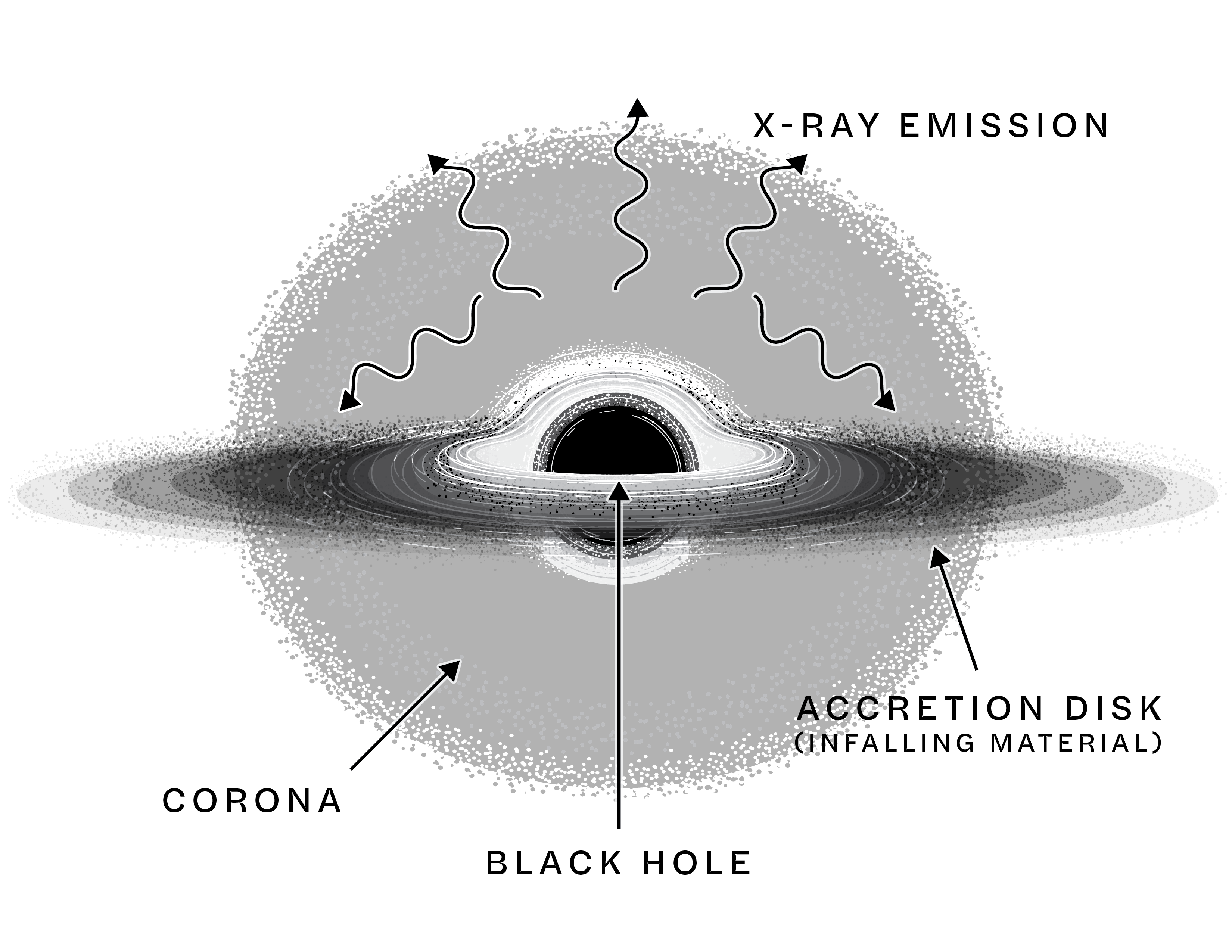}
\caption{\textbf{\textit{Cosmic neutrino cradle.}}
The core of an active galaxy combines the ingredients to produce neutrinos, with protons accelerated by the black hole and/or by the accretion disk and a target (corona) with a large optical depth in both gamma rays and hydrogen that converts them into neutrinos.
}
\label{fig:Black-Hole_Diagram}
\end{figure}

Dimensional analysis is sufficient to show that the observations point to the production of neutrinos in the obscured dense core surrounding the supermassive black holes at their centers.
A conceptual version of the neutrino source in~\cref{fig:Black-Hole_Diagram} shows the dense core of X-rays confined to a radius $\rm R$ that converts protons accelerated near the supermassive black hole into neutrinos. Astronomers refer to the gamma-ray-obscured core emitting X-rays as a ``corona."

The opacity of the X-ray target to protons is
\begin{equation}
\label{eq:pgammaopacity}
\rm \tau_{p\gamma}\simeq \frac{\kappa_{p\gamma} R_{target}}{\lambda_{p\gamma}} \simeq \kappa_{p\gamma} \rm R_{target}\, \sigma_{p\gamma} \,n_{\gamma},
\end{equation}
which is determined by how often the proton interacts in a target of size $\rm R_{target}$ given its interaction length $\lambda_{p\gamma}$. With each interaction, the proton loses a fraction $\kappa_{p\gamma}$ of its energy, the inelasticity.
The interaction length is determined by the number density of target X-ray photons $\rm n_{\gamma} \simeq n_X$ and the interaction cross section $\sigma_{p\gamma}$:
\begin{equation}
\rm n_X = \frac{u_X}{\epsilon_X} = \frac{1}{4 \pi R^2c}\, \frac{L_X}{\epsilon_X}\,,
\end{equation}
and therefore,
\begin{equation}
\rm \tau_{p\gamma} = \frac{\kappa_{p\gamma} \sigma_{p\gamma}}{4 \pi c}\, \frac{1}{R}\, \frac {L_X}{\epsilon_X} .
\label{taugammagamma}
\end{equation}
We here determined the target density in X-rays in a corona of size $R$ from the energy density of X-rays $\rm u_X$ divided by their energy $\epsilon_X$, and we subsequently identified the energy flux $\rm cu_X$ with the measured X-ray luminosity of the source $\rm 4 \pi R^2\, L_X$.
In the end, the opacity of the corona to protons is proportional to the density of target photons, i.e., to the X-ray luminosity $\rm L_X$, and inversely proportional to the radius $\rm R$. The above result can also be rewritten as
\begin{equation}
\label{eq:targetopacity}
\rm \tau_{p\gamma} \simeq 10^2\,\left[\frac{R_s}{R}\right]\,\left[\frac{\SI{1}\keV}{\epsilon_X}\right]\left[\frac{L_X}{L_{\rm edd}}\right] \simeq 10^{-3}\,\tau_{\gamma\gamma}\,,
\end{equation}
where $\rm R_s$ and $\rm L_{edd}$ are the Schwartzschild radius and the Eddington luminosity of the black hole. For a black hole mass of approximately $10^7$ solar masses for NGC 1068, $\rm L_X \sim 10^{43} \,\rm erg/s$ at keV energies~\cite{padovani2024supermassive} yields $\rm L_X \simeq 10^{-2}\, L_{\rm edd}$.

Even for a relative modest opacity of $\tau_{p\gamma} \gtrsim 0.1$, the above result requires that the X-rays be confined to a radius of $\rm R \simeq 10 \, R_s$. 
Importantly, this will automatically guarantee the suppression of the pionic photon flux by more than one order of magnitude relative to neutrinos by~\cref{opacitygamma}.

Detailed modeling of the accelerator and target reinforce the conclusion that neutrinos originate within $\rm R/R_s \sim 10^{-4}\,pc$ of the supermassive black hole~\cite{Murase:2022dog}.
Producing neutrinos, even at the base of the jet at a distance, will fail to yield the neutrino flux observed for NGC~1068 and suppress the appearance of a TeV gamma-ray flux at TeV energy and below.

Interestingly, high densities of both radiation~\cite{Bauer:2014rla,Marinucci:2015fqo,Ricci:2017dhj} and matter~\cite{Rosas:2021zbx,García-Burillo_2016} are associated with the core of NGC~1068, and therefore neutrinos are efficiently produced in both $p\gamma$ and $pp$ interactions.
A second look at the multiwavelength data shown in~\cref{fig:ngcdatatheory} leads us to conclude that $\tau_{pp} > \tau_{p\gamma}$ because the $p\gamma$ threshold for producing neutrinos on a target of keV-energy photons at the $\Delta$-resonance is at PeV energy.
Instead, the typical energy of the neutrinos observed by IceCube from the direction of NGC~1068 is closer to $\sim \SI{10}\TeV$; see~\cref{fig:ngcdatatheory}.
These neutrinos are predominantly produced in $pp$ interactions, with a reduced threshold for producing multiple pions, which results in neutrinos of lower energy.

From~\cref{eq:pgammaopacity}, we calculate the opacity to protons of the large matter  density close to the black hole
\begin{equation}
\rm \tau_{pp} \simeq \frac{\kappa_{pp}\, R} {\lambda_{pp}} \simeq \kappa_{pp}\, \rm R\, \sigma_{pp} \,n_p \simeq \kappa_{pp}\, \sigma_{pp}\, N_H,
\label{eq:gas}
\end{equation}
where we have introduced the line-of-sight density $\rm N_H = R\, n_p$, which is sufficiently large for NGC~1068 to achieve an opacity $\rm \tau_{pp} \gtrsim 1$.
In~\cref{fig:ngcdatatheory}, two such models are compared to the flux observed by IceCube from NGC~1068.
Accelerated protons interact in the corona with both the dense gas in $pp$ interaction and with the X-ray photons by $p\gamma \rightarrow pe^+e^-$, the Bethe-Heitler process.
The latter channels the energy stored in the flux of relativistic protons to relativistic pairs with extended distributions.
Contrary to accelerated leptons, whose maximum energy is limited by radiative losses, the maximal energy of pairs is determined by the kinematics of the process and, in this case, populates the MeV energy region.

From~\cref{eq:gas}, one may be tempted to conclude that, with the production of neutrinos in $pp$ interactions depending only on the line-of-sight density $N_H$, we can evade the constraint on the size of the production region $\rm R \lesssim 100 \, R_s$. 
This is not the case because a high density of X-rays in the corona is still required to suppress the flux of pionic gamma rays produced in $pp$ interactions to evade the strong upper limits from the MAGIC telescope shown in~\cref{fig:ngcdatatheory}.
The suppression of the $\gamma$-ray flux by over one order of magnitude cannot be achieved with interactions with protons only. 
Although produced by different mechanisms, the gamma rays accompanying the neutrinos still emerge at MeV energies, the characteristic energy at which the X-ray corona becomes transparent to photons.
For a more detailed review of the models, see Refs.~\cite{Murase:2022dog,Kheirandish:2021wkm,Eichmann:2022lxh,Anchordoqui:2021vms}.

Do cosmic neutrinos (and cosmic rays) originate in the obscured cores of active galaxies?
Some do, and with NGC~1068 and NGC~4151, we have identified the first sources of high-energy cosmic rays since their discovery more than a century ago. 
Instead, modeling the multiwavelength spectrum of TXS~0506+056 represents an unmet challenge.
The source is different: it emits neutrinos in bursts with Eddington luminosity characterized by a harder neutrino spectrum, close to $\rm E^{-2.2}$.
The gamma-ray-obscured spectra at the time of neutrino production~\cite{Halzen:2021xkf} may indicate that the neutrinos are produced in the vicinity of the core as is the case for NGC~1068~\cite{kun2024correlation}, but in association with catastrophic events hinted at by the observation of a strong optical flare in 2017~\cite{Lipunov:2020ptp}.

\vspace{0.2cm} \NibSolidRight \textit{While any conclusions regarding the origin of the highest energy cosmic rays may be premature, we feel that with more and larger instruments, the revelation of a multitude of sources and the resolution of the puzzle is within reach.
The emission of neutrinos from the few sources we have identified indicates that they are produced near supermassive black holes.
Thus, neutrino observations, and their comparison to gamma-ray observations, provide information about the matter and radiation density around the black hole.
}

\subsection{What will we learn about stellar collapse and neutrino physics from Galactic and extragalactic supernovae?\label{subsec:supernovae}}

A core-collapse supernova radiates the vast majority of the binding energy of the resulting compact remnant in the form of neutrinos of all flavors.
Information on the astrophysics of the collapse and subsequent explosion, and about the physics of the neutrinos themselves, is encoded in the time, energy, and flavor structure of the neutrino burst.
When supernova 1987A exploded, only a couple of tens of events provided sufficient information to not only probe the physics of the explosion but also to shed new light on neutrino properties, such as their mass, magnetic moment, and lifetime.
Neutrino oscillations produce significant modifications of the neutrino spectra.
Matter transitions in the expanding remnant modify the time- and energy-dependent profiles of the spectra of neutrinos and antineutrinos, which depend on the neutrino mixing parameters and the presently unknown neutrino mass ordering.
Additionally, depending on the type of supernova, ``collective'' effects resulting from neutrino-neutrino interactions can result in additional modifications of the spectra; for recent reviews, see Ref.~\cite{Sarmah:2024tis}.

Although designed as a high-energy Cherenkov detector with a nominal threshold of $\SI{100}\GeV$ ($\sim \SI{5}\GeV$ in IceCube's DeepCore~\cite{IceCube:2024xjj}, $\sim \SI{1}\GeV$ in the IceCube Upgrade~\cite{Ishihara:2019aao}), IceCube can identify with high significance the interactions of below-threshold MeV neutrinos produced by the passage of the flux resulting from a Galactic supernova~\cite{Abbasi:2011ss}.
It is observed as a collective increase in all photomultiplier counting rates over their very low dark noise in radioactivity-free ice.
The rate increase is caused by the Cherenkov light from shower particles produced by supernova electron antineutrinos interacting in the ice, predominantly by the inverse beta decay reaction.
The secondary positron tracks of about $\SI{0.6}\cm \times (E_\nu/{\rm MeV})$ length radiate $178 \times (E_{e^+}/{\rm MeV})$ Cherenkov photons in the $300-\SI{600}\nm$ wavelength range.
From the approximate $E^2$ dependence of the neutrino cross section and the linear energy dependence of the track length, the light yield per neutrino scales with $E^3$.
With absorption lengths exceeding $\SI{100}\m$, photons travel long distances in the ice, so each photomultipier tube effectively monitors several hundred cubic meters of ice.
A similar detection is also possible in water-based Cherenkov neutrino telescopes such as KM3NeT~\cite{KM3NeT:2021moe}, though these suffer from larger radiative backgrounds compared to the radioactivity-free ice at the South Pole.

Although the rate increase in a single light sensor, for this discussion IceCube's digital optical module or DOM, is not statistically significant, its significance increases when recorded collectively in multiple independent sensors.
It reaches over $10\sigma$ for a supernova at $\SI{10}\kpc$ generating one million photoelectrons, producing a detailed movie of the time evolution of the neutrino signal.
Whereas DOMs overwhelmingly record single photons, the frequency of two-photon coincidences can be used to measure the average energy of the neutrinos.
With a two-megaton effective volume for supernova neutrinos, IceCube is the most precise detector for analyzing their neutrino light curve~\cite{Abbasi:2011ss}.
This has been demonstrated by establishing a limit on the frequency of supernovae using more than ten years of data~\cite{IceCube:2023esf}.
While the run time of IceCube and other high-energy neutrino telescopes is shorter than Baksan~\cite{Novoseltsev:2019gdt} and LVD~\cite{Vigorito:2021sgy}, the larger tonnage of IceCube makes it sensitive to a broader range of progenitor masses and provides constraints that are more robust to the different emission models~\cite{IceCube:2023ogt}.
Thus, neutrino telescopes are complementary to these smaller, long-running experiments, which are sensitive only to nearby, high-luminosity supernovae.
Finally, IceCube participates in the SuperNova Early Warning System SNEWS~\cite{SNEWS:2020tbu}, which combines in real time potential signals recorded in participating neutrino experiments~\cite{Kara:2024xug}.

Besides continuously monitoring the highly stable noise rate in the DOMs~\cite{IceCube:2023xzp}, IceCube additionally buffers the raw data recorded by the detector for several days; i.e., the information recorded by every single photomultiplier is saved to disk~\cite{HeeremanvonZuydtwyck:2015mbs}. 
This data can be saved for a period of days but must subsequently be dumped because the size of the record becomes unmanageable.
It is, however, saved when an interesting astronomical event occurs~\cite{HeeremanvonZuydtwyck:2015mbs}.
The low-level data, saved when a supernova occurs, provides several advantages. The complete information is available over the duration of the supernova without cuts, and the data—buffered at an early stage of the data acquisition system on the so-called string hubs—will be available in the unlikely case that the data acquisition fails or saturates, for instance, in the case of an extremely close supernova.
The automatized ``HitSpooling" has been working reliably for several years, including the data transfer to the north in the case of potentially interesting alerts .

\begin{figure*}[htbp]
\includegraphics[width=0.9\linewidth]{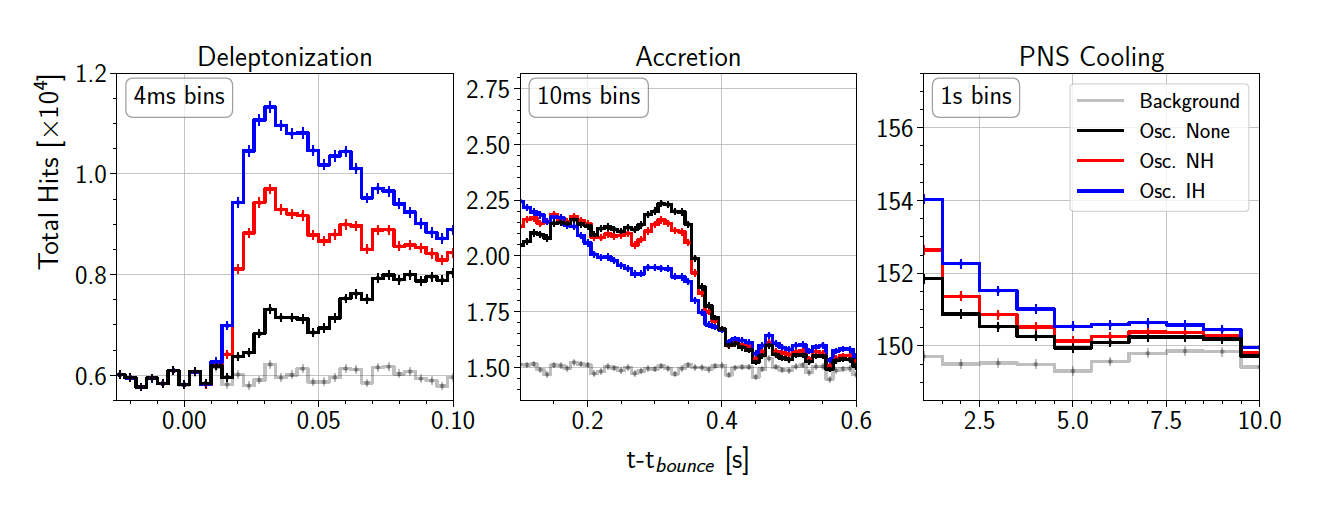}
\caption{
\textbf{\textit{Supernovae neutrino events in a neutrino observatory.}}
Simulated photoelectrons in IceCube from the interactions of electron antineutrinos produced by the collapse of a 13-solar-mass progenitor $\SI{10}\kpc$ from Earth~\cite{Nakazato:2012qf}.
Different bin sizes are chosen to illustrate time features in different phases of the star's collapse: (left) the deleptonization peak, (center) a $\sim 0.5$-second plateau as matter accretes onto the core forming a proto-neutron star, and (right) $\sim 10$~seconds of exponential decay as the proto-neutron cools.
Contrasted are the results assuming neutrino oscillations for normal and inverted hierarchy of the masses.}
\label{fig:supernova}
\end{figure*}

The excellent sensitivity to neutrino properties, such as the neutrino hierarchy, as well as the possibility of detecting the neutronization burst, a short outbreak of anti-\,$\nu_e$ released by electron capture on protons soon after collapse, are discussed in Refs.~\cite{Abbasi:2011ss,IceCube:2023xzp}.
In~\cref{fig:supernova}, we illustrate the sensitivity of IceCube to the neutrino mass hierarchy for a supernova at the most likely distance of $\SI{10}\kpc$.
On the astrophysics side, the characteristic ``SASI'' oscillations~\cite{Drago_2023} of the standing accretion shock will leave clear signatures in the high-statistics IceCube observations, as will the collapse of the star to a black hole.

IceCube supernova data will also represent an extraordinary opportunity for a multimessenger analysis with the gravitational wave signal~\cite{Takiwaki:2021dve}.
The gravitational wave strain of a supernova burst is proportional to the second time derivative of its moment of inertia $I$, which we can rewrite in terms of its mass, radius, and characteristic frequency
\begin{equation}
h \sim \rm \frac{d^2I}{dt^2} \sim MR^2 f^2.     
\end{equation}
An asymmetry in the breakout shock, required for a nonvanishing strain, develops because of the rotation of the star and also results from the SASI oscillations.
The same frequency is visible in the IceCube data and is revealed by transforming the high statistics time dependence data into the frequency domain.

IceCube observations also represent an opportunity to search for the superluminous extragalactic interaction-powered Type~IIn supernovae where particles can be accelerated to relativistic energies by the ejecta of the star interacting with the dense circumnuclear medium~\cite{IceCube:2023amf,Murase_2018}.

\vspace{0.2cm} \NibSolidRight \textit{Neutrino telescopes are in a position to make crucial and complementary observations to MeV–keV-scale neutrino detectors for Galactic core-collapse supernovae. Such observations can not only tell us about the supernovae themselves but also serve as a particle physics laboratory to measure phenomena such as neutrino mass hierarchy. Extragalactic Type~IIn supernovae are neutrino-bright source candidates in neutrino astronomy.
}

\subsection{What will we learn about the nature of dark matter with neutrino telescopes?}\label{subsec:DM}

A neutrino observatory is a powerful tool for searching for the particle nature of dark matter and provides unique capacities to study dark matter complementary to Earth-based experiments.
For example, if ``weakly" interacting massive particles (WIMPs) make up dark matter, they have been swept up by the Sun for billions of years as the Solar System moves about the Galactic halo~\cite{Jungman:1995df}.
Though interacting weakly, they will occasionally scatter elastically with nuclei in the Sun and lose enough momentum to become gravitationally bound; see Ref.~\cite{Maity:2023rez} for a recent review on the dark matter electron scattering scenario.
Over the Sun's lifetime, a sufficient density of WIMPs may accumulate in its center so that an equilibrium is established between their capture and annihilation.
The annihilation products of these WIMPs represent an indirect signature of halo dark matter, their presence revealed by neutrinos, which escape the Sun with minimal absorption for sub-100~GeV masses.
The neutrinos are, for instance, the decay products of heavy quarks and weak bosons resulting from the annihilation of  WIMPs into $\chi\chi\rightarrow \tau^{+}\tau^{-}$, $b\bar{b}$, or $W^+ W^-$.
Neutrino telescopes are sensitive to such neutrinos because of their relatively large neutrino energy, reflecting the mass of the parent WIMP, making them distinct from neutrinos produced in nuclear processes in the Sun.

\begin{figure}
 \begin{center}
\includegraphics[width=1.1\linewidth,angle=0]{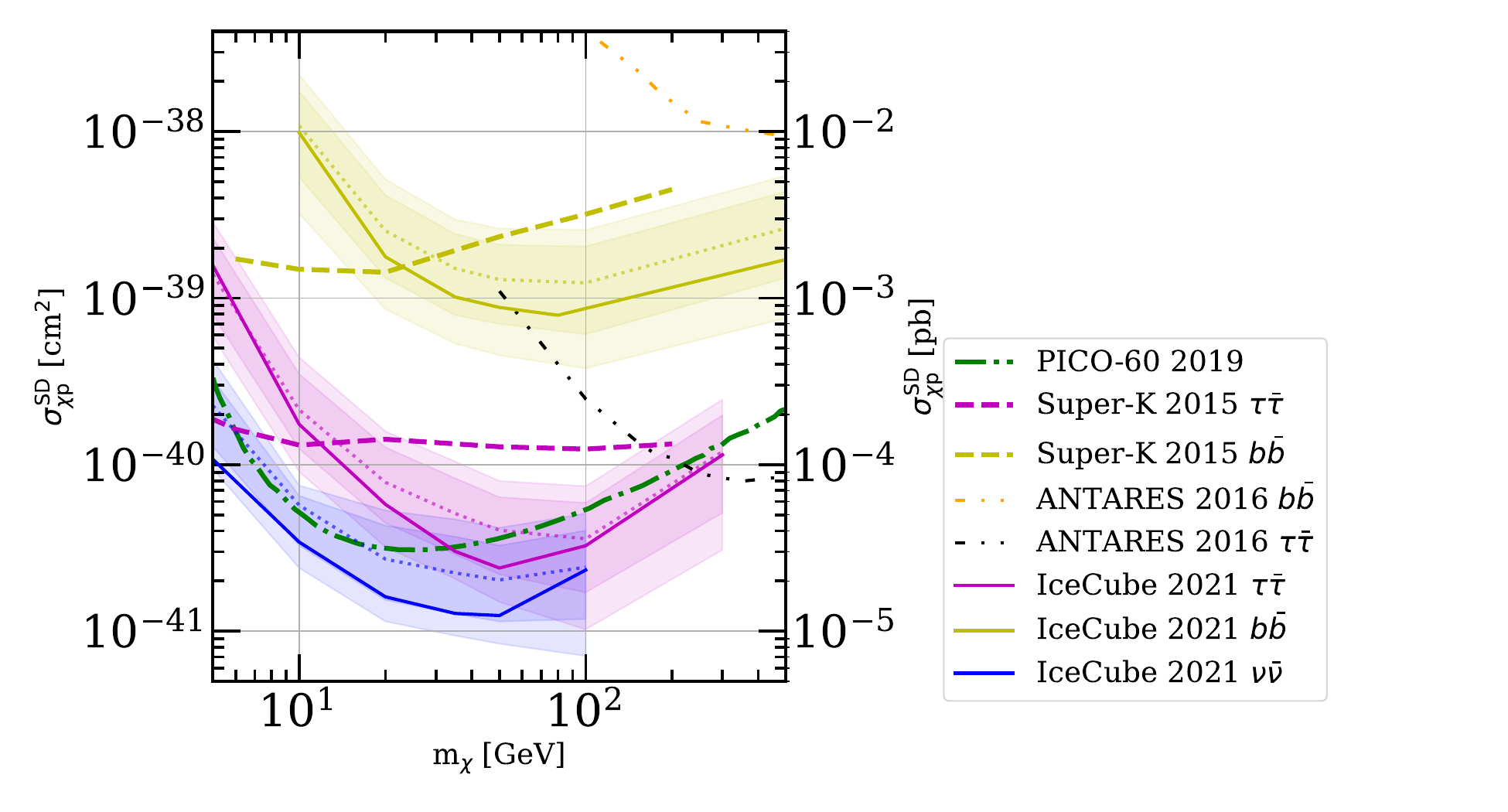}
\end{center}
\caption{\textit{\textbf{Upper limits at 90\% confidence level on the spin-dependent dark matter-proton cross section assuming that the neutrinos are produced by $\nu\bar\nu$, $b\bar{b}$ and $W^+W^-$ annihilation~\cite{IceCube:2021xzo}.}}
Limits from Super-Kamiokande~\cite{Super-Kamiokande:2015xms}, ANTARES~\cite{ANTARES:2016xuh}, and a direct-detection experiment (PICO)~\cite{PICO:2019vsc} are shown for comparison.
For higher-mass constraints using neutrino telescopes, see Refs.~\cite{Aartsen:2012kia,Lazar:2023ymm}.}
\label{fig:solar-dark-matter}
\end{figure}

The beauty of the indirect detection technique using neutrinos originating in the Sun is that the astrophysics of the problem is understood.
The source in the Sun has built up over solar time, sampling the dark matter throughout the galaxy; therefore, any unanticipated structure in the halo has been averaged out over time.
Other astrophysical uncertainties, such as the dark matter velocity distribution, have little or no impact on the capture rate of dark matter in the Sun~\cite{Choi:2013eda,Danninger:2014xza}.
Given a WIMP mass and cross section (and the assumption that the dark matter is not exclusively matter or antimatter), one can unambiguously predict the signal in a neutrino telescope.
If not observed, the model will be ruled out; see, e.g.,~\cref{fig:solar-dark-matter} for a recent result.
This is in contrast to indirect searches for photons from WIMP annihilation, whose sensitivity depends critically on the structure of halo dark matter. Observation requires cuspy structure near the Galactic Center or clustering on appropriate scales elsewhere, and observation necessitates not only appropriate WIMP properties but also favorable astrophysical circumstances.

Beyond WIMP dark matter, neutrino telescopes can also search for light dark matter through its effect in various astrophysical observables.
One well-motivated scenario of light dark matter is that of the axion, which is posed to solve the strong CP problem in QCD~\cite{Peccei:1977hh,Peccei:1977ur,Weinberg:1977ma,Wilczek:1977pj} but also appears in many realizations of string theory~\cite{Arvanitaki:2009fg}.
Neutrino telescopes are sensitive to axion-like dark matter through their effects in supernovae cooling~\cite{Fischer:2016cyd,Betranhandy:2022bvr,Mori:2023mjw}, modifying the duration of the Galactic neutrino signal in the neutrino telescope.
Additionally, neutrino telescopes can probe the interactions of ultralight dark matter with neutrinos through the imprint of this interaction on the flavor of astrophysical neutrinos~\cite{Reynoso:2016hjr,Brdar:2017kbt,Farzan:2018pnk,Farzan:2021gbx,Reynoso:2022vrn,Arguelles:2024cjj}.
The latter constraints are among the strongest constraints between ultralight dark matter and neutrinos.
In general, these astrophysical searches are complementary to laboratory searches for ultralight dark matter such as axion-like particles.

Finally, there are two other unique ways in which neutrino telescopes can search for dark matter.
First, if dark matter is very heavy, e.g., when its masses are above 100~TeV, the photons produced from the decay or annihilation of dark matter in our halo do not reach Earth unscattered. 
Instead, they interact with the extragalactic light or the cosmic microwave background~\cite{Berezinsky:2016feh,Skrzypek:2022hpy}.
In this scenario, neutrino telescopes provide the most direct and effective probe of very heavy dark matter annihilation and decay~\cite{Beacom:2006tt,Murase:2012rd,Murase:2012xs,Bhattacharya:2019ucd,Arguelles:2019ouk,Chianese:2021htv,Arguelles:2022nbl,Fiorillo:2023clw}.
Neutrino telescopes may find this signal by looking for neutrinos from concentrations of WIMPs in the Milky Way~\cite{Aartsen:2017ulx,Aartsen:2015xej,Aartsen:2016pfc,Abbasi:2011eq,Aartsen:2014hva} and nearby galaxies~\cite{Aartsen:2013dxa}.
These searches are especially important since neutrinos are the final channel to test the WIMP miracle in a model-independent way~\cite{Leane:2018kjk}, where dark matter interaction with the Standard Model is at the weak scale. 
Second, neutrinos can interact with dark matter from their sources to Earth.
Interactions between neutrinos and dark matter are motivated by the \textit{scotogenic} neutrino mass generation mechanism~\cite{Ma:2006km,Kubo:2006rm,Ma:2008ba,Ma:2013yga,Fraser:2014yha,Blennow:2019fhy}, where the neutrino mass is produced by these interactions, and recent cosmological observations~\cite{DiValentino:2017oaw,Hooper:2021rjc,Hooper:2022byl,Abdalla:2022yfr,Giare:2023qqn}.
The study of the opacity of the Universe to high-energy neutrinos yields constraints on the neutrino-dark matter interaction~\cite{Arguelles:2017atb,Kelly:2018tyg,Ferrer:2022kei,IceCube:2022clp} that are comparable with cosmological studies.
Additionally, the intensity of neutrino emission from the sources provides information about the hadronic environments where these neutrinos are produced.
This observation can place constraints on the proton-dark matter interaction, which is probing the same coupling as direct-detection experiments on Earth~\cite{Herrera:2023nww}. 

\vspace{0.2cm} \NibSolidRight \textit{Neutrino telescopes, given their broad energy range from tens of GeV to multiple PeV, offer a unique way to discover dark matter.
They are the most effective probes for very heavy dark matter, and they offer a way to test, in a model independent way, one of the last remaining channels of the WIMP miracle: dark matter annihilation to neutrinos.
Additionally, the Universe's transparency to high-energy neutrinos allows for the potential discovery of neutrino-dark matter interactions.}

\subsection{What will astrophysical neutrinos tell us about neutrino properties?}\label{subsec:SM}

The discovery of nonzero neutrino masses~\cite{Cleveland:1998nv,SNO:2001kpb,SNO:2002tuh,Super-Kamiokande:2004orf} has been one of the most significant observations of the last century~\cite{Davis:2003kh,McDonald:2016ixn,Kajita:2016cak}.
This measurement implies that the Standard Model particle content is incomplete and that new physics is responsible for the neutrino masses~\cite{Mohapatra:2005wg}.
So far, we have only been able to learn about neutrino masses by studying the change in neutrino flavors over long distances.
In fact, two naturally occurring neutrino sources were used to discover this phenomenon: solar neutrinos~\cite{Cleveland:1998nv,SNO:2001kpb,SNO:2002tuh} and neutrinos produced in cosmic-ray collisions in Earth’s atmosphere~\cite{Super-Kamiokande:2004orf}.
As we will discuss in this section, the study of high-energy astrophysical neutrinos with neutrino telescopes can address three fundamental questions in neutrino physics: the nature of neutrino masses, the lifetime of the neutrino, and the unitarity of the neutrino mixing matrix.
Of these three questions, the latter two require new physics beyond the existence of massive neutrinos to yield observable signatures in neutrino observatories; however, the first question is already invited by the observation of nonzero neutrino masses.

\textit{Shedding light into the origin of neutrino masses.---} With more than twenty years of measurements using natural and anthropogenic neutrino sources, the parameters that dictate neutrino flavor transition have now been measured to the few-percent level~\cite{Esteban:2020cvm,deSalas:2020pgw}.
Despite these improved measurements, we still do not know the mechanism responsible for neutrino masses.
Models that aim to explain neutrino masses~\cite{Mohapatra:2005wg} can be broadly organized into scenarios where neutrinos are Majorana particles and where they are predominantly Dirac fermions.
Much attention has been placed on the former scenario, where the Weinberg operator describes neutrino masses at low-energy scales.
This operator can be UV-completed to high-energy scales by introducing additional heavy neutrino states in what is known as the see-saw mechanism, where the smallness of neutrino masses is explained by the smallness of the Dirac-like mass term compared to the larger Majorana-like term associated with these heavy neutrinos.
A prediction of this scenario is that neutrino-less double beta decay should be observed, pending accidental cancellations due to neutrino Majorana phases~\cite{Agostini:2017jim,Denton:2023hkx}.
This has led to a vibrant experimental program aimed at observing this process~\cite{Agostini:2022zub}. 

The other possibility is that neutrinos are Dirac-like in nature.
In this case, new, right-handed neutrinos are added to the Standard Model particle content.
These can yield neutrino masses similar to the case of charged leptons through $y \bar L \tilde H N_R$, where $H$ is the Higgs double, $y$ the Yukawa couplings, and $N_R$ the new, right-handed neutrino field.
In this case, to achieve the right scale for the neutrino masses, the Yukawa couplings are distinctly small compared to those associated with other fermions.
The lack of an explanation for the smallness of the Yukawa in this scenario and the lack of experimental signatures to promptly confirm this have generally made this scenario less appealing.
However, recent advancements in quantum gravity~\cite{Ooguri:2016pdq,Gonzalo:2021zsp}, within the Swampland program, point to the fact that neutrinos should be Dirac-like in nature~\cite{Vafa:2024fpx}.
These results coupled with the recent discovery of high-energy astrophysical neutrino sources, which, as we will discuss, provides new avenues to discover Dirac-like neutrinos, makes revisiting this scenario timely.

If the masses of neutrinos are predominantly due to a Dirac-like mechanism, as suggested by recent work~\cite{Ooguri:2016pdq,Gonzalo:2021zsp}, or by plain similarity to how the other fermions acquire masses in the Standard Model, then, in the context of quantum gravity, we expect that neutrinos are quasi-Dirac particles~\cite{Valle:1983dk}.
This is because the value of the right-handed neutrino mass term is proportional to the amount of lepton number violation, which is expected to be broken at the Planck scale, as no global symmetries are expected to be preserved by quantum gravity.
In the scenario of a small Majorana mass term compared to the Dirac mass term, one produces ultrasmall mass splitting between the left- and right-handed neutrinos.
As we will discuss below, this can produce the disappearance of neutrinos, with oscillation frequencies that scale like $L/E$ on Galactic~\cite{Crocker:2001zs} or extragalactic scales~\cite{Keranen:2003xd,Beacom:2003eu,Esmaili:2009fk,Esmaili:2012ac,Shoemaker:2015qul, Carloni:2022cqz}.
Finally, even when we ignore the motivation of quantum gravity and pose that neutrinos are Dirac-like particles, studying extremely long baseline oscillations is of fundamental importance.
This is because, in the Dirac-like scenario, the right-handed neutrino mass term ($\frac{1}{2} m_R \bar N_R^c N_R$) presents a fundamental, unconstrained parameter of the new Standard Model.
Determining if this parameter is nonzero but small would have profound implications in fundamental physics as it proves new, ultrasmall energy scales.

As proposed in Ref.~\cite{Carloni:2022cqz}, one can use the observation of multiple high-energy neutrino sources to search for the disappearance of neutrinos in the Dirac-like scenario.
This technique has the advantage over prior proposals---see, e.g.,  Refs.~\cite{Beacom:2003eu,Esmaili:2009fk,Esmaili:2012ac,Shoemaker:2015qul}---in that it is within experimental capabilities and mitigates the astrophysical source modeling uncertainties by correlating neutrino observations from different sources at the same $L/E$.
The oscillation probability in this scenario is given by
\begin{equation}
P_{\alpha \beta}=\frac{1}{2} \sum_{j=1}^3\left|U_{\beta j}\right|^2\left|U_{\alpha j}\right|^2\left[1+\cos \left(\frac{\delta m_j^2 L_{\text {eff }}}{2 E_\nu}\right)\right],
\end{equation}
where $L_{\text {eff }}$ is the effective distance to the source after correcting for cosmic expansion, and $U$ is the mixing matrix measured in terrestrial experiments.
The disappearance is expected to be the same strength for all flavors in the minimal model scenario.
This has critical phenomenological implications since this implies that the effect should be present in tracks, produced predominantly by muon neutrinos, and cascades, produced by all flavors.

\begin{figure*}[htbp]
\includegraphics[width=0.4\linewidth]{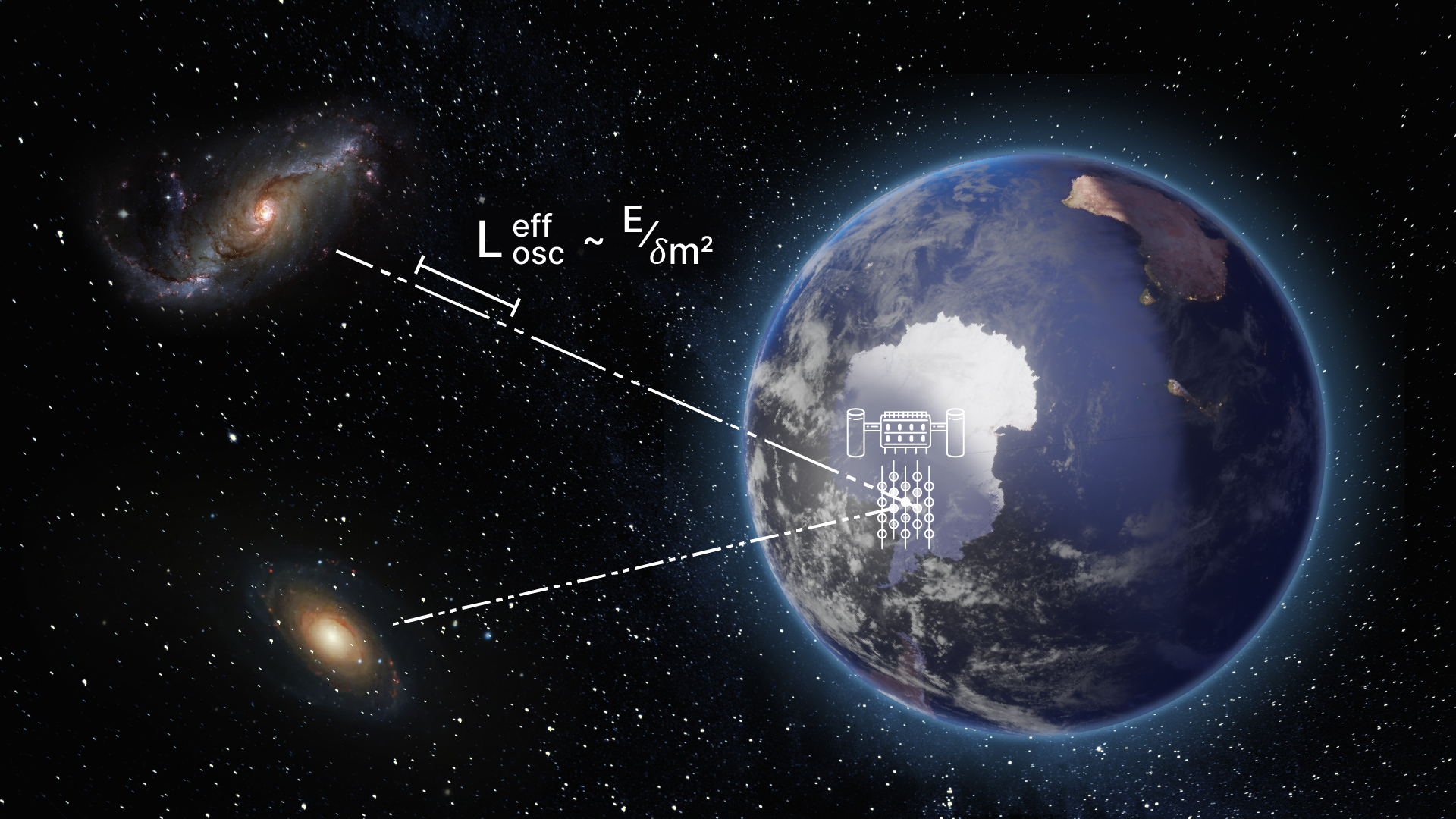} 
\includegraphics[width=0.4\linewidth]{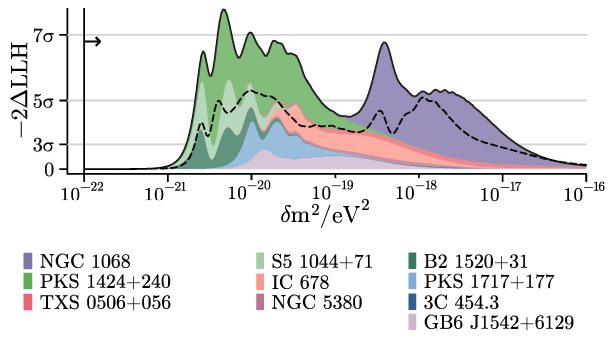}
\caption{\textbf{\textit{Neutrino oscillations at cosmic scales and the sensitivity to extremely small energy differences.}}
Left: Artistic rendition of ultralong baseline neutrino oscillations from astrophysical sources. Right: Expected sensitivity of IceCube-Gen2 to quasi-Dirac neutrinos. Figures reproduced from Ref.~\cite{Carloni:2022cqz}.}
\label{fig:pseudo-dirac-sensitivity}
\end{figure*}

Given the current experimental capabilities, this implies that searches for extraterrestrial sources will be restricted to searching for distortions in the muon spectra energy distribution since we do not currently expect to establish high-significance correlations between cascade and point-like neutrino sources.
The signature of this scenario is thus the disappearance of muon neutrinos at sources that share the common value of $L/E$; see~\cref{fig:pseudo-dirac-sensitivity} (left).
\Cref{fig:pseudo-dirac-sensitivity} (right) shows the sensitivity that is expected from the combination of multiple extraterrestrial neutrino sources as a function of the ultrasmall mass-square difference, $\delta m^2$, for IceCube-Gen2.
As can be noticed, the sensitivity can reach discovery-level significance due to the combination of multiple sources that probe the same values of $L/E$ even when the astrophysical fluxes are marginalized over all multiple shapes of the neutrino spectra; see the dashed line in the figure.
However, understanding the neutrino fluxes of these sources would yield significant improvement in the sensitivity.
For example, if the fluxes were to be modeled by an unbroken power law in energy, the sensitivity would scale to match the shaded region demarcated by the solid line.
Thus, the change in significance between the dashed and solid lines in this figure provides a proxy for the impact of extraterrestrial astrophysical source modeling in this scenario.
Additionally, as noted in Ref.~\cite{Crocker:2001zs}, high-energy neutrinos produced in the galaxy are also sensitive to mass-squared differences on the order of $10^{-14}{\rm eV^2}$, which offers a complementary source to look for this effect. 

\textit{Probing the neutrino lifetime and the unitarity of the neutrino mixing matrix.---} Other properties of the neutrino can be studied by measuring high-energy astrophysical neutrinos. 
These are the lifetime of the neutrino and the number of neutrino species.
The extremely long baselines that astrophysical neutrinos travel made them especially sensitive to the neutrino lifetime.
Given the unknowns in the order of neutrino mass states, i.e., our lack of understanding of which of the neutrinos is less massive, we can consider two scenarios.
In one scenario, the neutrino mass state with the most electron-neutrino content is the lightest; this is known as the normal ordering. In the other, the neutrino with the least electron fraction is the lightest, known as the abnormal or inverted ordering.
The neutrino ordering problem is expected to be resolved over the next several years as new data is gathered from terrestrial neutrino oscillation experiments.
In particular, neutrino telescopes are expected to be important in establishing neutrino ordering~\cite{Ribordy:2013xea,KM3NeT:2021ozk,Arguelles:2022hrt,IceCube:2023ins} by performing precise measurements of the oscillation parameters using large data sets.

The Standard Model expectation is that all known neutrinos are stable over cosmological distances~\cite{Hosotani:1981mq,Pal:1981rm,Nieves:1982bq}; however, new physics scenarios~\cite{Bahcall:1972my,Chikashige:1980qk} can significantly reduce the heavier neutrino lifetime, allowing them to decay into their lighter partners.
This directly affects the ratio of neutrino flavors detected at Earth~\cite{Beacom:2002vi,Bustamante:2016ciw,Denton:2018aml,Song:2020nfh,Abdullahi:2020rge,Liu:2023flr}.
The neutrino ordering above leads to two extreme scenarios, where the neutrino with either the most or the least electron content remains at Earth. 
These two scenarios lead to dramatically distinct fractions of electron neutrinos at Earth; in one case, one expects  $\left|U_{e 3}\right|^2 \simeq 0.67 \%$ of the neutrinos to be electron-flavored, while in the other scenario, only $\left|U_{e 1}\right|^2 \simeq 0.02 \%$ are electron-flavored~\cite{Bustamante:2016ciw}.
Current observations by IceCube~\cite{IceCube:2015gsk,IceCube:2020fpi} already disfavor the large-electron-fraction scenario at the $2 \sigma$ level, placing constraints on the lifetime of the neutrino that is significantly stronger than Earth-based experiments under the assumption of normal neutrino ordering, which is currently preferred by neutrino oscillation measurements~\cite{Esteban:2020cvm}.
Thus, further measurements of the fraction of electron neutrinos at Earth in the astrophysical beam and their energy dependence~\cite{Liu:2023flr} will help to further constrain scenarios that reduce the neutrino lifetime. 
Finally, the number of neutrino states can also affect the expected distribution of astrophysical neutrino flavors at Earth~\cite{Brdar:2016thq,Arguelles:2019tum}. 
Currently, terrestrial neutrino oscillation experiments constrain the unitarity of the neutrino mixing matrix to only $20\%$ at the $3\sigma$ level~\cite{Parke:2015goa,Ellis:2020hus}, leaving space for additional heavier neutrino states to exist. 
The existence of additional neutrino states has been suggested by anomalies in accelerator~\cite{LSND:1996ubh,MiniBooNE:2013uba}, radioactive source~\cite{SAGE:1998fvr,GALLEX:1997lja,Barinov:2021asz}, and neutrino reactor experiments~\cite{Serebrov:2022ajm}, though these have not been confirmed in other neutrino experiments~\cite{Arguelles:2021meu,MINOS:2017cae} with similar reach.
Beyond the experimental motivation, the search for additional neutrino states is of significant interest in fundamental physics as it is predicted in scenarios that aim to resolve the hierarchy problem, e.g., the dark dimension solution~\cite{Montero:2022prj}.
The current sensitivity of the astrophysical neutrino flavor ratio is insufficient to shed light on the existence of new neutrino states~\cite{Beacom:2003nh,Arguelles:2019tum,Song:2020nfh}, but future measurements could enable the indirect detection of these additional neutrinos.

\vspace{0.2cm} \NibSolidRight \textit{The observation of astrophysical neutrinos from Galactic and extragalactic sources can uncover the nature of neutrino masses.
By observing the transformation of neutrinos into new, invisible neutrino states, they can establish that neutrinos are Dirac-like particles, which would have profound implications in particle physics.
Their detailed study can yield new insight into the stability of neutrinos and additional neutrino states.}

\subsection{What new findings in fundamental and particle physics can be made with super-high-energy neutrino beams?\label{subsec:super-beam}}

Two important features of high-energy astrophysical neutrinos make them unique probes of new physics~\cite{Arguelles:2019rbn,Ackermann:2019cxh}: the extremely long lengths traversed from neutrino emission to detection and their high energy.

\begin{figure*}[htbp]
\includegraphics[width=0.3\linewidth]{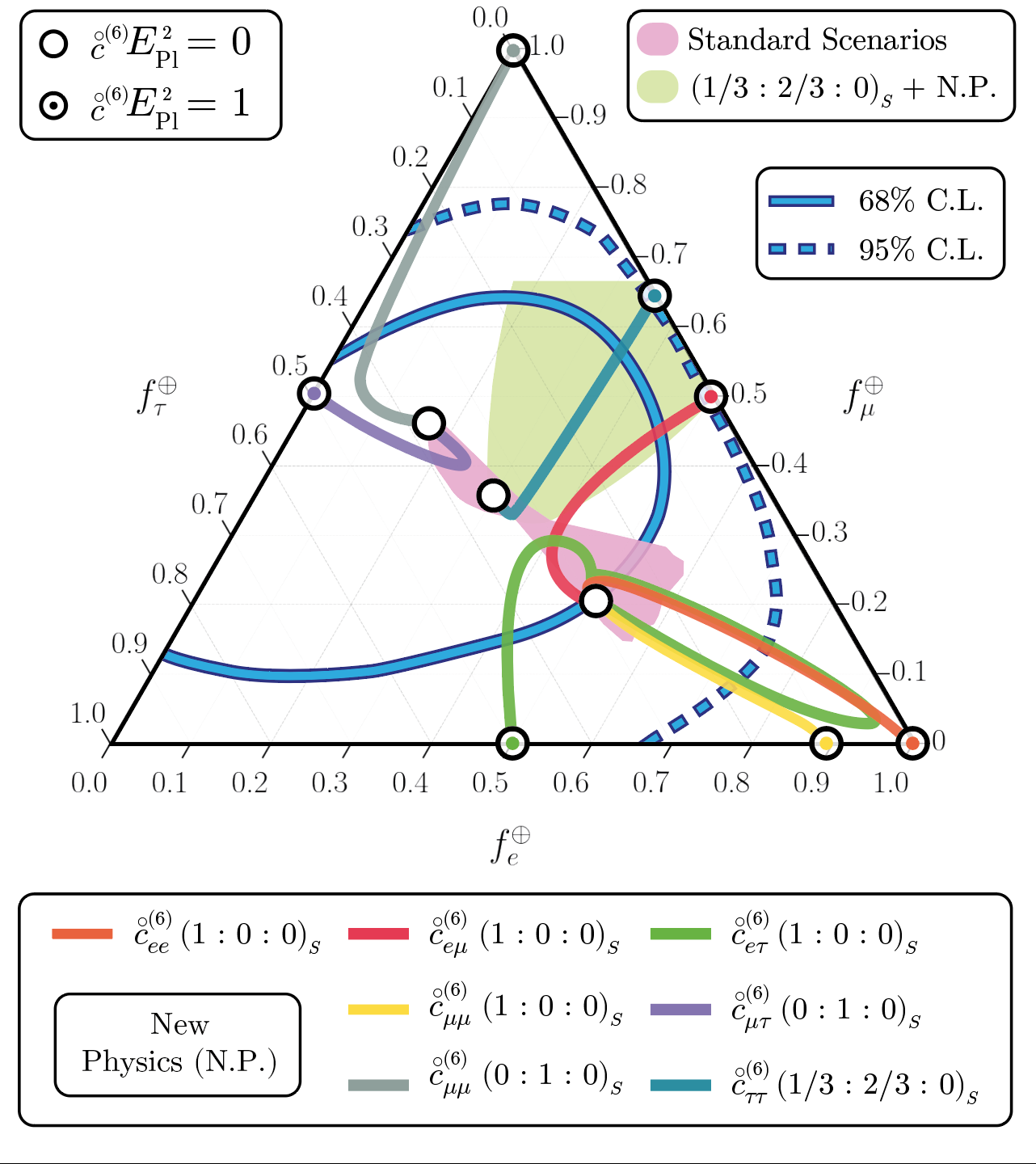} 
\includegraphics[width=0.4\linewidth]{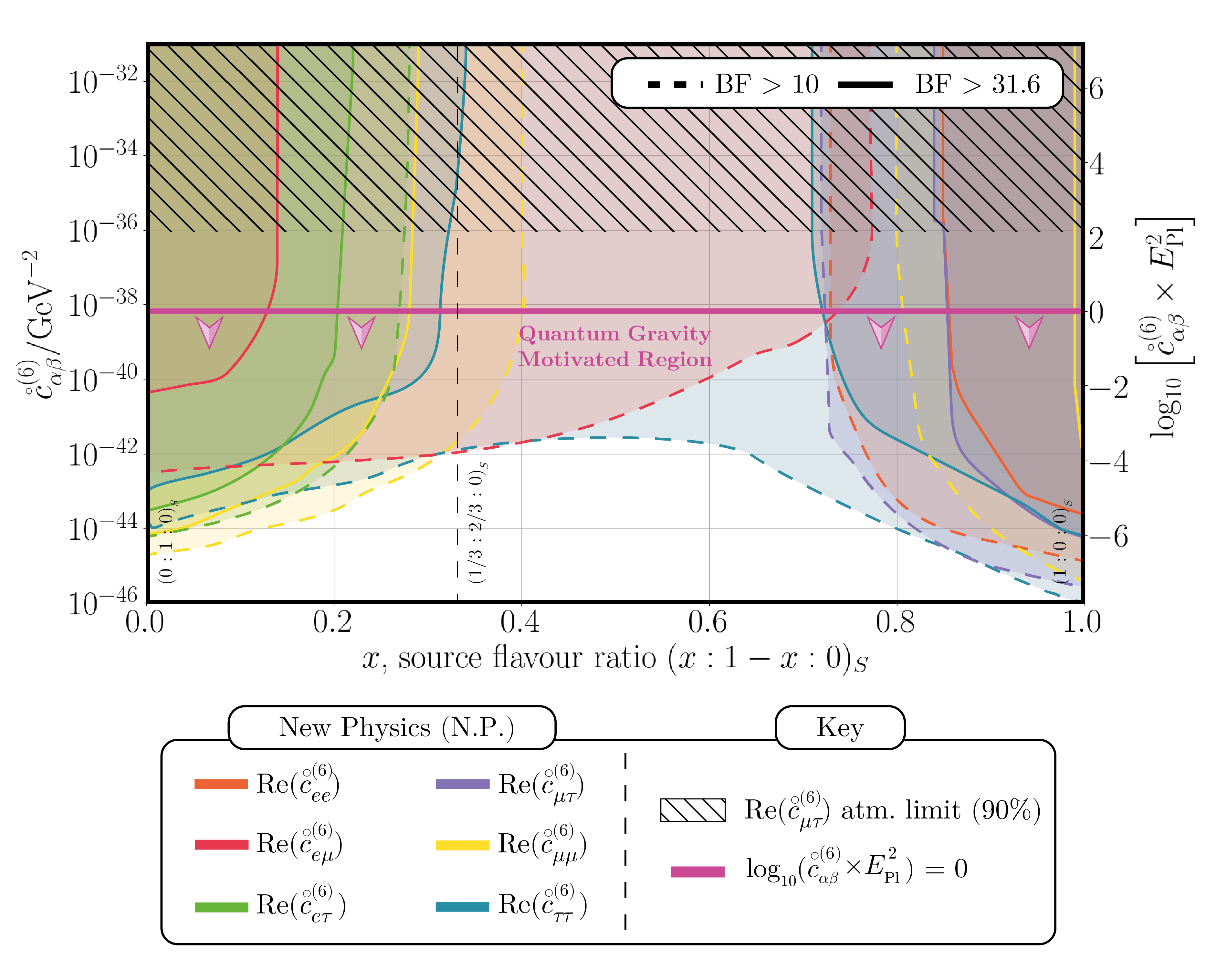}
\caption{\textbf{\textit{Quantum gravity effects on the flavor triangle and constraints from IceCube.}}
Left: Trajectories in the astrophysical neutrino flavor triangle for different dimension-six flavor operators.
Right: Constraints on dimension-six operators that parameterize the interaction of neutrinos with a Lorentz-violating field.
Figures reproduced from Ref.~\cite{IceCube:2021tdn}.}
\label{fig:quantum-gravity}
\end{figure*}

The extremely long distance from production to detection makes high-energy astrophysical neutrinos one of the longest naturally occurring ``interferometers''~\cite{Aartsen:2017ibm} available to search for new phenomena. 
This can be exploited by studying the relative ratio of astrophysical neutrino flavors~\cite{Learned:1994wg,Arguelles:2015dca,Bustamante:2015waa}.
First, it is an observable that is well predicted within the Standard Model with massive neutrinos and is robust under various production mechanisms, thus unambiguously revealing new physics when deviations are observed~\cite{Shoemaker:2015qul,Song:2020nfh}.
However, if the astrophysical environments where neutrinos are produced are very dense, the presence of strong neutrino-matter interactions can produce small, energy-dependent flavor ratios that deviate from democratic flavor composition~\cite{Dev:2023znd}. 
Second, the long lever arm from production to detection serves as an extremely long-lever-arm interferometer allowing for the observation of extremely small effects~\cite{Arguelles:2015dca,IceCube:2021tdn}. 
Namely, as neutrinos travel from their sources to Earth they can coherently interact with the medium they traverse, leaving an imprint on the observed flavor ratio at Earth.
For example, in some theories of quantum gravity, Lorentz symmetry is spontaneously broken at high scales, leaving a remnant Lorentz-violating field throughout space. 
Neutrinos can interact with this field modifying their flavor composition; see~\cref{fig:quantum-gravity} (left). 
The strength and type of this hypothetical interaction are unknown, but it is usually described by a theory known as the Standard Model extension, which includes effective Lorentz-violating terms~\cite{Colladay:1998fq}.
Within this framework, the interactions of neutrinos with this field are expected to be suppressed by powers of the Planck scale.
Currently, high-energy astrophysical neutrinos provide the best way to search for these minute interactions, which accumulate as the neutrinos travel from their source to Earth.
This yields constraints on the coupling strengths of these new, space-time forces that are well beyond the Planck scale~\cite{IceCube:2021tdn}; see ~\cref{fig:quantum-gravity} (right).
Although, for most of the parameters, they are dependent on the flavor structure of the interaction and the flavor composition at the source. 
Other effects that can be probed by studying astrophysical neutrino flavor are coherent interactions between neutrinos and dark matter~\cite{deSalas:2016svi,Capozzi:2018bps,Farzan:2018pnk} and extremely long-range forces~\cite{Bustamante:2018mzu}; see Ref.~\cite{Arguelles:2024cjj} for a recent recast of these constraints in various scenarios.
Additionally, the study of neutrino flavor can provide information about the dense environments in which neutrinos are produced~\cite{Dev:2023znd} and information on new, secret neutrino-nucleon interactions~\cite{Gonzalez-Garcia:2016gpq}.

The high energies and extremely large detectors required to measure high-energy astrophysical neutrinos provide a second opportunity for discovery.
These large detectors measure a large fraction of high-energy astrophysical neutrinos, which can achieve statistics on the order of hundreds of thousands~\cite{IceCube:2024kel,IceCube:2024uzv}.
These high-energy atmospheric neutrinos have been shown to be one of the most sensitive probes of light sterile neutrinos~\cite{IceCube:2024kel,IceCube:2024uzv}, Lorentz symmetry breaking~\cite{IceCube:2017qyp}, decoherence and fundamental tests of quantum mechanics~\cite{Coloma:2018idr,Stuttard:2020qfv,Jones:2024qfr,ICECUBE:2024fej}, and nonstandard neutrino-nucleon interactions~\cite{Salvado:2016uqu,IceCube:2022ubv}.
These searches are ultimately limited by uncertainties in the atmospheric neutrino flux~\cite{Yanez:2023lsy} and the unknown shape of the astrophysical neutrino spectra.
Additionally, neutrino interactions at energies above an EeV probe increasingly inelastic regimes where the proton structure function has yet to be studied.
These high-energy neutrinos probe regions within the proton where the number of partons is increasingly large, and color-shadowing is expected~\cite{Block:2011vz}, yet directly unobserved in collider experiments to date.
The so-called darkening of the proton at high energies is expected to reduce the neutrino-nucleon interaction cross section~\cite{Arguelles:2015wba}.
The deeply inelastic component of the neutrino-nucleon cross section can be measured by studying the transparency of Earth to neutrinos~\cite{IceCube:2017roe,Bustamante:2017xuy,Zhou:2019frk,IceCube:2020rnc,Bertone:2018dse,Esteban:2022uuw}, while the mostly elastic component can produce new morphological features~\cite{GarciaSoto:2022vlw}.
Finally, new particles can be discovered at these high energies by their impact in the neutrino cross section~\cite{Alvarez-Muniz:2002snq,Dey:2017ede,Babu:2019vff}.
The physics of some of these latter scenarios can degenerate with certain astrophysical uncertainties, such as the energy distribution of astrophysical neutrinos.
Although, if the astrophysical neutrino flux is dominated by an isotropic source emission, the angular distribution can be used to break this degeneracy to a certain extent~\cite {Esteban:2022uuw}.

\vspace{0.2cm} \NibSolidRight \textit{The study of the relative flavors of neutrinos at Earth allows us to measure the smallest interactions having sensitivity to Planck-scale physics, potentially enabling the observation of Planck-scale physics for the first time.
Additionally, the measurement of Earth's transparency to high-energy neutrinos through various layers of the Earth will provide insights into uncharted depths of the proton, potentially allowing us to observe gluon shadowing.
}

\section{Enabling technologies}\label{sec:technologies}
The fundamental challenge of neutrino astronomy is the enormous number of background particles, the neutrinos and muons created in the Earth's atmosphere, that the neutrino telescopes encounter.
The overburden of water or ice above the detector is usually on the order of a kilometer. Thus, atmospheric muons penetrate through to the detector to produce many orders of magnitude more events compared to neutrinos.
Atmospheric neutrinos, while lower in rate compared to muons, are background events that are larger in rate than astrophysical neutrinos and in almost all cases indistinguishable from them at an event-by-event level. 

To combat the disproportionate background rate, many statistical techniques are employed.
They are usually in the form of data filtering, which aims to cut out background events while retaining as many signal events as possible. While every event observed in a detector usually corresponds to a single particle detection, noise and multiple particles in the detector at the same time can make event classification challenging. Furthermore, classifying events as ``signal" or ``background" is a statistical process. It is usually not possible to definitively distinguish each event as one or the other, so only probabilistic statements can be made based on the event's characteristics. This can be as simple as using the particle's estimated energy as a characteristic to cut on, as it accounts for the fact that atmospheric components have a softer spectrum, or filtering for Earth-penetrating events to reduce muon rates.
More sophisticated techniques are also used, such as veto-based event selections, which aim to keep only events with neutrino interaction vertices contained within the detector, and event selections targeting unique interactions or flavors, such as showers, identifiable tau events, or Glashow events, as discussed in~\cref{subsec:gamma}.
Generally, a combination of many techniques leads to a final analysis-ready event sample.

In every case, though, the final event samples, while higher in signal-to-noise ratio, become statistically limiting in astrophysical neutrinos. Either they are still background dominated or become extremely reduced event samples with a handful of events each.
Thus, the main key to unlocking more results in neutrino astronomy in the coming decades is statistical power.
More astrophysical neutrinos are obtained by collecting more data, and more data are collected faster with more detectors around the globe.
This means more deployed photomultiplier tubes (PMTs) in water and ice.
Thus, the metric of success, to first order, for the next generation of detectors is more instrumented volume around the world.

Beyond this crude metric, optimization of new neutrino telescopes that are coming online and their complementarities to each other and existing telescopes becomes important.
While the original neutrino telescopes imagined decades ago aimed for one globally optimized design, in the last decade, with multiple discoveries emerging, it has become clear that different detector configurations are becoming important in targeting different fluxes of astrophysical neutrinos.
We have entered the era of divergent detector needs. This is to be celebrated as a sign of a maturing field. 

\subsection{Astrophysical Neutrino Fluxes\label{sec:astro-fluxes}}
\begin{figure}[ht]
\centering
\includegraphics[width=1.0\columnwidth]{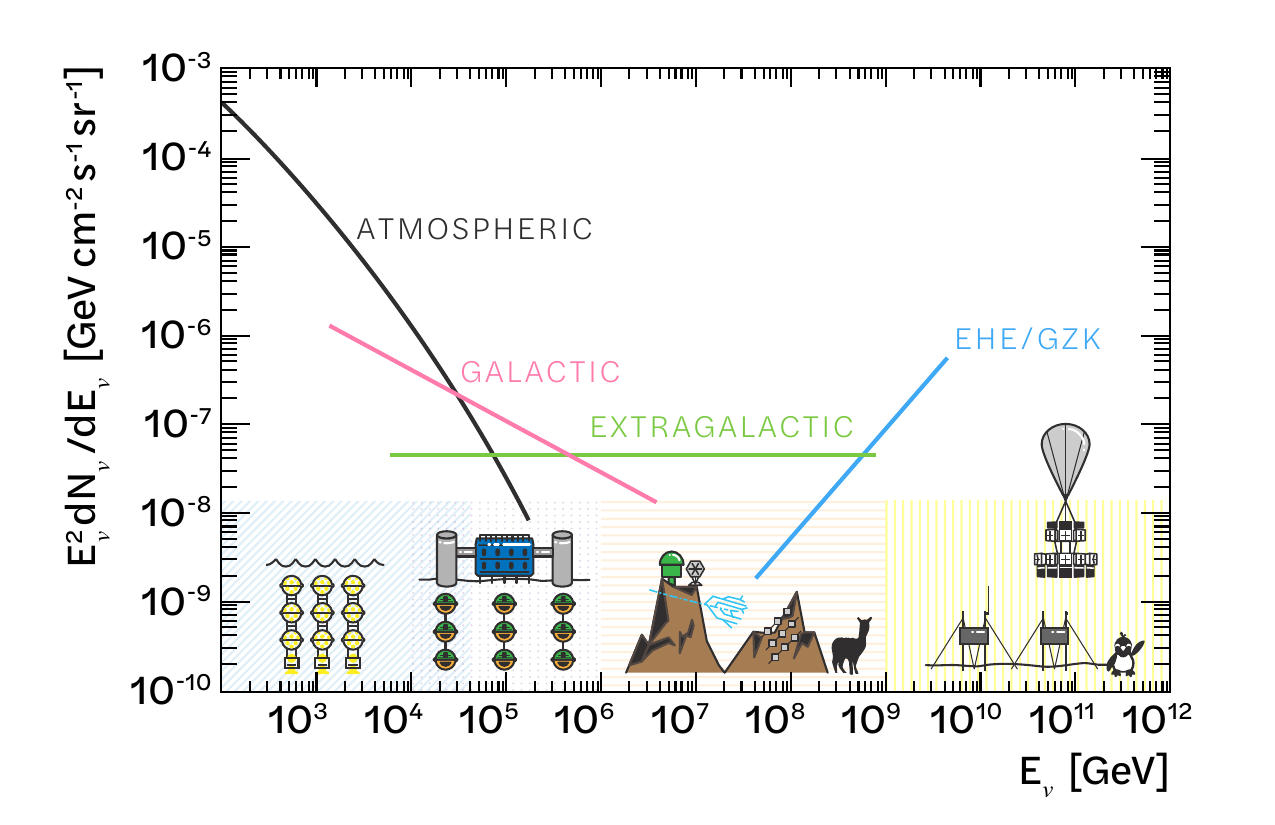}
\caption{\textbf{\textit{Different detector technologies and their dominant energy region.}}
The colored lines in this figure show the different neutrino components of interest as a function of energy, where the ranges should be taken as order of magnitude estimations.
The leading technologies for high-energy astrophysical neutrino detection are illustrated: water-Cherenkov, ice-Cherenkov, Earth-skimming, and radio. 
As discussed in this \textit{Perspective}, the former three detection methods are synergic as they provide complementary information on the same Galactic and extragalactic neutrino sources.
On the other hand, radio is expected to reach out towards new, unexplored energy domains.}
\label{fig:nu_fluxes}
\end{figure}
In the TeV to EeV range, the neutrino fluxes can be described by four principal components, as shown in Fig.~\ref{fig:nu_fluxes}.
Here, we reduce the discussion to describing each component as described in~\cref{eq:powerlaw}, a single power-law flux with spectral index $\gamma_{\rm astro}$, to focus on broader observational needs.
It is clear from recent observations that the diffuse fluxes and the sources that possibly contribute to the fluxes could have more complicated energy spectra. 

Starting from the lowest part of this energy range, the atmospheric neutrino flux dominates with a level that is many orders larger than any astrophysical fluxes.
This is the background to any astrophysical observations and has the steepest spectrum at a spectral index of $\sim3.7$. 

The next observable component in energy is the diffuse Galactic flux.
As described in Section~\ref{subsec:galactic-neutrinos}, it is possible that, within a narrow energy range, this flux dominates in the Galactic plane region of the sky.
While its spectral index is expected to be harder at $\sim2.7$, the discrimination of this flux against the atmospheric background using energy measurements of events will not be so effective since the fluxes have the closest spectral indices.
The crucial observational metric for targeting Galactic neutrinos is thus angular resolution. 
Localizing the diffuse plane or Galactic point sources with good angular resolution allows discrimination against a directionally uniform background, where energy has limited discrimination power.

Neutrinos of extragalactic origin are expected to have spectral indices of $\sim2$.
The flux is now much harder compared to the atmospheric background, so angular resolution requirements can be loosened by relying more on the observed energy of the events.
The distribution of energies of the signal events is sufficiently different to the background for identifying sources.

Finally, at the highest energies or extremely high-energy (EHE) region, cosmogenic neutrinos, also known as GZK neutrinos for Greisen–Zatsepin–Kuzmin~\cite{Greisen:1966jv,Zatsepin:1966jv}, are expected to dominate.
Because this flux is at the tail end of steeply falling fluxes in power law, cosmogenic neutrinos are exceedingly rare.

\subsection{Ice and water optical neutrino detectors}\label{sec:opt-det}
A neutrino telescope optimized to the three fluxes described above will naturally act as three different detectors.
In any steeply falling power-law flux, the higher the energy one targets, the lower the expected number of neutrinos.
Thus the size of the detector will need to be larger for higher energy fluxes.
For the GZK flux, this necessitates such a large detector volume that optical signals become challenging to instrument.
A neutrino telescope must have sensors at a smaller spacing than the attenuation length of the signal for requiring coincidence in multiple sensors.
Therefore, signals that travel further without attenuating, such as radio pulses, allow a larger detector volume to be instrumented by using the same number of sensors but spaced further apart.
Many future radio-based neutrino telescopes~\cite{Aguilar_2021,IceCube-Gen2:2020qha,Abarr_2021,GRAND:2018iaj,Wissel:2020sec} are planned, but this section will focus on optical-based detectors in keeping with the motivations in Section~\ref{sec:questinos}.

\begin{figure}[ht]
\centering
\includegraphics[width=1.0\columnwidth]{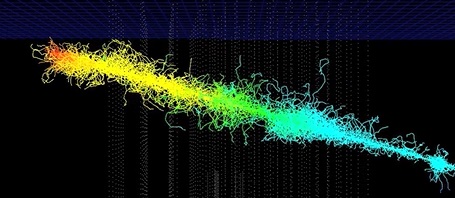}
\includegraphics[width=1.0\columnwidth]{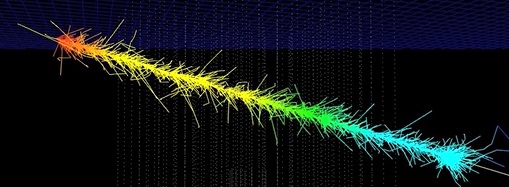}
\caption{\textbf{\textit{Difference between light emission in water and ice.}}
A muon-neutrino-induced muon traversing an ice detector (above) and a water detector (below).
Lines emitted from the muon track show the path of Cherenkov photons emitted from the muon.
Figure courtesy of C.~Kopper (FAU Erlangen-Nürnberg, ECAP).}
\label{fig:tracks}
\end{figure}

Neutrino telescopes with optical sensors collect Cherenkov photons from charged particles that result from neutrino interactions.
The optical properties of the detector media therefore naturally informs the optimization of its detector.
Water generally has a shorter attenuation length of optical and UV light, but a longer scattering length. 
Ice generally has the opposite, where the attenuation length is longer but the scattering length is shorter.
\Cref{fig:tracks} compares a high-energy muon traversing the entire detector length for detectors using ice and water as media. 
The Cherenkov photons are shown as lines coming out of the muon track.
Color denotes the time difference, with red indicating earlier and blue indicating later.
The photons in ice can be seen covering a larger volume around the track, owing to the longer absorption length.
However, the photons do not travel as straight as those in water because of the short scattering length.

The nondirect path of photons in ice creates the largest systematic effect in the reconstruction of the muon track and, thus, in the direction of the neutrino that created the muon.
Water, with all else held equal, is a superior medium for directional reconstruction of neutrinos in the sky.

\begin{figure}[htbp]
\centering
\includegraphics[width=\columnwidth]{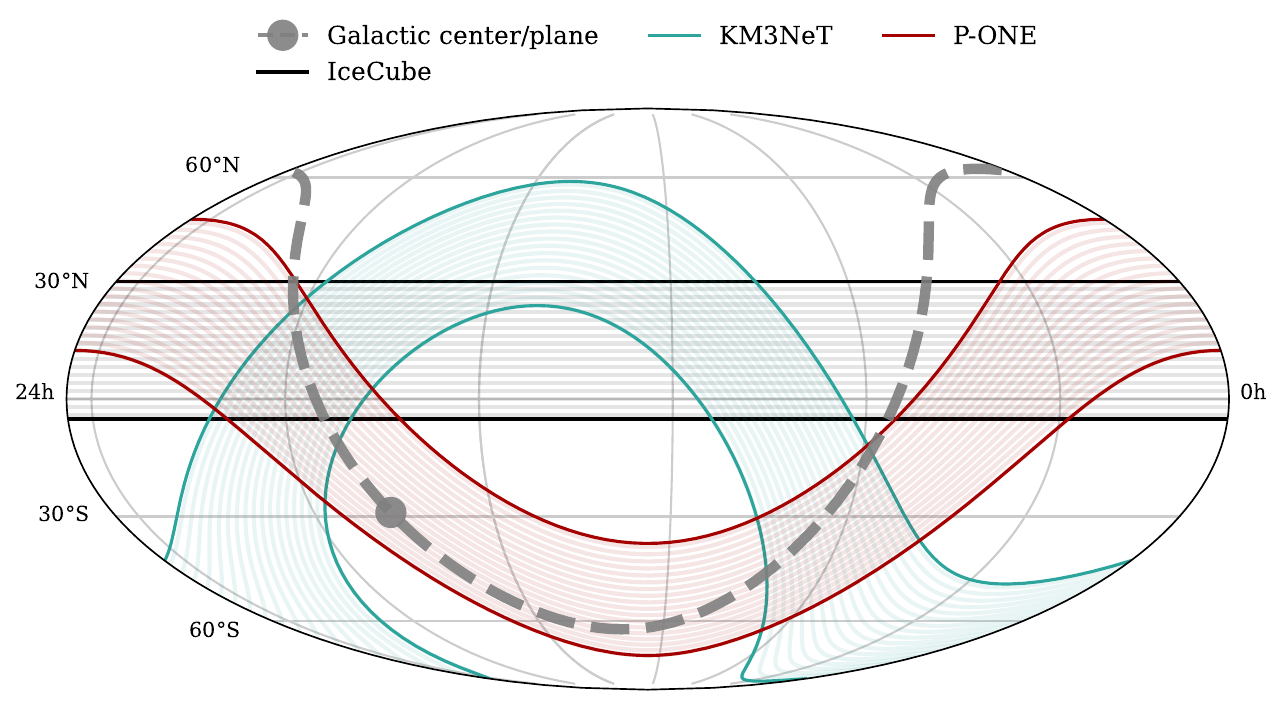}
\caption{\textbf{\textit{Instantaneous regions of highest detection efficiency for various existing and proposed water- and ice-Cherenkov neutrino observatories.}}
The color regions show the largest source discovery potential when assuming an unbroken $E^{-2}$ energy spectra.
The edges of the bands are defined by the discovery potential dropping by a factor of two with respect to its maximum value. This is an instantaneous snapshot, and the bands move as the Earth rotates. Figure courtesy of L. Schumacher (FAU Erlangen-Nürnberg, ECAP).
}
\label{fig:global}
\end{figure}

However, the distance photons travel in water is a disadvantage, especially when the event is not a high-energy muon track traversing the entire detector.
When the effective distance the photon traverses from the charged particle is limited, the sensors must be placed closer in order to obtain coincidence detection of many sensors.
This means that for the same number of sensors, if the same coincidence level is required, the water detector must be made smaller with shorter sensor distances.

This naturally points to ice-based detectors as better suited for higher energy fluxes, such as extragalactic fluxes.
The larger spacing of sensors allows a larger instrumented volume which is more suitable for detecting lower fluxes of neutrinos.
The less-optimized directional resolution of the neutrinos is better tolerated as the extragalactic flux is much harder than background, as previously described.
Conversely, the water-based detectors are better suited for softer fluxes, such as Galactic neutrinos and extragalactic sources with softer emission.
By leveraging the superior directional pointing, discriminating from the soft background becomes possible, and higher fluxes at lower energies compensate for the relatively smaller instrumented volume, again, as described previously.

\subsection{Complementarity in the Global Landscape}
The need for differently optimized telescopes to target different astrophysical neutrino fluxes is not surprising.
Much like how electromagnetic astronomy developed different telescopes targeting different wavelengths and has matured into multiwavelength astronomy, the field of neutrino astronomy is maturing rapidly.
Differently optimized detectors operating simultaneously is the second key to a continuing boom in neutrino astronomy in the coming decades, second only to the first metric of globally increased instrumented volume as discussed earlier.

One can argue then that the third key in global optimization is the locations of neutrino telescopes.
Multiple telescopes operating at different locations in the world, covering different areas of the sky at any given time, ready for real-time multimessenger astronomy, is not a new concept in astronomy.
This is indeed also the case for neutrino astronomy, as shown in~\cref{fig:global}.
While neutrino telescopes have $4\pi$ coverage, sensitivities depend heavily on the zenith angle of the source at the detector. 
The optimized regions of the sky for telescopes in the Mediterranean Sea and the Pacific Ocean as well as at the South Pole show complementary instantaneous coverages that together span the entire sky. 

A final optimization in the neutrino telescope ecosystem is the relative signal-to-background ratio.
Water and ice Cherenkov neutrino telescopes can only differentiate between atmospheric and astrophysical neutrinos for very specific events. 
For example, most of the muon neutrinos observed in these experiments cannot be claimed to be of astrophysical origin since many of them are produced in Earth's atmosphere.
This is not the case for tau neutrinos, where their observation implies that they have a high probability of being astrophysical in nature~\cite{Beacom:2001xn,Soto:2021vdc}. 
Thus, the dominant sample of neutrinos detected in these experiments has low astrophysical neutrino purity.
On the other hand, Earth-skimming tau-neutrino detectors have a high astrophysical-neutrino purity, though they suffer from high-energy thresholds and, thus, limit sample sizes.
Currently, lower-energy Earth-skimming experiments, such as TAMBO or TRINITY, are being developed as individual neutrino telescopes, which complement existing ultra-high-energy Earth-skimming experiments such as Auger~\cite{PierreAuger:2019ens}.
However, the vision of the authors is to think about neutrino telescope development within a network of telescopes rather than in isolation.
Thus, we envision arrays of Earth-skimming neutrino telescopes, whose energy threshold can match the energy range of current and future water/ice Cherenkov neutrino telescopes.  
These will provide a high-purity astrophysical sample that can be instrumental in finding neutrino sources.

If the last two decades of neutrino astronomy can be characterized as the race for becoming ``the'' globally optimized telescope, then the key to continued success in the next two decades is multiple complementary neutrino telescopes, each targeting different neutrino fluxes and flavors, using different media, in different locations around the world, all summing up to a substantial increased instrumented volume globally, operating simultaneously and cross-checking neutrino signals from the sky. Neutrino astronomy deserves to be a strong partner in multimessenger astronomy, as discussed in the science cases presented above.
One or two neutrino telescopes alone cannot accomplish this.
In contrast to the competition of the previous decades, specialization, cooperation, and many complementary telescopes are needed for a truly multimessenger era. 

\subsection{Additional Supporting Technologies}

The experiments discussed above, which are currently either being deployed or constructed, or are in design stages, will operate in an era where machine learning and artificial intelligence will have more than a decade of well-established techniques.
Already in the last three years, we have seen significant impacts on the scientific performance of neutrino telescopes such as IceCube~\cite{Abbasi:2021ryj}, Baikal-GVD~\cite{Kharuk:2023xnl}, and KM3NeT~\cite{Reck:2021zqw} due to the use of artificial intelligence.
For example, the improvement in cascade reconstruction and selection efficiency by means of convolutional neural networks transformed a statistical hint into a firm observation of the Galactic plane. 
These initial approaches were promptly superseded by approaches that allow for general detector geometries~\cite{IceCube:2022njh}, improve the execution speed~\cite{Yu:2023ehc,Capel:2023ijl}, or operate in power-constraint environments~\cite{Jin:2023xts}. 
A recent Kaggle challenge~\cite{Eller:2023myr} on IceCube simulation brought our community's problems to the machine-learning ecosystem, leading to improved event reconstructions. 
The advancement of these methods has also led to the synergistic development of open-source simulation tools~\cite{Lazar:2023rol} and common data formats among experiments. 
All of these improvements and community-building activities will have a significant impact in the development of next-generation neutrino telescope simulation and reconstruction. 
We expect that the speed of development will grow very rapidly.
The steady growth in the capacity of commercially available quantum computers heralds a new era of enhanced simulations and event reconstructions.
Demonstrated recently~\cite{Lazar:2024luq}, even today's quantum resources are capable of facilitating the study of neutrino events.

\section{Conclusion and parting words}\label{sec:conclusion}

Neutrino astronomy needs greater signal-above-background and more astrophysical sources.
A ``neutrino sources catalog'' is desperately needed in the near future to make the transformational discoveries that we believe are possible within the next decade.
The potential discoveries will provide insight into the Universe's more extreme environments and also promise to significantly impact neutrino physics and, more broadly, particle physics.
This requires differently optimized telescopes, in different parts of the world and using different approaches, to make simultaneous observations of sources, by combining data, that yield discoveries to push forward multimessenger astronomy and particle physics through astronomical neutrinos. 
Currently, neutrino telescope collaborations operate independently with limited crosstalk, such as joint conferences 
like VLVnT~\cite{VLVnT} and NuTel~\cite{NuTeL}
and the Global Neutrino Network (GNN)~\cite{GNN}, where ongoing efforts are presented. 
However, it does not make sense to reinvent techniques, tools, and observational strategies. 
The authors strongly believe that a fundamental partnership of many neutrino telescope collaborations, beyond what is imagined within the current efforts, is the path to a thriving field in the next decade. 

\acknowledgments

We thank Jean DeMerit and Julia Book Motzkin for carefully reading the manuscript.
We also thank Claudio Kopper and Lisa Schumacher for their help with figures eleven and twelve.
Finally, we thank Matheus Hostert, Kiara Carloni, and Pavel Zhelnin for their useful feedback on earlier version of this manuscript.
CAA are supported by the Faculty of Arts and Sciences of Harvard University, the National Science Foundation, the Research Corporation for Science Advancement, and the David \& Lucile Packard Foundation.
The research of FH is supported in part by the U.S. National Science Foundation under grants~PHY-2209445 and OPP-2042807. NK is supported by the U.S. National Science Foundation under grants~PHY-1847827 and PHY-2209445.

\bibliography{prx-nu-astro}


\end{document}

%% file: packages.tex
\usepackage{tikz}
\usepackage{amsmath}
\usepackage{graphicx}
\usepackage{fullpage}
\usepackage[colorlinks=true,allcolors=blue]{hyperref}
\usepackage[capitalise]{cleveref}
\usepackage[utf8]{inputenc}
\usepackage{siunitx}
\usepackage{url}
\usepackage{rotating}
\usepackage{epsfig,graphics,rotate}
\usepackage{wrapfig}
\usepackage[caption=false]{subfig}
\usepackage{bm}
\usepackage{amssymb}
\usepackage{amsmath}
\usepackage{amsfonts}
\usepackage{booktabs}
\usepackage{microtype}
\usepackage{multirow, makecell}
\usepackage{lipsum}
\usepackage{xspace}
\usepackage{comment}
\usepackage{floatrow}
\usepackage{ragged2e}

\usepackage{nicefrac, xfrac}
\usepackage{mathtools} 
\usepackage{relsize}   

\usepackage{float}

\usepackage{psvectorian}
\usepackage{bbding}

\DeclareSIUnit \s {\second}
\DeclareSIUnit \ns {\nano\second}
\DeclareSIUnit \mus {\micro\second}
\DeclareSIUnit \ms {\milli\second}
\DeclareSIUnit \MB {\mega\byte}
\DeclareSIUnit \GB {\giga\byte}
\DeclareSIUnit \TB {\tera\byte}
\DeclareSIUnit \PB {\peta\byte}
\DeclareSIUnit \Mbps {\mega\bit/\s}
\DeclareSIUnit \Gbps {\giga\bit/\s}
\DeclareSIUnit \Tbps {\tera\bit/\s}
\DeclareSIUnit \Pbps {\peta\bit/\s}
\DeclareSIUnit \kton {\kilo\tonne} 
\DeclareSIUnit \kt {\kilo\tonne}
\DeclareSIUnit \Mt {\mega\tonne}
\DeclareSIUnit \eV {\electronvolt}
\DeclareSIUnit \keV {\kilo\electronvolt}
\DeclareSIUnit \MeV {\mega\electronvolt}
\DeclareSIUnit \GeV {\giga\electronvolt}
\DeclareSIUnit \TeV {\tera\electronvolt}
\DeclareSIUnit \PeV {\peta\electronvolt}
\DeclareSIUnit \EeV {\exa\electronvolt}
\DeclareSIUnit \m {\meter}
\DeclareSIUnit \cm {\centi\meter}
\DeclareSIUnit \in {\inchcommand}
\DeclareSIUnit \km {\kilo\meter}
\DeclareSIUnit \kV {\kilo\volt}
\DeclareSIUnit \kW {\kilo\watt}
\DeclareSIUnit \MW {\mega\watt}
\DeclareSIUnit \MHz {\mega\hertz}
\DeclareSIUnit \mrad {\milli\radian}
\DeclareSIUnit \year {years}
\DeclareSIUnit \POT {POT}
\DeclareSIUnit \sig {$\sigma$}
\DeclareSIUnit\parsec{pc}
\DeclareSIUnit\lightyear{ly}
\DeclareSIUnit\foot{ft}
\DeclareSIUnit\ft{ft}
\DeclareSIUnit \ppb{ppb}
\DeclareSIUnit \ppt{ppt}
\DeclareSIUnit \samples{S}
\DeclareSIUnit \pe{PE}
\DeclareSIUnit \kpc{kpc}

%% file: main.bbl
\begin{thebibliography}{293}%
\makeatletter
\providecommand \@ifxundefined [1]{%
 \@ifx{#1\undefined}
}%
\providecommand \@ifnum [1]{%
 \ifnum #1\expandafter \@firstoftwo
 \else \expandafter \@secondoftwo
 \fi
}%
\providecommand \@ifx [1]{%
 \ifx #1\expandafter \@firstoftwo
 \else \expandafter \@secondoftwo
 \fi
}%
\providecommand \natexlab [1]{#1}%
\providecommand \enquote  [1]{``#1''}%
\providecommand \bibnamefont  [1]{#1}%
\providecommand \bibfnamefont [1]{#1}%
\providecommand \citenamefont [1]{#1}%
\providecommand \href@noop [0]{\@secondoftwo}%
\providecommand \href [0]{\begingroup \@sanitize@url \@href}%
\providecommand \@href[1]{\@@startlink{#1}\@@href}%
\providecommand \@@href[1]{\endgroup#1\@@endlink}%
\providecommand \@sanitize@url [0]{\catcode `\\12\catcode `\$12\catcode `\&12\catcode `\#12\catcode `\^12\catcode `\_12\catcode `\%12\relax}%
\providecommand \@@startlink[1]{}%
\providecommand \@@endlink[0]{}%
\providecommand \url  [0]{\begingroup\@sanitize@url \@url }%
\providecommand \@url [1]{\endgroup\@href {#1}{\urlprefix }}%
\providecommand \urlprefix  [0]{URL }%
\providecommand \Eprint [0]{\href }%
\providecommand \doibase [0]{http://dx.doi.org/}%
\providecommand \selectlanguage [0]{\@gobble}%
\providecommand \bibinfo  [0]{\@secondoftwo}%
\providecommand \bibfield  [0]{\@secondoftwo}%
\providecommand \translation [1]{[#1]}%
\providecommand \BibitemOpen [0]{}%
\providecommand \bibitemStop [0]{}%
\providecommand \bibitemNoStop [0]{.\EOS\space}%
\providecommand \EOS [0]{\spacefactor3000\relax}%
\providecommand \BibitemShut  [1]{\csname bibitem#1\endcsname}%
\let\auto@bib@innerbib\@empty
\bibitem [{\citenamefont {Hill}\ \emph {et~al.}(2018)\citenamefont {Hill}, \citenamefont {Masui},\ and\ \citenamefont {Scott}}]{Hill:2018trh}%
  \BibitemOpen
  \bibfield  {author} {\bibinfo {author} {\bibfnamefont {Ryley}\ \bibnamefont {Hill}}, \bibinfo {author} {\bibfnamefont {Kiyoshi~W.}\ \bibnamefont {Masui}}, \ and\ \bibinfo {author} {\bibfnamefont {Douglas}\ \bibnamefont {Scott}},\ }\bibfield  {title} {\enquote {\bibinfo {title} {{The Spectrum of the Universe}},}\ }\href {\doibase 10.1177/0003702818767133} {\bibfield  {journal} {\bibinfo  {journal} {Appl. Spectrosc.}\ }\textbf {\bibinfo {volume} {72}},\ \bibinfo {pages} {663--688} (\bibinfo {year} {2018})},\ \Eprint {http://arxiv.org/abs/1802.03694} {arXiv:1802.03694 [astro-ph.CO]} \BibitemShut {NoStop}%
\bibitem [{\citenamefont {Gould}\ and\ \citenamefont {Schr\'eder}(1966)}]{Gould:1966pza}%
  \BibitemOpen
  \bibfield  {author} {\bibinfo {author} {\bibfnamefont {Robert}\ \bibnamefont {Gould}}\ and\ \bibinfo {author} {\bibfnamefont {Gerald}\ \bibnamefont {Schr\'eder}},\ }\bibfield  {title} {\enquote {\bibinfo {title} {{Opacity of the Universe to High-Energy Photons}},}\ }\href {\doibase 10.1103/PhysRevLett.16.252} {\bibfield  {journal} {\bibinfo  {journal} {Phys. Rev. Lett.}\ }\textbf {\bibinfo {volume} {16}},\ \bibinfo {pages} {252--254} (\bibinfo {year} {1966})}\BibitemShut {NoStop}%
\bibitem [{\citenamefont {Olive}\ \emph {et~al.}(2014)\citenamefont {Olive} \emph {et~al.}}]{ParticleDataGroup:2014cgo}%
  \BibitemOpen
  \bibfield  {author} {\bibinfo {author} {\bibfnamefont {K.~A.}\ \bibnamefont {Olive}} \emph {et~al.} (\bibinfo {collaboration} {Particle Data Group}),\ }\bibfield  {title} {\enquote {\bibinfo {title} {{Review of Particle Physics}},}\ }\href {\doibase 10.1088/1674-1137/38/9/090001} {\bibfield  {journal} {\bibinfo  {journal} {Chin. Phys. C}\ }\textbf {\bibinfo {volume} {38}},\ \bibinfo {pages} {090001} (\bibinfo {year} {2014})}\BibitemShut {NoStop}%
\bibitem [{\citenamefont {Spiering}(2012)}]{Spiering:2012xe}%
  \BibitemOpen
  \bibfield  {author} {\bibinfo {author} {\bibfnamefont {Christian}\ \bibnamefont {Spiering}},\ }\bibfield  {title} {\enquote {\bibinfo {title} {{Towards High-Energy Neutrino Astronomy. A Historical Review}},}\ }\href {\doibase 10.1140/epjh/e2012-30014-2} {\bibfield  {journal} {\bibinfo  {journal} {Eur. Phys. J. H}\ }\textbf {\bibinfo {volume} {37}},\ \bibinfo {pages} {515--565} (\bibinfo {year} {2012})},\ \Eprint {http://arxiv.org/abs/1207.4952} {arXiv:1207.4952 [astro-ph.IM]} \BibitemShut {NoStop}%
\bibitem [{\citenamefont {Aartsen}\ \emph {et~al.}(2017{\natexlab{a}})\citenamefont {Aartsen} \emph {et~al.}}]{IceCube:2016zyt}%
  \BibitemOpen
  \bibfield  {author} {\bibinfo {author} {\bibfnamefont {M.~G.}\ \bibnamefont {Aartsen}} \emph {et~al.} (\bibinfo {collaboration} {IceCube}),\ }\bibfield  {title} {\enquote {\bibinfo {title} {{The IceCube Neutrino Observatory: Instrumentation and Online Systems}},}\ }\href {\doibase 10.1088/1748-0221/12/03/P03012} {\bibfield  {journal} {\bibinfo  {journal} {JINST}\ }\textbf {\bibinfo {volume} {12}},\ \bibinfo {pages} {P03012} (\bibinfo {year} {2017}{\natexlab{a}})},\ \bibinfo {note} {[Erratum: JINST 19, E05001 (2024)]},\ \Eprint {http://arxiv.org/abs/1612.05093} {arXiv:1612.05093 [astro-ph.IM]} \BibitemShut {NoStop}%
\bibitem [{\citenamefont {Aartsen}\ \emph {et~al.}(2013{\natexlab{a}})\citenamefont {Aartsen} \emph {et~al.}}]{Aartsen:2013jdh}%
  \BibitemOpen
  \bibfield  {author} {\bibinfo {author} {\bibfnamefont {M.~G.}\ \bibnamefont {Aartsen}} \emph {et~al.} (\bibinfo {collaboration} {IceCube}),\ }\bibfield  {title} {\enquote {\bibinfo {title} {{Evidence for High-Energy Extraterrestrial Neutrinos at the IceCube Detector}},}\ }\href {\doibase 10.1126/science.1242856} {\bibfield  {journal} {\bibinfo  {journal} {Science}\ }\textbf {\bibinfo {volume} {342}},\ \bibinfo {pages} {1242856} (\bibinfo {year} {2013}{\natexlab{a}})},\ \Eprint {http://arxiv.org/abs/1311.5238} {arXiv:1311.5238 [astro-ph.HE]} \BibitemShut {NoStop}%
\bibitem [{\citenamefont {Aartsen}\ \emph {et~al.}(2014)\citenamefont {Aartsen} \emph {et~al.}}]{IceCube:2014stg}%
  \BibitemOpen
  \bibfield  {author} {\bibinfo {author} {\bibfnamefont {M.~G.}\ \bibnamefont {Aartsen}} \emph {et~al.} (\bibinfo {collaboration} {IceCube}),\ }\bibfield  {title} {\enquote {\bibinfo {title} {{Observation of High-Energy Astrophysical Neutrinos in Three Years of IceCube Data}},}\ }\href {\doibase 10.1103/PhysRevLett.113.101101} {\bibfield  {journal} {\bibinfo  {journal} {Phys. Rev. Lett.}\ }\textbf {\bibinfo {volume} {113}},\ \bibinfo {pages} {101101} (\bibinfo {year} {2014})},\ \Eprint {http://arxiv.org/abs/1405.5303} {arXiv:1405.5303 [astro-ph.HE]} \BibitemShut {NoStop}%
\bibitem [{\citenamefont {Fang}\ \emph {et~al.}(2022)\citenamefont {Fang}, \citenamefont {Gallagher},\ and\ \citenamefont {Halzen}}]{Fang:2022trf}%
  \BibitemOpen
  \bibfield  {author} {\bibinfo {author} {\bibfnamefont {Ke}~\bibnamefont {Fang}}, \bibinfo {author} {\bibfnamefont {John~S.}\ \bibnamefont {Gallagher}}, \ and\ \bibinfo {author} {\bibfnamefont {Francis}\ \bibnamefont {Halzen}},\ }\bibfield  {title} {\enquote {\bibinfo {title} {{The TeV Diffuse Cosmic Neutrino Spectrum and the Nature of Astrophysical Neutrino Sources}},}\ }\href {\doibase 10.3847/1538-4357/ac7649} {\bibfield  {journal} {\bibinfo  {journal} {Astrophys. J.}\ }\textbf {\bibinfo {volume} {933}},\ \bibinfo {pages} {190} (\bibinfo {year} {2022})},\ \Eprint {http://arxiv.org/abs/2205.03740} {arXiv:2205.03740 [astro-ph.HE]} \BibitemShut {NoStop}%
\bibitem [{\citenamefont {Abbasi}\ \emph {et~al.}(2023{\natexlab{a}})\citenamefont {Abbasi} \emph {et~al.}}]{IceCube:2023ame}%
  \BibitemOpen
  \bibfield  {author} {\bibinfo {author} {\bibfnamefont {R.}~\bibnamefont {Abbasi}} \emph {et~al.} (\bibinfo {collaboration} {IceCube}),\ }\bibfield  {title} {\enquote {\bibinfo {title} {{Observation of high-energy neutrinos from the Galactic plane}},}\ }\href {\doibase 10.1126/science.adc9818} {\bibfield  {journal} {\bibinfo  {journal} {Science}\ }\textbf {\bibinfo {volume} {380}},\ \bibinfo {pages} {adc9818} (\bibinfo {year} {2023}{\natexlab{a}})},\ \Eprint {http://arxiv.org/abs/2307.04427} {arXiv:2307.04427 [astro-ph.HE]} \BibitemShut {NoStop}%
\bibitem [{\citenamefont {Fang}\ \emph {et~al.}(2024)\citenamefont {Fang}, \citenamefont {Gallagher},\ and\ \citenamefont {Halzen}}]{Fang:2023azx}%
  \BibitemOpen
  \bibfield  {author} {\bibinfo {author} {\bibfnamefont {Ke}~\bibnamefont {Fang}}, \bibinfo {author} {\bibfnamefont {John~S.}\ \bibnamefont {Gallagher}}, \ and\ \bibinfo {author} {\bibfnamefont {Francis}\ \bibnamefont {Halzen}},\ }\bibfield  {title} {\enquote {\bibinfo {title} {{The Milky Way revealed to be a neutrino desert by the IceCube Galactic plane observation}},}\ }\href {\doibase 10.1038/s41550-023-02128-0} {\bibfield  {journal} {\bibinfo  {journal} {Nature Astron.}\ }\textbf {\bibinfo {volume} {8}},\ \bibinfo {pages} {241--246} (\bibinfo {year} {2024})},\ \Eprint {http://arxiv.org/abs/2306.17275} {arXiv:2306.17275 [astro-ph.HE]} \BibitemShut {NoStop}%
\bibitem [{\citenamefont {Abbasi}\ \emph {et~al.}(2022{\natexlab{a}})\citenamefont {Abbasi} \emph {et~al.}}]{IceCube:2022der}%
  \BibitemOpen
  \bibfield  {author} {\bibinfo {author} {\bibfnamefont {R.}~\bibnamefont {Abbasi}} \emph {et~al.} (\bibinfo {collaboration} {IceCube}),\ }\bibfield  {title} {\enquote {\bibinfo {title} {{Evidence for neutrino emission from the nearby active galaxy NGC 1068}},}\ }\href {\doibase 10.1126/science.abg3395} {\bibfield  {journal} {\bibinfo  {journal} {Science}\ }\textbf {\bibinfo {volume} {378}},\ \bibinfo {pages} {538--543} (\bibinfo {year} {2022}{\natexlab{a}})},\ \Eprint {http://arxiv.org/abs/2211.09972} {arXiv:2211.09972 [astro-ph.HE]} \BibitemShut {NoStop}%
\bibitem [{\citenamefont {Privon}\ \emph {et~al.}(2023)\citenamefont {Privon} \emph {et~al.}}]{IceCube:2023jds}%
  \BibitemOpen
  \bibfield  {author} {\bibinfo {author} {\bibfnamefont {George~C.}\ \bibnamefont {Privon}} \emph {et~al.} (\bibinfo {collaboration} {IceCube}),\ }\bibfield  {title} {\enquote {\bibinfo {title} {{Search for high-energy neutrino emission from hard X-ray AGN with IceCube}},}\ }\href {\doibase 10.22323/1.444.1032} {\bibfield  {journal} {\bibinfo  {journal} {PoS}\ }\textbf {\bibinfo {volume} {ICRC2023}},\ \bibinfo {pages} {1032} (\bibinfo {year} {2023})},\ \Eprint {http://arxiv.org/abs/2307.15349} {arXiv:2307.15349 [astro-ph.HE]} \BibitemShut {NoStop}%
\bibitem [{\citenamefont {Aartsen}\ \emph {et~al.}(2018{\natexlab{a}})\citenamefont {Aartsen} \emph {et~al.}}]{IceCube:2018dnn}%
  \BibitemOpen
  \bibfield  {author} {\bibinfo {author} {\bibfnamefont {M.~G.}\ \bibnamefont {Aartsen}} \emph {et~al.} (\bibinfo {collaboration} {IceCube, Fermi-LAT, MAGIC, AGILE, ASAS-SN, HAWC, H.E.S.S., INTEGRAL, Kanata, Kiso, Kapteyn, Liverpool Telescope, Subaru, Swift NuSTAR, VERITAS, VLA/17B-403}),\ }\bibfield  {title} {\enquote {\bibinfo {title} {{Multimessenger observations of a flaring blazar coincident with high-energy neutrino IceCube-170922A}},}\ }\href {\doibase 10.1126/science.aat1378} {\bibfield  {journal} {\bibinfo  {journal} {Science}\ }\textbf {\bibinfo {volume} {361}},\ \bibinfo {pages} {eaat1378} (\bibinfo {year} {2018}{\natexlab{a}})},\ \Eprint {http://arxiv.org/abs/1807.08816} {arXiv:1807.08816 [astro-ph.HE]} \BibitemShut {NoStop}%
\bibitem [{\citenamefont {Aartsen}\ \emph {et~al.}(2018{\natexlab{b}})\citenamefont {Aartsen} \emph {et~al.}}]{IceCube:2018cha}%
  \BibitemOpen
  \bibfield  {author} {\bibinfo {author} {\bibfnamefont {M.~G.}\ \bibnamefont {Aartsen}} \emph {et~al.} (\bibinfo {collaboration} {IceCube}),\ }\bibfield  {title} {\enquote {\bibinfo {title} {{Neutrino emission from the direction of the blazar TXS 0506+056 prior to the IceCube-170922A alert}},}\ }\href {\doibase 10.1126/science.aat2890} {\bibfield  {journal} {\bibinfo  {journal} {Science}\ }\textbf {\bibinfo {volume} {361}},\ \bibinfo {pages} {147--151} (\bibinfo {year} {2018}{\natexlab{b}})},\ \Eprint {http://arxiv.org/abs/1807.08794} {arXiv:1807.08794 [astro-ph.HE]} \BibitemShut {NoStop}%
\bibitem [{\citenamefont {Murase}(2022)}]{Murase:2022dog}%
  \BibitemOpen
  \bibfield  {author} {\bibinfo {author} {\bibfnamefont {Kohta}\ \bibnamefont {Murase}},\ }\bibfield  {title} {\enquote {\bibinfo {title} {{Hidden Hearts of Neutrino Active Galaxies}},}\ }\href {\doibase 10.3847/2041-8213/aca53c} {\bibfield  {journal} {\bibinfo  {journal} {Astrophys. J. Lett.}\ }\textbf {\bibinfo {volume} {941}},\ \bibinfo {pages} {L17} (\bibinfo {year} {2022})},\ \Eprint {http://arxiv.org/abs/2211.04460} {arXiv:2211.04460 [astro-ph.HE]} \BibitemShut {NoStop}%
\bibitem [{\citenamefont {Aynutdinov}\ \emph {et~al.}(2023)\citenamefont {Aynutdinov} \emph {et~al.}}]{Baikal-GVD:2023qib}%
  \BibitemOpen
  \bibfield  {author} {\bibinfo {author} {\bibfnamefont {V.~M.}\ \bibnamefont {Aynutdinov}} \emph {et~al.} (\bibinfo {collaboration} {Baikal-GVD}),\ }\bibfield  {title} {\enquote {\bibinfo {title} {{Baikal-GVD Astrophysical Neutrino Candidate near the Blazar TXS\textasciitilde{}0506+056}},}\ }in\ \href@noop {} {\emph {\bibinfo {booktitle} {{38th International Cosmic Ray Conference}}}}\ (\bibinfo {year} {2023})\ \Eprint {http://arxiv.org/abs/2308.13686} {arXiv:2308.13686 [astro-ph.HE]} \BibitemShut {NoStop}%
\bibitem [{\citenamefont {Albert}\ \emph {et~al.}(2018{\natexlab{a}})\citenamefont {Albert} \emph {et~al.}}]{ANTARES:2018osx}%
  \BibitemOpen
  \bibfield  {author} {\bibinfo {author} {\bibfnamefont {A.}~\bibnamefont {Albert}} \emph {et~al.} (\bibinfo {collaboration} {ANTARES}),\ }\bibfield  {title} {\enquote {\bibinfo {title} {{The Search for Neutrinos from TXS 0506+056 with the ANTARES Telescope}},}\ }\href {\doibase 10.3847/2041-8213/aad8c0} {\bibfield  {journal} {\bibinfo  {journal} {Astrophys. J. Lett.}\ }\textbf {\bibinfo {volume} {863}},\ \bibinfo {pages} {L30} (\bibinfo {year} {2018}{\natexlab{a}})},\ \Eprint {http://arxiv.org/abs/1807.04309} {arXiv:1807.04309 [astro-ph.HE]} \BibitemShut {NoStop}%
\bibitem [{\citenamefont {Ishihara}(2021)}]{Ishihara:2019aao}%
  \BibitemOpen
  \bibfield  {author} {\bibinfo {author} {\bibfnamefont {Aya}\ \bibnamefont {Ishihara}} (\bibinfo {collaboration} {IceCube}),\ }\bibfield  {title} {\enquote {\bibinfo {title} {{The IceCube Upgrade - Design and Science Goals}},}\ }\href {\doibase 10.22323/1.358.1031} {\bibfield  {journal} {\bibinfo  {journal} {PoS}\ }\textbf {\bibinfo {volume} {ICRC2019}},\ \bibinfo {pages} {1031} (\bibinfo {year} {2021})},\ \Eprint {http://arxiv.org/abs/1908.09441} {arXiv:1908.09441 [astro-ph.HE]} \BibitemShut {NoStop}%
\bibitem [{\citenamefont {Andres}\ \emph {et~al.}(2000)\citenamefont {Andres} \emph {et~al.}}]{Andres:1999hm}%
  \BibitemOpen
  \bibfield  {author} {\bibinfo {author} {\bibfnamefont {E.}~\bibnamefont {Andres}} \emph {et~al.},\ }\bibfield  {title} {\enquote {\bibinfo {title} {{The AMANDA neutrino telescope: Principle of operation and first results}},}\ }\href {\doibase 10.1016/S0927-6505(99)00092-4} {\bibfield  {journal} {\bibinfo  {journal} {Astropart. Phys.}\ }\textbf {\bibinfo {volume} {13}},\ \bibinfo {pages} {1--20} (\bibinfo {year} {2000})},\ \Eprint {http://arxiv.org/abs/astro-ph/9906203} {arXiv:astro-ph/9906203} \BibitemShut {NoStop}%
\bibitem [{\citenamefont {Aartsen}\ \emph {et~al.}(2021{\natexlab{a}})\citenamefont {Aartsen} \emph {et~al.}}]{IceCube-Gen2:2020qha}%
  \BibitemOpen
  \bibfield  {author} {\bibinfo {author} {\bibfnamefont {M.~G.}\ \bibnamefont {Aartsen}} \emph {et~al.} (\bibinfo {collaboration} {IceCube-Gen2}),\ }\bibfield  {title} {\enquote {\bibinfo {title} {{IceCube-Gen2: the window to the extreme Universe}},}\ }\href {\doibase 10.1088/1361-6471/abbd48} {\bibfield  {journal} {\bibinfo  {journal} {J. Phys. G}\ }\textbf {\bibinfo {volume} {48}},\ \bibinfo {pages} {060501} (\bibinfo {year} {2021}{\natexlab{a}})},\ \Eprint {http://arxiv.org/abs/2008.04323} {arXiv:2008.04323 [astro-ph.HE]} \BibitemShut {NoStop}%
\bibitem [{\citenamefont {{National Academies of Sciences, Engineering, and Medicine}}(2023)}]{NAP26141}%
  \BibitemOpen
  \bibfield  {author} {\bibinfo {author} {\bibnamefont {{National Academies of Sciences, Engineering, and Medicine}}},\ }\href {\doibase 10.17226/26141} {\emph {\bibinfo {title} {Pathways to Discovery in Astronomy and Astrophysics for the 2020s}}}\ (\bibinfo  {publisher} {The National Academies Press},\ \bibinfo {address} {Washington, DC},\ \bibinfo {year} {2023})\BibitemShut {NoStop}%
\bibitem [{\citenamefont {Asai}\ \emph {et~al.}(2023)\citenamefont {Asai} \emph {et~al.}}]{P5:2023wyd}%
  \BibitemOpen
  \bibfield  {author} {\bibinfo {author} {\bibfnamefont {Shoji}\ \bibnamefont {Asai}} \emph {et~al.} (\bibinfo {collaboration} {P5}),\ }\bibfield  {title} {\enquote {\bibinfo {title} {{Exploring the Quantum Universe: Pathways to Innovation and Discovery in Particle Physics}},}\ }\href {\doibase 10.2172/2368847} {\  (\bibinfo {year} {2023}),\ 10.2172/2368847},\ \Eprint {http://arxiv.org/abs/2407.19176} {arXiv:2407.19176 [hep-ex]} \BibitemShut {NoStop}%
\bibitem [{\citenamefont {Shoibonov}(2019)}]{Shoibonov:2019gfj}%
  \BibitemOpen
  \bibfield  {author} {\bibinfo {author} {\bibfnamefont {Bair}\ \bibnamefont {Shoibonov}} (\bibinfo {collaboration} {Baikal}),\ }\bibfield  {title} {\enquote {\bibinfo {title} {{Baikal-GVD - the Next Generation Neutrino Telescope in Lake Baikal}},}\ }\href {\doibase 10.1088/1742-6596/1263/1/012005} {\bibfield  {journal} {\bibinfo  {journal} {J. Phys. Conf. Ser.}\ }\textbf {\bibinfo {volume} {1263}},\ \bibinfo {pages} {012005} (\bibinfo {year} {2019})}\BibitemShut {NoStop}%
\bibitem [{\citenamefont {Sanguineti}(2022)}]{Sanguineti:2023qfa}%
  \BibitemOpen
  \bibfield  {author} {\bibinfo {author} {\bibfnamefont {Matteo}\ \bibnamefont {Sanguineti}} (\bibinfo {collaboration} {KM3NeT}),\ }\bibfield  {title} {\enquote {\bibinfo {title} {{Status and physics results of the KM3NeT experiment}},}\ }\href {\doibase 10.1393/ncc/i2023-23004-3} {\bibfield  {journal} {\bibinfo  {journal} {Nuovo Cim. C}\ }\textbf {\bibinfo {volume} {46}},\ \bibinfo {pages} {4} (\bibinfo {year} {2022})}\BibitemShut {NoStop}%
\bibitem [{\citenamefont {Bailly}\ \emph {et~al.}(2021)\citenamefont {Bailly} \emph {et~al.}}]{Bailly:2021dxn}%
  \BibitemOpen
  \bibfield  {author} {\bibinfo {author} {\bibfnamefont {Nicolai}\ \bibnamefont {Bailly}} \emph {et~al.},\ }\bibfield  {title} {\enquote {\bibinfo {title} {{Two-year optical site characterization for the Pacific Ocean Neutrino Experiment (P-ONE) in the Cascadia Basin}},}\ }\href {\doibase 10.1140/epjc/s10052-021-09872-5} {\bibfield  {journal} {\bibinfo  {journal} {Eur. Phys. J. C}\ }\textbf {\bibinfo {volume} {81}},\ \bibinfo {pages} {1071} (\bibinfo {year} {2021})},\ \Eprint {http://arxiv.org/abs/2108.04961} {arXiv:2108.04961 [astro-ph.IM]} \BibitemShut {NoStop}%
\bibitem [{\citenamefont {Roberts}(1992)}]{Roberts:1992re}%
  \BibitemOpen
  \bibfield  {author} {\bibinfo {author} {\bibfnamefont {A.}~\bibnamefont {Roberts}},\ }\bibfield  {title} {\enquote {\bibinfo {title} {{The Birth of high-energy neutrino astronomy: A Personal history of the DUMAND project}},}\ }\href {\doibase 10.1103/RevModPhys.64.259} {\bibfield  {journal} {\bibinfo  {journal} {Rev. Mod. Phys.}\ }\textbf {\bibinfo {volume} {64}},\ \bibinfo {pages} {259--312} (\bibinfo {year} {1992})}\BibitemShut {NoStop}%
\bibitem [{\citenamefont {Ageron}\ \emph {et~al.}(2011)\citenamefont {Ageron} \emph {et~al.}}]{ANTARES:2011hfw}%
  \BibitemOpen
  \bibfield  {author} {\bibinfo {author} {\bibfnamefont {M.}~\bibnamefont {Ageron}} \emph {et~al.} (\bibinfo {collaboration} {ANTARES}),\ }\bibfield  {title} {\enquote {\bibinfo {title} {{ANTARES: the first undersea neutrino telescope}},}\ }\href {\doibase 10.1016/j.nima.2011.06.103} {\bibfield  {journal} {\bibinfo  {journal} {Nucl. Instrum. Meth. A}\ }\textbf {\bibinfo {volume} {656}},\ \bibinfo {pages} {11--38} (\bibinfo {year} {2011})},\ \Eprint {http://arxiv.org/abs/1104.1607} {arXiv:1104.1607 [astro-ph.IM]} \BibitemShut {NoStop}%
\bibitem [{\citenamefont {Albert}\ \emph {et~al.}(2020)\citenamefont {Albert} \emph {et~al.}}]{ANTARES:2020srt}%
  \BibitemOpen
  \bibfield  {author} {\bibinfo {author} {\bibfnamefont {A.}~\bibnamefont {Albert}} \emph {et~al.} (\bibinfo {collaboration} {ANTARES, IceCube}),\ }\bibfield  {title} {\enquote {\bibinfo {title} {{ANTARES and IceCube Combined Search for Neutrino Point-like and Extended Sources in the Southern Sky}},}\ }\href {\doibase 10.3847/1538-4357/ab7afb} {\bibfield  {journal} {\bibinfo  {journal} {Astrophys. J.}\ }\textbf {\bibinfo {volume} {892}},\ \bibinfo {pages} {92} (\bibinfo {year} {2020})},\ \Eprint {http://arxiv.org/abs/2001.04412} {arXiv:2001.04412 [astro-ph.HE]} \BibitemShut {NoStop}%
\bibitem [{\citenamefont {Albert}\ \emph {et~al.}(2018{\natexlab{b}})\citenamefont {Albert} \emph {et~al.}}]{ANTARES:2018nyb}%
  \BibitemOpen
  \bibfield  {author} {\bibinfo {author} {\bibfnamefont {A.}~\bibnamefont {Albert}} \emph {et~al.} (\bibinfo {collaboration} {ANTARES, IceCube}),\ }\bibfield  {title} {\enquote {\bibinfo {title} {{Joint Constraints on Galactic Diffuse Neutrino Emission from the ANTARES and IceCube Neutrino Telescopes}},}\ }\href {\doibase 10.3847/2041-8213/aaeecf} {\bibfield  {journal} {\bibinfo  {journal} {Astrophys. J. Lett.}\ }\textbf {\bibinfo {volume} {868}},\ \bibinfo {pages} {L20} (\bibinfo {year} {2018}{\natexlab{b}})},\ \Eprint {http://arxiv.org/abs/1808.03531} {arXiv:1808.03531 [astro-ph.HE]} \BibitemShut {NoStop}%
\bibitem [{\citenamefont {Ye}\ \emph {et~al.}(2023)\citenamefont {Ye} \emph {et~al.}}]{Ye:2023dch}%
  \BibitemOpen
  \bibfield  {author} {\bibinfo {author} {\bibfnamefont {Z.~P.}\ \bibnamefont {Ye}} \emph {et~al.},\ }\bibfield  {title} {\enquote {\bibinfo {title} {{A multi-cubic-kilometre neutrino telescope in the western Pacific Ocean}},}\ }\href {\doibase 10.1038/s41550-023-02087-6} {\bibfield  {journal} {\bibinfo  {journal} {Nature Astron.}\ }\textbf {\bibinfo {volume} {7}},\ \bibinfo {pages} {1497--1505} (\bibinfo {year} {2023})}\BibitemShut {NoStop}%
\bibitem [{\citenamefont {Huang}\ \emph {et~al.}(2023)\citenamefont {Huang}, \citenamefont {Cao}, \citenamefont {Chen}, \citenamefont {Liu}, \citenamefont {Wang}, \citenamefont {You},\ and\ \citenamefont {Qi}}]{Huang:2023mzt}%
  \BibitemOpen
  \bibfield  {author} {\bibinfo {author} {\bibfnamefont {Tian-Qi}\ \bibnamefont {Huang}}, \bibinfo {author} {\bibfnamefont {Zhen}\ \bibnamefont {Cao}}, \bibinfo {author} {\bibfnamefont {Mingjun}\ \bibnamefont {Chen}}, \bibinfo {author} {\bibfnamefont {Jiali}\ \bibnamefont {Liu}}, \bibinfo {author} {\bibfnamefont {Zike}\ \bibnamefont {Wang}}, \bibinfo {author} {\bibfnamefont {Xiaohao}\ \bibnamefont {You}}, \ and\ \bibinfo {author} {\bibfnamefont {Ying}\ \bibnamefont {Qi}},\ }\bibfield  {title} {\enquote {\bibinfo {title} {{Proposal for the High Energy Neutrino Telescope}},}\ }\href {\doibase 10.22323/1.444.1080} {\bibfield  {journal} {\bibinfo  {journal} {PoS}\ }\textbf {\bibinfo {volume} {ICRC2023}},\ \bibinfo {pages} {1080} (\bibinfo {year} {2023})}\BibitemShut {NoStop}%
\bibitem [{\citenamefont {et~al}(2022)}]{ACKERMANN202255}%
  \BibitemOpen
  \bibfield  {author} {\bibinfo {author} {\bibfnamefont {Markus~Ackermann}\ \bibnamefont {et~al}},\ }\bibfield  {title} {\enquote {\bibinfo {title} {High-energy and ultra-high-energy neutrinos: A snowmass white paper},}\ }\href {\doibase https://doi.org/10.1016/j.jheap.2022.08.001} {\bibfield  {journal} {\bibinfo  {journal} {Journal of High Energy Astrophysics}\ }\textbf {\bibinfo {volume} {36}},\ \bibinfo {pages} {55--110} (\bibinfo {year} {2022})}\BibitemShut {NoStop}%
\bibitem [{\citenamefont {Otte}\ \emph {et~al.}(2023)\citenamefont {Otte} \emph {et~al.}}]{Otte:2023osf}%
  \BibitemOpen
  \bibfield  {author} {\bibinfo {author} {\bibfnamefont {A.~Nepomuk}\ \bibnamefont {Otte}} \emph {et~al.},\ }\bibfield  {title} {\enquote {\bibinfo {title} {{Trinity: The PeV Neutrino Observatory}},}\ }\href {\doibase 10.22323/1.444.1170} {\bibfield  {journal} {\bibinfo  {journal} {PoS}\ }\textbf {\bibinfo {volume} {ICRC2023}},\ \bibinfo {pages} {1170} (\bibinfo {year} {2023})}\BibitemShut {NoStop}%
\bibitem [{\citenamefont {Thompson}\ \emph {et~al.}(2023)\citenamefont {Thompson} \emph {et~al.}}]{TAMBO:2023plw}%
  \BibitemOpen
  \bibfield  {author} {\bibinfo {author} {\bibfnamefont {Will}\ \bibnamefont {Thompson}} \emph {et~al.} (\bibinfo {collaboration} {TAMBO}),\ }\bibfield  {title} {\enquote {\bibinfo {title} {{TAMBO: Searching for Tau Neutrinos in the Peruvian Andes}},}\ }\href {\doibase 10.22323/1.444.1109} {\bibfield  {journal} {\bibinfo  {journal} {PoS}\ }\textbf {\bibinfo {volume} {ICRC2023}},\ \bibinfo {pages} {1109} (\bibinfo {year} {2023})}\BibitemShut {NoStop}%
\bibitem [{\citenamefont {Aasi}\ \emph {et~al.}(2015)\citenamefont {Aasi} \emph {et~al.}}]{LIGOScientific:2014pky}%
  \BibitemOpen
  \bibfield  {author} {\bibinfo {author} {\bibfnamefont {J.}~\bibnamefont {Aasi}} \emph {et~al.} (\bibinfo {collaboration} {LIGO Scientific}),\ }\bibfield  {title} {\enquote {\bibinfo {title} {{Advanced LIGO}},}\ }\href {\doibase 10.1088/0264-9381/32/7/074001} {\bibfield  {journal} {\bibinfo  {journal} {Class. Quant. Grav.}\ }\textbf {\bibinfo {volume} {32}},\ \bibinfo {pages} {074001} (\bibinfo {year} {2015})},\ \Eprint {http://arxiv.org/abs/1411.4547} {arXiv:1411.4547 [gr-qc]} \BibitemShut {NoStop}%
\bibitem [{\citenamefont {Acernese}\ \emph {et~al.}(2015)\citenamefont {Acernese} \emph {et~al.}}]{VIRGO:2014yos}%
  \BibitemOpen
  \bibfield  {author} {\bibinfo {author} {\bibfnamefont {F.}~\bibnamefont {Acernese}} \emph {et~al.} (\bibinfo {collaboration} {VIRGO}),\ }\bibfield  {title} {\enquote {\bibinfo {title} {{Advanced Virgo: a second-generation interferometric gravitational wave detector}},}\ }\href {\doibase 10.1088/0264-9381/32/2/024001} {\bibfield  {journal} {\bibinfo  {journal} {Class. Quant. Grav.}\ }\textbf {\bibinfo {volume} {32}},\ \bibinfo {pages} {024001} (\bibinfo {year} {2015})},\ \Eprint {http://arxiv.org/abs/1408.3978} {arXiv:1408.3978 [gr-qc]} \BibitemShut {NoStop}%
\bibitem [{\citenamefont {Akutsu}\ \emph {et~al.}(2021)\citenamefont {Akutsu} \emph {et~al.}}]{KAGRA:2020tym}%
  \BibitemOpen
  \bibfield  {author} {\bibinfo {author} {\bibfnamefont {T.}~\bibnamefont {Akutsu}} \emph {et~al.} (\bibinfo {collaboration} {KAGRA}),\ }\bibfield  {title} {\enquote {\bibinfo {title} {{Overview of KAGRA: Detector design and construction history}},}\ }\href {\doibase 10.1093/ptep/ptaa125} {\bibfield  {journal} {\bibinfo  {journal} {PTEP}\ }\textbf {\bibinfo {volume} {2021}},\ \bibinfo {pages} {05A101} (\bibinfo {year} {2021})},\ \Eprint {http://arxiv.org/abs/2005.05574} {arXiv:2005.05574 [physics.ins-det]} \BibitemShut {NoStop}%
\bibitem [{\citenamefont {Abbott}\ \emph {et~al.}(2016)\citenamefont {Abbott} \emph {et~al.}}]{KAGRA:2013rdx}%
  \BibitemOpen
  \bibfield  {author} {\bibinfo {author} {\bibfnamefont {B.~P.}\ \bibnamefont {Abbott}} \emph {et~al.} (\bibinfo {collaboration} {KAGRA, LIGO Scientific, Virgo}),\ }\bibfield  {title} {\enquote {\bibinfo {title} {{Prospects for observing and localizing gravitational-wave transients with Advanced LIGO, Advanced Virgo and KAGRA}},}\ }\href {\doibase 10.1007/s41114-020-00026-9} {\bibfield  {journal} {\bibinfo  {journal} {Living Rev. Rel.}\ }\textbf {\bibinfo {volume} {19}},\ \bibinfo {pages} {1} (\bibinfo {year} {2016})},\ \Eprint {http://arxiv.org/abs/1304.0670} {arXiv:1304.0670 [gr-qc]} \BibitemShut {NoStop}%
\bibitem [{\citenamefont {Abbasi}\ \emph {et~al.}(2023{\natexlab{b}})\citenamefont {Abbasi} \emph {et~al.}}]{IceCube:2022mma}%
  \BibitemOpen
  \bibfield  {author} {\bibinfo {author} {\bibfnamefont {R.}~\bibnamefont {Abbasi}} \emph {et~al.} (\bibinfo {collaboration} {IceCube}),\ }\bibfield  {title} {\enquote {\bibinfo {title} {{IceCube Search for Neutrinos Coincident with Gravitational Wave Events from LIGO/Virgo Run O3}},}\ }\href {\doibase 10.3847/1538-4357/aca5fc} {\bibfield  {journal} {\bibinfo  {journal} {Astrophys. J.}\ }\textbf {\bibinfo {volume} {944}},\ \bibinfo {pages} {80} (\bibinfo {year} {2023}{\natexlab{b}})},\ \Eprint {http://arxiv.org/abs/2208.09532} {arXiv:2208.09532 [astro-ph.HE]} \BibitemShut {NoStop}%
\bibitem [{\citenamefont {\'Alvarez-Mu\~niz}\ \emph {et~al.}(2020)\citenamefont {\'Alvarez-Mu\~niz} \emph {et~al.}}]{GRAND:2018iaj}%
  \BibitemOpen
  \bibfield  {author} {\bibinfo {author} {\bibfnamefont {Jaime}\ \bibnamefont {\'Alvarez-Mu\~niz}} \emph {et~al.} (\bibinfo {collaboration} {GRAND}),\ }\bibfield  {title} {\enquote {\bibinfo {title} {{The Giant Radio Array for Neutrino Detection (GRAND): Science and Design}},}\ }\href {\doibase 10.1007/s11433-018-9385-7} {\bibfield  {journal} {\bibinfo  {journal} {Sci. China Phys. Mech. Astron.}\ }\textbf {\bibinfo {volume} {63}},\ \bibinfo {pages} {219501} (\bibinfo {year} {2020})},\ \Eprint {http://arxiv.org/abs/1810.09994} {arXiv:1810.09994 [astro-ph.HE]} \BibitemShut {NoStop}%
\bibitem [{\citenamefont {Olinto}\ \emph {et~al.}(2021)\citenamefont {Olinto} \emph {et~al.}}]{POEMMA:2020ykm}%
  \BibitemOpen
  \bibfield  {author} {\bibinfo {author} {\bibfnamefont {A.~V.}\ \bibnamefont {Olinto}} \emph {et~al.} (\bibinfo {collaboration} {POEMMA}),\ }\bibfield  {title} {\enquote {\bibinfo {title} {{The POEMMA (Probe of Extreme Multi-Messenger Astrophysics) observatory}},}\ }\href {\doibase 10.1088/1475-7516/2021/06/007} {\bibfield  {journal} {\bibinfo  {journal} {JCAP}\ }\textbf {\bibinfo {volume} {06}},\ \bibinfo {pages} {007} (\bibinfo {year} {2021})},\ \Eprint {http://arxiv.org/abs/2012.07945} {arXiv:2012.07945 [astro-ph.IM]} \BibitemShut {NoStop}%
\bibitem [{\citenamefont {Abarr}\ \emph {et~al.}(2021)\citenamefont {Abarr} \emph {et~al.}}]{PUEO:2020bnn}%
  \BibitemOpen
  \bibfield  {author} {\bibinfo {author} {\bibfnamefont {Q.}~\bibnamefont {Abarr}} \emph {et~al.} (\bibinfo {collaboration} {PUEO}),\ }\bibfield  {title} {\enquote {\bibinfo {title} {{The Payload for Ultrahigh Energy Observations (PUEO): a white paper}},}\ }\href {\doibase 10.1088/1748-0221/16/08/P08035} {\bibfield  {journal} {\bibinfo  {journal} {JINST}\ }\textbf {\bibinfo {volume} {16}},\ \bibinfo {pages} {P08035} (\bibinfo {year} {2021})},\ \Eprint {http://arxiv.org/abs/2010.02892} {arXiv:2010.02892 [astro-ph.IM]} \BibitemShut {NoStop}%
\bibitem [{\citenamefont {Aguilar}\ \emph {et~al.}(2021{\natexlab{a}})\citenamefont {Aguilar} \emph {et~al.}}]{RNO-G:2020rmc}%
  \BibitemOpen
  \bibfield  {author} {\bibinfo {author} {\bibfnamefont {J.~A.}\ \bibnamefont {Aguilar}} \emph {et~al.} (\bibinfo {collaboration} {RNO-G}),\ }\bibfield  {title} {\enquote {\bibinfo {title} {{Design and Sensitivity of the Radio Neutrino Observatory in Greenland (RNO-G)}},}\ }\href {\doibase 10.1088/1748-0221/16/03/P03025} {\bibfield  {journal} {\bibinfo  {journal} {JINST}\ }\textbf {\bibinfo {volume} {16}},\ \bibinfo {pages} {P03025} (\bibinfo {year} {2021}{\natexlab{a}})},\ \bibinfo {note} {[Erratum: JINST 18, E03001 (2023)]},\ \Eprint {http://arxiv.org/abs/2010.12279} {arXiv:2010.12279 [astro-ph.IM]} \BibitemShut {NoStop}%
\bibitem [{\citenamefont {Prohira}\ \emph {et~al.}(2020)\citenamefont {Prohira} \emph {et~al.}}]{Prohira:2019glh}%
  \BibitemOpen
  \bibfield  {author} {\bibinfo {author} {\bibfnamefont {S.}~\bibnamefont {Prohira}} \emph {et~al.},\ }\bibfield  {title} {\enquote {\bibinfo {title} {{Observation of Radar Echoes From High-Energy Particle Cascades}},}\ }\href {\doibase 10.1103/PhysRevLett.124.091101} {\bibfield  {journal} {\bibinfo  {journal} {Phys. Rev. Lett.}\ }\textbf {\bibinfo {volume} {124}},\ \bibinfo {pages} {091101} (\bibinfo {year} {2020})},\ \Eprint {http://arxiv.org/abs/1910.12830} {arXiv:1910.12830 [astro-ph.HE]} \BibitemShut {NoStop}%
\bibitem [{\citenamefont {Allison}\ \emph {et~al.}(2019)\citenamefont {Allison} \emph {et~al.}}]{Allison:2018ynt}%
  \BibitemOpen
  \bibfield  {author} {\bibinfo {author} {\bibfnamefont {P.}~\bibnamefont {Allison}} \emph {et~al.},\ }\bibfield  {title} {\enquote {\bibinfo {title} {{Design and performance of an interferometric trigger array for radio detection of high-energy neutrinos}},}\ }\href {\doibase 10.1016/j.nima.2019.01.067} {\bibfield  {journal} {\bibinfo  {journal} {Nucl. Instrum. Meth. A}\ }\textbf {\bibinfo {volume} {930}},\ \bibinfo {pages} {112--125} (\bibinfo {year} {2019})},\ \Eprint {http://arxiv.org/abs/1809.04573} {arXiv:1809.04573 [astro-ph.IM]} \BibitemShut {NoStop}%
\bibitem [{\citenamefont {Adams}\ \emph {et~al.}(2017)\citenamefont {Adams} \emph {et~al.}}]{Adams:2017fjh}%
  \BibitemOpen
  \bibfield  {author} {\bibinfo {author} {\bibfnamefont {James~H.}\ \bibnamefont {Adams}} \emph {et~al.},\ }\bibfield  {title} {\enquote {\bibinfo {title} {{White paper on EUSO-SPB2}},}\ }\href@noop {} {\  (\bibinfo {year} {2017})},\ \Eprint {http://arxiv.org/abs/1703.04513} {arXiv:1703.04513 [astro-ph.HE]} \BibitemShut {NoStop}%
\bibitem [{\citenamefont {Huege}\ and\ \citenamefont {Besson}(2017)}]{Huege:2017khw}%
  \BibitemOpen
  \bibfield  {author} {\bibinfo {author} {\bibfnamefont {Tim}\ \bibnamefont {Huege}}\ and\ \bibinfo {author} {\bibfnamefont {Dave}\ \bibnamefont {Besson}},\ }\bibfield  {title} {\enquote {\bibinfo {title} {{Radio-wave detection of ultra-high-energy neutrinos and cosmic rays}},}\ }\href {\doibase 10.1093/ptep/ptx009} {\bibfield  {journal} {\bibinfo  {journal} {PTEP}\ }\textbf {\bibinfo {volume} {2017}},\ \bibinfo {pages} {12A106} (\bibinfo {year} {2017})},\ \Eprint {http://arxiv.org/abs/1701.02987} {arXiv:1701.02987 [astro-ph.IM]} \BibitemShut {NoStop}%
\bibitem [{\citenamefont {Ahlers}\ and\ \citenamefont {Halzen}(2015)}]{Ahlers:2015lln}%
  \BibitemOpen
  \bibfield  {author} {\bibinfo {author} {\bibfnamefont {Markus}\ \bibnamefont {Ahlers}}\ and\ \bibinfo {author} {\bibfnamefont {Francis}\ \bibnamefont {Halzen}},\ }\bibfield  {title} {\enquote {\bibinfo {title} {{High-energy cosmic neutrino puzzle: a review}},}\ }\href {\doibase 10.1088/0034-4885/78/12/126901} {\bibfield  {journal} {\bibinfo  {journal} {Rept. Prog. Phys.}\ }\textbf {\bibinfo {volume} {78}},\ \bibinfo {pages} {126901} (\bibinfo {year} {2015})}\BibitemShut {NoStop}%
\bibitem [{\citenamefont {Berezinsky}\ and\ \citenamefont {Kalashev}(2016)}]{Berezinsky:2016feh}%
  \BibitemOpen
  \bibfield  {author} {\bibinfo {author} {\bibfnamefont {V.}~\bibnamefont {Berezinsky}}\ and\ \bibinfo {author} {\bibfnamefont {O.}~\bibnamefont {Kalashev}},\ }\bibfield  {title} {\enquote {\bibinfo {title} {{High energy electromagnetic cascades in extragalactic space: physics and features}},}\ }\href {\doibase 10.1103/PhysRevD.94.023007} {\bibfield  {journal} {\bibinfo  {journal} {Phys. Rev. D}\ }\textbf {\bibinfo {volume} {94}},\ \bibinfo {pages} {023007} (\bibinfo {year} {2016})},\ \Eprint {http://arxiv.org/abs/1603.03989} {arXiv:1603.03989 [astro-ph.HE]} \BibitemShut {NoStop}%
\bibitem [{\citenamefont {Aartsen}\ \emph {et~al.}(2015{\natexlab{a}})\citenamefont {Aartsen} \emph {et~al.}}]{IceCube:2015qii}%
  \BibitemOpen
  \bibfield  {author} {\bibinfo {author} {\bibfnamefont {M.~G.}\ \bibnamefont {Aartsen}} \emph {et~al.} (\bibinfo {collaboration} {IceCube}),\ }\bibfield  {title} {\enquote {\bibinfo {title} {{Evidence for Astrophysical Muon Neutrinos from the Northern Sky with IceCube}},}\ }\href {\doibase 10.1103/PhysRevLett.115.081102} {\bibfield  {journal} {\bibinfo  {journal} {Phys. Rev. Lett.}\ }\textbf {\bibinfo {volume} {115}},\ \bibinfo {pages} {081102} (\bibinfo {year} {2015}{\natexlab{a}})},\ \Eprint {http://arxiv.org/abs/1507.04005} {arXiv:1507.04005 [astro-ph.HE]} \BibitemShut {NoStop}%
\bibitem [{\citenamefont {Abbasi}\ \emph {et~al.}(2022{\natexlab{b}})\citenamefont {Abbasi} \emph {et~al.}}]{Abbasi:2021qfz}%
  \BibitemOpen
  \bibfield  {author} {\bibinfo {author} {\bibfnamefont {R.}~\bibnamefont {Abbasi}} \emph {et~al.},\ }\bibfield  {title} {\enquote {\bibinfo {title} {{Improved Characterization of the Astrophysical Muon\textendash{}neutrino Flux with 9.5 Years of IceCube Data}},}\ }\href {\doibase 10.3847/1538-4357/ac4d29} {\bibfield  {journal} {\bibinfo  {journal} {Astrophys. J.}\ }\textbf {\bibinfo {volume} {928}},\ \bibinfo {pages} {50} (\bibinfo {year} {2022}{\natexlab{b}})},\ \Eprint {http://arxiv.org/abs/2111.10299} {arXiv:2111.10299 [astro-ph.HE]} \BibitemShut {NoStop}%
\bibitem [{\citenamefont {Albert}\ \emph {et~al.}(2024)\citenamefont {Albert} \emph {et~al.}}]{ANTARES:2024ihw}%
  \BibitemOpen
  \bibfield  {author} {\bibinfo {author} {\bibfnamefont {A.}~\bibnamefont {Albert}} \emph {et~al.} (\bibinfo {collaboration} {ANTARES}),\ }\bibfield  {title} {\enquote {\bibinfo {title} {{Constraints on the energy spectrum of the diffuse cosmic neutrino flux from the ANTARES neutrino telescope}},}\ }\href {\doibase 10.1088/1475-7516/2024/08/038} {\bibfield  {journal} {\bibinfo  {journal} {JCAP}\ }\textbf {\bibinfo {volume} {08}},\ \bibinfo {pages} {038} (\bibinfo {year} {2024})},\ \Eprint {http://arxiv.org/abs/2407.00328} {arXiv:2407.00328 [astro-ph.HE]} \BibitemShut {NoStop}%
\bibitem [{\citenamefont {Allakhverdyan}\ \emph {et~al.}(2023)\citenamefont {Allakhverdyan} \emph {et~al.}}]{Baikal-GVD:2022fis}%
  \BibitemOpen
  \bibfield  {author} {\bibinfo {author} {\bibfnamefont {V.~A.}\ \bibnamefont {Allakhverdyan}} \emph {et~al.} (\bibinfo {collaboration} {Baikal-GVD}),\ }\bibfield  {title} {\enquote {\bibinfo {title} {{Diffuse neutrino flux measurements with the Baikal-GVD neutrino telescope}},}\ }\href {\doibase 10.1103/PhysRevD.107.042005} {\bibfield  {journal} {\bibinfo  {journal} {Phys. Rev. D}\ }\textbf {\bibinfo {volume} {107}},\ \bibinfo {pages} {042005} (\bibinfo {year} {2023})},\ \Eprint {http://arxiv.org/abs/2211.09447} {arXiv:2211.09447 [astro-ph.HE]} \BibitemShut {NoStop}%
\bibitem [{\citenamefont {Aiello}\ \emph {et~al.}(2024)\citenamefont {Aiello} \emph {et~al.}}]{KM3NeT:2024uhg}%
  \BibitemOpen
  \bibfield  {author} {\bibinfo {author} {\bibfnamefont {S.}~\bibnamefont {Aiello}} \emph {et~al.} (\bibinfo {collaboration} {KM3NeT}),\ }\bibfield  {title} {\enquote {\bibinfo {title} {{Differential Sensitivity of the KM3NeT/ARCA detector to a diffuse neutrino flux and to point-like source emission: Exploring the case of the Starburst Galaxies}},}\ }\href {\doibase 10.1016/j.astropartphys.2024.102990} {\bibfield  {journal} {\bibinfo  {journal} {Astropart. Phys.}\ }\textbf {\bibinfo {volume} {162}},\ \bibinfo {pages} {102990} (\bibinfo {year} {2024})},\ \Eprint {http://arxiv.org/abs/2402.09088} {arXiv:2402.09088 [astro-ph.HE]} \BibitemShut {NoStop}%
\bibitem [{\citenamefont {Aartsen}\ \emph {et~al.}(2017{\natexlab{b}})\citenamefont {Aartsen} \emph {et~al.}}]{Aartsen:2017mau}%
  \BibitemOpen
  \bibfield  {author} {\bibinfo {author} {\bibfnamefont {M.~G.}\ \bibnamefont {Aartsen}} \emph {et~al.} (\bibinfo {collaboration} {IceCube}),\ }\bibfield  {title} {\enquote {\bibinfo {title} {{The IceCube Neutrino Observatory - Contributions to ICRC 2017 Part II: Properties of the Atmospheric and Astrophysical Neutrino Flux}},}\ }\href@noop {} {\  (\bibinfo {year} {2017}{\natexlab{b}})},\ \Eprint {http://arxiv.org/abs/1710.01191} {arXiv:1710.01191 [astro-ph.HE]} \BibitemShut {NoStop}%
\bibitem [{\citenamefont {Aartsen}\ \emph {et~al.}(2020)\citenamefont {Aartsen} \emph {et~al.}}]{IceCube:2020acn}%
  \BibitemOpen
  \bibfield  {author} {\bibinfo {author} {\bibfnamefont {M.~G.}\ \bibnamefont {Aartsen}} \emph {et~al.} (\bibinfo {collaboration} {IceCube}),\ }\bibfield  {title} {\enquote {\bibinfo {title} {{Characteristics of the diffuse astrophysical electron and tau neutrino flux with six years of IceCube high energy cascade data}},}\ }\href {\doibase 10.1103/PhysRevLett.125.121104} {\bibfield  {journal} {\bibinfo  {journal} {Phys. Rev. Lett.}\ }\textbf {\bibinfo {volume} {125}},\ \bibinfo {pages} {121104} (\bibinfo {year} {2020})},\ \Eprint {http://arxiv.org/abs/2001.09520} {arXiv:2001.09520 [astro-ph.HE]} \BibitemShut {NoStop}%
\bibitem [{\citenamefont {Abbasi}\ \emph {et~al.}(2022{\natexlab{c}})\citenamefont {Abbasi} \emph {et~al.}}]{IceCube:2020fpi}%
  \BibitemOpen
  \bibfield  {author} {\bibinfo {author} {\bibfnamefont {R.}~\bibnamefont {Abbasi}} \emph {et~al.} (\bibinfo {collaboration} {IceCube}),\ }\bibfield  {title} {\enquote {\bibinfo {title} {{Detection of astrophysical tau neutrino candidates in IceCube}},}\ }\href {\doibase 10.1140/epjc/s10052-022-10795-y} {\bibfield  {journal} {\bibinfo  {journal} {Eur. Phys. J. C}\ }\textbf {\bibinfo {volume} {82}},\ \bibinfo {pages} {1031} (\bibinfo {year} {2022}{\natexlab{c}})},\ \Eprint {http://arxiv.org/abs/2011.03561} {arXiv:2011.03561 [hep-ex]} \BibitemShut {NoStop}%
\bibitem [{\citenamefont {Schonert}\ \emph {et~al.}(2009)\citenamefont {Schonert}, \citenamefont {Gaisser}, \citenamefont {Resconi},\ and\ \citenamefont {Schulz}}]{Schonert:2008is}%
  \BibitemOpen
  \bibfield  {author} {\bibinfo {author} {\bibfnamefont {Stefan}\ \bibnamefont {Schonert}}, \bibinfo {author} {\bibfnamefont {Thomas~K.}\ \bibnamefont {Gaisser}}, \bibinfo {author} {\bibfnamefont {Elisa}\ \bibnamefont {Resconi}}, \ and\ \bibinfo {author} {\bibfnamefont {Olaf}\ \bibnamefont {Schulz}},\ }\bibfield  {title} {\enquote {\bibinfo {title} {{Vetoing atmospheric neutrinos in a high energy neutrino telescope}},}\ }\href {\doibase 10.1103/PhysRevD.79.043009} {\bibfield  {journal} {\bibinfo  {journal} {Phys. Rev. D}\ }\textbf {\bibinfo {volume} {79}},\ \bibinfo {pages} {043009} (\bibinfo {year} {2009})},\ \Eprint {http://arxiv.org/abs/0812.4308} {arXiv:0812.4308 [astro-ph]} \BibitemShut {NoStop}%
\bibitem [{\citenamefont {Gaisser}\ \emph {et~al.}(2014)\citenamefont {Gaisser}, \citenamefont {Jero}, \citenamefont {Karle},\ and\ \citenamefont {van Santen}}]{Gaisser:2014bja}%
  \BibitemOpen
  \bibfield  {author} {\bibinfo {author} {\bibfnamefont {Thomas~K.}\ \bibnamefont {Gaisser}}, \bibinfo {author} {\bibfnamefont {Kyle}\ \bibnamefont {Jero}}, \bibinfo {author} {\bibfnamefont {Albrecht}\ \bibnamefont {Karle}}, \ and\ \bibinfo {author} {\bibfnamefont {Jakob}\ \bibnamefont {van Santen}},\ }\bibfield  {title} {\enquote {\bibinfo {title} {{Generalized self-veto probability for atmospheric neutrinos}},}\ }\href {\doibase 10.1103/PhysRevD.90.023009} {\bibfield  {journal} {\bibinfo  {journal} {Phys. Rev. D}\ }\textbf {\bibinfo {volume} {90}},\ \bibinfo {pages} {023009} (\bibinfo {year} {2014})},\ \Eprint {http://arxiv.org/abs/1405.0525} {arXiv:1405.0525 [astro-ph.HE]} \BibitemShut {NoStop}%
\bibitem [{\citenamefont {Arg\"uelles}\ \emph {et~al.}(2018)\citenamefont {Arg\"uelles}, \citenamefont {Palomares-Ruiz}, \citenamefont {Schneider}, \citenamefont {Wille},\ and\ \citenamefont {Yuan}}]{Arguelles:2018awr}%
  \BibitemOpen
  \bibfield  {author} {\bibinfo {author} {\bibfnamefont {Carlos~A.}\ \bibnamefont {Arg\"uelles}}, \bibinfo {author} {\bibfnamefont {Sergio}\ \bibnamefont {Palomares-Ruiz}}, \bibinfo {author} {\bibfnamefont {Austin}\ \bibnamefont {Schneider}}, \bibinfo {author} {\bibfnamefont {Logan}\ \bibnamefont {Wille}}, \ and\ \bibinfo {author} {\bibfnamefont {Tianlu}\ \bibnamefont {Yuan}},\ }\bibfield  {title} {\enquote {\bibinfo {title} {{Unified atmospheric neutrino passing fractions for large-scale neutrino telescopes}},}\ }\href {\doibase 10.1088/1475-7516/2018/07/047} {\bibfield  {journal} {\bibinfo  {journal} {JCAP}\ }\textbf {\bibinfo {volume} {07}},\ \bibinfo {pages} {047} (\bibinfo {year} {2018})},\ \Eprint {http://arxiv.org/abs/1805.11003} {arXiv:1805.11003 [hep-ph]} \BibitemShut {NoStop}%
\bibitem [{\citenamefont {Beacom}\ and\ \citenamefont {Candia}(2004)}]{Beacom:2004jb}%
  \BibitemOpen
  \bibfield  {author} {\bibinfo {author} {\bibfnamefont {John~F.}\ \bibnamefont {Beacom}}\ and\ \bibinfo {author} {\bibfnamefont {Julian}\ \bibnamefont {Candia}},\ }\bibfield  {title} {\enquote {\bibinfo {title} {{Shower power: Isolating the prompt atmospheric neutrino flux using electron neutrinos}},}\ }\href {\doibase 10.1088/1475-7516/2004/11/009} {\bibfield  {journal} {\bibinfo  {journal} {JCAP}\ }\textbf {\bibinfo {volume} {11}},\ \bibinfo {pages} {009} (\bibinfo {year} {2004})},\ \Eprint {http://arxiv.org/abs/hep-ph/0409046} {arXiv:hep-ph/0409046} \BibitemShut {NoStop}%
\bibitem [{\citenamefont {Aartsen}\ \emph {et~al.}(2013{\natexlab{b}})\citenamefont {Aartsen} \emph {et~al.}}]{IceCube:2013low}%
  \BibitemOpen
  \bibfield  {author} {\bibinfo {author} {\bibfnamefont {M.~G.}\ \bibnamefont {Aartsen}} \emph {et~al.} (\bibinfo {collaboration} {IceCube}),\ }\bibfield  {title} {\enquote {\bibinfo {title} {{Evidence for High-Energy Extraterrestrial Neutrinos at the IceCube Detector}},}\ }\href {\doibase 10.1126/science.1242856} {\bibfield  {journal} {\bibinfo  {journal} {Science}\ }\textbf {\bibinfo {volume} {342}},\ \bibinfo {pages} {1242856} (\bibinfo {year} {2013}{\natexlab{b}})},\ \Eprint {http://arxiv.org/abs/1311.5238} {arXiv:1311.5238 [astro-ph.HE]} \BibitemShut {NoStop}%
\bibitem [{\citenamefont {Abbasi}\ \emph {et~al.}(2021{\natexlab{a}})\citenamefont {Abbasi} \emph {et~al.}}]{IceCube:2020wum}%
  \BibitemOpen
  \bibfield  {author} {\bibinfo {author} {\bibfnamefont {R.}~\bibnamefont {Abbasi}} \emph {et~al.} (\bibinfo {collaboration} {IceCube}),\ }\bibfield  {title} {\enquote {\bibinfo {title} {{The IceCube high-energy starting event sample: Description and flux characterization with 7.5 years of data}},}\ }\href {\doibase 10.1103/PhysRevD.104.022002} {\bibfield  {journal} {\bibinfo  {journal} {Phys. Rev. D}\ }\textbf {\bibinfo {volume} {104}},\ \bibinfo {pages} {022002} (\bibinfo {year} {2021}{\natexlab{a}})},\ \Eprint {http://arxiv.org/abs/2011.03545} {arXiv:2011.03545 [astro-ph.HE]} \BibitemShut {NoStop}%
\bibitem [{\citenamefont {Aartsen}\ \emph {et~al.}(2019)\citenamefont {Aartsen} \emph {et~al.}}]{IceCube:2018pgc}%
  \BibitemOpen
  \bibfield  {author} {\bibinfo {author} {\bibfnamefont {M.~G.}\ \bibnamefont {Aartsen}} \emph {et~al.} (\bibinfo {collaboration} {IceCube}),\ }\bibfield  {title} {\enquote {\bibinfo {title} {{Measurements using the inelasticity distribution of multi-TeV neutrino interactions in IceCube}},}\ }\href {\doibase 10.1103/PhysRevD.99.032004} {\bibfield  {journal} {\bibinfo  {journal} {Phys. Rev. D}\ }\textbf {\bibinfo {volume} {99}},\ \bibinfo {pages} {032004} (\bibinfo {year} {2019})},\ \Eprint {http://arxiv.org/abs/1808.07629} {arXiv:1808.07629 [hep-ex]} \BibitemShut {NoStop}%
\bibitem [{\citenamefont {Aartsen}\ \emph {et~al.}(2021{\natexlab{b}})\citenamefont {Aartsen} \emph {et~al.}}]{IceCube:2021rpz}%
  \BibitemOpen
  \bibfield  {author} {\bibinfo {author} {\bibfnamefont {M.~G.}\ \bibnamefont {Aartsen}} \emph {et~al.} (\bibinfo {collaboration} {IceCube}),\ }\bibfield  {title} {\enquote {\bibinfo {title} {{Detection of a particle shower at the Glashow resonance with IceCube}},}\ }\href {\doibase 10.1038/s41586-021-03256-1} {\bibfield  {journal} {\bibinfo  {journal} {Nature}\ }\textbf {\bibinfo {volume} {591}},\ \bibinfo {pages} {220--224} (\bibinfo {year} {2021}{\natexlab{b}})},\ \bibinfo {note} {[Erratum: Nature 592, E11 (2021)]},\ \Eprint {http://arxiv.org/abs/2110.15051} {arXiv:2110.15051 [hep-ex]} \BibitemShut {NoStop}%
\bibitem [{\citenamefont {Abbasi}\ \emph {et~al.}(2022{\natexlab{d}})\citenamefont {Abbasi} \emph {et~al.}}]{Abbasi:2020zmr}%
  \BibitemOpen
  \bibfield  {author} {\bibinfo {author} {\bibfnamefont {R.}~\bibnamefont {Abbasi}} \emph {et~al.} (\bibinfo {collaboration} {IceCube}),\ }\bibfield  {title} {\enquote {\bibinfo {title} {{Detection of astrophysical tau neutrino candidates in IceCube}},}\ }\href {\doibase 10.1140/epjc/s10052-022-10795-y} {\bibfield  {journal} {\bibinfo  {journal} {Eur. Phys. J. C}\ }\textbf {\bibinfo {volume} {82}},\ \bibinfo {pages} {1031} (\bibinfo {year} {2022}{\natexlab{d}})},\ \Eprint {http://arxiv.org/abs/2011.03561} {arXiv:2011.03561 [hep-ex]} \BibitemShut {NoStop}%
\bibitem [{\citenamefont {Lu}(2018)}]{Lu:2017nti}%
  \BibitemOpen
  \bibfield  {author} {\bibinfo {author} {\bibfnamefont {Lu}~\bibnamefont {Lu}} (\bibinfo {collaboration} {IceCube}),\ }\bibfield  {title} {\enquote {\bibinfo {title} {{Multi-flavour PeV neutrino search with IceCube}},}\ }\href {\doibase 10.22323/1.301.1002} {\bibfield  {journal} {\bibinfo  {journal} {PoS}\ }\textbf {\bibinfo {volume} {ICRC2017}},\ \bibinfo {pages} {1002} (\bibinfo {year} {2018})}\BibitemShut {NoStop}%
\bibitem [{\citenamefont {Abbasi}\ \emph {et~al.}(2022{\natexlab{e}})\citenamefont {Abbasi} \emph {et~al.}}]{IceCube:2021uhz}%
  \BibitemOpen
  \bibfield  {author} {\bibinfo {author} {\bibfnamefont {R.}~\bibnamefont {Abbasi}} \emph {et~al.} (\bibinfo {collaboration} {IceCube}),\ }\bibfield  {title} {\enquote {\bibinfo {title} {{Improved Characterization of the Astrophysical Muon\textendash{}neutrino Flux with 9.5 Years of IceCube Data}},}\ }\href {\doibase 10.3847/1538-4357/ac4d29} {\bibfield  {journal} {\bibinfo  {journal} {Astrophys. J.}\ }\textbf {\bibinfo {volume} {928}},\ \bibinfo {pages} {50} (\bibinfo {year} {2022}{\natexlab{e}})},\ \Eprint {http://arxiv.org/abs/2111.10299} {arXiv:2111.10299 [astro-ph.HE]} \BibitemShut {NoStop}%
\bibitem [{\citenamefont {Hopkins}\ and\ \citenamefont {Beacom}(2006)}]{Hopkins:2006bw}%
  \BibitemOpen
  \bibfield  {author} {\bibinfo {author} {\bibfnamefont {Andrew~M.}\ \bibnamefont {Hopkins}}\ and\ \bibinfo {author} {\bibfnamefont {John~F.}\ \bibnamefont {Beacom}},\ }\bibfield  {title} {\enquote {\bibinfo {title} {{On the normalisation of the cosmic star formation history}},}\ }\href {\doibase 10.1086/506610} {\bibfield  {journal} {\bibinfo  {journal} {Astrophys. J.}\ }\textbf {\bibinfo {volume} {651}},\ \bibinfo {pages} {142--154} (\bibinfo {year} {2006})},\ \Eprint {http://arxiv.org/abs/astro-ph/0601463} {arXiv:astro-ph/0601463} \BibitemShut {NoStop}%
\bibitem [{\citenamefont {Aartsen}\ \emph {et~al.}(2017{\natexlab{c}})\citenamefont {Aartsen} \emph {et~al.}}]{IceCube:2017zho}%
  \BibitemOpen
  \bibfield  {author} {\bibinfo {author} {\bibfnamefont {M.~G.}\ \bibnamefont {Aartsen}} \emph {et~al.} (\bibinfo {collaboration} {IceCube}),\ }\bibfield  {title} {\enquote {\bibinfo {title} {{The IceCube Neutrino Observatory - Contributions to ICRC 2017 Part II: Properties of the Atmospheric and Astrophysical Neutrino Flux}},}\ }\href@noop {} {\  (\bibinfo {year} {2017}{\natexlab{c}})},\ \Eprint {http://arxiv.org/abs/1710.01191} {arXiv:1710.01191 [astro-ph.HE]} \BibitemShut {NoStop}%
\bibitem [{\citenamefont {Ackermann}\ \emph {et~al.}(2015)\citenamefont {Ackermann} \emph {et~al.}}]{Fermi-LAT:2014ryh}%
  \BibitemOpen
  \bibfield  {author} {\bibinfo {author} {\bibfnamefont {M.}~\bibnamefont {Ackermann}} \emph {et~al.} (\bibinfo {collaboration} {Fermi-LAT}),\ }\bibfield  {title} {\enquote {\bibinfo {title} {{The spectrum of isotropic diffuse gamma-ray emission between 100 MeV and 820 GeV}},}\ }\href {\doibase 10.1088/0004-637X/799/1/86} {\bibfield  {journal} {\bibinfo  {journal} {Astrophys. J.}\ }\textbf {\bibinfo {volume} {799}},\ \bibinfo {pages} {86} (\bibinfo {year} {2015})},\ \Eprint {http://arxiv.org/abs/1410.3696} {arXiv:1410.3696 [astro-ph.HE]} \BibitemShut {NoStop}%
\bibitem [{\citenamefont {{Weidenspointner}}(1999)}]{1999PhDT.......284W}%
  \BibitemOpen
  \bibfield  {author} {\bibinfo {author} {\bibfnamefont {Georg}\ \bibnamefont {{Weidenspointner}}},\ }\emph {\bibinfo {title} {{The origin of the cosmic gamma-ray background in the COMPTEL energy range}}},\ \href@noop {} {Ph.D. thesis},\ \bibinfo  {school} {Munich University of Technology, Germany} (\bibinfo {year} {1999})\BibitemShut {NoStop}%
\bibitem [{\citenamefont {Watanabe}\ \emph {et~al.}(1999)\citenamefont {Watanabe}, \citenamefont {Hartmann}, \citenamefont {Leising},\ and\ \citenamefont {The}}]{Watanabe:1998ds}%
  \BibitemOpen
  \bibfield  {author} {\bibinfo {author} {\bibfnamefont {K.}~\bibnamefont {Watanabe}}, \bibinfo {author} {\bibfnamefont {D.~H.}\ \bibnamefont {Hartmann}}, \bibinfo {author} {\bibfnamefont {M.~D.}\ \bibnamefont {Leising}}, \ and\ \bibinfo {author} {\bibfnamefont {L.~S.}\ \bibnamefont {The}},\ }\bibfield  {title} {\enquote {\bibinfo {title} {{The Diffuse gamma-ray background from supernovae}},}\ }\href {\doibase 10.1086/307110} {\bibfield  {journal} {\bibinfo  {journal} {Astrophys. J.}\ }\textbf {\bibinfo {volume} {516}},\ \bibinfo {pages} {285--296} (\bibinfo {year} {1999})},\ \Eprint {http://arxiv.org/abs/astro-ph/9809197} {arXiv:astro-ph/9809197} \BibitemShut {NoStop}%
\bibitem [{\citenamefont {Strong}\ \emph {et~al.}(2004)\citenamefont {Strong}, \citenamefont {Moskalenko},\ and\ \citenamefont {Reimer}}]{Strong:2004ry}%
  \BibitemOpen
  \bibfield  {author} {\bibinfo {author} {\bibfnamefont {A.~W.}\ \bibnamefont {Strong}}, \bibinfo {author} {\bibfnamefont {I.~V.}\ \bibnamefont {Moskalenko}}, \ and\ \bibinfo {author} {\bibfnamefont {O.}~\bibnamefont {Reimer}},\ }\bibfield  {title} {\enquote {\bibinfo {title} {{A new determination of the extragalactic diffuse gamma-ray background from egret data}},}\ }\href {\doibase 10.1086/423196} {\bibfield  {journal} {\bibinfo  {journal} {Astrophys. J.}\ }\textbf {\bibinfo {volume} {613}},\ \bibinfo {pages} {956--961} (\bibinfo {year} {2004})},\ \Eprint {http://arxiv.org/abs/astro-ph/0405441} {arXiv:astro-ph/0405441} \BibitemShut {NoStop}%
\bibitem [{\citenamefont {Atwood}\ \emph {et~al.}(2009)\citenamefont {Atwood} \emph {et~al.}}]{Fermi-LAT:2009ihh}%
  \BibitemOpen
  \bibfield  {author} {\bibinfo {author} {\bibfnamefont {W.~B.}\ \bibnamefont {Atwood}} \emph {et~al.} (\bibinfo {collaboration} {Fermi-LAT}),\ }\bibfield  {title} {\enquote {\bibinfo {title} {{The Large Area Telescope on the Fermi Gamma-ray Space Telescope Mission}},}\ }\href {\doibase 10.1088/0004-637X/697/2/1071} {\bibfield  {journal} {\bibinfo  {journal} {Astrophys. J.}\ }\textbf {\bibinfo {volume} {697}},\ \bibinfo {pages} {1071--1102} (\bibinfo {year} {2009})},\ \Eprint {http://arxiv.org/abs/0902.1089} {arXiv:0902.1089 [astro-ph.IM]} \BibitemShut {NoStop}%
\bibitem [{\citenamefont {Aartsen}\ \emph {et~al.}(2017{\natexlab{d}})\citenamefont {Aartsen} \emph {et~al.}}]{IceCube:2016qvd}%
  \BibitemOpen
  \bibfield  {author} {\bibinfo {author} {\bibfnamefont {M.~G.}\ \bibnamefont {Aartsen}} \emph {et~al.} (\bibinfo {collaboration} {IceCube}),\ }\bibfield  {title} {\enquote {\bibinfo {title} {{The contribution of Fermi-2LAC blazars to the diffuse TeV-PeV neutrino flux}},}\ }\href {\doibase 10.3847/1538-4357/835/1/45} {\bibfield  {journal} {\bibinfo  {journal} {Astrophys. J.}\ }\textbf {\bibinfo {volume} {835}},\ \bibinfo {pages} {45} (\bibinfo {year} {2017}{\natexlab{d}})},\ \Eprint {http://arxiv.org/abs/1611.03874} {arXiv:1611.03874 [astro-ph.HE]} \BibitemShut {NoStop}%
\bibitem [{\citenamefont {Murase}\ \emph {et~al.}(2013)\citenamefont {Murase}, \citenamefont {Ahlers},\ and\ \citenamefont {Lacki}}]{Murase:2013rfa}%
  \BibitemOpen
  \bibfield  {author} {\bibinfo {author} {\bibfnamefont {Kohta}\ \bibnamefont {Murase}}, \bibinfo {author} {\bibfnamefont {Markus}\ \bibnamefont {Ahlers}}, \ and\ \bibinfo {author} {\bibfnamefont {Brian~C.}\ \bibnamefont {Lacki}},\ }\bibfield  {title} {\enquote {\bibinfo {title} {{Testing the Hadronuclear Origin of PeV Neutrinos Observed with IceCube}},}\ }\href {\doibase 10.1103/PhysRevD.88.121301} {\bibfield  {journal} {\bibinfo  {journal} {Phys. Rev. D}\ }\textbf {\bibinfo {volume} {88}},\ \bibinfo {pages} {121301} (\bibinfo {year} {2013})},\ \Eprint {http://arxiv.org/abs/1306.3417} {arXiv:1306.3417 [astro-ph.HE]} \BibitemShut {NoStop}%
\bibitem [{\citenamefont {Murase}\ \emph {et~al.}(2016)\citenamefont {Murase}, \citenamefont {Guetta},\ and\ \citenamefont {Ahlers}}]{Murase:2015xka}%
  \BibitemOpen
  \bibfield  {author} {\bibinfo {author} {\bibfnamefont {Kohta}\ \bibnamefont {Murase}}, \bibinfo {author} {\bibfnamefont {Dafne}\ \bibnamefont {Guetta}}, \ and\ \bibinfo {author} {\bibfnamefont {Markus}\ \bibnamefont {Ahlers}},\ }\bibfield  {title} {\enquote {\bibinfo {title} {{Hidden Cosmic-Ray Accelerators as an Origin of TeV-PeV Cosmic Neutrinos}},}\ }\href {\doibase 10.1103/PhysRevLett.116.071101} {\bibfield  {journal} {\bibinfo  {journal} {Phys. Rev. Lett.}\ }\textbf {\bibinfo {volume} {116}},\ \bibinfo {pages} {071101} (\bibinfo {year} {2016})},\ \Eprint {http://arxiv.org/abs/1509.00805} {arXiv:1509.00805 [astro-ph.HE]} \BibitemShut {NoStop}%
\bibitem [{\citenamefont {Capanema}\ \emph {et~al.}(2020)\citenamefont {Capanema}, \citenamefont {Esmaili},\ and\ \citenamefont {Murase}}]{Capanema:2020rjj}%
  \BibitemOpen
  \bibfield  {author} {\bibinfo {author} {\bibfnamefont {Antonio}\ \bibnamefont {Capanema}}, \bibinfo {author} {\bibfnamefont {Arman}\ \bibnamefont {Esmaili}}, \ and\ \bibinfo {author} {\bibfnamefont {Kohta}\ \bibnamefont {Murase}},\ }\bibfield  {title} {\enquote {\bibinfo {title} {{New constraints on the origin of medium-energy neutrinos observed by IceCube}},}\ }\href {\doibase 10.1103/PhysRevD.101.103012} {\bibfield  {journal} {\bibinfo  {journal} {Phys. Rev. D}\ }\textbf {\bibinfo {volume} {101}},\ \bibinfo {pages} {103012} (\bibinfo {year} {2020})},\ \Eprint {http://arxiv.org/abs/2002.07192} {arXiv:2002.07192 [hep-ph]} \BibitemShut {NoStop}%
\bibitem [{\citenamefont {Capanema}\ \emph {et~al.}(2021)\citenamefont {Capanema}, \citenamefont {Esmaili},\ and\ \citenamefont {Serpico}}]{Capanema:2020oet}%
  \BibitemOpen
  \bibfield  {author} {\bibinfo {author} {\bibfnamefont {Antonio}\ \bibnamefont {Capanema}}, \bibinfo {author} {\bibfnamefont {Arman}\ \bibnamefont {Esmaili}}, \ and\ \bibinfo {author} {\bibfnamefont {Pasquale~Dario}\ \bibnamefont {Serpico}},\ }\bibfield  {title} {\enquote {\bibinfo {title} {{Where do IceCube neutrinos come from? Hints from the diffuse gamma-ray flux}},}\ }\href {\doibase 10.1088/1475-7516/2021/02/037} {\bibfield  {journal} {\bibinfo  {journal} {JCAP}\ }\textbf {\bibinfo {volume} {02}},\ \bibinfo {pages} {037} (\bibinfo {year} {2021})},\ \Eprint {http://arxiv.org/abs/2007.07911} {arXiv:2007.07911 [hep-ph]} \BibitemShut {NoStop}%
\bibitem [{\citenamefont {Ambrosone}(2024)}]{Ambrosone:2024zrf}%
  \BibitemOpen
  \bibfield  {author} {\bibinfo {author} {\bibfnamefont {Antonio}\ \bibnamefont {Ambrosone}},\ }\bibfield  {title} {\enquote {\bibinfo {title} {{Berezinsky Hidden Sources: An Emergent Tension in the High-Energy Neutrino Sky?}}}\ }\href {\doibase 10.1088/1475-7516/2024/09/075} {\bibfield  {journal} {\bibinfo  {journal} {JCAP}\ }\textbf {\bibinfo {volume} {09}},\ \bibinfo {pages} {075} (\bibinfo {year} {2024})},\ \Eprint {http://arxiv.org/abs/2406.13336} {arXiv:2406.13336 [astro-ph.HE]} \BibitemShut {NoStop}%
\bibitem [{\citenamefont {Svensson}(1987)}]{Svensson:1987nlx}%
  \BibitemOpen
  \bibfield  {author} {\bibinfo {author} {\bibfnamefont {Roland}\ \bibnamefont {Svensson}},\ }\bibfield  {title} {\enquote {\bibinfo {title} {{Non-thermal pair production in compact X-ray sources: first-order Compton cascades in soft radiation fields}},}\ }\href {\doibase 10.1093/mnras/227.2.403} {\bibfield  {journal} {\bibinfo  {journal} {Mon. Not. Roy. Astron. Soc.}\ }\textbf {\bibinfo {volume} {227}},\ \bibinfo {pages} {403--451} (\bibinfo {year} {1987})}\BibitemShut {NoStop}%
\bibitem [{\citenamefont {Moskalenko}\ \emph {et~al.}()\citenamefont {Moskalenko}, \citenamefont {Jóhannesson},\ and\ \citenamefont {Porter}}]{GALPROP}%
  \BibitemOpen
  \bibfield  {author} {\bibinfo {author} {\bibfnamefont {I.}~\bibnamefont {Moskalenko}}, \bibinfo {author} {\bibfnamefont {G.}~\bibnamefont {Jóhannesson}}, \ and\ \bibinfo {author} {\bibfnamefont {T.}~\bibnamefont {Porter}},\ }\href@noop {} {\enquote {\bibinfo {title} {Galprop code for cosmic-ray transport and diffuse emission production},}\ }\bibinfo {howpublished} {\url{https://galprop.stanford.edu/}},\ \bibinfo {note} {2024}\BibitemShut {NoStop}%
\bibitem [{\citenamefont {Halzen}\ \emph {et~al.}(2017)\citenamefont {Halzen}, \citenamefont {Kheirandish},\ and\ \citenamefont {Niro}}]{Halzen:2016seh}%
  \BibitemOpen
  \bibfield  {author} {\bibinfo {author} {\bibfnamefont {Francis}\ \bibnamefont {Halzen}}, \bibinfo {author} {\bibfnamefont {Ali}\ \bibnamefont {Kheirandish}}, \ and\ \bibinfo {author} {\bibfnamefont {Viviana}\ \bibnamefont {Niro}},\ }\bibfield  {title} {\enquote {\bibinfo {title} {{Prospects for Detecting Galactic Sources of Cosmic Neutrinos with IceCube: An Update}},}\ }\href {\doibase 10.1016/j.astropartphys.2016.11.004} {\bibfield  {journal} {\bibinfo  {journal} {Astropart. Phys.}\ }\textbf {\bibinfo {volume} {86}},\ \bibinfo {pages} {46--56} (\bibinfo {year} {2017})},\ \Eprint {http://arxiv.org/abs/1609.03072} {arXiv:1609.03072 [astro-ph.HE]} \BibitemShut {NoStop}%
\bibitem [{\citenamefont {Albert}\ \emph {et~al.}(2017)\citenamefont {Albert} \emph {et~al.}}]{ANTARES:2017nlh}%
  \BibitemOpen
  \bibfield  {author} {\bibinfo {author} {\bibfnamefont {A.}~\bibnamefont {Albert}} \emph {et~al.} (\bibinfo {collaboration} {ANTARES}),\ }\bibfield  {title} {\enquote {\bibinfo {title} {{New constraints on all flavor Galactic diffuse neutrino emission with the ANTARES telescope}},}\ }\href {\doibase 10.1103/PhysRevD.96.062001} {\bibfield  {journal} {\bibinfo  {journal} {Phys. Rev. D}\ }\textbf {\bibinfo {volume} {96}},\ \bibinfo {pages} {062001} (\bibinfo {year} {2017})},\ \Eprint {http://arxiv.org/abs/1705.00497} {arXiv:1705.00497 [astro-ph.HE]} \BibitemShut {NoStop}%
\bibitem [{\citenamefont {Aartsen}\ \emph {et~al.}(2017{\natexlab{e}})\citenamefont {Aartsen} \emph {et~al.}}]{IceCube:2017trr}%
  \BibitemOpen
  \bibfield  {author} {\bibinfo {author} {\bibfnamefont {M.~G.}\ \bibnamefont {Aartsen}} \emph {et~al.} (\bibinfo {collaboration} {IceCube}),\ }\bibfield  {title} {\enquote {\bibinfo {title} {{Constraints on Galactic Neutrino Emission with Seven Years of IceCube Data}},}\ }\href {\doibase 10.3847/1538-4357/aa8dfb} {\bibfield  {journal} {\bibinfo  {journal} {Astrophys. J.}\ }\textbf {\bibinfo {volume} {849}},\ \bibinfo {pages} {67} (\bibinfo {year} {2017}{\natexlab{e}})},\ \Eprint {http://arxiv.org/abs/1707.03416} {arXiv:1707.03416 [astro-ph.HE]} \BibitemShut {NoStop}%
\bibitem [{\citenamefont {Bustamante}(2023)}]{Bustamante_2023}%
  \BibitemOpen
  \bibfield  {author} {\bibinfo {author} {\bibfnamefont {M.}~\bibnamefont {Bustamante}},\ }\bibfield  {title} {\enquote {\bibinfo {title} {The milky way shines in high-energy neutrinos},}\ }\href {\doibase 10.1038/s42254-023-00679-9} {\bibfield  {journal} {\bibinfo  {journal} {Nature Reviews Physics}\ }\textbf {\bibinfo {volume} {6}},\ \bibinfo {pages} {8–10} (\bibinfo {year} {2023})}\BibitemShut {NoStop}%
\bibitem [{\citenamefont {Ambrosone}\ \emph {et~al.}(2024)\citenamefont {Ambrosone}, \citenamefont {Groth}, \citenamefont {Peretti},\ and\ \citenamefont {Ahlers}}]{Ambrosone_2024}%
  \BibitemOpen
  \bibfield  {author} {\bibinfo {author} {\bibfnamefont {Antonio}\ \bibnamefont {Ambrosone}}, \bibinfo {author} {\bibfnamefont {Kathrine~Mørch}\ \bibnamefont {Groth}}, \bibinfo {author} {\bibfnamefont {Enrico}\ \bibnamefont {Peretti}}, \ and\ \bibinfo {author} {\bibfnamefont {Markus}\ \bibnamefont {Ahlers}},\ }\bibfield  {title} {\enquote {\bibinfo {title} {Galactic diffuse neutrino emission from sources beyond the discovery horizon},}\ }\href {\doibase 10.1103/physrevd.109.043007} {\bibfield  {journal} {\bibinfo  {journal} {Physical Review D}\ }\textbf {\bibinfo {volume} {109}} (\bibinfo {year} {2024}),\ 10.1103/physrevd.109.043007}\BibitemShut {NoStop}%
\bibitem [{\citenamefont {Aiello}\ \emph {et~al.}(2019)\citenamefont {Aiello} \emph {et~al.}}]{KM3NeT:2018wnd}%
  \BibitemOpen
  \bibfield  {author} {\bibinfo {author} {\bibfnamefont {S.}~\bibnamefont {Aiello}} \emph {et~al.} (\bibinfo {collaboration} {KM3NeT}),\ }\bibfield  {title} {\enquote {\bibinfo {title} {{Sensitivity of the KM3NeT/ARCA neutrino telescope to point-like neutrino sources}},}\ }\href {\doibase 10.1016/j.astropartphys.2019.04.002} {\bibfield  {journal} {\bibinfo  {journal} {Astropart. Phys.}\ }\textbf {\bibinfo {volume} {111}},\ \bibinfo {pages} {100--110} (\bibinfo {year} {2019})},\ \Eprint {http://arxiv.org/abs/1810.08499} {arXiv:1810.08499 [astro-ph.HE]} \BibitemShut {NoStop}%
\bibitem [{\citenamefont {Ackermann}\ \emph {et~al.}(2012)\citenamefont {Ackermann} \emph {et~al.}}]{fermi-diffusegp}%
  \BibitemOpen
  \bibfield  {author} {\bibinfo {author} {\bibfnamefont {M.}~\bibnamefont {Ackermann}} \emph {et~al.} (\bibinfo {collaboration} {\textit{Fermi}-LAT}),\ }\bibfield  {title} {\enquote {\bibinfo {title} {Fermi-lat observations of the diffuse $\gamma$-ray emission: Implications for cosmic rays and the interstellar medium},}\ }\href {\doibase 10.1088/0004-637x/750/1/3} {\bibfield  {journal} {\bibinfo  {journal} {Astrophys. J.}\ }\textbf {\bibinfo {volume} {750}},\ \bibinfo {pages} {3} (\bibinfo {year} {2012})}\BibitemShut {NoStop}%
\bibitem [{\citenamefont {Gaggero}\ \emph {et~al.}(2015)\citenamefont {Gaggero} \emph {et~al.}}]{kra}%
  \BibitemOpen
  \bibfield  {author} {\bibinfo {author} {\bibfnamefont {Daniele}\ \bibnamefont {Gaggero}} \emph {et~al.},\ }\bibfield  {title} {\enquote {\bibinfo {title} {The gamma-ray and neutrino sky: A consistent picture of fermi -lat, milagro, and icecube results},}\ }\href {\doibase 10.1088/2041-8205/815/2/l25} {\bibfield  {journal} {\bibinfo  {journal} {Astrophys. J.}\ }\textbf {\bibinfo {volume} {815}},\ \bibinfo {pages} {L25} (\bibinfo {year} {2015})}\BibitemShut {NoStop}%
\bibitem [{\citenamefont {Fang}\ \emph {et~al.}(2023)\citenamefont {Fang}, \citenamefont {Rodriguez}, \citenamefont {Halzen},\ and\ \citenamefont {Gallagher}}]{Fang:2023vdg}%
  \BibitemOpen
  \bibfield  {author} {\bibinfo {author} {\bibfnamefont {Ke}~\bibnamefont {Fang}}, \bibinfo {author} {\bibfnamefont {Enrique~Lopez}\ \bibnamefont {Rodriguez}}, \bibinfo {author} {\bibfnamefont {Francis}\ \bibnamefont {Halzen}}, \ and\ \bibinfo {author} {\bibfnamefont {John~S.}\ \bibnamefont {Gallagher}},\ }\bibfield  {title} {\enquote {\bibinfo {title} {{High-energy Neutrinos from the Inner Circumnuclear Region of NGC 1068}},}\ }\href {\doibase 10.3847/1538-4357/acee70} {\bibfield  {journal} {\bibinfo  {journal} {Astrophys. J.}\ }\textbf {\bibinfo {volume} {956}},\ \bibinfo {pages} {8} (\bibinfo {year} {2023})},\ \Eprint {http://arxiv.org/abs/2307.07121} {arXiv:2307.07121 [astro-ph.HE]} \BibitemShut {NoStop}%
\bibitem [{\citenamefont {Ajello}\ \emph {et~al.}(2020)\citenamefont {Ajello} \emph {et~al.}}]{Fermi-LAT:2019pir}%
  \BibitemOpen
  \bibfield  {author} {\bibinfo {author} {\bibfnamefont {M.}~\bibnamefont {Ajello}} \emph {et~al.} (\bibinfo {collaboration} {Fermi-LAT}),\ }\bibfield  {title} {\enquote {\bibinfo {title} {{The Fourth Catalog of Active Galactic Nuclei Detected by the Fermi Large Area Telescope}},}\ }\href {\doibase 10.3847/1538-4357/ab791e} {\bibfield  {journal} {\bibinfo  {journal} {Astrophys. J.}\ }\textbf {\bibinfo {volume} {892}},\ \bibinfo {pages} {105} (\bibinfo {year} {2020})},\ \Eprint {http://arxiv.org/abs/1905.10771} {arXiv:1905.10771 [astro-ph.HE]} \BibitemShut {NoStop}%
\bibitem [{\citenamefont {Abdollahi}\ \emph {et~al.}(2020)\citenamefont {Abdollahi} \emph {et~al.}}]{Fermi-LAT:2019yla}%
  \BibitemOpen
  \bibfield  {author} {\bibinfo {author} {\bibfnamefont {S.}~\bibnamefont {Abdollahi}} \emph {et~al.} (\bibinfo {collaboration} {Fermi-LAT}),\ }\bibfield  {title} {\enquote {\bibinfo {title} {{$Fermi$ Large Area Telescope Fourth Source Catalog}},}\ }\href {\doibase 10.3847/1538-4365/ab6bcb} {\bibfield  {journal} {\bibinfo  {journal} {Astrophys. J. Suppl.}\ }\textbf {\bibinfo {volume} {247}},\ \bibinfo {pages} {33} (\bibinfo {year} {2020})},\ \Eprint {http://arxiv.org/abs/1902.10045} {arXiv:1902.10045 [astro-ph.HE]} \BibitemShut {NoStop}%
\bibitem [{\citenamefont {Acciari}\ \emph {et~al.}(2019)\citenamefont {Acciari} \emph {et~al.}}]{Acciari:2019raw}%
  \BibitemOpen
  \bibfield  {author} {\bibinfo {author} {\bibfnamefont {V.~A.}\ \bibnamefont {Acciari}} \emph {et~al.} (\bibinfo {collaboration} {MAGIC}),\ }\bibfield  {title} {\enquote {\bibinfo {title} {{Constraints on gamma-ray and neutrino emission from NGC 1068 with the MAGIC telescopes}},}\ }\href {\doibase 10.3847/1538-4357/ab3a51} {\bibfield  {journal} {\bibinfo  {journal} {Astrophys. J.}\ }\textbf {\bibinfo {volume} {883}},\ \bibinfo {pages} {135} (\bibinfo {year} {2019})},\ \Eprint {http://arxiv.org/abs/1906.10954} {arXiv:1906.10954 [astro-ph.HE]} \BibitemShut {NoStop}%
\bibitem [{\citenamefont {Inoue}\ \emph {et~al.}(2019)\citenamefont {Inoue}, \citenamefont {Khangulyan}, \citenamefont {Inoue},\ and\ \citenamefont {Doi}}]{Inoue:2019fil}%
  \BibitemOpen
  \bibfield  {author} {\bibinfo {author} {\bibfnamefont {Yoshiyuki}\ \bibnamefont {Inoue}}, \bibinfo {author} {\bibfnamefont {Dmitry}\ \bibnamefont {Khangulyan}}, \bibinfo {author} {\bibfnamefont {Susumu}\ \bibnamefont {Inoue}}, \ and\ \bibinfo {author} {\bibfnamefont {Akihiro}\ \bibnamefont {Doi}},\ }\bibfield  {title} {\enquote {\bibinfo {title} {{On high-energy particles in accretion disk coronae of supermassive black holes: implications for MeV gamma rays and high-energy neutrinos from AGN cores}},}\ }\href {\doibase 10.3847/1538-4357/ab2715} {\  (\bibinfo {year} {2019}),\ 10.3847/1538-4357/ab2715},\ \Eprint {http://arxiv.org/abs/1904.00554} {arXiv:1904.00554 [astro-ph.HE]} \BibitemShut {NoStop}%
\bibitem [{\citenamefont {Inoue}\ \emph {et~al.}(2020)\citenamefont {Inoue}, \citenamefont {Khangulyan},\ and\ \citenamefont {Doi}}]{Inoue:2019yfs}%
  \BibitemOpen
  \bibfield  {author} {\bibinfo {author} {\bibfnamefont {Yoshiyuki}\ \bibnamefont {Inoue}}, \bibinfo {author} {\bibfnamefont {Dmitry}\ \bibnamefont {Khangulyan}}, \ and\ \bibinfo {author} {\bibfnamefont {Akihiro}\ \bibnamefont {Doi}},\ }\bibfield  {title} {\enquote {\bibinfo {title} {{On the Origin of High-energy Neutrinos from NGC 1068: The Role of Nonthermal Coronal Activity}},}\ }\href {\doibase 10.3847/2041-8213/ab7661} {\bibfield  {journal} {\bibinfo  {journal} {Astrophys. J. Lett.}\ }\textbf {\bibinfo {volume} {891}},\ \bibinfo {pages} {L33} (\bibinfo {year} {2020})},\ \Eprint {http://arxiv.org/abs/1909.02239} {arXiv:1909.02239 [astro-ph.HE]} \BibitemShut {NoStop}%
\bibitem [{\citenamefont {Murase}\ \emph {et~al.}(2020)\citenamefont {Murase}, \citenamefont {Kimura},\ and\ \citenamefont {Meszaros}}]{Murase:2019vdl}%
  \BibitemOpen
  \bibfield  {author} {\bibinfo {author} {\bibfnamefont {Kohta}\ \bibnamefont {Murase}}, \bibinfo {author} {\bibfnamefont {Shigeo~S.}\ \bibnamefont {Kimura}}, \ and\ \bibinfo {author} {\bibfnamefont {Peter}\ \bibnamefont {Meszaros}},\ }\bibfield  {title} {\enquote {\bibinfo {title} {{Hidden Cores of Active Galactic Nuclei as the Origin of Medium-Energy Neutrinos: Critical Tests with the MeV Gamma-Ray Connection}},}\ }\href {\doibase 10.1103/PhysRevLett.125.011101} {\bibfield  {journal} {\bibinfo  {journal} {Phys. Rev. Lett.}\ }\textbf {\bibinfo {volume} {125}},\ \bibinfo {pages} {011101} (\bibinfo {year} {2020})},\ \Eprint {http://arxiv.org/abs/1904.04226} {arXiv:1904.04226 [astro-ph.HE]} \BibitemShut {NoStop}%
\bibitem [{\citenamefont {Padovani}\ \emph {et~al.}(2024)\citenamefont {Padovani}, \citenamefont {Resconi}, \citenamefont {Ajello}, \citenamefont {Bellenghi}, \citenamefont {Bianchi}, \citenamefont {Blasi}, \citenamefont {Huang}, \citenamefont {Gabici}, \citenamefont {Rosas}, \citenamefont {Niederhausen} \emph {et~al.}}]{padovani2024supermassive}%
  \BibitemOpen
  \bibfield  {author} {\bibinfo {author} {\bibfnamefont {P}~\bibnamefont {Padovani}}, \bibinfo {author} {\bibfnamefont {E}~\bibnamefont {Resconi}}, \bibinfo {author} {\bibfnamefont {M}~\bibnamefont {Ajello}}, \bibinfo {author} {\bibfnamefont {C}~\bibnamefont {Bellenghi}}, \bibinfo {author} {\bibfnamefont {S}~\bibnamefont {Bianchi}}, \bibinfo {author} {\bibfnamefont {P}~\bibnamefont {Blasi}}, \bibinfo {author} {\bibfnamefont {K-Y}\ \bibnamefont {Huang}}, \bibinfo {author} {\bibfnamefont {S}~\bibnamefont {Gabici}}, \bibinfo {author} {\bibfnamefont {V~G{\'a}mez}\ \bibnamefont {Rosas}}, \bibinfo {author} {\bibfnamefont {H}~\bibnamefont {Niederhausen}},  \emph {et~al.},\ }\bibfield  {title} {\enquote {\bibinfo {title} {Supermassive black holes and very high-energy neutrinos: the case of ngc 1068},}\ }\href@noop {} {\bibfield  {journal} {\bibinfo  {journal} {arXiv preprint arXiv:2405.20146}\ } (\bibinfo {year} {2024})}\BibitemShut {NoStop}%
\bibitem [{\citenamefont {Bauer}\ \emph {et~al.}(2015)\citenamefont {Bauer} \emph {et~al.}}]{Bauer:2014rla}%
  \BibitemOpen
  \bibfield  {author} {\bibinfo {author} {\bibfnamefont {Franz~E.}\ \bibnamefont {Bauer}} \emph {et~al.},\ }\bibfield  {title} {\enquote {\bibinfo {title} {{NuSTAR Spectroscopy of Multi-Component X-ray Reflection from NGC 1068}},}\ }\href {\doibase 10.1088/0004-637X/812/2/116} {\bibfield  {journal} {\bibinfo  {journal} {Astrophys. J.}\ }\textbf {\bibinfo {volume} {812}},\ \bibinfo {pages} {116} (\bibinfo {year} {2015})},\ \Eprint {http://arxiv.org/abs/1411.0670} {arXiv:1411.0670 [astro-ph.HE]} \BibitemShut {NoStop}%
\bibitem [{\citenamefont {Marinucci}\ \emph {et~al.}(2016)\citenamefont {Marinucci} \emph {et~al.}}]{Marinucci:2015fqo}%
  \BibitemOpen
  \bibfield  {author} {\bibinfo {author} {\bibfnamefont {A.}~\bibnamefont {Marinucci}} \emph {et~al.},\ }\bibfield  {title} {\enquote {\bibinfo {title} {{NuSTAR catches the unveiling nucleus of NGC 1068}},}\ }\href {\doibase 10.1093/mnrasl/slv178} {\bibfield  {journal} {\bibinfo  {journal} {Mon. Not. Roy. Astron. Soc.}\ }\textbf {\bibinfo {volume} {456}},\ \bibinfo {pages} {L94--L98} (\bibinfo {year} {2016})},\ \Eprint {http://arxiv.org/abs/1511.03503} {arXiv:1511.03503 [astro-ph.HE]} \BibitemShut {NoStop}%
\bibitem [{\citenamefont {Ricci}\ \emph {et~al.}(2017)\citenamefont {Ricci} \emph {et~al.}}]{Ricci:2017dhj}%
  \BibitemOpen
  \bibfield  {author} {\bibinfo {author} {\bibfnamefont {Claudio}\ \bibnamefont {Ricci}} \emph {et~al.},\ }\bibfield  {title} {\enquote {\bibinfo {title} {{BAT AGN Spectroscopic Survey - V. X-ray properties of the Swift/BAT 70-month AGN catalog}},}\ }\href {\doibase 10.3847/1538-4365/aa96ad} {\bibfield  {journal} {\bibinfo  {journal} {Astrophys. J. Suppl.}\ }\textbf {\bibinfo {volume} {233}},\ \bibinfo {pages} {17} (\bibinfo {year} {2017})},\ \Eprint {http://arxiv.org/abs/1709.03989} {arXiv:1709.03989 [astro-ph.HE]} \BibitemShut {NoStop}%
\bibitem [{\citenamefont {Rosas}\ \emph {et~al.}(2022)\citenamefont {Rosas} \emph {et~al.}}]{Rosas:2021zbx}%
  \BibitemOpen
  \bibfield  {author} {\bibinfo {author} {\bibfnamefont {Violeta~Gamez}\ \bibnamefont {Rosas}} \emph {et~al.},\ }\bibfield  {title} {\enquote {\bibinfo {title} {{Thermal imaging of dust hiding the black hole in NGC 1068}},}\ }\href {\doibase 10.1038/s41586-021-04311-7} {\bibfield  {journal} {\bibinfo  {journal} {Nature}\ }\textbf {\bibinfo {volume} {602}},\ \bibinfo {pages} {403--407} (\bibinfo {year} {2022})},\ \Eprint {http://arxiv.org/abs/2112.13694} {arXiv:2112.13694 [astro-ph.GA]} \BibitemShut {NoStop}%
\bibitem [{\citenamefont {et~al.}(2016)}]{García-Burillo_2016}%
  \BibitemOpen
  \bibfield  {author} {\bibinfo {author} {\bibfnamefont {S.~García-Burillo}\ \bibnamefont {et~al.}},\ }\bibfield  {title} {\enquote {\bibinfo {title} {Alma resolves the torus of ngc 1068: Continuum and molecular line emission},}\ }\href {\doibase 10.3847/2041-8205/823/1/L12} {\bibfield  {journal} {\bibinfo  {journal} {The Astrophysical Journal Letters}\ }\textbf {\bibinfo {volume} {823}},\ \bibinfo {pages} {L12} (\bibinfo {year} {2016})}\BibitemShut {NoStop}%
\bibitem [{\citenamefont {Kheirandish}\ \emph {et~al.}(2021)\citenamefont {Kheirandish}, \citenamefont {Murase},\ and\ \citenamefont {Kimura}}]{Kheirandish:2021wkm}%
  \BibitemOpen
  \bibfield  {author} {\bibinfo {author} {\bibfnamefont {Ali}\ \bibnamefont {Kheirandish}}, \bibinfo {author} {\bibfnamefont {Kohta}\ \bibnamefont {Murase}}, \ and\ \bibinfo {author} {\bibfnamefont {Shigeo~S.}\ \bibnamefont {Kimura}},\ }\bibfield  {title} {\enquote {\bibinfo {title} {{High-energy Neutrinos from Magnetized Coronae of Active Galactic Nuclei and Prospects for Identification of Seyfert Galaxies and Quasars in Neutrino Telescopes}},}\ }\href {\doibase 10.3847/1538-4357/ac1c77} {\bibfield  {journal} {\bibinfo  {journal} {Astrophys. J.}\ }\textbf {\bibinfo {volume} {922}},\ \bibinfo {pages} {45} (\bibinfo {year} {2021})},\ \Eprint {http://arxiv.org/abs/2102.04475} {arXiv:2102.04475 [astro-ph.HE]} \BibitemShut {NoStop}%
\bibitem [{\citenamefont {Eichmann}\ \emph {et~al.}(2022)\citenamefont {Eichmann}, \citenamefont {Oikonomou}, \citenamefont {Salvatore}, \citenamefont {Dettmar},\ and\ \citenamefont {Becker~Tjus}}]{Eichmann:2022lxh}%
  \BibitemOpen
  \bibfield  {author} {\bibinfo {author} {\bibfnamefont {Bj\"orn}\ \bibnamefont {Eichmann}}, \bibinfo {author} {\bibfnamefont {Foteini}\ \bibnamefont {Oikonomou}}, \bibinfo {author} {\bibfnamefont {Silvia}\ \bibnamefont {Salvatore}}, \bibinfo {author} {\bibfnamefont {Ralf-J\"urgen}\ \bibnamefont {Dettmar}}, \ and\ \bibinfo {author} {\bibfnamefont {Julia}\ \bibnamefont {Becker~Tjus}},\ }\bibfield  {title} {\enquote {\bibinfo {title} {{Solving the Multimessenger Puzzle of the AGN-starburst Composite Galaxy NGC 1068}},}\ }\href {\doibase 10.3847/1538-4357/ac9588} {\bibfield  {journal} {\bibinfo  {journal} {Astrophys. J.}\ }\textbf {\bibinfo {volume} {939}},\ \bibinfo {pages} {43} (\bibinfo {year} {2022})},\ \Eprint {http://arxiv.org/abs/2207.00102} {arXiv:2207.00102 [astro-ph.HE]} \BibitemShut {NoStop}%
\bibitem [{\citenamefont {Anchordoqui}\ \emph {et~al.}(2021)\citenamefont {Anchordoqui}, \citenamefont {Krizmanic},\ and\ \citenamefont {Stecker}}]{Anchordoqui:2021vms}%
  \BibitemOpen
  \bibfield  {author} {\bibinfo {author} {\bibfnamefont {Luis~A.}\ \bibnamefont {Anchordoqui}}, \bibinfo {author} {\bibfnamefont {John~F.}\ \bibnamefont {Krizmanic}}, \ and\ \bibinfo {author} {\bibfnamefont {Floyd~W.}\ \bibnamefont {Stecker}},\ }\bibfield  {title} {\enquote {\bibinfo {title} {{High-Energy Neutrinos from NGC 1068}},}\ }\href {\doibase 10.22323/1.395.0993} {\bibfield  {journal} {\bibinfo  {journal} {PoS}\ }\textbf {\bibinfo {volume} {ICRC2021}},\ \bibinfo {pages} {993} (\bibinfo {year} {2021})},\ \Eprint {http://arxiv.org/abs/2102.12409} {arXiv:2102.12409 [astro-ph.HE]} \BibitemShut {NoStop}%
\bibitem [{\citenamefont {Halzen}(2021)}]{Halzen:2021xkf}%
  \BibitemOpen
  \bibfield  {author} {\bibinfo {author} {\bibfnamefont {Francis}\ \bibnamefont {Halzen}},\ }\bibfield  {title} {\enquote {\bibinfo {title} {{High-Energy Neutrinos from the Cosmos}},}\ }\href {\doibase 10.1002/andp.202100309} {\bibfield  {journal} {\bibinfo  {journal} {Annalen Phys.}\ }\textbf {\bibinfo {volume} {533}},\ \bibinfo {pages} {2100309} (\bibinfo {year} {2021})}\BibitemShut {NoStop}%
\bibitem [{\citenamefont {Kun}\ \emph {et~al.}(2024)\citenamefont {Kun}, \citenamefont {Bartos}, \citenamefont {Tjus}, \citenamefont {Biermann}, \citenamefont {Franckowiak}, \citenamefont {Halzen}, \citenamefont {del Palacio},\ and\ \citenamefont {Woo}}]{kun2024correlation}%
  \BibitemOpen
  \bibfield  {author} {\bibinfo {author} {\bibfnamefont {Emma}\ \bibnamefont {Kun}}, \bibinfo {author} {\bibfnamefont {Imre}\ \bibnamefont {Bartos}}, \bibinfo {author} {\bibfnamefont {Julia~Becker}\ \bibnamefont {Tjus}}, \bibinfo {author} {\bibfnamefont {Peter~L}\ \bibnamefont {Biermann}}, \bibinfo {author} {\bibfnamefont {Anna}\ \bibnamefont {Franckowiak}}, \bibinfo {author} {\bibfnamefont {Francis}\ \bibnamefont {Halzen}}, \bibinfo {author} {\bibfnamefont {Santiago}\ \bibnamefont {del Palacio}}, \ and\ \bibinfo {author} {\bibfnamefont {Jooyun}\ \bibnamefont {Woo}},\ }\bibfield  {title} {\enquote {\bibinfo {title} {A correlation between hard x-rays and neutrinos in radio-loud and radio-quiet agn},}\ }\href@noop {} {\bibfield  {journal} {\bibinfo  {journal} {arXiv preprint arXiv:2404.06867}\ } (\bibinfo {year} {2024})}\BibitemShut {NoStop}%
\bibitem [{\citenamefont {Lipunov}\ \emph {et~al.}(2020)\citenamefont {Lipunov} \emph {et~al.}}]{Lipunov:2020ptp}%
  \BibitemOpen
  \bibfield  {author} {\bibinfo {author} {\bibfnamefont {V.~M.}\ \bibnamefont {Lipunov}} \emph {et~al.},\ }\bibfield  {title} {\enquote {\bibinfo {title} {{Optical Observations Reveal Strong Evidence for High Energy Neutrino Progenitor}},}\ }\href {\doibase 10.3847/2041-8213/ab96ba} {\  (\bibinfo {year} {2020}),\ 10.3847/2041-8213/ab96ba},\ \Eprint {http://arxiv.org/abs/2006.04918} {arXiv:2006.04918 [astro-ph.HE]} \BibitemShut {NoStop}%
\bibitem [{\citenamefont {Sarmah}\ \emph {et~al.}(2024)\citenamefont {Sarmah}, \citenamefont {Chakraborty}, \citenamefont {Tamborra},\ and\ \citenamefont {Auchettl}}]{Sarmah:2024tis}%
  \BibitemOpen
  \bibfield  {author} {\bibinfo {author} {\bibfnamefont {Prantik}\ \bibnamefont {Sarmah}}, \bibinfo {author} {\bibfnamefont {Sovan}\ \bibnamefont {Chakraborty}}, \bibinfo {author} {\bibfnamefont {Irene}\ \bibnamefont {Tamborra}}, \ and\ \bibinfo {author} {\bibfnamefont {Katie}\ \bibnamefont {Auchettl}},\ }\bibfield  {title} {\enquote {\bibinfo {title} {{Gamma-Rays and~High Energy Neutrinos from~Young Supernovae}},}\ }\href {\doibase 10.1007/978-981-97-0289-3_25} {\bibfield  {journal} {\bibinfo  {journal} {Springer Proc. Phys.}\ }\textbf {\bibinfo {volume} {304}},\ \bibinfo {pages} {120--123} (\bibinfo {year} {2024})}\BibitemShut {NoStop}%
\bibitem [{\citenamefont {Abbasi}\ \emph {et~al.}(2024{\natexlab{a}})\citenamefont {Abbasi} \emph {et~al.}}]{IceCube:2024xjj}%
  \BibitemOpen
  \bibfield  {author} {\bibinfo {author} {\bibfnamefont {R.}~\bibnamefont {Abbasi}} \emph {et~al.} (\bibinfo {collaboration} {IceCube}),\ }\bibfield  {title} {\enquote {\bibinfo {title} {{Measurement of atmospheric neutrino oscillation parameters using convolutional neural networks with 9.3 years of data in IceCube DeepCore}},}\ }\href@noop {} {\  (\bibinfo {year} {2024}{\natexlab{a}})},\ \Eprint {http://arxiv.org/abs/2405.02163} {arXiv:2405.02163 [hep-ex]} \BibitemShut {NoStop}%
\bibitem [{\citenamefont {Abbasi}\ \emph {et~al.}(2011{\natexlab{a}})\citenamefont {Abbasi} \emph {et~al.}}]{Abbasi:2011ss}%
  \BibitemOpen
  \bibfield  {author} {\bibinfo {author} {\bibfnamefont {R.}~\bibnamefont {Abbasi}} \emph {et~al.} (\bibinfo {collaboration} {IceCube}),\ }\bibfield  {title} {\enquote {\bibinfo {title} {{IceCube Sensitivity for Low-Energy Neutrinos from Nearby Supernovae}},}\ }\href {\doibase 10.1051/0004-6361/201117810e} {\bibfield  {journal} {\bibinfo  {journal} {Astron. Astrophys.}\ }\textbf {\bibinfo {volume} {535}},\ \bibinfo {pages} {A109} (\bibinfo {year} {2011}{\natexlab{a}})},\ \bibinfo {note} {[Erratum: Astron.Astrophys. 563, C1 (2014)]},\ \Eprint {http://arxiv.org/abs/1108.0171} {arXiv:1108.0171 [astro-ph.HE]} \BibitemShut {NoStop}%
\bibitem [{\citenamefont {Aiello}\ \emph {et~al.}(2021)\citenamefont {Aiello} \emph {et~al.}}]{KM3NeT:2021moe}%
  \BibitemOpen
  \bibfield  {author} {\bibinfo {author} {\bibfnamefont {S.}~\bibnamefont {Aiello}} \emph {et~al.} (\bibinfo {collaboration} {KM3NeT}),\ }\bibfield  {title} {\enquote {\bibinfo {title} {{The KM3NeT potential for the next core-collapse supernova observation with neutrinos}},}\ }\href {\doibase 10.1140/epjc/s10052-021-09187-5} {\bibfield  {journal} {\bibinfo  {journal} {Eur. Phys. J. C}\ }\textbf {\bibinfo {volume} {81}},\ \bibinfo {pages} {445} (\bibinfo {year} {2021})},\ \Eprint {http://arxiv.org/abs/2102.05977} {arXiv:2102.05977 [astro-ph.HE]} \BibitemShut {NoStop}%
\bibitem [{\citenamefont {Abbasi}\ \emph {et~al.}(2023{\natexlab{c}})\citenamefont {Abbasi} \emph {et~al.}}]{IceCube:2023esf}%
  \BibitemOpen
  \bibfield  {author} {\bibinfo {author} {\bibfnamefont {R.}~\bibnamefont {Abbasi}} \emph {et~al.} (\bibinfo {collaboration} {IceCube}),\ }\bibfield  {title} {\enquote {\bibinfo {title} {{Constraining High-energy Neutrino Emission from Supernovae with IceCube}},}\ }\href {\doibase 10.3847/2041-8213/acd2c9} {\bibfield  {journal} {\bibinfo  {journal} {Astrophys. J. Lett.}\ }\textbf {\bibinfo {volume} {949}},\ \bibinfo {pages} {L12} (\bibinfo {year} {2023}{\natexlab{c}})},\ \Eprint {http://arxiv.org/abs/2303.03316} {arXiv:2303.03316 [astro-ph.HE]} \BibitemShut {NoStop}%
\bibitem [{\citenamefont {Novoseltsev}\ \emph {et~al.}(2020)\citenamefont {Novoseltsev}, \citenamefont {Boliev}, \citenamefont {Dzaparova}, \citenamefont {Kochkarov}, \citenamefont {Kurenya}, \citenamefont {Novoseltseva}, \citenamefont {Petkov}, \citenamefont {Striganov},\ and\ \citenamefont {Yanin}}]{Novoseltsev:2019gdt}%
  \BibitemOpen
  \bibfield  {author} {\bibinfo {author} {\bibfnamefont {Yu~F.}\ \bibnamefont {Novoseltsev}}, \bibinfo {author} {\bibfnamefont {M.~M.}\ \bibnamefont {Boliev}}, \bibinfo {author} {\bibfnamefont {I.~M.}\ \bibnamefont {Dzaparova}}, \bibinfo {author} {\bibfnamefont {M.~M.}\ \bibnamefont {Kochkarov}}, \bibinfo {author} {\bibfnamefont {A.~N.}\ \bibnamefont {Kurenya}}, \bibinfo {author} {\bibfnamefont {R.~V.}\ \bibnamefont {Novoseltseva}}, \bibinfo {author} {\bibfnamefont {V.~B.}\ \bibnamefont {Petkov}}, \bibinfo {author} {\bibfnamefont {P.~S.}\ \bibnamefont {Striganov}}, \ and\ \bibinfo {author} {\bibfnamefont {A.~F.}\ \bibnamefont {Yanin}},\ }\bibfield  {title} {\enquote {\bibinfo {title} {{Supernova Neutrino Burst Monitor at the Baksan Underground Scintillation Telescope}},}\ }\href {\doibase 10.1016/j.astropartphys.2019.102404} {\bibfield  {journal} {\bibinfo  {journal} {Astropart. Phys.}\ }\textbf {\bibinfo {volume} {117}},\ \bibinfo {pages} {102404} (\bibinfo {year} {2020})},\ \Eprint
  {http://arxiv.org/abs/1907.03019} {arXiv:1907.03019 [physics.ins-det]} \BibitemShut {NoStop}%
\bibitem [{\citenamefont {Vigorito}\ \emph {et~al.}(2021)\citenamefont {Vigorito}, \citenamefont {Bruno}, \citenamefont {Fulgione},\ and\ \citenamefont {Molinario}}]{Vigorito:2021sgy}%
  \BibitemOpen
  \bibfield  {author} {\bibinfo {author} {\bibfnamefont {Carlo~Francesco}\ \bibnamefont {Vigorito}}, \bibinfo {author} {\bibfnamefont {Gianmarco}\ \bibnamefont {Bruno}}, \bibinfo {author} {\bibfnamefont {Walter}\ \bibnamefont {Fulgione}}, \ and\ \bibinfo {author} {\bibfnamefont {Andrea}\ \bibnamefont {Molinario}} (\bibinfo {collaboration} {LVD}),\ }\bibfield  {title} {\enquote {\bibinfo {title} {{Update of the supernova neutrinos monitoring with the LVD experiment}},}\ }\href {\doibase 10.22323/1.395.1111} {\bibfield  {journal} {\bibinfo  {journal} {PoS}\ }\textbf {\bibinfo {volume} {ICRC2021}},\ \bibinfo {pages} {1111} (\bibinfo {year} {2021})}\BibitemShut {NoStop}%
\bibitem [{\citenamefont {Abbasi}\ \emph {et~al.}(2024{\natexlab{b}})\citenamefont {Abbasi} \emph {et~al.}}]{IceCube:2023ogt}%
  \BibitemOpen
  \bibfield  {author} {\bibinfo {author} {\bibfnamefont {R.}~\bibnamefont {Abbasi}} \emph {et~al.} (\bibinfo {collaboration} {IceCube}),\ }\bibfield  {title} {\enquote {\bibinfo {title} {{Search for Galactic Core-collapse Supernovae in a Decade of Data Taken with the IceCube Neutrino Observatory}},}\ }\href {\doibase 10.3847/1538-4357/ad07d1} {\bibfield  {journal} {\bibinfo  {journal} {Astrophys. J.}\ }\textbf {\bibinfo {volume} {961}},\ \bibinfo {pages} {84} (\bibinfo {year} {2024}{\natexlab{b}})},\ \Eprint {http://arxiv.org/abs/2308.01172} {arXiv:2308.01172 [astro-ph.HE]} \BibitemShut {NoStop}%
\bibitem [{\citenamefont {Al~Kharusi}\ \emph {et~al.}(2021)\citenamefont {Al~Kharusi} \emph {et~al.}}]{SNEWS:2020tbu}%
  \BibitemOpen
  \bibfield  {author} {\bibinfo {author} {\bibfnamefont {S.}~\bibnamefont {Al~Kharusi}} \emph {et~al.} (\bibinfo {collaboration} {SNEWS}),\ }\bibfield  {title} {\enquote {\bibinfo {title} {{SNEWS 2.0: a next-generation supernova early warning system for multi-messenger astronomy}},}\ }\href {\doibase 10.1088/1367-2630/abde33} {\bibfield  {journal} {\bibinfo  {journal} {New J. Phys.}\ }\textbf {\bibinfo {volume} {23}},\ \bibinfo {pages} {031201} (\bibinfo {year} {2021})},\ \Eprint {http://arxiv.org/abs/2011.00035} {arXiv:2011.00035 [astro-ph.HE]} \BibitemShut {NoStop}%
\bibitem [{\citenamefont {Kara}\ \emph {et~al.}(2024)\citenamefont {Kara} \emph {et~al.}}]{Kara:2024xug}%
  \BibitemOpen
  \bibfield  {author} {\bibinfo {author} {\bibfnamefont {M.}~\bibnamefont {Kara}} \emph {et~al.},\ }\bibfield  {title} {\enquote {\bibinfo {title} {{The SNEWS 2.0 Alert Software for the Coincident Detection of Neutrinos from Core-Collapse Supernovae}},}\ }\href {\doibase 10.1088/1748-0221/19/10/P10017} {\bibfield  {journal} {\bibinfo  {journal} {JINST}\ }\textbf {\bibinfo {volume} {19}},\ \bibinfo {pages} {P10017} (\bibinfo {year} {2024})},\ \Eprint {http://arxiv.org/abs/2406.17743} {arXiv:2406.17743 [astro-ph.IM]} \BibitemShut {NoStop}%
\bibitem [{\citenamefont {Abbasi}\ \emph {et~al.}(2023{\natexlab{d}})\citenamefont {Abbasi} \emph {et~al.}}]{IceCube:2023xzp}%
  \BibitemOpen
  \bibfield  {author} {\bibinfo {author} {\bibfnamefont {Rasha}\ \bibnamefont {Abbasi}} \emph {et~al.} (\bibinfo {collaboration} {IceCube}),\ }\bibfield  {title} {\enquote {\bibinfo {title} {{Galactic Core-Collapse Supernovae at IceCube: \textquotedblleft{}Fire Drill\textquotedblright{} Data Challenges and follow-up}},}\ }\href {\doibase 10.22323/1.444.1111} {\bibfield  {journal} {\bibinfo  {journal} {PoS}\ }\textbf {\bibinfo {volume} {ICRC2023}},\ \bibinfo {pages} {1111} (\bibinfo {year} {2023}{\natexlab{d}})}\BibitemShut {NoStop}%
\bibitem [{\citenamefont {Heereman~von Zuydtwyck}(2015)}]{HeeremanvonZuydtwyck:2015mbs}%
  \BibitemOpen
  \bibfield  {author} {\bibinfo {author} {\bibfnamefont {David~Freiherr}\ \bibnamefont {Heereman~von Zuydtwyck}},\ }\emph {\bibinfo {title} {{HitSpooling}: {An Improvement for the Supernova Neutrino Detection System in IceCube}}},\ \href@noop {} {Ph.D. thesis},\ \bibinfo  {school} {University of Brussels} (\bibinfo {year} {2015})\BibitemShut {NoStop}%
\bibitem [{\citenamefont {Drago}\ \emph {et~al.}(2023)\citenamefont {Drago}, \citenamefont {Andresen}, \citenamefont {Di~Palma}, \citenamefont {Tamborra},\ and\ \citenamefont {Torres-Forné}}]{Drago_2023}%
  \BibitemOpen
  \bibfield  {author} {\bibinfo {author} {\bibfnamefont {Marco}\ \bibnamefont {Drago}}, \bibinfo {author} {\bibfnamefont {Haakon}\ \bibnamefont {Andresen}}, \bibinfo {author} {\bibfnamefont {Irene}\ \bibnamefont {Di~Palma}}, \bibinfo {author} {\bibfnamefont {Irene}\ \bibnamefont {Tamborra}}, \ and\ \bibinfo {author} {\bibfnamefont {Alejandro}\ \bibnamefont {Torres-Forné}},\ }\bibfield  {title} {\enquote {\bibinfo {title} {Multimessenger observations of core-collapse supernovae: Exploiting the standing accretion shock instability},}\ }\href {\doibase 10.1103/physrevd.108.103036} {\bibfield  {journal} {\bibinfo  {journal} {Physical Review D}\ }\textbf {\bibinfo {volume} {108}} (\bibinfo {year} {2023}),\ 10.1103/physrevd.108.103036}\BibitemShut {NoStop}%
\bibitem [{\citenamefont {Nakazato}\ \emph {et~al.}(2013)\citenamefont {Nakazato}, \citenamefont {Sumiyoshi}, \citenamefont {Suzuki}, \citenamefont {Totani}, \citenamefont {Umeda},\ and\ \citenamefont {Yamada}}]{Nakazato:2012qf}%
  \BibitemOpen
  \bibfield  {author} {\bibinfo {author} {\bibfnamefont {Ken'ichiro}\ \bibnamefont {Nakazato}}, \bibinfo {author} {\bibfnamefont {Kohsuke}\ \bibnamefont {Sumiyoshi}}, \bibinfo {author} {\bibfnamefont {Hideyuki}\ \bibnamefont {Suzuki}}, \bibinfo {author} {\bibfnamefont {Tomonori}\ \bibnamefont {Totani}}, \bibinfo {author} {\bibfnamefont {Hideyuki}\ \bibnamefont {Umeda}}, \ and\ \bibinfo {author} {\bibfnamefont {Shoichi}\ \bibnamefont {Yamada}},\ }\bibfield  {title} {\enquote {\bibinfo {title} {{Supernova Neutrino Light Curves and Spectra for Various Progenitor Stars: From Core Collapse to Proto-neutron Star Cooling}},}\ }\href {\doibase 10.1088/0067-0049/205/1/2} {\bibfield  {journal} {\bibinfo  {journal} {Astrophys. J. Suppl.}\ }\textbf {\bibinfo {volume} {205}},\ \bibinfo {pages} {2} (\bibinfo {year} {2013})},\ \Eprint {http://arxiv.org/abs/1210.6841} {arXiv:1210.6841 [astro-ph.HE]} \BibitemShut {NoStop}%
\bibitem [{\citenamefont {Takiwaki}\ \emph {et~al.}(2021)\citenamefont {Takiwaki}, \citenamefont {Kotake},\ and\ \citenamefont {Foglizzo}}]{Takiwaki:2021dve}%
  \BibitemOpen
  \bibfield  {author} {\bibinfo {author} {\bibfnamefont {Tomoya}\ \bibnamefont {Takiwaki}}, \bibinfo {author} {\bibfnamefont {Kei}\ \bibnamefont {Kotake}}, \ and\ \bibinfo {author} {\bibfnamefont {Thierry}\ \bibnamefont {Foglizzo}},\ }\bibfield  {title} {\enquote {\bibinfo {title} {{Insights into non-axisymmetric instabilities in three-dimensional rotating supernova models with neutrino and gravitational-wave signatures}},}\ }\href {\doibase 10.1093/mnras/stab2607} {\bibfield  {journal} {\bibinfo  {journal} {Mon. Not. Roy. Astron. Soc.}\ }\textbf {\bibinfo {volume} {508}},\ \bibinfo {pages} {966--985} (\bibinfo {year} {2021})},\ \Eprint {http://arxiv.org/abs/2107.02933} {arXiv:2107.02933 [astro-ph.HE]} \BibitemShut {NoStop}%
\bibitem [{\citenamefont {Lincetto}\ \emph {et~al.}(2023)\citenamefont {Lincetto} \emph {et~al.}}]{IceCube:2023amf}%
  \BibitemOpen
  \bibfield  {author} {\bibinfo {author} {\bibfnamefont {Massimiliano}\ \bibnamefont {Lincetto}} \emph {et~al.} (\bibinfo {collaboration} {IceCube}),\ }\bibfield  {title} {\enquote {\bibinfo {title} {{Searching for high-energy neutrinos from shock-interaction powered supernovae with the IceCube Neutrino Observatory}},}\ }\href {\doibase 10.22323/1.444.1105} {\bibfield  {journal} {\bibinfo  {journal} {PoS}\ }\textbf {\bibinfo {volume} {ICRC2023}},\ \bibinfo {pages} {1105} (\bibinfo {year} {2023})},\ \Eprint {http://arxiv.org/abs/2308.01047} {arXiv:2308.01047 [astro-ph.HE]} \BibitemShut {NoStop}%
\bibitem [{\citenamefont {Murase}(2018)}]{Murase_2018}%
  \BibitemOpen
  \bibfield  {author} {\bibinfo {author} {\bibfnamefont {Kohta}\ \bibnamefont {Murase}},\ }\bibfield  {title} {\enquote {\bibinfo {title} {New prospects for detecting high-energy neutrinos from nearby supernovae},}\ }\href {\doibase 10.1103/physrevd.97.081301} {\bibfield  {journal} {\bibinfo  {journal} {Physical Review D}\ }\textbf {\bibinfo {volume} {97}} (\bibinfo {year} {2018}),\ 10.1103/physrevd.97.081301}\BibitemShut {NoStop}%
\bibitem [{\citenamefont {Jungman}\ \emph {et~al.}(1996)\citenamefont {Jungman}, \citenamefont {Kamionkowski},\ and\ \citenamefont {Griest}}]{Jungman:1995df}%
  \BibitemOpen
  \bibfield  {author} {\bibinfo {author} {\bibfnamefont {Gerard}\ \bibnamefont {Jungman}}, \bibinfo {author} {\bibfnamefont {Marc}\ \bibnamefont {Kamionkowski}}, \ and\ \bibinfo {author} {\bibfnamefont {Kim}\ \bibnamefont {Griest}},\ }\bibfield  {title} {\enquote {\bibinfo {title} {{Supersymmetric dark matter}},}\ }\href {\doibase 10.1016/0370-1573(95)00058-5} {\bibfield  {journal} {\bibinfo  {journal} {Phys. Rept.}\ }\textbf {\bibinfo {volume} {267}},\ \bibinfo {pages} {195--373} (\bibinfo {year} {1996})},\ \Eprint {http://arxiv.org/abs/hep-ph/9506380} {arXiv:hep-ph/9506380} \BibitemShut {NoStop}%
\bibitem [{\citenamefont {Maity}\ \emph {et~al.}(2023)\citenamefont {Maity}, \citenamefont {Saha}, \citenamefont {Mondal},\ and\ \citenamefont {Laha}}]{Maity:2023rez}%
  \BibitemOpen
  \bibfield  {author} {\bibinfo {author} {\bibfnamefont {Tarak~Nath}\ \bibnamefont {Maity}}, \bibinfo {author} {\bibfnamefont {Akash~Kumar}\ \bibnamefont {Saha}}, \bibinfo {author} {\bibfnamefont {Sagnik}\ \bibnamefont {Mondal}}, \ and\ \bibinfo {author} {\bibfnamefont {Ranjan}\ \bibnamefont {Laha}},\ }\bibfield  {title} {\enquote {\bibinfo {title} {{Neutrinos from the Sun can discover dark matter-electron scattering}},}\ }\href@noop {} {\  (\bibinfo {year} {2023})},\ \Eprint {http://arxiv.org/abs/2308.12336} {arXiv:2308.12336 [hep-ph]} \BibitemShut {NoStop}%
\bibitem [{\citenamefont {Choi}\ \emph {et~al.}(2014)\citenamefont {Choi}, \citenamefont {Rott},\ and\ \citenamefont {Itow}}]{Choi:2013eda}%
  \BibitemOpen
  \bibfield  {author} {\bibinfo {author} {\bibfnamefont {K.}~\bibnamefont {Choi}}, \bibinfo {author} {\bibfnamefont {Carsten}\ \bibnamefont {Rott}}, \ and\ \bibinfo {author} {\bibfnamefont {Yoshitaka}\ \bibnamefont {Itow}},\ }\bibfield  {title} {\enquote {\bibinfo {title} {{Impact of the dark matter velocity distribution on capture rates in the Sun}},}\ }\href {\doibase 10.1088/1475-7516/2014/05/049} {\bibfield  {journal} {\bibinfo  {journal} {JCAP}\ }\textbf {\bibinfo {volume} {05}},\ \bibinfo {pages} {049} (\bibinfo {year} {2014})},\ \Eprint {http://arxiv.org/abs/1312.0273} {arXiv:1312.0273 [astro-ph.HE]} \BibitemShut {NoStop}%
\bibitem [{\citenamefont {Danninger}\ and\ \citenamefont {Rott}(2014)}]{Danninger:2014xza}%
  \BibitemOpen
  \bibfield  {author} {\bibinfo {author} {\bibfnamefont {Matthias}\ \bibnamefont {Danninger}}\ and\ \bibinfo {author} {\bibfnamefont {Carsten}\ \bibnamefont {Rott}},\ }\bibfield  {title} {\enquote {\bibinfo {title} {{Solar WIMPs unravelled: Experiments, astrophysical uncertainties, and interactive tools}},}\ }\href {\doibase 10.1016/j.dark.2014.10.002} {\bibfield  {journal} {\bibinfo  {journal} {Phys. Dark Univ.}\ }\textbf {\bibinfo {volume} {5-6}},\ \bibinfo {pages} {35--44} (\bibinfo {year} {2014})},\ \Eprint {http://arxiv.org/abs/1509.08230} {arXiv:1509.08230 [astro-ph.HE]} \BibitemShut {NoStop}%
\bibitem [{\citenamefont {Peccei}\ and\ \citenamefont {Quinn}(1977{\natexlab{a}})}]{Peccei:1977hh}%
  \BibitemOpen
  \bibfield  {author} {\bibinfo {author} {\bibfnamefont {R.~D.}\ \bibnamefont {Peccei}}\ and\ \bibinfo {author} {\bibfnamefont {Helen~R.}\ \bibnamefont {Quinn}},\ }\bibfield  {title} {\enquote {\bibinfo {title} {{CP Conservation in the Presence of Instantons}},}\ }\href {\doibase 10.1103/PhysRevLett.38.1440} {\bibfield  {journal} {\bibinfo  {journal} {Phys. Rev. Lett.}\ }\textbf {\bibinfo {volume} {38}},\ \bibinfo {pages} {1440--1443} (\bibinfo {year} {1977}{\natexlab{a}})}\BibitemShut {NoStop}%
\bibitem [{\citenamefont {Peccei}\ and\ \citenamefont {Quinn}(1977{\natexlab{b}})}]{Peccei:1977ur}%
  \BibitemOpen
  \bibfield  {author} {\bibinfo {author} {\bibfnamefont {R.~D.}\ \bibnamefont {Peccei}}\ and\ \bibinfo {author} {\bibfnamefont {Helen~R.}\ \bibnamefont {Quinn}},\ }\bibfield  {title} {\enquote {\bibinfo {title} {{Constraints Imposed by CP Conservation in the Presence of Instantons}},}\ }\href {\doibase 10.1103/PhysRevD.16.1791} {\bibfield  {journal} {\bibinfo  {journal} {Phys. Rev. D}\ }\textbf {\bibinfo {volume} {16}},\ \bibinfo {pages} {1791--1797} (\bibinfo {year} {1977}{\natexlab{b}})}\BibitemShut {NoStop}%
\bibitem [{\citenamefont {Weinberg}(1978)}]{Weinberg:1977ma}%
  \BibitemOpen
  \bibfield  {author} {\bibinfo {author} {\bibfnamefont {Steven}\ \bibnamefont {Weinberg}},\ }\bibfield  {title} {\enquote {\bibinfo {title} {{A New Light Boson?}}}\ }\href {\doibase 10.1103/PhysRevLett.40.223} {\bibfield  {journal} {\bibinfo  {journal} {Phys. Rev. Lett.}\ }\textbf {\bibinfo {volume} {40}},\ \bibinfo {pages} {223--226} (\bibinfo {year} {1978})}\BibitemShut {NoStop}%
\bibitem [{\citenamefont {Wilczek}(1978)}]{Wilczek:1977pj}%
  \BibitemOpen
  \bibfield  {author} {\bibinfo {author} {\bibfnamefont {Frank}\ \bibnamefont {Wilczek}},\ }\bibfield  {title} {\enquote {\bibinfo {title} {{Problem of Strong $P$ and $T$ Invariance in the Presence of Instantons}},}\ }\href {\doibase 10.1103/PhysRevLett.40.279} {\bibfield  {journal} {\bibinfo  {journal} {Phys. Rev. Lett.}\ }\textbf {\bibinfo {volume} {40}},\ \bibinfo {pages} {279--282} (\bibinfo {year} {1978})}\BibitemShut {NoStop}%
\bibitem [{\citenamefont {Arvanitaki}\ \emph {et~al.}(2010)\citenamefont {Arvanitaki}, \citenamefont {Dimopoulos}, \citenamefont {Dubovsky}, \citenamefont {Kaloper},\ and\ \citenamefont {March-Russell}}]{Arvanitaki:2009fg}%
  \BibitemOpen
  \bibfield  {author} {\bibinfo {author} {\bibfnamefont {Asimina}\ \bibnamefont {Arvanitaki}}, \bibinfo {author} {\bibfnamefont {Savas}\ \bibnamefont {Dimopoulos}}, \bibinfo {author} {\bibfnamefont {Sergei}\ \bibnamefont {Dubovsky}}, \bibinfo {author} {\bibfnamefont {Nemanja}\ \bibnamefont {Kaloper}}, \ and\ \bibinfo {author} {\bibfnamefont {John}\ \bibnamefont {March-Russell}},\ }\bibfield  {title} {\enquote {\bibinfo {title} {{String Axiverse}},}\ }\href {\doibase 10.1103/PhysRevD.81.123530} {\bibfield  {journal} {\bibinfo  {journal} {Phys. Rev. D}\ }\textbf {\bibinfo {volume} {81}},\ \bibinfo {pages} {123530} (\bibinfo {year} {2010})},\ \Eprint {http://arxiv.org/abs/0905.4720} {arXiv:0905.4720 [hep-th]} \BibitemShut {NoStop}%
\bibitem [{\citenamefont {Fischer}\ \emph {et~al.}(2016)\citenamefont {Fischer}, \citenamefont {Chakraborty}, \citenamefont {Giannotti}, \citenamefont {Mirizzi}, \citenamefont {Payez},\ and\ \citenamefont {Ringwald}}]{Fischer:2016cyd}%
  \BibitemOpen
  \bibfield  {author} {\bibinfo {author} {\bibfnamefont {Tobias}\ \bibnamefont {Fischer}}, \bibinfo {author} {\bibfnamefont {Sovan}\ \bibnamefont {Chakraborty}}, \bibinfo {author} {\bibfnamefont {Maurizio}\ \bibnamefont {Giannotti}}, \bibinfo {author} {\bibfnamefont {Alessandro}\ \bibnamefont {Mirizzi}}, \bibinfo {author} {\bibfnamefont {Alexandre}\ \bibnamefont {Payez}}, \ and\ \bibinfo {author} {\bibfnamefont {Andreas}\ \bibnamefont {Ringwald}},\ }\bibfield  {title} {\enquote {\bibinfo {title} {{Probing axions with the neutrino signal from the next galactic supernova}},}\ }\href {\doibase 10.1103/PhysRevD.94.085012} {\bibfield  {journal} {\bibinfo  {journal} {Phys. Rev. D}\ }\textbf {\bibinfo {volume} {94}},\ \bibinfo {pages} {085012} (\bibinfo {year} {2016})},\ \Eprint {http://arxiv.org/abs/1605.08780} {arXiv:1605.08780 [astro-ph.HE]} \BibitemShut {NoStop}%
\bibitem [{\citenamefont {Betranhandy}\ and\ \citenamefont {O'Connor}(2022)}]{Betranhandy:2022bvr}%
  \BibitemOpen
  \bibfield  {author} {\bibinfo {author} {\bibfnamefont {Aurore}\ \bibnamefont {Betranhandy}}\ and\ \bibinfo {author} {\bibfnamefont {Evan}\ \bibnamefont {O'Connor}},\ }\bibfield  {title} {\enquote {\bibinfo {title} {{Neutrino driven explosions aided by axion cooling in multidimensional simulations of core-collapse supernovae}},}\ }\href {\doibase 10.1103/PhysRevD.106.063019} {\bibfield  {journal} {\bibinfo  {journal} {Phys. Rev. D}\ }\textbf {\bibinfo {volume} {106}},\ \bibinfo {pages} {063019} (\bibinfo {year} {2022})},\ \Eprint {http://arxiv.org/abs/2204.00503} {arXiv:2204.00503 [astro-ph.HE]} \BibitemShut {NoStop}%
\bibitem [{\citenamefont {Mori}\ \emph {et~al.}(2023)\citenamefont {Mori}, \citenamefont {Takiwaki}, \citenamefont {Kotake},\ and\ \citenamefont {Horiuchi}}]{Mori:2023mjw}%
  \BibitemOpen
  \bibfield  {author} {\bibinfo {author} {\bibfnamefont {Kanji}\ \bibnamefont {Mori}}, \bibinfo {author} {\bibfnamefont {Tomoya}\ \bibnamefont {Takiwaki}}, \bibinfo {author} {\bibfnamefont {Kei}\ \bibnamefont {Kotake}}, \ and\ \bibinfo {author} {\bibfnamefont {Shunsaku}\ \bibnamefont {Horiuchi}},\ }\bibfield  {title} {\enquote {\bibinfo {title} {{Multimessenger signals of heavy axionlike particles in core-collapse supernovae: Two-dimensional simulations}},}\ }\href {\doibase 10.1103/PhysRevD.108.063027} {\bibfield  {journal} {\bibinfo  {journal} {Phys. Rev. D}\ }\textbf {\bibinfo {volume} {108}},\ \bibinfo {pages} {063027} (\bibinfo {year} {2023})},\ \Eprint {http://arxiv.org/abs/2304.11360} {arXiv:2304.11360 [astro-ph.HE]} \BibitemShut {NoStop}%
\bibitem [{\citenamefont {Reynoso}\ and\ \citenamefont {Sampayo}(2016)}]{Reynoso:2016hjr}%
  \BibitemOpen
  \bibfield  {author} {\bibinfo {author} {\bibfnamefont {Mat\'\i{}as~M.}\ \bibnamefont {Reynoso}}\ and\ \bibinfo {author} {\bibfnamefont {Oscar~A.}\ \bibnamefont {Sampayo}},\ }\bibfield  {title} {\enquote {\bibinfo {title} {{Propagation of high-energy neutrinos in a background of ultralight scalar dark matter}},}\ }\href {\doibase 10.1016/j.astropartphys.2016.05.004} {\bibfield  {journal} {\bibinfo  {journal} {Astropart. Phys.}\ }\textbf {\bibinfo {volume} {82}},\ \bibinfo {pages} {10--20} (\bibinfo {year} {2016})},\ \Eprint {http://arxiv.org/abs/1605.09671} {arXiv:1605.09671 [hep-ph]} \BibitemShut {NoStop}%
\bibitem [{\citenamefont {Brdar}\ \emph {et~al.}(2018)\citenamefont {Brdar}, \citenamefont {Kopp}, \citenamefont {Liu}, \citenamefont {Prass},\ and\ \citenamefont {Wang}}]{Brdar:2017kbt}%
  \BibitemOpen
  \bibfield  {author} {\bibinfo {author} {\bibfnamefont {Vedran}\ \bibnamefont {Brdar}}, \bibinfo {author} {\bibfnamefont {Joachim}\ \bibnamefont {Kopp}}, \bibinfo {author} {\bibfnamefont {Jia}\ \bibnamefont {Liu}}, \bibinfo {author} {\bibfnamefont {Pascal}\ \bibnamefont {Prass}}, \ and\ \bibinfo {author} {\bibfnamefont {Xiao-Ping}\ \bibnamefont {Wang}},\ }\bibfield  {title} {\enquote {\bibinfo {title} {{Fuzzy dark matter and nonstandard neutrino interactions}},}\ }\href {\doibase 10.1103/PhysRevD.97.043001} {\bibfield  {journal} {\bibinfo  {journal} {Phys. Rev. D}\ }\textbf {\bibinfo {volume} {97}},\ \bibinfo {pages} {043001} (\bibinfo {year} {2018})},\ \Eprint {http://arxiv.org/abs/1705.09455} {arXiv:1705.09455 [hep-ph]} \BibitemShut {NoStop}%
\bibitem [{\citenamefont {Farzan}\ and\ \citenamefont {Palomares-Ruiz}(2019)}]{Farzan:2018pnk}%
  \BibitemOpen
  \bibfield  {author} {\bibinfo {author} {\bibfnamefont {Yasaman}\ \bibnamefont {Farzan}}\ and\ \bibinfo {author} {\bibfnamefont {Sergio}\ \bibnamefont {Palomares-Ruiz}},\ }\bibfield  {title} {\enquote {\bibinfo {title} {{Flavor of cosmic neutrinos preserved by ultralight dark matter}},}\ }\href {\doibase 10.1103/PhysRevD.99.051702} {\bibfield  {journal} {\bibinfo  {journal} {Phys. Rev. D}\ }\textbf {\bibinfo {volume} {99}},\ \bibinfo {pages} {051702} (\bibinfo {year} {2019})},\ \Eprint {http://arxiv.org/abs/1810.00892} {arXiv:1810.00892 [hep-ph]} \BibitemShut {NoStop}%
\bibitem [{\citenamefont {Farzan}(2021)}]{Farzan:2021gbx}%
  \BibitemOpen
  \bibfield  {author} {\bibinfo {author} {\bibfnamefont {Yasaman}\ \bibnamefont {Farzan}},\ }\bibfield  {title} {\enquote {\bibinfo {title} {{On the \ensuremath{\tau} flavor of the cosmic neutrino flux}},}\ }\href {\doibase 10.1007/JHEP07(2021)174} {\bibfield  {journal} {\bibinfo  {journal} {JHEP}\ }\textbf {\bibinfo {volume} {07}},\ \bibinfo {pages} {174} (\bibinfo {year} {2021})},\ \Eprint {http://arxiv.org/abs/2105.03272} {arXiv:2105.03272 [hep-ph]} \BibitemShut {NoStop}%
\bibitem [{\citenamefont {Reynoso}\ \emph {et~al.}(2022)\citenamefont {Reynoso}, \citenamefont {Sampayo},\ and\ \citenamefont {Carulli}}]{Reynoso:2022vrn}%
  \BibitemOpen
  \bibfield  {author} {\bibinfo {author} {\bibfnamefont {Mat\'\i{}as~M.}\ \bibnamefont {Reynoso}}, \bibinfo {author} {\bibfnamefont {Oscar~A.}\ \bibnamefont {Sampayo}}, \ and\ \bibinfo {author} {\bibfnamefont {Agust\'\i{}n~M.}\ \bibnamefont {Carulli}},\ }\bibfield  {title} {\enquote {\bibinfo {title} {{Neutrino interactions with ultralight axion-like dark matter}},}\ }\href {\doibase 10.1140/epjc/s10052-022-10228-w} {\bibfield  {journal} {\bibinfo  {journal} {Eur. Phys. J. C}\ }\textbf {\bibinfo {volume} {82}},\ \bibinfo {pages} {274} (\bibinfo {year} {2022})},\ \Eprint {http://arxiv.org/abs/2203.11642} {arXiv:2203.11642 [hep-ph]} \BibitemShut {NoStop}%
\bibitem [{\citenamefont {Arg\"uelles}\ \emph {et~al.}(2024)\citenamefont {Arg\"uelles}, \citenamefont {Farrag},\ and\ \citenamefont {Katori}}]{Arguelles:2024cjj}%
  \BibitemOpen
  \bibfield  {author} {\bibinfo {author} {\bibfnamefont {Carlos~A.}\ \bibnamefont {Arg\"uelles}}, \bibinfo {author} {\bibfnamefont {Kareem}\ \bibnamefont {Farrag}}, \ and\ \bibinfo {author} {\bibfnamefont {Teppei}\ \bibnamefont {Katori}},\ }\bibfield  {title} {\enquote {\bibinfo {title} {{Ultra-light Dark Matter Limits from Astrophysical Neutrino Flavor}},}\ }\href@noop {} {\  (\bibinfo {year} {2024})},\ \Eprint {http://arxiv.org/abs/2404.10926} {arXiv:2404.10926 [hep-ph]} \BibitemShut {NoStop}%
\bibitem [{\citenamefont {Skrzypek}\ \emph {et~al.}(2023)\citenamefont {Skrzypek}, \citenamefont {Chianese}, \citenamefont {Arg\"uelles},\ and\ \citenamefont {Delgado~Arg\"uelles}}]{Skrzypek:2022hpy}%
  \BibitemOpen
  \bibfield  {author} {\bibinfo {author} {\bibfnamefont {Barbara}\ \bibnamefont {Skrzypek}}, \bibinfo {author} {\bibfnamefont {Marco}\ \bibnamefont {Chianese}}, \bibinfo {author} {\bibfnamefont {C.~A.}\ \bibnamefont {Arg\"uelles}}, \ and\ \bibinfo {author} {\bibfnamefont {Carlos}\ \bibnamefont {Delgado~Arg\"uelles}},\ }\bibfield  {title} {\enquote {\bibinfo {title} {{Multi-messenger high-energy signatures of decaying dark matter and the effect of background light}},}\ }\href {\doibase 10.1088/1475-7516/2023/01/037} {\bibfield  {journal} {\bibinfo  {journal} {JCAP}\ }\textbf {\bibinfo {volume} {01}},\ \bibinfo {pages} {037} (\bibinfo {year} {2023})},\ \Eprint {http://arxiv.org/abs/2205.03416} {arXiv:2205.03416 [astro-ph.HE]} \BibitemShut {NoStop}%
\bibitem [{\citenamefont {Beacom}\ \emph {et~al.}(2007)\citenamefont {Beacom}, \citenamefont {Bell},\ and\ \citenamefont {Mack}}]{Beacom:2006tt}%
  \BibitemOpen
  \bibfield  {author} {\bibinfo {author} {\bibfnamefont {John~F.}\ \bibnamefont {Beacom}}, \bibinfo {author} {\bibfnamefont {Nicole~F.}\ \bibnamefont {Bell}}, \ and\ \bibinfo {author} {\bibfnamefont {Gregory~D.}\ \bibnamefont {Mack}},\ }\bibfield  {title} {\enquote {\bibinfo {title} {{General Upper Bound on the Dark Matter Total Annihilation Cross Section}},}\ }\href {\doibase 10.1103/PhysRevLett.99.231301} {\bibfield  {journal} {\bibinfo  {journal} {Phys. Rev. Lett.}\ }\textbf {\bibinfo {volume} {99}},\ \bibinfo {pages} {231301} (\bibinfo {year} {2007})},\ \Eprint {http://arxiv.org/abs/astro-ph/0608090} {arXiv:astro-ph/0608090} \BibitemShut {NoStop}%
\bibitem [{\citenamefont {Murase}\ and\ \citenamefont {Beacom}(2013)}]{Murase:2012rd}%
  \BibitemOpen
  \bibfield  {author} {\bibinfo {author} {\bibfnamefont {Kohta}\ \bibnamefont {Murase}}\ and\ \bibinfo {author} {\bibfnamefont {John~F.}\ \bibnamefont {Beacom}},\ }\bibfield  {title} {\enquote {\bibinfo {title} {{Galaxy Clusters as Reservoirs of Heavy Dark Matter and High-Energy Cosmic Rays: Constraints from Neutrino Observations}},}\ }\href {\doibase 10.1088/1475-7516/2013/02/028} {\bibfield  {journal} {\bibinfo  {journal} {JCAP}\ }\textbf {\bibinfo {volume} {02}},\ \bibinfo {pages} {028} (\bibinfo {year} {2013})},\ \Eprint {http://arxiv.org/abs/1209.0225} {arXiv:1209.0225 [astro-ph.HE]} \BibitemShut {NoStop}%
\bibitem [{\citenamefont {Murase}\ and\ \citenamefont {Beacom}(2012)}]{Murase:2012xs}%
  \BibitemOpen
  \bibfield  {author} {\bibinfo {author} {\bibfnamefont {Kohta}\ \bibnamefont {Murase}}\ and\ \bibinfo {author} {\bibfnamefont {John~F.}\ \bibnamefont {Beacom}},\ }\bibfield  {title} {\enquote {\bibinfo {title} {{Constraining Very Heavy Dark Matter Using Diffuse Backgrounds of Neutrinos and Cascaded Gamma Rays}},}\ }\href {\doibase 10.1088/1475-7516/2012/10/043} {\bibfield  {journal} {\bibinfo  {journal} {JCAP}\ }\textbf {\bibinfo {volume} {10}},\ \bibinfo {pages} {043} (\bibinfo {year} {2012})},\ \Eprint {http://arxiv.org/abs/1206.2595} {arXiv:1206.2595 [hep-ph]} \BibitemShut {NoStop}%
\bibitem [{\citenamefont {Bhattacharya}\ \emph {et~al.}(2019)\citenamefont {Bhattacharya}, \citenamefont {Esmaili}, \citenamefont {Palomares-Ruiz},\ and\ \citenamefont {Sarcevic}}]{Bhattacharya:2019ucd}%
  \BibitemOpen
  \bibfield  {author} {\bibinfo {author} {\bibfnamefont {Atri}\ \bibnamefont {Bhattacharya}}, \bibinfo {author} {\bibfnamefont {Arman}\ \bibnamefont {Esmaili}}, \bibinfo {author} {\bibfnamefont {Sergio}\ \bibnamefont {Palomares-Ruiz}}, \ and\ \bibinfo {author} {\bibfnamefont {Ina}\ \bibnamefont {Sarcevic}},\ }\bibfield  {title} {\enquote {\bibinfo {title} {{Update on decaying and annihilating heavy dark matter with the 6-year IceCube HESE data}},}\ }\href {\doibase 10.1088/1475-7516/2019/05/051} {\bibfield  {journal} {\bibinfo  {journal} {JCAP}\ }\textbf {\bibinfo {volume} {05}},\ \bibinfo {pages} {051} (\bibinfo {year} {2019})},\ \Eprint {http://arxiv.org/abs/1903.12623} {arXiv:1903.12623 [hep-ph]} \BibitemShut {NoStop}%
\bibitem [{\citenamefont {Arg\"uelles}\ \emph {et~al.}(2021)\citenamefont {Arg\"uelles}, \citenamefont {Diaz}, \citenamefont {Kheirandish}, \citenamefont {Olivares-Del-Campo}, \citenamefont {Safa},\ and\ \citenamefont {Vincent}}]{Arguelles:2019ouk}%
  \BibitemOpen
  \bibfield  {author} {\bibinfo {author} {\bibfnamefont {Carlos~A.}\ \bibnamefont {Arg\"uelles}}, \bibinfo {author} {\bibfnamefont {Alejandro}\ \bibnamefont {Diaz}}, \bibinfo {author} {\bibfnamefont {Ali}\ \bibnamefont {Kheirandish}}, \bibinfo {author} {\bibfnamefont {Andr\'es}\ \bibnamefont {Olivares-Del-Campo}}, \bibinfo {author} {\bibfnamefont {Ibrahim}\ \bibnamefont {Safa}}, \ and\ \bibinfo {author} {\bibfnamefont {Aaron~C.}\ \bibnamefont {Vincent}},\ }\bibfield  {title} {\enquote {\bibinfo {title} {{Dark matter annihilation to neutrinos}},}\ }\href {\doibase 10.1103/RevModPhys.93.035007} {\bibfield  {journal} {\bibinfo  {journal} {Rev. Mod. Phys.}\ }\textbf {\bibinfo {volume} {93}},\ \bibinfo {pages} {035007} (\bibinfo {year} {2021})},\ \Eprint {http://arxiv.org/abs/1912.09486} {arXiv:1912.09486 [hep-ph]} \BibitemShut {NoStop}%
\bibitem [{\citenamefont {Chianese}\ \emph {et~al.}(2021)\citenamefont {Chianese}, \citenamefont {Fiorillo}, \citenamefont {Hajjar}, \citenamefont {Miele}, \citenamefont {Morisi},\ and\ \citenamefont {Saviano}}]{Chianese:2021htv}%
  \BibitemOpen
  \bibfield  {author} {\bibinfo {author} {\bibfnamefont {Marco}\ \bibnamefont {Chianese}}, \bibinfo {author} {\bibfnamefont {Damiano F.~G.}\ \bibnamefont {Fiorillo}}, \bibinfo {author} {\bibfnamefont {Rasmi}\ \bibnamefont {Hajjar}}, \bibinfo {author} {\bibfnamefont {Gennaro}\ \bibnamefont {Miele}}, \bibinfo {author} {\bibfnamefont {Stefano}\ \bibnamefont {Morisi}}, \ and\ \bibinfo {author} {\bibfnamefont {Ninetta}\ \bibnamefont {Saviano}},\ }\bibfield  {title} {\enquote {\bibinfo {title} {{Heavy decaying dark matter at future neutrino radio telescopes}},}\ }\href {\doibase 10.1088/1475-7516/2021/05/074} {\bibfield  {journal} {\bibinfo  {journal} {JCAP}\ }\textbf {\bibinfo {volume} {05}},\ \bibinfo {pages} {074} (\bibinfo {year} {2021})},\ \Eprint {http://arxiv.org/abs/2103.03254} {arXiv:2103.03254 [hep-ph]} \BibitemShut {NoStop}%
\bibitem [{\citenamefont {Arg\"uelles}\ \emph {et~al.}(2023{\natexlab{a}})\citenamefont {Arg\"uelles}, \citenamefont {Delgado}, \citenamefont {Friedlander}, \citenamefont {Kheirandish}, \citenamefont {Safa}, \citenamefont {Vincent},\ and\ \citenamefont {White}}]{Arguelles:2022nbl}%
  \BibitemOpen
  \bibfield  {author} {\bibinfo {author} {\bibfnamefont {Carlos~A.}\ \bibnamefont {Arg\"uelles}}, \bibinfo {author} {\bibfnamefont {Diyaselis}\ \bibnamefont {Delgado}}, \bibinfo {author} {\bibfnamefont {Avi}\ \bibnamefont {Friedlander}}, \bibinfo {author} {\bibfnamefont {Ali}\ \bibnamefont {Kheirandish}}, \bibinfo {author} {\bibfnamefont {Ibrahim}\ \bibnamefont {Safa}}, \bibinfo {author} {\bibfnamefont {Aaron~C.}\ \bibnamefont {Vincent}}, \ and\ \bibinfo {author} {\bibfnamefont {Henry}\ \bibnamefont {White}},\ }\bibfield  {title} {\enquote {\bibinfo {title} {{Dark matter decay to neutrinos}},}\ }\href {\doibase 10.1103/PhysRevD.108.123021} {\bibfield  {journal} {\bibinfo  {journal} {Phys. Rev. D}\ }\textbf {\bibinfo {volume} {108}},\ \bibinfo {pages} {123021} (\bibinfo {year} {2023}{\natexlab{a}})},\ \Eprint {http://arxiv.org/abs/2210.01303} {arXiv:2210.01303 [hep-ph]} \BibitemShut {NoStop}%
\bibitem [{\citenamefont {Fiorillo}\ \emph {et~al.}(2023)\citenamefont {Fiorillo}, \citenamefont {Valera}, \citenamefont {Bustamante},\ and\ \citenamefont {Winter}}]{Fiorillo:2023clw}%
  \BibitemOpen
  \bibfield  {author} {\bibinfo {author} {\bibfnamefont {Damiano F.~G.}\ \bibnamefont {Fiorillo}}, \bibinfo {author} {\bibfnamefont {V\'\i{}ctor~B.}\ \bibnamefont {Valera}}, \bibinfo {author} {\bibfnamefont {Mauricio}\ \bibnamefont {Bustamante}}, \ and\ \bibinfo {author} {\bibfnamefont {Walter}\ \bibnamefont {Winter}},\ }\bibfield  {title} {\enquote {\bibinfo {title} {{Searches for dark matter decay with ultrahigh-energy neutrinos endure backgrounds}},}\ }\href {\doibase 10.1103/PhysRevD.108.103012} {\bibfield  {journal} {\bibinfo  {journal} {Phys. Rev. D}\ }\textbf {\bibinfo {volume} {108}},\ \bibinfo {pages} {103012} (\bibinfo {year} {2023})},\ \Eprint {http://arxiv.org/abs/2307.02538} {arXiv:2307.02538 [astro-ph.HE]} \BibitemShut {NoStop}%
\bibitem [{\citenamefont {Aartsen}\ \emph {et~al.}(2017{\natexlab{f}})\citenamefont {Aartsen} \emph {et~al.}}]{Aartsen:2017ulx}%
  \BibitemOpen
  \bibfield  {author} {\bibinfo {author} {\bibfnamefont {M.~G.}\ \bibnamefont {Aartsen}} \emph {et~al.} (\bibinfo {collaboration} {IceCube}),\ }\bibfield  {title} {\enquote {\bibinfo {title} {{Search for Neutrinos from Dark Matter Self-Annihilations in the center of the Milky Way with 3 years of IceCube/DeepCore}},}\ }\href {\doibase 10.1140/epjc/s10052-017-5213-y} {\bibfield  {journal} {\bibinfo  {journal} {Eur. Phys. J. C}\ }\textbf {\bibinfo {volume} {77}},\ \bibinfo {pages} {627} (\bibinfo {year} {2017}{\natexlab{f}})},\ \Eprint {http://arxiv.org/abs/1705.08103} {arXiv:1705.08103 [hep-ex]} \BibitemShut {NoStop}%
\bibitem [{\citenamefont {Aartsen}\ \emph {et~al.}(2015{\natexlab{b}})\citenamefont {Aartsen} \emph {et~al.}}]{Aartsen:2015xej}%
  \BibitemOpen
  \bibfield  {author} {\bibinfo {author} {\bibfnamefont {M.~G.}\ \bibnamefont {Aartsen}} \emph {et~al.} (\bibinfo {collaboration} {IceCube}),\ }\bibfield  {title} {\enquote {\bibinfo {title} {{Search for Dark Matter Annihilation in the Galactic Center with IceCube-79}},}\ }\href {\doibase 10.1140/epjc/s10052-015-3713-1} {\bibfield  {journal} {\bibinfo  {journal} {Eur. Phys. J. C}\ }\textbf {\bibinfo {volume} {75}},\ \bibinfo {pages} {492} (\bibinfo {year} {2015}{\natexlab{b}})},\ \Eprint {http://arxiv.org/abs/1505.07259} {arXiv:1505.07259 [astro-ph.HE]} \BibitemShut {NoStop}%
\bibitem [{\citenamefont {Aartsen}\ \emph {et~al.}(2016)\citenamefont {Aartsen} \emph {et~al.}}]{Aartsen:2016pfc}%
  \BibitemOpen
  \bibfield  {author} {\bibinfo {author} {\bibfnamefont {M.~G.}\ \bibnamefont {Aartsen}} \emph {et~al.} (\bibinfo {collaboration} {IceCube}),\ }\bibfield  {title} {\enquote {\bibinfo {title} {{All-flavour Search for Neutrinos from Dark Matter Annihilations in the Milky Way with IceCube/DeepCore}},}\ }\href {\doibase 10.1140/epjc/s10052-016-4375-3} {\bibfield  {journal} {\bibinfo  {journal} {Eur. Phys. J. C}\ }\textbf {\bibinfo {volume} {76}},\ \bibinfo {pages} {531} (\bibinfo {year} {2016})},\ \Eprint {http://arxiv.org/abs/1606.00209} {arXiv:1606.00209 [astro-ph.HE]} \BibitemShut {NoStop}%
\bibitem [{\citenamefont {Abbasi}\ \emph {et~al.}(2011{\natexlab{b}})\citenamefont {Abbasi} \emph {et~al.}}]{Abbasi:2011eq}%
  \BibitemOpen
  \bibfield  {author} {\bibinfo {author} {\bibfnamefont {R.}~\bibnamefont {Abbasi}} \emph {et~al.} (\bibinfo {collaboration} {IceCube}),\ }\bibfield  {title} {\enquote {\bibinfo {title} {{Search for dark matter from the Galactic halo with the IceCube Neutrino Telescope}},}\ }\href {\doibase 10.1103/PhysRevD.84.022004} {\bibfield  {journal} {\bibinfo  {journal} {Phys. Rev. D}\ }\textbf {\bibinfo {volume} {84}},\ \bibinfo {pages} {022004} (\bibinfo {year} {2011}{\natexlab{b}})},\ \Eprint {http://arxiv.org/abs/1101.3349} {arXiv:1101.3349 [astro-ph.HE]} \BibitemShut {NoStop}%
\bibitem [{\citenamefont {Aartsen}\ \emph {et~al.}(2015{\natexlab{c}})\citenamefont {Aartsen} \emph {et~al.}}]{Aartsen:2014hva}%
  \BibitemOpen
  \bibfield  {author} {\bibinfo {author} {\bibfnamefont {M.~G.}\ \bibnamefont {Aartsen}} \emph {et~al.} (\bibinfo {collaboration} {IceCube}),\ }\bibfield  {title} {\enquote {\bibinfo {title} {{Multipole analysis of IceCube data to search for dark matter accumulated in the Galactic halo}},}\ }\href {\doibase 10.1140/epjc/s10052-014-3224-5} {\bibfield  {journal} {\bibinfo  {journal} {Eur. Phys. J. C}\ }\textbf {\bibinfo {volume} {75}},\ \bibinfo {pages} {20} (\bibinfo {year} {2015}{\natexlab{c}})},\ \Eprint {http://arxiv.org/abs/1406.6868} {arXiv:1406.6868 [astro-ph.HE]} \BibitemShut {NoStop}%
\bibitem [{\citenamefont {Aartsen}\ \emph {et~al.}(2013{\natexlab{c}})\citenamefont {Aartsen} \emph {et~al.}}]{Aartsen:2013dxa}%
  \BibitemOpen
  \bibfield  {author} {\bibinfo {author} {\bibfnamefont {M.~G.}\ \bibnamefont {Aartsen}} \emph {et~al.} (\bibinfo {collaboration} {IceCube}),\ }\bibfield  {title} {\enquote {\bibinfo {title} {{IceCube Search for Dark Matter Annihilation in nearby Galaxies and Galaxy Clusters}},}\ }\href {\doibase 10.1103/PhysRevD.88.122001} {\bibfield  {journal} {\bibinfo  {journal} {Phys. Rev. D}\ }\textbf {\bibinfo {volume} {88}},\ \bibinfo {pages} {122001} (\bibinfo {year} {2013}{\natexlab{c}})},\ \Eprint {http://arxiv.org/abs/1307.3473} {arXiv:1307.3473 [astro-ph.HE]} \BibitemShut {NoStop}%
\bibitem [{\citenamefont {Leane}\ \emph {et~al.}(2018)\citenamefont {Leane}, \citenamefont {Slatyer}, \citenamefont {Beacom},\ and\ \citenamefont {Ng}}]{Leane:2018kjk}%
  \BibitemOpen
  \bibfield  {author} {\bibinfo {author} {\bibfnamefont {Rebecca~K.}\ \bibnamefont {Leane}}, \bibinfo {author} {\bibfnamefont {Tracy~R.}\ \bibnamefont {Slatyer}}, \bibinfo {author} {\bibfnamefont {John~F.}\ \bibnamefont {Beacom}}, \ and\ \bibinfo {author} {\bibfnamefont {Kenny C.~Y.}\ \bibnamefont {Ng}},\ }\bibfield  {title} {\enquote {\bibinfo {title} {{GeV-scale thermal WIMPs: Not even slightly ruled out}},}\ }\href {\doibase 10.1103/PhysRevD.98.023016} {\bibfield  {journal} {\bibinfo  {journal} {Phys. Rev. D}\ }\textbf {\bibinfo {volume} {98}},\ \bibinfo {pages} {023016} (\bibinfo {year} {2018})},\ \Eprint {http://arxiv.org/abs/1805.10305} {arXiv:1805.10305 [hep-ph]} \BibitemShut {NoStop}%
\bibitem [{\citenamefont {Ma}(2006)}]{Ma:2006km}%
  \BibitemOpen
  \bibfield  {author} {\bibinfo {author} {\bibfnamefont {Ernest}\ \bibnamefont {Ma}},\ }\bibfield  {title} {\enquote {\bibinfo {title} {{Verifiable radiative seesaw mechanism of neutrino mass and dark matter}},}\ }\href {\doibase 10.1103/PhysRevD.73.077301} {\bibfield  {journal} {\bibinfo  {journal} {Phys. Rev. D}\ }\textbf {\bibinfo {volume} {73}},\ \bibinfo {pages} {077301} (\bibinfo {year} {2006})},\ \Eprint {http://arxiv.org/abs/hep-ph/0601225} {arXiv:hep-ph/0601225} \BibitemShut {NoStop}%
\bibitem [{\citenamefont {Kubo}\ and\ \citenamefont {Suematsu}(2006)}]{Kubo:2006rm}%
  \BibitemOpen
  \bibfield  {author} {\bibinfo {author} {\bibfnamefont {Jisuke}\ \bibnamefont {Kubo}}\ and\ \bibinfo {author} {\bibfnamefont {Daijiro}\ \bibnamefont {Suematsu}},\ }\bibfield  {title} {\enquote {\bibinfo {title} {{Neutrino masses and CDM in a non-supersymmetric model}},}\ }\href {\doibase 10.1016/j.physletb.2006.11.005} {\bibfield  {journal} {\bibinfo  {journal} {Phys. Lett. B}\ }\textbf {\bibinfo {volume} {643}},\ \bibinfo {pages} {336--341} (\bibinfo {year} {2006})},\ \Eprint {http://arxiv.org/abs/hep-ph/0610006} {arXiv:hep-ph/0610006} \BibitemShut {NoStop}%
\bibitem [{\citenamefont {Ma}(2008)}]{Ma:2008ba}%
  \BibitemOpen
  \bibfield  {author} {\bibinfo {author} {\bibfnamefont {Ernest}\ \bibnamefont {Ma}},\ }\bibfield  {title} {\enquote {\bibinfo {title} {{Supersymmetric U(1) Gauge Realization of the Dark Scalar Doublet Model of Radiative Neutrino Mass}},}\ }\href {\doibase 10.1142/S0217732308026753} {\bibfield  {journal} {\bibinfo  {journal} {Mod. Phys. Lett. A}\ }\textbf {\bibinfo {volume} {23}},\ \bibinfo {pages} {721--725} (\bibinfo {year} {2008})},\ \Eprint {http://arxiv.org/abs/0801.2545} {arXiv:0801.2545 [hep-ph]} \BibitemShut {NoStop}%
\bibitem [{\citenamefont {Ma}\ \emph {et~al.}(2013)\citenamefont {Ma}, \citenamefont {Picek},\ and\ \citenamefont {Radov\v{c}i\'c}}]{Ma:2013yga}%
  \BibitemOpen
  \bibfield  {author} {\bibinfo {author} {\bibfnamefont {Ernest}\ \bibnamefont {Ma}}, \bibinfo {author} {\bibfnamefont {Ivica}\ \bibnamefont {Picek}}, \ and\ \bibinfo {author} {\bibfnamefont {Branimir}\ \bibnamefont {Radov\v{c}i\'c}},\ }\bibfield  {title} {\enquote {\bibinfo {title} {{New Scotogenic Model of Neutrino Mass with $U(1)_D$ Gauge Interaction}},}\ }\href {\doibase 10.1016/j.physletb.2013.09.049} {\bibfield  {journal} {\bibinfo  {journal} {Phys. Lett. B}\ }\textbf {\bibinfo {volume} {726}},\ \bibinfo {pages} {744--746} (\bibinfo {year} {2013})},\ \Eprint {http://arxiv.org/abs/1308.5313} {arXiv:1308.5313 [hep-ph]} \BibitemShut {NoStop}%
\bibitem [{\citenamefont {Fraser}\ \emph {et~al.}(2014)\citenamefont {Fraser}, \citenamefont {Ma},\ and\ \citenamefont {Popov}}]{Fraser:2014yha}%
  \BibitemOpen
  \bibfield  {author} {\bibinfo {author} {\bibfnamefont {Sean}\ \bibnamefont {Fraser}}, \bibinfo {author} {\bibfnamefont {Ernest}\ \bibnamefont {Ma}}, \ and\ \bibinfo {author} {\bibfnamefont {Oleg}\ \bibnamefont {Popov}},\ }\bibfield  {title} {\enquote {\bibinfo {title} {{Scotogenic Inverse Seesaw Model of Neutrino Mass}},}\ }\href {\doibase 10.1016/j.physletb.2014.08.069} {\bibfield  {journal} {\bibinfo  {journal} {Phys. Lett. B}\ }\textbf {\bibinfo {volume} {737}},\ \bibinfo {pages} {280--282} (\bibinfo {year} {2014})},\ \Eprint {http://arxiv.org/abs/1408.4785} {arXiv:1408.4785 [hep-ph]} \BibitemShut {NoStop}%
\bibitem [{\citenamefont {Blennow}\ \emph {et~al.}(2019)\citenamefont {Blennow}, \citenamefont {Fernandez-Martinez}, \citenamefont {Olivares-Del~Campo}, \citenamefont {Pascoli}, \citenamefont {Rosauro-Alcaraz},\ and\ \citenamefont {Titov}}]{Blennow:2019fhy}%
  \BibitemOpen
  \bibfield  {author} {\bibinfo {author} {\bibfnamefont {M.}~\bibnamefont {Blennow}}, \bibinfo {author} {\bibfnamefont {E.}~\bibnamefont {Fernandez-Martinez}}, \bibinfo {author} {\bibfnamefont {A.}~\bibnamefont {Olivares-Del~Campo}}, \bibinfo {author} {\bibfnamefont {S.}~\bibnamefont {Pascoli}}, \bibinfo {author} {\bibfnamefont {S.}~\bibnamefont {Rosauro-Alcaraz}}, \ and\ \bibinfo {author} {\bibfnamefont {A.~V.}\ \bibnamefont {Titov}},\ }\bibfield  {title} {\enquote {\bibinfo {title} {{Neutrino Portals to Dark Matter}},}\ }\href {\doibase 10.1140/epjc/s10052-019-7060-5} {\bibfield  {journal} {\bibinfo  {journal} {Eur. Phys. J. C}\ }\textbf {\bibinfo {volume} {79}},\ \bibinfo {pages} {555} (\bibinfo {year} {2019})},\ \Eprint {http://arxiv.org/abs/1903.00006} {arXiv:1903.00006 [hep-ph]} \BibitemShut {NoStop}%
\bibitem [{\citenamefont {Di~Valentino}\ \emph {et~al.}(2018)\citenamefont {Di~Valentino}, \citenamefont {B\o{}ehm}, \citenamefont {Hivon},\ and\ \citenamefont {Bouchet}}]{DiValentino:2017oaw}%
  \BibitemOpen
  \bibfield  {author} {\bibinfo {author} {\bibfnamefont {Eleonora}\ \bibnamefont {Di~Valentino}}, \bibinfo {author} {\bibfnamefont {C\'eline}\ \bibnamefont {B\o{}ehm}}, \bibinfo {author} {\bibfnamefont {Eric}\ \bibnamefont {Hivon}}, \ and\ \bibinfo {author} {\bibfnamefont {Fran\c{c}ois~R.}\ \bibnamefont {Bouchet}},\ }\bibfield  {title} {\enquote {\bibinfo {title} {{Reducing the $H_0$ and $\sigma_8$ tensions with Dark Matter-neutrino interactions}},}\ }\href {\doibase 10.1103/PhysRevD.97.043513} {\bibfield  {journal} {\bibinfo  {journal} {Phys. Rev. D}\ }\textbf {\bibinfo {volume} {97}},\ \bibinfo {pages} {043513} (\bibinfo {year} {2018})},\ \Eprint {http://arxiv.org/abs/1710.02559} {arXiv:1710.02559 [astro-ph.CO]} \BibitemShut {NoStop}%
\bibitem [{\citenamefont {Hooper}\ and\ \citenamefont {Lucca}(2022)}]{Hooper:2021rjc}%
  \BibitemOpen
  \bibfield  {author} {\bibinfo {author} {\bibfnamefont {Deanna~C.}\ \bibnamefont {Hooper}}\ and\ \bibinfo {author} {\bibfnamefont {Matteo}\ \bibnamefont {Lucca}},\ }\bibfield  {title} {\enquote {\bibinfo {title} {{Hints of dark matter-neutrino interactions in Lyman-\ensuremath{\alpha} data}},}\ }\href {\doibase 10.1103/PhysRevD.105.103504} {\bibfield  {journal} {\bibinfo  {journal} {Phys. Rev. D}\ }\textbf {\bibinfo {volume} {105}},\ \bibinfo {pages} {103504} (\bibinfo {year} {2022})},\ \Eprint {http://arxiv.org/abs/2110.04024} {arXiv:2110.04024 [astro-ph.CO]} \BibitemShut {NoStop}%
\bibitem [{\citenamefont {Hooper}\ \emph {et~al.}(2022)\citenamefont {Hooper}, \citenamefont {Sch\"oneberg}, \citenamefont {Murgia}, \citenamefont {Archidiacono}, \citenamefont {Lesgourgues},\ and\ \citenamefont {Viel}}]{Hooper:2022byl}%
  \BibitemOpen
  \bibfield  {author} {\bibinfo {author} {\bibfnamefont {Deanna~C.}\ \bibnamefont {Hooper}}, \bibinfo {author} {\bibfnamefont {Nils}\ \bibnamefont {Sch\"oneberg}}, \bibinfo {author} {\bibfnamefont {Riccardo}\ \bibnamefont {Murgia}}, \bibinfo {author} {\bibfnamefont {Maria}\ \bibnamefont {Archidiacono}}, \bibinfo {author} {\bibfnamefont {Julien}\ \bibnamefont {Lesgourgues}}, \ and\ \bibinfo {author} {\bibfnamefont {Matteo}\ \bibnamefont {Viel}},\ }\bibfield  {title} {\enquote {\bibinfo {title} {{One likelihood to bind them all: Lyman-\ensuremath{\alpha} constraints on non-standard dark matter}},}\ }\href {\doibase 10.1088/1475-7516/2022/10/032} {\bibfield  {journal} {\bibinfo  {journal} {JCAP}\ }\textbf {\bibinfo {volume} {10}},\ \bibinfo {pages} {032} (\bibinfo {year} {2022})},\ \Eprint {http://arxiv.org/abs/2206.08188} {arXiv:2206.08188 [astro-ph.CO]} \BibitemShut {NoStop}%
\bibitem [{\citenamefont {Abdalla}\ \emph {et~al.}(2022)\citenamefont {Abdalla} \emph {et~al.}}]{Abdalla:2022yfr}%
  \BibitemOpen
  \bibfield  {author} {\bibinfo {author} {\bibfnamefont {Elcio}\ \bibnamefont {Abdalla}} \emph {et~al.},\ }\bibfield  {title} {\enquote {\bibinfo {title} {{Cosmology intertwined: A review of the particle physics, astrophysics, and cosmology associated with the cosmological tensions and anomalies}},}\ }\href {\doibase 10.1016/j.jheap.2022.04.002} {\bibfield  {journal} {\bibinfo  {journal} {JHEAp}\ }\textbf {\bibinfo {volume} {34}},\ \bibinfo {pages} {49--211} (\bibinfo {year} {2022})},\ \Eprint {http://arxiv.org/abs/2203.06142} {arXiv:2203.06142 [astro-ph.CO]} \BibitemShut {NoStop}%
\bibitem [{\citenamefont {Giar\`e}\ \emph {et~al.}(2024)\citenamefont {Giar\`e}, \citenamefont {G\'omez-Valent}, \citenamefont {Di~Valentino},\ and\ \citenamefont {van~de Bruck}}]{Giare:2023qqn}%
  \BibitemOpen
  \bibfield  {author} {\bibinfo {author} {\bibfnamefont {William}\ \bibnamefont {Giar\`e}}, \bibinfo {author} {\bibfnamefont {Adri\`a}\ \bibnamefont {G\'omez-Valent}}, \bibinfo {author} {\bibfnamefont {Eleonora}\ \bibnamefont {Di~Valentino}}, \ and\ \bibinfo {author} {\bibfnamefont {Carsten}\ \bibnamefont {van~de Bruck}},\ }\bibfield  {title} {\enquote {\bibinfo {title} {{Hints of neutrino dark matter scattering in the CMB? Constraints from the marginalized and profile distributions}},}\ }\href {\doibase 10.1103/PhysRevD.109.063516} {\bibfield  {journal} {\bibinfo  {journal} {Phys. Rev. D}\ }\textbf {\bibinfo {volume} {109}},\ \bibinfo {pages} {063516} (\bibinfo {year} {2024})},\ \Eprint {http://arxiv.org/abs/2311.09116} {arXiv:2311.09116 [astro-ph.CO]} \BibitemShut {NoStop}%
\bibitem [{\citenamefont {Arg\"uelles}\ \emph {et~al.}(2017)\citenamefont {Arg\"uelles}, \citenamefont {Kheirandish},\ and\ \citenamefont {Vincent}}]{Arguelles:2017atb}%
  \BibitemOpen
  \bibfield  {author} {\bibinfo {author} {\bibfnamefont {Carlos~A.}\ \bibnamefont {Arg\"uelles}}, \bibinfo {author} {\bibfnamefont {Ali}\ \bibnamefont {Kheirandish}}, \ and\ \bibinfo {author} {\bibfnamefont {Aaron~C.}\ \bibnamefont {Vincent}},\ }\bibfield  {title} {\enquote {\bibinfo {title} {{Imaging Galactic Dark Matter with High-Energy Cosmic Neutrinos}},}\ }\href {\doibase 10.1103/PhysRevLett.119.201801} {\bibfield  {journal} {\bibinfo  {journal} {Phys. Rev. Lett.}\ }\textbf {\bibinfo {volume} {119}},\ \bibinfo {pages} {201801} (\bibinfo {year} {2017})},\ \Eprint {http://arxiv.org/abs/1703.00451} {arXiv:1703.00451 [hep-ph]} \BibitemShut {NoStop}%
\bibitem [{\citenamefont {Kelly}\ and\ \citenamefont {Machado}(2018)}]{Kelly:2018tyg}%
  \BibitemOpen
  \bibfield  {author} {\bibinfo {author} {\bibfnamefont {Kevin~J.}\ \bibnamefont {Kelly}}\ and\ \bibinfo {author} {\bibfnamefont {Pedro A.~N.}\ \bibnamefont {Machado}},\ }\bibfield  {title} {\enquote {\bibinfo {title} {{Multimessenger Astronomy and New Neutrino Physics}},}\ }\href {\doibase 10.1088/1475-7516/2018/10/048} {\bibfield  {journal} {\bibinfo  {journal} {JCAP}\ }\textbf {\bibinfo {volume} {10}},\ \bibinfo {pages} {048} (\bibinfo {year} {2018})},\ \Eprint {http://arxiv.org/abs/1808.02889} {arXiv:1808.02889 [hep-ph]} \BibitemShut {NoStop}%
\bibitem [{\citenamefont {Ferrer}\ \emph {et~al.}(2023)\citenamefont {Ferrer}, \citenamefont {Herrera},\ and\ \citenamefont {Ibarra}}]{Ferrer:2022kei}%
  \BibitemOpen
  \bibfield  {author} {\bibinfo {author} {\bibfnamefont {Francesc}\ \bibnamefont {Ferrer}}, \bibinfo {author} {\bibfnamefont {Gonzalo}\ \bibnamefont {Herrera}}, \ and\ \bibinfo {author} {\bibfnamefont {Alejandro}\ \bibnamefont {Ibarra}},\ }\bibfield  {title} {\enquote {\bibinfo {title} {{New constraints on the dark matter-neutrino and dark matter-photon scattering cross sections from TXS 0506+056}},}\ }\href {\doibase 10.1088/1475-7516/2023/05/057} {\bibfield  {journal} {\bibinfo  {journal} {JCAP}\ }\textbf {\bibinfo {volume} {05}},\ \bibinfo {pages} {057} (\bibinfo {year} {2023})},\ \Eprint {http://arxiv.org/abs/2209.06339} {arXiv:2209.06339 [hep-ph]} \BibitemShut {NoStop}%
\bibitem [{\citenamefont {Abbasi}\ \emph {et~al.}(2023{\natexlab{e}})\citenamefont {Abbasi} \emph {et~al.}}]{IceCube:2022clp}%
  \BibitemOpen
  \bibfield  {author} {\bibinfo {author} {\bibfnamefont {R.}~\bibnamefont {Abbasi}} \emph {et~al.} (\bibinfo {collaboration} {IceCube}),\ }\bibfield  {title} {\enquote {\bibinfo {title} {{Searches for connections between dark matter and high-energy neutrinos with IceCube}},}\ }\href {\doibase 10.1088/1475-7516/2023/10/003} {\bibfield  {journal} {\bibinfo  {journal} {JCAP}\ }\textbf {\bibinfo {volume} {10}},\ \bibinfo {pages} {003} (\bibinfo {year} {2023}{\natexlab{e}})},\ \Eprint {http://arxiv.org/abs/2205.12950} {arXiv:2205.12950 [hep-ex]} \BibitemShut {NoStop}%
\bibitem [{\citenamefont {Herrera}\ and\ \citenamefont {Murase}(2023)}]{Herrera:2023nww}%
  \BibitemOpen
  \bibfield  {author} {\bibinfo {author} {\bibfnamefont {Gonzalo}\ \bibnamefont {Herrera}}\ and\ \bibinfo {author} {\bibfnamefont {Kohta}\ \bibnamefont {Murase}},\ }\bibfield  {title} {\enquote {\bibinfo {title} {{Probing Light Dark Matter through Cosmic-Ray Cooling in Active Galactic Nuclei}},}\ }\href {\doibase 10.1103/PhysRevD.110.L011701} {\bibfield  {journal} {\bibinfo  {journal} {Phys. Rev. D}\ }\textbf {\bibinfo {volume} {110}},\ \bibinfo {pages} {L011701} (\bibinfo {year} {2023})},\ \Eprint {http://arxiv.org/abs/2307.09460} {arXiv:2307.09460 [hep-ph]} \BibitemShut {NoStop}%
\bibitem [{\citenamefont {Abbasi}\ \emph {et~al.}(2022{\natexlab{f}})\citenamefont {Abbasi} \emph {et~al.}}]{IceCube:2021xzo}%
  \BibitemOpen
  \bibfield  {author} {\bibinfo {author} {\bibfnamefont {R.}~\bibnamefont {Abbasi}} \emph {et~al.} (\bibinfo {collaboration} {IceCube}),\ }\bibfield  {title} {\enquote {\bibinfo {title} {{Search for GeV-scale dark matter annihilation in the Sun with IceCube DeepCore}},}\ }\href {\doibase 10.1103/PhysRevD.105.062004} {\bibfield  {journal} {\bibinfo  {journal} {Phys. Rev. D}\ }\textbf {\bibinfo {volume} {105}},\ \bibinfo {pages} {062004} (\bibinfo {year} {2022}{\natexlab{f}})},\ \Eprint {http://arxiv.org/abs/2111.09970} {arXiv:2111.09970 [astro-ph.HE]} \BibitemShut {NoStop}%
\bibitem [{\citenamefont {Choi}\ \emph {et~al.}(2015)\citenamefont {Choi} \emph {et~al.}}]{Super-Kamiokande:2015xms}%
  \BibitemOpen
  \bibfield  {author} {\bibinfo {author} {\bibfnamefont {K.}~\bibnamefont {Choi}} \emph {et~al.} (\bibinfo {collaboration} {Super-Kamiokande}),\ }\bibfield  {title} {\enquote {\bibinfo {title} {{Search for neutrinos from annihilation of captured low-mass dark matter particles in the Sun by Super-Kamiokande}},}\ }\href {\doibase 10.1103/PhysRevLett.114.141301} {\bibfield  {journal} {\bibinfo  {journal} {Phys. Rev. Lett.}\ }\textbf {\bibinfo {volume} {114}},\ \bibinfo {pages} {141301} (\bibinfo {year} {2015})},\ \Eprint {http://arxiv.org/abs/1503.04858} {arXiv:1503.04858 [hep-ex]} \BibitemShut {NoStop}%
\bibitem [{\citenamefont {Adrian-Martinez}\ \emph {et~al.}(2016)\citenamefont {Adrian-Martinez} \emph {et~al.}}]{ANTARES:2016xuh}%
  \BibitemOpen
  \bibfield  {author} {\bibinfo {author} {\bibfnamefont {S.}~\bibnamefont {Adrian-Martinez}} \emph {et~al.} (\bibinfo {collaboration} {ANTARES}),\ }\bibfield  {title} {\enquote {\bibinfo {title} {{Limits on Dark Matter Annihilation in the Sun using the ANTARES Neutrino Telescope}},}\ }\href {\doibase 10.1016/j.physletb.2016.05.019} {\bibfield  {journal} {\bibinfo  {journal} {Phys. Lett. B}\ }\textbf {\bibinfo {volume} {759}},\ \bibinfo {pages} {69--74} (\bibinfo {year} {2016})},\ \Eprint {http://arxiv.org/abs/1603.02228} {arXiv:1603.02228 [astro-ph.HE]} \BibitemShut {NoStop}%
\bibitem [{\citenamefont {Amole}\ \emph {et~al.}(2019)\citenamefont {Amole} \emph {et~al.}}]{PICO:2019vsc}%
  \BibitemOpen
  \bibfield  {author} {\bibinfo {author} {\bibfnamefont {C.}~\bibnamefont {Amole}} \emph {et~al.} (\bibinfo {collaboration} {PICO}),\ }\bibfield  {title} {\enquote {\bibinfo {title} {{Dark Matter Search Results from the Complete Exposure of the PICO-60 C$_3$F$_8$ Bubble Chamber}},}\ }\href {\doibase 10.1103/PhysRevD.100.022001} {\bibfield  {journal} {\bibinfo  {journal} {Phys. Rev. D}\ }\textbf {\bibinfo {volume} {100}},\ \bibinfo {pages} {022001} (\bibinfo {year} {2019})},\ \Eprint {http://arxiv.org/abs/1902.04031} {arXiv:1902.04031 [astro-ph.CO]} \BibitemShut {NoStop}%
\bibitem [{\citenamefont {Aartsen}\ \emph {et~al.}(2013{\natexlab{d}})\citenamefont {Aartsen} \emph {et~al.}}]{Aartsen:2012kia}%
  \BibitemOpen
  \bibfield  {author} {\bibinfo {author} {\bibfnamefont {M.~G.}\ \bibnamefont {Aartsen}} \emph {et~al.} (\bibinfo {collaboration} {IceCube}),\ }\bibfield  {title} {\enquote {\bibinfo {title} {{Search for dark matter annihilations in the Sun with the 79-string IceCube detector}},}\ }\href {\doibase 10.1103/PhysRevLett.110.131302} {\bibfield  {journal} {\bibinfo  {journal} {Phys. Rev. Lett.}\ }\textbf {\bibinfo {volume} {110}},\ \bibinfo {pages} {131302} (\bibinfo {year} {2013}{\natexlab{d}})},\ \Eprint {http://arxiv.org/abs/1212.4097} {arXiv:1212.4097 [astro-ph.HE]} \BibitemShut {NoStop}%
\bibitem [{\citenamefont {Lazar}(2023)}]{Lazar:2023ymm}%
  \BibitemOpen
  \bibfield  {author} {\bibinfo {author} {\bibfnamefont {Jeffrey~P.}\ \bibnamefont {Lazar}},\ }\emph {\bibinfo {title} {{Searching for solar neutrinos and building an open-source future for neutrino astronomy}}},\ \href@noop {} {Ph.D. thesis},\ \bibinfo  {school} {U. Wisconsin, Madison (main)} (\bibinfo {year} {2023})\BibitemShut {NoStop}%
\bibitem [{\citenamefont {Cleveland}\ \emph {et~al.}(1998)\citenamefont {Cleveland}, \citenamefont {Daily}, \citenamefont {Davis}, \citenamefont {Distel}, \citenamefont {Lande}, \citenamefont {Lee}, \citenamefont {Wildenhain},\ and\ \citenamefont {Ullman}}]{Cleveland:1998nv}%
  \BibitemOpen
  \bibfield  {author} {\bibinfo {author} {\bibfnamefont {B.~T.}\ \bibnamefont {Cleveland}}, \bibinfo {author} {\bibfnamefont {Timothy}\ \bibnamefont {Daily}}, \bibinfo {author} {\bibfnamefont {Raymond}\ \bibnamefont {Davis}, \bibfnamefont {Jr.}}, \bibinfo {author} {\bibfnamefont {James~R.}\ \bibnamefont {Distel}}, \bibinfo {author} {\bibfnamefont {Kenneth}\ \bibnamefont {Lande}}, \bibinfo {author} {\bibfnamefont {C.~K.}\ \bibnamefont {Lee}}, \bibinfo {author} {\bibfnamefont {Paul~S.}\ \bibnamefont {Wildenhain}}, \ and\ \bibinfo {author} {\bibfnamefont {Jack}\ \bibnamefont {Ullman}},\ }\bibfield  {title} {\enquote {\bibinfo {title} {{Measurement of the solar electron neutrino flux with the Homestake chlorine detector}},}\ }\href {\doibase 10.1086/305343} {\bibfield  {journal} {\bibinfo  {journal} {Astrophys. J.}\ }\textbf {\bibinfo {volume} {496}},\ \bibinfo {pages} {505--526} (\bibinfo {year} {1998})}\BibitemShut {NoStop}%
\bibitem [{\citenamefont {Ahmad}\ \emph {et~al.}(2001)\citenamefont {Ahmad} \emph {et~al.}}]{SNO:2001kpb}%
  \BibitemOpen
  \bibfield  {author} {\bibinfo {author} {\bibfnamefont {Q.~R.}\ \bibnamefont {Ahmad}} \emph {et~al.} (\bibinfo {collaboration} {SNO}),\ }\bibfield  {title} {\enquote {\bibinfo {title} {{Measurement of the rate of $\nu_e+d \to p+p+e^-$ interactions produced by $^8$B solar neutrinos at the Sudbury Neutrino Observatory}},}\ }\href {\doibase 10.1103/PhysRevLett.87.071301} {\bibfield  {journal} {\bibinfo  {journal} {Phys. Rev. Lett.}\ }\textbf {\bibinfo {volume} {87}},\ \bibinfo {pages} {071301} (\bibinfo {year} {2001})},\ \Eprint {http://arxiv.org/abs/nucl-ex/0106015} {arXiv:nucl-ex/0106015} \BibitemShut {NoStop}%
\bibitem [{\citenamefont {Ahmad}\ \emph {et~al.}(2002)\citenamefont {Ahmad} \emph {et~al.}}]{SNO:2002tuh}%
  \BibitemOpen
  \bibfield  {author} {\bibinfo {author} {\bibfnamefont {Q.~R.}\ \bibnamefont {Ahmad}} \emph {et~al.} (\bibinfo {collaboration} {SNO}),\ }\bibfield  {title} {\enquote {\bibinfo {title} {{Direct evidence for neutrino flavor transformation from neutral current interactions in the Sudbury Neutrino Observatory}},}\ }\href {\doibase 10.1103/PhysRevLett.89.011301} {\bibfield  {journal} {\bibinfo  {journal} {Phys. Rev. Lett.}\ }\textbf {\bibinfo {volume} {89}},\ \bibinfo {pages} {011301} (\bibinfo {year} {2002})},\ \Eprint {http://arxiv.org/abs/nucl-ex/0204008} {arXiv:nucl-ex/0204008} \BibitemShut {NoStop}%
\bibitem [{\citenamefont {Ashie}\ \emph {et~al.}(2004)\citenamefont {Ashie} \emph {et~al.}}]{Super-Kamiokande:2004orf}%
  \BibitemOpen
  \bibfield  {author} {\bibinfo {author} {\bibfnamefont {Y.}~\bibnamefont {Ashie}} \emph {et~al.} (\bibinfo {collaboration} {Super-Kamiokande}),\ }\bibfield  {title} {\enquote {\bibinfo {title} {{Evidence for an oscillatory signature in atmospheric neutrino oscillation}},}\ }\href {\doibase 10.1103/PhysRevLett.93.101801} {\bibfield  {journal} {\bibinfo  {journal} {Phys. Rev. Lett.}\ }\textbf {\bibinfo {volume} {93}},\ \bibinfo {pages} {101801} (\bibinfo {year} {2004})},\ \Eprint {http://arxiv.org/abs/hep-ex/0404034} {arXiv:hep-ex/0404034} \BibitemShut {NoStop}%
\bibitem [{\citenamefont {Davis}(2003)}]{Davis:2003kh}%
  \BibitemOpen
  \bibfield  {author} {\bibinfo {author} {\bibfnamefont {R.}~\bibnamefont {Davis}},\ }\bibfield  {title} {\enquote {\bibinfo {title} {{Nobel Lecture: A half-century with solar neutrinos}},}\ }\href {\doibase 10.1103/RevModPhys.75.985} {\bibfield  {journal} {\bibinfo  {journal} {Rev. Mod. Phys.}\ }\textbf {\bibinfo {volume} {75}},\ \bibinfo {pages} {985--994} (\bibinfo {year} {2003})}\BibitemShut {NoStop}%
\bibitem [{\citenamefont {McDonald}(2016)}]{McDonald:2016ixn}%
  \BibitemOpen
  \bibfield  {author} {\bibinfo {author} {\bibfnamefont {Arthur~B.}\ \bibnamefont {McDonald}},\ }\bibfield  {title} {\enquote {\bibinfo {title} {{Nobel Lecture: The Sudbury Neutrino Observatory: Observation of flavor change for solar neutrinos}},}\ }\href {\doibase 10.1103/RevModPhys.88.030502} {\bibfield  {journal} {\bibinfo  {journal} {Rev. Mod. Phys.}\ }\textbf {\bibinfo {volume} {88}},\ \bibinfo {pages} {030502} (\bibinfo {year} {2016})}\BibitemShut {NoStop}%
\bibitem [{\citenamefont {Kajita}(2016)}]{Kajita:2016cak}%
  \BibitemOpen
  \bibfield  {author} {\bibinfo {author} {\bibfnamefont {Takaaki}\ \bibnamefont {Kajita}},\ }\bibfield  {title} {\enquote {\bibinfo {title} {{Nobel Lecture: Discovery of atmospheric neutrino oscillations}},}\ }\href {\doibase 10.1103/RevModPhys.88.030501} {\bibfield  {journal} {\bibinfo  {journal} {Rev. Mod. Phys.}\ }\textbf {\bibinfo {volume} {88}},\ \bibinfo {pages} {030501} (\bibinfo {year} {2016})}\BibitemShut {NoStop}%
\bibitem [{\citenamefont {Mohapatra}\ \emph {et~al.}(2007)\citenamefont {Mohapatra} \emph {et~al.}}]{Mohapatra:2005wg}%
  \BibitemOpen
  \bibfield  {author} {\bibinfo {author} {\bibfnamefont {R.~N.}\ \bibnamefont {Mohapatra}} \emph {et~al.},\ }\bibfield  {title} {\enquote {\bibinfo {title} {{Theory of neutrinos: A White paper}},}\ }\href {\doibase 10.1088/0034-4885/70/11/R02} {\bibfield  {journal} {\bibinfo  {journal} {Rept. Prog. Phys.}\ }\textbf {\bibinfo {volume} {70}},\ \bibinfo {pages} {1757--1867} (\bibinfo {year} {2007})},\ \Eprint {http://arxiv.org/abs/hep-ph/0510213} {arXiv:hep-ph/0510213} \BibitemShut {NoStop}%
\bibitem [{\citenamefont {Esteban}\ \emph {et~al.}(2020)\citenamefont {Esteban}, \citenamefont {Gonzalez-Garcia}, \citenamefont {Maltoni}, \citenamefont {Schwetz},\ and\ \citenamefont {Zhou}}]{Esteban:2020cvm}%
  \BibitemOpen
  \bibfield  {author} {\bibinfo {author} {\bibfnamefont {Ivan}\ \bibnamefont {Esteban}}, \bibinfo {author} {\bibfnamefont {M.~C.}\ \bibnamefont {Gonzalez-Garcia}}, \bibinfo {author} {\bibfnamefont {Michele}\ \bibnamefont {Maltoni}}, \bibinfo {author} {\bibfnamefont {Thomas}\ \bibnamefont {Schwetz}}, \ and\ \bibinfo {author} {\bibfnamefont {Albert}\ \bibnamefont {Zhou}},\ }\bibfield  {title} {\enquote {\bibinfo {title} {{The fate of hints: updated global analysis of three-flavor neutrino oscillations}},}\ }\href {\doibase 10.1007/JHEP09(2020)178} {\bibfield  {journal} {\bibinfo  {journal} {JHEP}\ }\textbf {\bibinfo {volume} {09}},\ \bibinfo {pages} {178} (\bibinfo {year} {2020})},\ \Eprint {http://arxiv.org/abs/2007.14792} {arXiv:2007.14792 [hep-ph]} \BibitemShut {NoStop}%
\bibitem [{\citenamefont {de~Salas}\ \emph {et~al.}(2021)\citenamefont {de~Salas}, \citenamefont {Forero}, \citenamefont {Gariazzo}, \citenamefont {Mart\'\i{}nez-Mirav\'e}, \citenamefont {Mena}, \citenamefont {Ternes}, \citenamefont {T\'ortola},\ and\ \citenamefont {Valle}}]{deSalas:2020pgw}%
  \BibitemOpen
  \bibfield  {author} {\bibinfo {author} {\bibfnamefont {P.~F.}\ \bibnamefont {de~Salas}}, \bibinfo {author} {\bibfnamefont {D.~V.}\ \bibnamefont {Forero}}, \bibinfo {author} {\bibfnamefont {S.}~\bibnamefont {Gariazzo}}, \bibinfo {author} {\bibfnamefont {P.}~\bibnamefont {Mart\'\i{}nez-Mirav\'e}}, \bibinfo {author} {\bibfnamefont {O.}~\bibnamefont {Mena}}, \bibinfo {author} {\bibfnamefont {C.~A.}\ \bibnamefont {Ternes}}, \bibinfo {author} {\bibfnamefont {M.}~\bibnamefont {T\'ortola}}, \ and\ \bibinfo {author} {\bibfnamefont {J.~W.~F.}\ \bibnamefont {Valle}},\ }\bibfield  {title} {\enquote {\bibinfo {title} {{2020 global reassessment of the neutrino oscillation picture}},}\ }\href {\doibase 10.1007/JHEP02(2021)071} {\bibfield  {journal} {\bibinfo  {journal} {JHEP}\ }\textbf {\bibinfo {volume} {02}},\ \bibinfo {pages} {071} (\bibinfo {year} {2021})},\ \Eprint {http://arxiv.org/abs/2006.11237} {arXiv:2006.11237 [hep-ph]} \BibitemShut {NoStop}%
\bibitem [{\citenamefont {Agostini}\ \emph {et~al.}(2017)\citenamefont {Agostini}, \citenamefont {Benato},\ and\ \citenamefont {Detwiler}}]{Agostini:2017jim}%
  \BibitemOpen
  \bibfield  {author} {\bibinfo {author} {\bibfnamefont {Matteo}\ \bibnamefont {Agostini}}, \bibinfo {author} {\bibfnamefont {Giovanni}\ \bibnamefont {Benato}}, \ and\ \bibinfo {author} {\bibfnamefont {Jason}\ \bibnamefont {Detwiler}},\ }\bibfield  {title} {\enquote {\bibinfo {title} {{Discovery probability of next-generation neutrinoless double- \ensuremath{\beta} decay experiments}},}\ }\href {\doibase 10.1103/PhysRevD.96.053001} {\bibfield  {journal} {\bibinfo  {journal} {Phys. Rev. D}\ }\textbf {\bibinfo {volume} {96}},\ \bibinfo {pages} {053001} (\bibinfo {year} {2017})},\ \Eprint {http://arxiv.org/abs/1705.02996} {arXiv:1705.02996 [hep-ex]} \BibitemShut {NoStop}%
\bibitem [{\citenamefont {Denton}\ and\ \citenamefont {Gehrlein}(2024)}]{Denton:2023hkx}%
  \BibitemOpen
  \bibfield  {author} {\bibinfo {author} {\bibfnamefont {Peter~B.}\ \bibnamefont {Denton}}\ and\ \bibinfo {author} {\bibfnamefont {Julia}\ \bibnamefont {Gehrlein}},\ }\bibfield  {title} {\enquote {\bibinfo {title} {{Survey of neutrino flavor predictions and the neutrinoless double beta decay funnel}},}\ }\href {\doibase 10.1103/PhysRevD.109.055028} {\bibfield  {journal} {\bibinfo  {journal} {Phys. Rev. D}\ }\textbf {\bibinfo {volume} {109}},\ \bibinfo {pages} {055028} (\bibinfo {year} {2024})},\ \Eprint {http://arxiv.org/abs/2308.09737} {arXiv:2308.09737 [hep-ph]} \BibitemShut {NoStop}%
\bibitem [{\citenamefont {Agostini}\ \emph {et~al.}(2023)\citenamefont {Agostini}, \citenamefont {Benato}, \citenamefont {Detwiler}, \citenamefont {Men\'endez},\ and\ \citenamefont {Vissani}}]{Agostini:2022zub}%
  \BibitemOpen
  \bibfield  {author} {\bibinfo {author} {\bibfnamefont {Matteo}\ \bibnamefont {Agostini}}, \bibinfo {author} {\bibfnamefont {Giovanni}\ \bibnamefont {Benato}}, \bibinfo {author} {\bibfnamefont {Jason~A.}\ \bibnamefont {Detwiler}}, \bibinfo {author} {\bibfnamefont {Javier}\ \bibnamefont {Men\'endez}}, \ and\ \bibinfo {author} {\bibfnamefont {Francesco}\ \bibnamefont {Vissani}},\ }\bibfield  {title} {\enquote {\bibinfo {title} {{Toward the discovery of matter creation with neutrinoless \ensuremath{\beta}\ensuremath{\beta} decay}},}\ }\href {\doibase 10.1103/RevModPhys.95.025002} {\bibfield  {journal} {\bibinfo  {journal} {Rev. Mod. Phys.}\ }\textbf {\bibinfo {volume} {95}},\ \bibinfo {pages} {025002} (\bibinfo {year} {2023})},\ \Eprint {http://arxiv.org/abs/2202.01787} {arXiv:2202.01787 [hep-ex]} \BibitemShut {NoStop}%
\bibitem [{\citenamefont {Ooguri}\ and\ \citenamefont {Vafa}(2017)}]{Ooguri:2016pdq}%
  \BibitemOpen
  \bibfield  {author} {\bibinfo {author} {\bibfnamefont {Hirosi}\ \bibnamefont {Ooguri}}\ and\ \bibinfo {author} {\bibfnamefont {Cumrun}\ \bibnamefont {Vafa}},\ }\bibfield  {title} {\enquote {\bibinfo {title} {{Non-supersymmetric AdS and the Swampland}},}\ }\href {\doibase 10.4310/ATMP.2017.v21.n7.a8} {\bibfield  {journal} {\bibinfo  {journal} {Adv. Theor. Math. Phys.}\ }\textbf {\bibinfo {volume} {21}},\ \bibinfo {pages} {1787--1801} (\bibinfo {year} {2017})},\ \Eprint {http://arxiv.org/abs/1610.01533} {arXiv:1610.01533 [hep-th]} \BibitemShut {NoStop}%
\bibitem [{\citenamefont {Gonzalo}\ \emph {et~al.}(2022)\citenamefont {Gonzalo}, \citenamefont {Ib\'a\~nez},\ and\ \citenamefont {Valenzuela}}]{Gonzalo:2021zsp}%
  \BibitemOpen
  \bibfield  {author} {\bibinfo {author} {\bibfnamefont {Eduardo}\ \bibnamefont {Gonzalo}}, \bibinfo {author} {\bibfnamefont {L.~E.}\ \bibnamefont {Ib\'a\~nez}}, \ and\ \bibinfo {author} {\bibfnamefont {I.}~\bibnamefont {Valenzuela}},\ }\bibfield  {title} {\enquote {\bibinfo {title} {{Swampland constraints on neutrino masses}},}\ }\href {\doibase 10.1007/JHEP02(2022)088} {\bibfield  {journal} {\bibinfo  {journal} {JHEP}\ }\textbf {\bibinfo {volume} {02}},\ \bibinfo {pages} {088} (\bibinfo {year} {2022})},\ \Eprint {http://arxiv.org/abs/2109.10961} {arXiv:2109.10961 [hep-th]} \BibitemShut {NoStop}%
\bibitem [{\citenamefont {Vafa}(2024)}]{Vafa:2024fpx}%
  \BibitemOpen
  \bibfield  {author} {\bibinfo {author} {\bibfnamefont {Cumrun}\ \bibnamefont {Vafa}},\ }\bibfield  {title} {\enquote {\bibinfo {title} {{Swamplandish Unification of the Dark Sector}},}\ }\href@noop {} {\  (\bibinfo {year} {2024})},\ \Eprint {http://arxiv.org/abs/2402.00981} {arXiv:2402.00981 [hep-ph]} \BibitemShut {NoStop}%
\bibitem [{\citenamefont {Valle}\ and\ \citenamefont {Singer}(1983)}]{Valle:1983dk}%
  \BibitemOpen
  \bibfield  {author} {\bibinfo {author} {\bibfnamefont {J.~W.~F.}\ \bibnamefont {Valle}}\ and\ \bibinfo {author} {\bibfnamefont {M.}~\bibnamefont {Singer}},\ }\bibfield  {title} {\enquote {\bibinfo {title} {{Lepton Number Violation With Quasi Dirac Neutrinos}},}\ }\href {\doibase 10.1103/PhysRevD.28.540} {\bibfield  {journal} {\bibinfo  {journal} {Phys. Rev. D}\ }\textbf {\bibinfo {volume} {28}},\ \bibinfo {pages} {540} (\bibinfo {year} {1983})}\BibitemShut {NoStop}%
\bibitem [{\citenamefont {Crocker}\ \emph {et~al.}(2002)\citenamefont {Crocker}, \citenamefont {Melia},\ and\ \citenamefont {Volkas}}]{Crocker:2001zs}%
  \BibitemOpen
  \bibfield  {author} {\bibinfo {author} {\bibfnamefont {Roland~M.}\ \bibnamefont {Crocker}}, \bibinfo {author} {\bibfnamefont {Fulvio}\ \bibnamefont {Melia}}, \ and\ \bibinfo {author} {\bibfnamefont {Raymond~R.}\ \bibnamefont {Volkas}},\ }\bibfield  {title} {\enquote {\bibinfo {title} {{Searching for long wavelength neutrino oscillations in the distorted neutrino spectrum of galactic supernova remnants}},}\ }\href {\doibase 10.1086/340278} {\bibfield  {journal} {\bibinfo  {journal} {Astrophys. J. Suppl.}\ }\textbf {\bibinfo {volume} {141}},\ \bibinfo {pages} {147--155} (\bibinfo {year} {2002})},\ \Eprint {http://arxiv.org/abs/astro-ph/0106090} {arXiv:astro-ph/0106090} \BibitemShut {NoStop}%
\bibitem [{\citenamefont {Keranen}\ \emph {et~al.}(2003)\citenamefont {Keranen}, \citenamefont {Maalampi}, \citenamefont {Myyrylainen},\ and\ \citenamefont {Riittinen}}]{Keranen:2003xd}%
  \BibitemOpen
  \bibfield  {author} {\bibinfo {author} {\bibfnamefont {P.}~\bibnamefont {Keranen}}, \bibinfo {author} {\bibfnamefont {J.}~\bibnamefont {Maalampi}}, \bibinfo {author} {\bibfnamefont {M.}~\bibnamefont {Myyrylainen}}, \ and\ \bibinfo {author} {\bibfnamefont {J.}~\bibnamefont {Riittinen}},\ }\bibfield  {title} {\enquote {\bibinfo {title} {{Effects of sterile neutrinos on the ultrahigh-energy cosmic neutrino flux}},}\ }\href {\doibase 10.1016/j.physletb.2003.09.006} {\bibfield  {journal} {\bibinfo  {journal} {Phys. Lett. B}\ }\textbf {\bibinfo {volume} {574}},\ \bibinfo {pages} {162--168} (\bibinfo {year} {2003})},\ \Eprint {http://arxiv.org/abs/hep-ph/0307041} {arXiv:hep-ph/0307041} \BibitemShut {NoStop}%
\bibitem [{\citenamefont {Beacom}\ \emph {et~al.}(2004)\citenamefont {Beacom}, \citenamefont {Bell}, \citenamefont {Hooper}, \citenamefont {Learned}, \citenamefont {Pakvasa},\ and\ \citenamefont {Weiler}}]{Beacom:2003eu}%
  \BibitemOpen
  \bibfield  {author} {\bibinfo {author} {\bibfnamefont {John~F.}\ \bibnamefont {Beacom}}, \bibinfo {author} {\bibfnamefont {Nicole~F.}\ \bibnamefont {Bell}}, \bibinfo {author} {\bibfnamefont {Dan}\ \bibnamefont {Hooper}}, \bibinfo {author} {\bibfnamefont {John~G.}\ \bibnamefont {Learned}}, \bibinfo {author} {\bibfnamefont {Sandip}\ \bibnamefont {Pakvasa}}, \ and\ \bibinfo {author} {\bibfnamefont {Thomas~J.}\ \bibnamefont {Weiler}},\ }\bibfield  {title} {\enquote {\bibinfo {title} {{PseudoDirac neutrinos: A Challenge for neutrino telescopes}},}\ }\href {\doibase 10.1103/PhysRevLett.92.011101} {\bibfield  {journal} {\bibinfo  {journal} {Phys. Rev. Lett.}\ }\textbf {\bibinfo {volume} {92}},\ \bibinfo {pages} {011101} (\bibinfo {year} {2004})},\ \Eprint {http://arxiv.org/abs/hep-ph/0307151} {arXiv:hep-ph/0307151} \BibitemShut {NoStop}%
\bibitem [{\citenamefont {Esmaili}(2010)}]{Esmaili:2009fk}%
  \BibitemOpen
  \bibfield  {author} {\bibinfo {author} {\bibfnamefont {Arman}\ \bibnamefont {Esmaili}},\ }\bibfield  {title} {\enquote {\bibinfo {title} {{Pseudo-Dirac Neutrino Scenario: Cosmic Neutrinos at Neutrino Telescopes}},}\ }\href {\doibase 10.1103/PhysRevD.81.013006} {\bibfield  {journal} {\bibinfo  {journal} {Phys. Rev. D}\ }\textbf {\bibinfo {volume} {81}},\ \bibinfo {pages} {013006} (\bibinfo {year} {2010})},\ \Eprint {http://arxiv.org/abs/0909.5410} {arXiv:0909.5410 [hep-ph]} \BibitemShut {NoStop}%
\bibitem [{\citenamefont {Esmaili}\ and\ \citenamefont {Farzan}(2012)}]{Esmaili:2012ac}%
  \BibitemOpen
  \bibfield  {author} {\bibinfo {author} {\bibfnamefont {Arman}\ \bibnamefont {Esmaili}}\ and\ \bibinfo {author} {\bibfnamefont {Yasaman}\ \bibnamefont {Farzan}},\ }\bibfield  {title} {\enquote {\bibinfo {title} {{Implications of the Pseudo-Dirac Scenario for Ultra High Energy Neutrinos from GRBs}},}\ }\href {\doibase 10.1088/1475-7516/2012/12/014} {\bibfield  {journal} {\bibinfo  {journal} {JCAP}\ }\textbf {\bibinfo {volume} {12}},\ \bibinfo {pages} {014} (\bibinfo {year} {2012})},\ \Eprint {http://arxiv.org/abs/1208.6012} {arXiv:1208.6012 [hep-ph]} \BibitemShut {NoStop}%
\bibitem [{\citenamefont {Shoemaker}\ and\ \citenamefont {Murase}(2016)}]{Shoemaker:2015qul}%
  \BibitemOpen
  \bibfield  {author} {\bibinfo {author} {\bibfnamefont {Ian~M.}\ \bibnamefont {Shoemaker}}\ and\ \bibinfo {author} {\bibfnamefont {Kohta}\ \bibnamefont {Murase}},\ }\bibfield  {title} {\enquote {\bibinfo {title} {{Probing BSM Neutrino Physics with Flavor and Spectral Distortions: Prospects for Future High-Energy Neutrino Telescopes}},}\ }\href {\doibase 10.1103/PhysRevD.93.085004} {\bibfield  {journal} {\bibinfo  {journal} {Phys. Rev. D}\ }\textbf {\bibinfo {volume} {93}},\ \bibinfo {pages} {085004} (\bibinfo {year} {2016})},\ \Eprint {http://arxiv.org/abs/1512.07228} {arXiv:1512.07228 [astro-ph.HE]} \BibitemShut {NoStop}%
\bibitem [{\citenamefont {Carloni}\ \emph {et~al.}(2024)\citenamefont {Carloni}, \citenamefont {Mart\'\i{}nez-Soler}, \citenamefont {Arguelles}, \citenamefont {Babu},\ and\ \citenamefont {Dev}}]{Carloni:2022cqz}%
  \BibitemOpen
  \bibfield  {author} {\bibinfo {author} {\bibfnamefont {Kiara}\ \bibnamefont {Carloni}}, \bibinfo {author} {\bibfnamefont {Ivan}\ \bibnamefont {Mart\'\i{}nez-Soler}}, \bibinfo {author} {\bibfnamefont {Carlos~A.}\ \bibnamefont {Arguelles}}, \bibinfo {author} {\bibfnamefont {K.~S.}\ \bibnamefont {Babu}}, \ and\ \bibinfo {author} {\bibfnamefont {P.~S.~Bhupal}\ \bibnamefont {Dev}},\ }\bibfield  {title} {\enquote {\bibinfo {title} {{Probing pseudo-Dirac neutrinos with astrophysical sources at IceCube}},}\ }\href {\doibase 10.1103/PhysRevD.109.L051702} {\bibfield  {journal} {\bibinfo  {journal} {Phys. Rev. D}\ }\textbf {\bibinfo {volume} {109}},\ \bibinfo {pages} {L051702} (\bibinfo {year} {2024})},\ \Eprint {http://arxiv.org/abs/2212.00737} {arXiv:2212.00737 [astro-ph.HE]} \BibitemShut {NoStop}%
\bibitem [{\citenamefont {Ribordy}\ and\ \citenamefont {Smirnov}(2013)}]{Ribordy:2013xea}%
  \BibitemOpen
  \bibfield  {author} {\bibinfo {author} {\bibfnamefont {Mathieu}\ \bibnamefont {Ribordy}}\ and\ \bibinfo {author} {\bibfnamefont {Alexei~Yu}\ \bibnamefont {Smirnov}},\ }\bibfield  {title} {\enquote {\bibinfo {title} {{Improving the neutrino mass hierarchy identification with inelasticity measurement in PINGU and ORCA}},}\ }\href {\doibase 10.1103/PhysRevD.87.113007} {\bibfield  {journal} {\bibinfo  {journal} {Phys. Rev. D}\ }\textbf {\bibinfo {volume} {87}},\ \bibinfo {pages} {113007} (\bibinfo {year} {2013})},\ \Eprint {http://arxiv.org/abs/1303.0758} {arXiv:1303.0758 [hep-ph]} \BibitemShut {NoStop}%
\bibitem [{\citenamefont {Aiello}\ \emph {et~al.}(2022)\citenamefont {Aiello} \emph {et~al.}}]{KM3NeT:2021ozk}%
  \BibitemOpen
  \bibfield  {author} {\bibinfo {author} {\bibfnamefont {S.}~\bibnamefont {Aiello}} \emph {et~al.} (\bibinfo {collaboration} {KM3NeT}),\ }\bibfield  {title} {\enquote {\bibinfo {title} {{Determining the neutrino mass ordering and oscillation parameters with KM3NeT/ORCA}},}\ }\href {\doibase 10.1140/epjc/s10052-021-09893-0} {\bibfield  {journal} {\bibinfo  {journal} {Eur. Phys. J. C}\ }\textbf {\bibinfo {volume} {82}},\ \bibinfo {pages} {26} (\bibinfo {year} {2022})},\ \Eprint {http://arxiv.org/abs/2103.09885} {arXiv:2103.09885 [hep-ex]} \BibitemShut {NoStop}%
\bibitem [{\citenamefont {Arg\"uelles}\ \emph {et~al.}(2023{\natexlab{b}})\citenamefont {Arg\"uelles}, \citenamefont {Fern\'andez}, \citenamefont {Mart\'\i{}nez-Soler},\ and\ \citenamefont {Jin}}]{Arguelles:2022hrt}%
  \BibitemOpen
  \bibfield  {author} {\bibinfo {author} {\bibfnamefont {C.~A.}\ \bibnamefont {Arg\"uelles}}, \bibinfo {author} {\bibfnamefont {P.}~\bibnamefont {Fern\'andez}}, \bibinfo {author} {\bibfnamefont {I.}~\bibnamefont {Mart\'\i{}nez-Soler}}, \ and\ \bibinfo {author} {\bibfnamefont {M.}~\bibnamefont {Jin}},\ }\bibfield  {title} {\enquote {\bibinfo {title} {{Measuring Oscillations with a Million Atmospheric Neutrinos}},}\ }\href {\doibase 10.1103/PhysRevX.13.041055} {\bibfield  {journal} {\bibinfo  {journal} {Phys. Rev. X}\ }\textbf {\bibinfo {volume} {13}},\ \bibinfo {pages} {041055} (\bibinfo {year} {2023}{\natexlab{b}})},\ \Eprint {http://arxiv.org/abs/2211.02666} {arXiv:2211.02666 [hep-ph]} \BibitemShut {NoStop}%
\bibitem [{\citenamefont {Eller}\ \emph {et~al.}(2023)\citenamefont {Eller} \emph {et~al.}}]{IceCube:2023ins}%
  \BibitemOpen
  \bibfield  {author} {\bibinfo {author} {\bibfnamefont {Philipp}\ \bibnamefont {Eller}} \emph {et~al.} (\bibinfo {collaboration} {IceCube}),\ }\bibfield  {title} {\enquote {\bibinfo {title} {{Sensitivity of the IceCube Upgrade to Atmospheric Neutrino Oscillations}},}\ }\href {\doibase 10.22323/1.444.1036} {\bibfield  {journal} {\bibinfo  {journal} {PoS}\ }\textbf {\bibinfo {volume} {ICRC2023}},\ \bibinfo {pages} {1036} (\bibinfo {year} {2023})},\ \Eprint {http://arxiv.org/abs/2307.15295} {arXiv:2307.15295 [astro-ph.HE]} \BibitemShut {NoStop}%
\bibitem [{\citenamefont {Hosotani}(1981)}]{Hosotani:1981mq}%
  \BibitemOpen
  \bibfield  {author} {\bibinfo {author} {\bibfnamefont {Yutaka}\ \bibnamefont {Hosotani}},\ }\bibfield  {title} {\enquote {\bibinfo {title} {{Majorana Masses, Photon Gas Heating and Cosmological Constraints on Neutrinos}},}\ }\href {\doibase 10.1016/0550-3213(82)90460-6} {\bibfield  {journal} {\bibinfo  {journal} {Nucl. Phys. B}\ }\textbf {\bibinfo {volume} {191}},\ \bibinfo {pages} {411} (\bibinfo {year} {1981})},\ \bibinfo {note} {[Erratum: Nucl.Phys.B 197, 546 (1982)]}\BibitemShut {NoStop}%
\bibitem [{\citenamefont {Pal}\ and\ \citenamefont {Wolfenstein}(1982)}]{Pal:1981rm}%
  \BibitemOpen
  \bibfield  {author} {\bibinfo {author} {\bibfnamefont {Palash~B.}\ \bibnamefont {Pal}}\ and\ \bibinfo {author} {\bibfnamefont {Lincoln}\ \bibnamefont {Wolfenstein}},\ }\bibfield  {title} {\enquote {\bibinfo {title} {{Radiative Decays of Massive Neutrinos}},}\ }\href {\doibase 10.1103/PhysRevD.25.766} {\bibfield  {journal} {\bibinfo  {journal} {Phys. Rev. D}\ }\textbf {\bibinfo {volume} {25}},\ \bibinfo {pages} {766} (\bibinfo {year} {1982})}\BibitemShut {NoStop}%
\bibitem [{\citenamefont {Nieves}(1983)}]{Nieves:1982bq}%
  \BibitemOpen
  \bibfield  {author} {\bibinfo {author} {\bibfnamefont {Jose~F.}\ \bibnamefont {Nieves}},\ }\bibfield  {title} {\enquote {\bibinfo {title} {{Two Photon Decays of Heavy Neutrinos}},}\ }\href {\doibase 10.1103/PhysRevD.28.1664} {\bibfield  {journal} {\bibinfo  {journal} {Phys. Rev. D}\ }\textbf {\bibinfo {volume} {28}},\ \bibinfo {pages} {1664} (\bibinfo {year} {1983})}\BibitemShut {NoStop}%
\bibitem [{\citenamefont {Bahcall}\ \emph {et~al.}(1972)\citenamefont {Bahcall}, \citenamefont {Cabibbo},\ and\ \citenamefont {Yahil}}]{Bahcall:1972my}%
  \BibitemOpen
  \bibfield  {author} {\bibinfo {author} {\bibfnamefont {John~N.}\ \bibnamefont {Bahcall}}, \bibinfo {author} {\bibfnamefont {N.}~\bibnamefont {Cabibbo}}, \ and\ \bibinfo {author} {\bibfnamefont {A.}~\bibnamefont {Yahil}},\ }\bibfield  {title} {\enquote {\bibinfo {title} {{Are neutrinos stable particles?}}}\ }\href {\doibase 10.1103/PhysRevLett.28.316} {\bibfield  {journal} {\bibinfo  {journal} {Phys. Rev. Lett.}\ }\textbf {\bibinfo {volume} {28}},\ \bibinfo {pages} {316--318} (\bibinfo {year} {1972})}\BibitemShut {NoStop}%
\bibitem [{\citenamefont {Chikashige}\ \emph {et~al.}(1980)\citenamefont {Chikashige}, \citenamefont {Mohapatra},\ and\ \citenamefont {Peccei}}]{Chikashige:1980qk}%
  \BibitemOpen
  \bibfield  {author} {\bibinfo {author} {\bibfnamefont {Y.}~\bibnamefont {Chikashige}}, \bibinfo {author} {\bibfnamefont {Rabindra~N.}\ \bibnamefont {Mohapatra}}, \ and\ \bibinfo {author} {\bibfnamefont {R.~D.}\ \bibnamefont {Peccei}},\ }\bibfield  {title} {\enquote {\bibinfo {title} {{Spontaneously Broken Lepton Number and Cosmological Constraints on the Neutrino Mass Spectrum}},}\ }\href {\doibase 10.1103/PhysRevLett.45.1926} {\bibfield  {journal} {\bibinfo  {journal} {Phys. Rev. Lett.}\ }\textbf {\bibinfo {volume} {45}},\ \bibinfo {pages} {1926} (\bibinfo {year} {1980})}\BibitemShut {NoStop}%
\bibitem [{\citenamefont {Beacom}\ \emph {et~al.}(2003{\natexlab{a}})\citenamefont {Beacom}, \citenamefont {Bell}, \citenamefont {Hooper}, \citenamefont {Pakvasa},\ and\ \citenamefont {Weiler}}]{Beacom:2002vi}%
  \BibitemOpen
  \bibfield  {author} {\bibinfo {author} {\bibfnamefont {John~F.}\ \bibnamefont {Beacom}}, \bibinfo {author} {\bibfnamefont {Nicole~F.}\ \bibnamefont {Bell}}, \bibinfo {author} {\bibfnamefont {Dan}\ \bibnamefont {Hooper}}, \bibinfo {author} {\bibfnamefont {Sandip}\ \bibnamefont {Pakvasa}}, \ and\ \bibinfo {author} {\bibfnamefont {Thomas~J.}\ \bibnamefont {Weiler}},\ }\bibfield  {title} {\enquote {\bibinfo {title} {{Decay of High-Energy Astrophysical Neutrinos}},}\ }\href {\doibase 10.1103/PhysRevLett.90.181301} {\bibfield  {journal} {\bibinfo  {journal} {Phys. Rev. Lett.}\ }\textbf {\bibinfo {volume} {90}},\ \bibinfo {pages} {181301} (\bibinfo {year} {2003}{\natexlab{a}})},\ \Eprint {http://arxiv.org/abs/hep-ph/0211305} {arXiv:hep-ph/0211305} \BibitemShut {NoStop}%
\bibitem [{\citenamefont {Bustamante}\ \emph {et~al.}(2017)\citenamefont {Bustamante}, \citenamefont {Beacom},\ and\ \citenamefont {Murase}}]{Bustamante:2016ciw}%
  \BibitemOpen
  \bibfield  {author} {\bibinfo {author} {\bibfnamefont {Mauricio}\ \bibnamefont {Bustamante}}, \bibinfo {author} {\bibfnamefont {John~F.}\ \bibnamefont {Beacom}}, \ and\ \bibinfo {author} {\bibfnamefont {Kohta}\ \bibnamefont {Murase}},\ }\bibfield  {title} {\enquote {\bibinfo {title} {{Testing decay of astrophysical neutrinos with incomplete information}},}\ }\href {\doibase 10.1103/PhysRevD.95.063013} {\bibfield  {journal} {\bibinfo  {journal} {Phys. Rev. D}\ }\textbf {\bibinfo {volume} {95}},\ \bibinfo {pages} {063013} (\bibinfo {year} {2017})},\ \Eprint {http://arxiv.org/abs/1610.02096} {arXiv:1610.02096 [astro-ph.HE]} \BibitemShut {NoStop}%
\bibitem [{\citenamefont {Denton}\ and\ \citenamefont {Tamborra}(2018)}]{Denton:2018aml}%
  \BibitemOpen
  \bibfield  {author} {\bibinfo {author} {\bibfnamefont {Peter~B.}\ \bibnamefont {Denton}}\ and\ \bibinfo {author} {\bibfnamefont {Irene}\ \bibnamefont {Tamborra}},\ }\bibfield  {title} {\enquote {\bibinfo {title} {{Invisible Neutrino Decay Could Resolve IceCube\textquoteright{}s Track and Cascade Tension}},}\ }\href {\doibase 10.1103/PhysRevLett.121.121802} {\bibfield  {journal} {\bibinfo  {journal} {Phys. Rev. Lett.}\ }\textbf {\bibinfo {volume} {121}},\ \bibinfo {pages} {121802} (\bibinfo {year} {2018})},\ \Eprint {http://arxiv.org/abs/1805.05950} {arXiv:1805.05950 [hep-ph]} \BibitemShut {NoStop}%
\bibitem [{\citenamefont {Song}\ \emph {et~al.}(2021)\citenamefont {Song}, \citenamefont {Li}, \citenamefont {Arg\"uelles}, \citenamefont {Bustamante},\ and\ \citenamefont {Vincent}}]{Song:2020nfh}%
  \BibitemOpen
  \bibfield  {author} {\bibinfo {author} {\bibfnamefont {Ningqiang}\ \bibnamefont {Song}}, \bibinfo {author} {\bibfnamefont {Shirley~Weishi}\ \bibnamefont {Li}}, \bibinfo {author} {\bibfnamefont {Carlos~A.}\ \bibnamefont {Arg\"uelles}}, \bibinfo {author} {\bibfnamefont {Mauricio}\ \bibnamefont {Bustamante}}, \ and\ \bibinfo {author} {\bibfnamefont {Aaron~C.}\ \bibnamefont {Vincent}},\ }\bibfield  {title} {\enquote {\bibinfo {title} {{The Future of High-Energy Astrophysical Neutrino Flavor Measurements}},}\ }\href {\doibase 10.1088/1475-7516/2021/04/054} {\bibfield  {journal} {\bibinfo  {journal} {JCAP}\ }\textbf {\bibinfo {volume} {04}},\ \bibinfo {pages} {054} (\bibinfo {year} {2021})},\ \Eprint {http://arxiv.org/abs/2012.12893} {arXiv:2012.12893 [hep-ph]} \BibitemShut {NoStop}%
\bibitem [{\citenamefont {Abdullahi}\ and\ \citenamefont {Denton}(2020)}]{Abdullahi:2020rge}%
  \BibitemOpen
  \bibfield  {author} {\bibinfo {author} {\bibfnamefont {Asli}\ \bibnamefont {Abdullahi}}\ and\ \bibinfo {author} {\bibfnamefont {Peter~B.}\ \bibnamefont {Denton}},\ }\bibfield  {title} {\enquote {\bibinfo {title} {{Visible Decay of Astrophysical Neutrinos at IceCube}},}\ }\href {\doibase 10.1103/PhysRevD.102.023018} {\bibfield  {journal} {\bibinfo  {journal} {Phys. Rev. D}\ }\textbf {\bibinfo {volume} {102}},\ \bibinfo {pages} {023018} (\bibinfo {year} {2020})},\ \Eprint {http://arxiv.org/abs/2005.07200} {arXiv:2005.07200 [hep-ph]} \BibitemShut {NoStop}%
\bibitem [{\citenamefont {Liu}\ \emph {et~al.}(2023)\citenamefont {Liu}, \citenamefont {Fiorillo}, \citenamefont {Arg\"uelles}, \citenamefont {Bustamante}, \citenamefont {Song},\ and\ \citenamefont {Vincent}}]{Liu:2023flr}%
  \BibitemOpen
  \bibfield  {author} {\bibinfo {author} {\bibfnamefont {Qinrui}\ \bibnamefont {Liu}}, \bibinfo {author} {\bibfnamefont {Damiano F.~G.}\ \bibnamefont {Fiorillo}}, \bibinfo {author} {\bibfnamefont {Carlos~A.}\ \bibnamefont {Arg\"uelles}}, \bibinfo {author} {\bibfnamefont {Mauricio}\ \bibnamefont {Bustamante}}, \bibinfo {author} {\bibfnamefont {Ningqiang}\ \bibnamefont {Song}}, \ and\ \bibinfo {author} {\bibfnamefont {Aaron~C.}\ \bibnamefont {Vincent}},\ }\bibfield  {title} {\enquote {\bibinfo {title} {{Identifying Energy-Dependent Flavor Transitions in High-Energy Astrophysical Neutrino Measurements}},}\ }\href@noop {} {\  (\bibinfo {year} {2023})},\ \Eprint {http://arxiv.org/abs/2312.07649} {arXiv:2312.07649 [astro-ph.HE]} \BibitemShut {NoStop}%
\bibitem [{\citenamefont {Aartsen}\ \emph {et~al.}(2015{\natexlab{d}})\citenamefont {Aartsen} \emph {et~al.}}]{IceCube:2015gsk}%
  \BibitemOpen
  \bibfield  {author} {\bibinfo {author} {\bibfnamefont {M.~G.}\ \bibnamefont {Aartsen}} \emph {et~al.} (\bibinfo {collaboration} {IceCube}),\ }\bibfield  {title} {\enquote {\bibinfo {title} {{A combined maximum-likelihood analysis of the high-energy astrophysical neutrino flux measured with IceCube}},}\ }\href {\doibase 10.1088/0004-637X/809/1/98} {\bibfield  {journal} {\bibinfo  {journal} {Astrophys. J.}\ }\textbf {\bibinfo {volume} {809}},\ \bibinfo {pages} {98} (\bibinfo {year} {2015}{\natexlab{d}})},\ \Eprint {http://arxiv.org/abs/1507.03991} {arXiv:1507.03991 [astro-ph.HE]} \BibitemShut {NoStop}%
\bibitem [{\citenamefont {Brdar}\ \emph {et~al.}(2017)\citenamefont {Brdar}, \citenamefont {Kopp},\ and\ \citenamefont {Wang}}]{Brdar:2016thq}%
  \BibitemOpen
  \bibfield  {author} {\bibinfo {author} {\bibfnamefont {Vedran}\ \bibnamefont {Brdar}}, \bibinfo {author} {\bibfnamefont {Joachim}\ \bibnamefont {Kopp}}, \ and\ \bibinfo {author} {\bibfnamefont {Xiao-Ping}\ \bibnamefont {Wang}},\ }\bibfield  {title} {\enquote {\bibinfo {title} {{Sterile Neutrinos and Flavor Ratios in IceCube}},}\ }\href {\doibase 10.1088/1475-7516/2017/01/026} {\bibfield  {journal} {\bibinfo  {journal} {JCAP}\ }\textbf {\bibinfo {volume} {01}},\ \bibinfo {pages} {026} (\bibinfo {year} {2017})},\ \Eprint {http://arxiv.org/abs/1611.04598} {arXiv:1611.04598 [hep-ph]} \BibitemShut {NoStop}%
\bibitem [{\citenamefont {Arg\"uelles}\ \emph {et~al.}(2020{\natexlab{a}})\citenamefont {Arg\"uelles}, \citenamefont {Farrag}, \citenamefont {Katori}, \citenamefont {Khandelwal}, \citenamefont {Mandalia},\ and\ \citenamefont {Salvado}}]{Arguelles:2019tum}%
  \BibitemOpen
  \bibfield  {author} {\bibinfo {author} {\bibfnamefont {Carlos~A.}\ \bibnamefont {Arg\"uelles}}, \bibinfo {author} {\bibfnamefont {Kareem}\ \bibnamefont {Farrag}}, \bibinfo {author} {\bibfnamefont {Teppei}\ \bibnamefont {Katori}}, \bibinfo {author} {\bibfnamefont {Rishabh}\ \bibnamefont {Khandelwal}}, \bibinfo {author} {\bibfnamefont {Shivesh}\ \bibnamefont {Mandalia}}, \ and\ \bibinfo {author} {\bibfnamefont {Jordi}\ \bibnamefont {Salvado}},\ }\bibfield  {title} {\enquote {\bibinfo {title} {{Sterile neutrinos in astrophysical neutrino flavor}},}\ }\href {\doibase 10.1088/1475-7516/2020/02/015} {\bibfield  {journal} {\bibinfo  {journal} {JCAP}\ }\textbf {\bibinfo {volume} {02}},\ \bibinfo {pages} {015} (\bibinfo {year} {2020}{\natexlab{a}})},\ \Eprint {http://arxiv.org/abs/1909.05341} {arXiv:1909.05341 [hep-ph]} \BibitemShut {NoStop}%
\bibitem [{\citenamefont {Parke}\ and\ \citenamefont {Ross-Lonergan}(2016)}]{Parke:2015goa}%
  \BibitemOpen
  \bibfield  {author} {\bibinfo {author} {\bibfnamefont {Stephen}\ \bibnamefont {Parke}}\ and\ \bibinfo {author} {\bibfnamefont {Mark}\ \bibnamefont {Ross-Lonergan}},\ }\bibfield  {title} {\enquote {\bibinfo {title} {{Unitarity and the three flavor neutrino mixing matrix}},}\ }\href {\doibase 10.1103/PhysRevD.93.113009} {\bibfield  {journal} {\bibinfo  {journal} {Phys. Rev. D}\ }\textbf {\bibinfo {volume} {93}},\ \bibinfo {pages} {113009} (\bibinfo {year} {2016})},\ \Eprint {http://arxiv.org/abs/1508.05095} {arXiv:1508.05095 [hep-ph]} \BibitemShut {NoStop}%
\bibitem [{\citenamefont {Ellis}\ \emph {et~al.}(2020)\citenamefont {Ellis}, \citenamefont {Kelly},\ and\ \citenamefont {Li}}]{Ellis:2020hus}%
  \BibitemOpen
  \bibfield  {author} {\bibinfo {author} {\bibfnamefont {Sebastian A.~R.}\ \bibnamefont {Ellis}}, \bibinfo {author} {\bibfnamefont {Kevin~J.}\ \bibnamefont {Kelly}}, \ and\ \bibinfo {author} {\bibfnamefont {Shirley~Weishi}\ \bibnamefont {Li}},\ }\bibfield  {title} {\enquote {\bibinfo {title} {{Current and Future Neutrino Oscillation Constraints on Leptonic Unitarity}},}\ }\href {\doibase 10.1007/JHEP12(2020)068} {\bibfield  {journal} {\bibinfo  {journal} {JHEP}\ }\textbf {\bibinfo {volume} {12}},\ \bibinfo {pages} {068} (\bibinfo {year} {2020})},\ \Eprint {http://arxiv.org/abs/2008.01088} {arXiv:2008.01088 [hep-ph]} \BibitemShut {NoStop}%
\bibitem [{\citenamefont {Athanassopoulos}\ \emph {et~al.}(1996)\citenamefont {Athanassopoulos} \emph {et~al.}}]{LSND:1996ubh}%
  \BibitemOpen
  \bibfield  {author} {\bibinfo {author} {\bibfnamefont {C.}~\bibnamefont {Athanassopoulos}} \emph {et~al.} (\bibinfo {collaboration} {LSND}),\ }\bibfield  {title} {\enquote {\bibinfo {title} {{Evidence for anti-muon-neutrino ---\ensuremath{>} anti-electron-neutrino oscillations from the LSND experiment at LAMPF}},}\ }\href {\doibase 10.1103/PhysRevLett.77.3082} {\bibfield  {journal} {\bibinfo  {journal} {Phys. Rev. Lett.}\ }\textbf {\bibinfo {volume} {77}},\ \bibinfo {pages} {3082--3085} (\bibinfo {year} {1996})},\ \Eprint {http://arxiv.org/abs/nucl-ex/9605003} {arXiv:nucl-ex/9605003} \BibitemShut {NoStop}%
\bibitem [{\citenamefont {Aguilar-Arevalo}\ \emph {et~al.}(2013)\citenamefont {Aguilar-Arevalo} \emph {et~al.}}]{MiniBooNE:2013uba}%
  \BibitemOpen
  \bibfield  {author} {\bibinfo {author} {\bibfnamefont {A.~A.}\ \bibnamefont {Aguilar-Arevalo}} \emph {et~al.} (\bibinfo {collaboration} {MiniBooNE}),\ }\bibfield  {title} {\enquote {\bibinfo {title} {{Improved Search for $\bar \nu_\mu \rightarrow \bar \nu_e$ Oscillations in the MiniBooNE Experiment}},}\ }\href {\doibase 10.1103/PhysRevLett.110.161801} {\bibfield  {journal} {\bibinfo  {journal} {Phys. Rev. Lett.}\ }\textbf {\bibinfo {volume} {110}},\ \bibinfo {pages} {161801} (\bibinfo {year} {2013})},\ \Eprint {http://arxiv.org/abs/1303.2588} {arXiv:1303.2588 [hep-ex]} \BibitemShut {NoStop}%
\bibitem [{\citenamefont {Abdurashitov}\ \emph {et~al.}(1999)\citenamefont {Abdurashitov} \emph {et~al.}}]{SAGE:1998fvr}%
  \BibitemOpen
  \bibfield  {author} {\bibinfo {author} {\bibfnamefont {J.~N.}\ \bibnamefont {Abdurashitov}} \emph {et~al.} (\bibinfo {collaboration} {SAGE}),\ }\bibfield  {title} {\enquote {\bibinfo {title} {{Measurement of the response of the Russian-American gallium experiment to neutrinos from a Cr-51 source}},}\ }\href {\doibase 10.1103/PhysRevC.59.2246} {\bibfield  {journal} {\bibinfo  {journal} {Phys. Rev. C}\ }\textbf {\bibinfo {volume} {59}},\ \bibinfo {pages} {2246--2263} (\bibinfo {year} {1999})},\ \Eprint {http://arxiv.org/abs/hep-ph/9803418} {arXiv:hep-ph/9803418} \BibitemShut {NoStop}%
\bibitem [{\citenamefont {Hampel}\ \emph {et~al.}(1998)\citenamefont {Hampel} \emph {et~al.}}]{GALLEX:1997lja}%
  \BibitemOpen
  \bibfield  {author} {\bibinfo {author} {\bibfnamefont {W.}~\bibnamefont {Hampel}} \emph {et~al.} (\bibinfo {collaboration} {GALLEX}),\ }\bibfield  {title} {\enquote {\bibinfo {title} {{Final results of the Cr-51 neutrino source experiments in GALLEX}},}\ }\href {\doibase 10.1016/S0370-2693(97)01562-1} {\bibfield  {journal} {\bibinfo  {journal} {Phys. Lett. B}\ }\textbf {\bibinfo {volume} {420}},\ \bibinfo {pages} {114--126} (\bibinfo {year} {1998})}\BibitemShut {NoStop}%
\bibitem [{\citenamefont {Barinov}\ \emph {et~al.}(2022)\citenamefont {Barinov} \emph {et~al.}}]{Barinov:2021asz}%
  \BibitemOpen
  \bibfield  {author} {\bibinfo {author} {\bibfnamefont {V.~V.}\ \bibnamefont {Barinov}} \emph {et~al.},\ }\bibfield  {title} {\enquote {\bibinfo {title} {{Results from the Baksan Experiment on Sterile Transitions (BEST)}},}\ }\href {\doibase 10.1103/PhysRevLett.128.232501} {\bibfield  {journal} {\bibinfo  {journal} {Phys. Rev. Lett.}\ }\textbf {\bibinfo {volume} {128}},\ \bibinfo {pages} {232501} (\bibinfo {year} {2022})},\ \Eprint {http://arxiv.org/abs/2109.11482} {arXiv:2109.11482 [nucl-ex]} \BibitemShut {NoStop}%
\bibitem [{\citenamefont {Serebrov}\ \emph {et~al.}(2022)\citenamefont {Serebrov}, \citenamefont {Samoilov}, \citenamefont {Chaikovskii},\ and\ \citenamefont {Zherebtsov}}]{Serebrov:2022ajm}%
  \BibitemOpen
  \bibfield  {author} {\bibinfo {author} {\bibfnamefont {A.~P.}\ \bibnamefont {Serebrov}}, \bibinfo {author} {\bibfnamefont {R.~M.}\ \bibnamefont {Samoilov}}, \bibinfo {author} {\bibfnamefont {M.~E.}\ \bibnamefont {Chaikovskii}}, \ and\ \bibinfo {author} {\bibfnamefont {O.~M.}\ \bibnamefont {Zherebtsov}},\ }\bibfield  {title} {\enquote {\bibinfo {title} {{Result of the Neutrino-4 Experiment and the Cosmological Constraints on the Sterile Neutrino (Brief Review)}},}\ }\href {\doibase 10.1134/S002136402260224X} {\bibfield  {journal} {\bibinfo  {journal} {JETP Lett.}\ }\textbf {\bibinfo {volume} {116}},\ \bibinfo {pages} {669--682} (\bibinfo {year} {2022})},\ \Eprint {http://arxiv.org/abs/2203.09401} {arXiv:2203.09401 [hep-ph]} \BibitemShut {NoStop}%
\bibitem [{\citenamefont {Arg\"uelles}\ \emph {et~al.}(2022)\citenamefont {Arg\"uelles}, \citenamefont {Esteban}, \citenamefont {Hostert}, \citenamefont {Kelly}, \citenamefont {Kopp}, \citenamefont {Machado}, \citenamefont {Martinez-Soler},\ and\ \citenamefont {Perez-Gonzalez}}]{Arguelles:2021meu}%
  \BibitemOpen
  \bibfield  {author} {\bibinfo {author} {\bibfnamefont {C.~A.}\ \bibnamefont {Arg\"uelles}}, \bibinfo {author} {\bibfnamefont {I.}~\bibnamefont {Esteban}}, \bibinfo {author} {\bibfnamefont {M.}~\bibnamefont {Hostert}}, \bibinfo {author} {\bibfnamefont {Kevin~J.}\ \bibnamefont {Kelly}}, \bibinfo {author} {\bibfnamefont {J.}~\bibnamefont {Kopp}}, \bibinfo {author} {\bibfnamefont {P.~A.~N.}\ \bibnamefont {Machado}}, \bibinfo {author} {\bibfnamefont {I.}~\bibnamefont {Martinez-Soler}}, \ and\ \bibinfo {author} {\bibfnamefont {Y.~F.}\ \bibnamefont {Perez-Gonzalez}},\ }\bibfield  {title} {\enquote {\bibinfo {title} {{MicroBooNE and the \ensuremath{\nu}e Interpretation of the MiniBooNE Low-Energy Excess}},}\ }\href {\doibase 10.1103/PhysRevLett.128.241802} {\bibfield  {journal} {\bibinfo  {journal} {Phys. Rev. Lett.}\ }\textbf {\bibinfo {volume} {128}},\ \bibinfo {pages} {241802} (\bibinfo {year} {2022})},\ \Eprint {http://arxiv.org/abs/2111.10359} {arXiv:2111.10359 [hep-ph]} \BibitemShut {NoStop}%
\bibitem [{\citenamefont {Adamson}\ \emph {et~al.}(2019)\citenamefont {Adamson} \emph {et~al.}}]{MINOS:2017cae}%
  \BibitemOpen
  \bibfield  {author} {\bibinfo {author} {\bibfnamefont {P.}~\bibnamefont {Adamson}} \emph {et~al.} (\bibinfo {collaboration} {MINOS+}),\ }\bibfield  {title} {\enquote {\bibinfo {title} {{Search for sterile neutrinos in MINOS and MINOS+ using a two-detector fit}},}\ }\href {\doibase 10.1103/PhysRevLett.122.091803} {\bibfield  {journal} {\bibinfo  {journal} {Phys. Rev. Lett.}\ }\textbf {\bibinfo {volume} {122}},\ \bibinfo {pages} {091803} (\bibinfo {year} {2019})},\ \Eprint {http://arxiv.org/abs/1710.06488} {arXiv:1710.06488 [hep-ex]} \BibitemShut {NoStop}%
\bibitem [{\citenamefont {Montero}\ \emph {et~al.}(2023)\citenamefont {Montero}, \citenamefont {Vafa},\ and\ \citenamefont {Valenzuela}}]{Montero:2022prj}%
  \BibitemOpen
  \bibfield  {author} {\bibinfo {author} {\bibfnamefont {Miguel}\ \bibnamefont {Montero}}, \bibinfo {author} {\bibfnamefont {Cumrun}\ \bibnamefont {Vafa}}, \ and\ \bibinfo {author} {\bibfnamefont {Irene}\ \bibnamefont {Valenzuela}},\ }\bibfield  {title} {\enquote {\bibinfo {title} {{The dark dimension and the Swampland}},}\ }\href {\doibase 10.1007/JHEP02(2023)022} {\bibfield  {journal} {\bibinfo  {journal} {JHEP}\ }\textbf {\bibinfo {volume} {02}},\ \bibinfo {pages} {022} (\bibinfo {year} {2023})},\ \Eprint {http://arxiv.org/abs/2205.12293} {arXiv:2205.12293 [hep-th]} \BibitemShut {NoStop}%
\bibitem [{\citenamefont {Beacom}\ \emph {et~al.}(2003{\natexlab{b}})\citenamefont {Beacom}, \citenamefont {Bell}, \citenamefont {Hooper}, \citenamefont {Pakvasa},\ and\ \citenamefont {Weiler}}]{Beacom:2003nh}%
  \BibitemOpen
  \bibfield  {author} {\bibinfo {author} {\bibfnamefont {John~F.}\ \bibnamefont {Beacom}}, \bibinfo {author} {\bibfnamefont {Nicole~F.}\ \bibnamefont {Bell}}, \bibinfo {author} {\bibfnamefont {Dan}\ \bibnamefont {Hooper}}, \bibinfo {author} {\bibfnamefont {Sandip}\ \bibnamefont {Pakvasa}}, \ and\ \bibinfo {author} {\bibfnamefont {Thomas~J.}\ \bibnamefont {Weiler}},\ }\bibfield  {title} {\enquote {\bibinfo {title} {{Measuring flavor ratios of high-energy astrophysical neutrinos}},}\ }\href {\doibase 10.1103/PhysRevD.68.093005} {\bibfield  {journal} {\bibinfo  {journal} {Phys. Rev. D}\ }\textbf {\bibinfo {volume} {68}},\ \bibinfo {pages} {093005} (\bibinfo {year} {2003}{\natexlab{b}})},\ \bibinfo {note} {[Erratum: Phys.Rev.D 72, 019901 (2005)]},\ \Eprint {http://arxiv.org/abs/hep-ph/0307025} {arXiv:hep-ph/0307025} \BibitemShut {NoStop}%
\bibitem [{\citenamefont {Arg\"uelles}\ \emph {et~al.}(2020{\natexlab{b}})\citenamefont {Arg\"uelles}, \citenamefont {Bustamante}, \citenamefont {Kheirandish}, \citenamefont {Palomares-Ruiz}, \citenamefont {Salvado},\ and\ \citenamefont {Vincent}}]{Arguelles:2019rbn}%
  \BibitemOpen
  \bibfield  {author} {\bibinfo {author} {\bibfnamefont {Carlos~A.}\ \bibnamefont {Arg\"uelles}}, \bibinfo {author} {\bibfnamefont {Mauricio}\ \bibnamefont {Bustamante}}, \bibinfo {author} {\bibfnamefont {Ali}\ \bibnamefont {Kheirandish}}, \bibinfo {author} {\bibfnamefont {Sergio}\ \bibnamefont {Palomares-Ruiz}}, \bibinfo {author} {\bibfnamefont {Jordi}\ \bibnamefont {Salvado}}, \ and\ \bibinfo {author} {\bibfnamefont {Aaron~C.}\ \bibnamefont {Vincent}},\ }\bibfield  {title} {\enquote {\bibinfo {title} {{Fundamental physics with high-energy cosmic neutrinos today and in the future}},}\ }\href {\doibase 10.22323/1.358.0849} {\bibfield  {journal} {\bibinfo  {journal} {PoS}\ }\textbf {\bibinfo {volume} {ICRC2019}},\ \bibinfo {pages} {849} (\bibinfo {year} {2020}{\natexlab{b}})},\ \Eprint {http://arxiv.org/abs/1907.08690} {arXiv:1907.08690 [astro-ph.HE]} \BibitemShut {NoStop}%
\bibitem [{\citenamefont {Ackermann}\ \emph {et~al.}(2019)\citenamefont {Ackermann} \emph {et~al.}}]{Ackermann:2019cxh}%
  \BibitemOpen
  \bibfield  {author} {\bibinfo {author} {\bibfnamefont {Markus}\ \bibnamefont {Ackermann}} \emph {et~al.},\ }\bibfield  {title} {\enquote {\bibinfo {title} {{Fundamental Physics with High-Energy Cosmic Neutrinos}},}\ }\href@noop {} {\bibfield  {journal} {\bibinfo  {journal} {Bull. Am. Astron. Soc.}\ }\textbf {\bibinfo {volume} {51}},\ \bibinfo {pages} {215} (\bibinfo {year} {2019})},\ \Eprint {http://arxiv.org/abs/1903.04333} {arXiv:1903.04333 [astro-ph.HE]} \BibitemShut {NoStop}%
\bibitem [{\citenamefont {Abbasi}\ \emph {et~al.}(2022{\natexlab{g}})\citenamefont {Abbasi} \emph {et~al.}}]{IceCube:2021tdn}%
  \BibitemOpen
  \bibfield  {author} {\bibinfo {author} {\bibfnamefont {R.}~\bibnamefont {Abbasi}} \emph {et~al.} (\bibinfo {collaboration} {IceCube}),\ }\bibfield  {title} {\enquote {\bibinfo {title} {{Search for quantum gravity using astrophysical neutrino flavour with IceCube}},}\ }\href {\doibase 10.1038/s41567-022-01762-1} {\bibfield  {journal} {\bibinfo  {journal} {Nature Phys.}\ }\textbf {\bibinfo {volume} {18}},\ \bibinfo {pages} {1287--1292} (\bibinfo {year} {2022}{\natexlab{g}})},\ \Eprint {http://arxiv.org/abs/2111.04654} {arXiv:2111.04654 [hep-ex]} \BibitemShut {NoStop}%
\bibitem [{\citenamefont {Aartsen}\ \emph {et~al.}(2018{\natexlab{c}})\citenamefont {Aartsen} \emph {et~al.}}]{Aartsen:2017ibm}%
  \BibitemOpen
  \bibfield  {author} {\bibinfo {author} {\bibfnamefont {M.~G.}\ \bibnamefont {Aartsen}} \emph {et~al.} (\bibinfo {collaboration} {IceCube}),\ }\bibfield  {title} {\enquote {\bibinfo {title} {{Neutrino Interferometry for High-Precision Tests of Lorentz Symmetry with IceCube}},}\ }\href {\doibase 10.1038/s41567-018-0172-2} {\bibfield  {journal} {\bibinfo  {journal} {Nature Phys.}\ }\textbf {\bibinfo {volume} {14}},\ \bibinfo {pages} {961--966} (\bibinfo {year} {2018}{\natexlab{c}})},\ \Eprint {http://arxiv.org/abs/1709.03434} {arXiv:1709.03434 [hep-ex]} \BibitemShut {NoStop}%
\bibitem [{\citenamefont {Learned}\ and\ \citenamefont {Pakvasa}(1995)}]{Learned:1994wg}%
  \BibitemOpen
  \bibfield  {author} {\bibinfo {author} {\bibfnamefont {John~G.}\ \bibnamefont {Learned}}\ and\ \bibinfo {author} {\bibfnamefont {Sandip}\ \bibnamefont {Pakvasa}},\ }\bibfield  {title} {\enquote {\bibinfo {title} {{Detecting tau-neutrino oscillations at PeV energies}},}\ }\href {\doibase 10.1016/0927-6505(94)00043-3} {\bibfield  {journal} {\bibinfo  {journal} {Astropart. Phys.}\ }\textbf {\bibinfo {volume} {3}},\ \bibinfo {pages} {267--274} (\bibinfo {year} {1995})},\ \Eprint {http://arxiv.org/abs/hep-ph/9405296} {arXiv:hep-ph/9405296} \BibitemShut {NoStop}%
\bibitem [{\citenamefont {Arg\"uelles}\ \emph {et~al.}(2015{\natexlab{a}})\citenamefont {Arg\"uelles}, \citenamefont {Katori},\ and\ \citenamefont {Salvado}}]{Arguelles:2015dca}%
  \BibitemOpen
  \bibfield  {author} {\bibinfo {author} {\bibfnamefont {Carlos~A.}\ \bibnamefont {Arg\"uelles}}, \bibinfo {author} {\bibfnamefont {Teppei}\ \bibnamefont {Katori}}, \ and\ \bibinfo {author} {\bibfnamefont {Jordi}\ \bibnamefont {Salvado}},\ }\bibfield  {title} {\enquote {\bibinfo {title} {{New Physics in Astrophysical Neutrino Flavor}},}\ }\href {\doibase 10.1103/PhysRevLett.115.161303} {\bibfield  {journal} {\bibinfo  {journal} {Phys. Rev. Lett.}\ }\textbf {\bibinfo {volume} {115}},\ \bibinfo {pages} {161303} (\bibinfo {year} {2015}{\natexlab{a}})},\ \Eprint {http://arxiv.org/abs/1506.02043} {arXiv:1506.02043 [hep-ph]} \BibitemShut {NoStop}%
\bibitem [{\citenamefont {Bustamante}\ \emph {et~al.}(2015)\citenamefont {Bustamante}, \citenamefont {Beacom},\ and\ \citenamefont {Winter}}]{Bustamante:2015waa}%
  \BibitemOpen
  \bibfield  {author} {\bibinfo {author} {\bibfnamefont {Mauricio}\ \bibnamefont {Bustamante}}, \bibinfo {author} {\bibfnamefont {John~F.}\ \bibnamefont {Beacom}}, \ and\ \bibinfo {author} {\bibfnamefont {Walter}\ \bibnamefont {Winter}},\ }\bibfield  {title} {\enquote {\bibinfo {title} {{Theoretically palatable flavor combinations of astrophysical neutrinos}},}\ }\href {\doibase 10.1103/PhysRevLett.115.161302} {\bibfield  {journal} {\bibinfo  {journal} {Phys. Rev. Lett.}\ }\textbf {\bibinfo {volume} {115}},\ \bibinfo {pages} {161302} (\bibinfo {year} {2015})},\ \Eprint {http://arxiv.org/abs/1506.02645} {arXiv:1506.02645 [astro-ph.HE]} \BibitemShut {NoStop}%
\bibitem [{\citenamefont {Dev}\ \emph {et~al.}(2023)\citenamefont {Dev}, \citenamefont {Jana},\ and\ \citenamefont {Porto}}]{Dev:2023znd}%
  \BibitemOpen
  \bibfield  {author} {\bibinfo {author} {\bibfnamefont {P.~S.~Bhupal}\ \bibnamefont {Dev}}, \bibinfo {author} {\bibfnamefont {Sudip}\ \bibnamefont {Jana}}, \ and\ \bibinfo {author} {\bibfnamefont {Yago}\ \bibnamefont {Porto}},\ }\bibfield  {title} {\enquote {\bibinfo {title} {{Flavor Matters, but Matter Flavors: Matter Effects on Flavor Composition of Astrophysical Neutrinos}},}\ }\href@noop {} {\  (\bibinfo {year} {2023})},\ \Eprint {http://arxiv.org/abs/2312.17315} {arXiv:2312.17315 [hep-ph]} \BibitemShut {NoStop}%
\bibitem [{\citenamefont {Colladay}\ and\ \citenamefont {Kostelecky}(1998)}]{Colladay:1998fq}%
  \BibitemOpen
  \bibfield  {author} {\bibinfo {author} {\bibfnamefont {Don}\ \bibnamefont {Colladay}}\ and\ \bibinfo {author} {\bibfnamefont {V.~Alan}\ \bibnamefont {Kostelecky}},\ }\bibfield  {title} {\enquote {\bibinfo {title} {{Lorentz violating extension of the standard model}},}\ }\href {\doibase 10.1103/PhysRevD.58.116002} {\bibfield  {journal} {\bibinfo  {journal} {Phys. Rev. D}\ }\textbf {\bibinfo {volume} {58}},\ \bibinfo {pages} {116002} (\bibinfo {year} {1998})},\ \Eprint {http://arxiv.org/abs/hep-ph/9809521} {arXiv:hep-ph/9809521} \BibitemShut {NoStop}%
\bibitem [{\citenamefont {de~Salas}\ \emph {et~al.}(2016)\citenamefont {de~Salas}, \citenamefont {Lineros},\ and\ \citenamefont {T\'ortola}}]{deSalas:2016svi}%
  \BibitemOpen
  \bibfield  {author} {\bibinfo {author} {\bibfnamefont {P.~F.}\ \bibnamefont {de~Salas}}, \bibinfo {author} {\bibfnamefont {R.~A.}\ \bibnamefont {Lineros}}, \ and\ \bibinfo {author} {\bibfnamefont {M.}~\bibnamefont {T\'ortola}},\ }\bibfield  {title} {\enquote {\bibinfo {title} {{Neutrino propagation in the galactic dark matter halo}},}\ }\href {\doibase 10.1103/PhysRevD.94.123001} {\bibfield  {journal} {\bibinfo  {journal} {Phys. Rev. D}\ }\textbf {\bibinfo {volume} {94}},\ \bibinfo {pages} {123001} (\bibinfo {year} {2016})},\ \Eprint {http://arxiv.org/abs/1601.05798} {arXiv:1601.05798 [astro-ph.HE]} \BibitemShut {NoStop}%
\bibitem [{\citenamefont {Capozzi}\ \emph {et~al.}(2018)\citenamefont {Capozzi}, \citenamefont {Shoemaker},\ and\ \citenamefont {Vecchi}}]{Capozzi:2018bps}%
  \BibitemOpen
  \bibfield  {author} {\bibinfo {author} {\bibfnamefont {Francesco}\ \bibnamefont {Capozzi}}, \bibinfo {author} {\bibfnamefont {Ian~M.}\ \bibnamefont {Shoemaker}}, \ and\ \bibinfo {author} {\bibfnamefont {Luca}\ \bibnamefont {Vecchi}},\ }\bibfield  {title} {\enquote {\bibinfo {title} {{Neutrino Oscillations in Dark Backgrounds}},}\ }\href {\doibase 10.1088/1475-7516/2018/07/004} {\bibfield  {journal} {\bibinfo  {journal} {JCAP}\ }\textbf {\bibinfo {volume} {07}},\ \bibinfo {pages} {004} (\bibinfo {year} {2018})},\ \Eprint {http://arxiv.org/abs/1804.05117} {arXiv:1804.05117 [hep-ph]} \BibitemShut {NoStop}%
\bibitem [{\citenamefont {Bustamante}\ and\ \citenamefont {Agarwalla}(2019)}]{Bustamante:2018mzu}%
  \BibitemOpen
  \bibfield  {author} {\bibinfo {author} {\bibfnamefont {Mauricio}\ \bibnamefont {Bustamante}}\ and\ \bibinfo {author} {\bibfnamefont {Sanjib~Kumar}\ \bibnamefont {Agarwalla}},\ }\bibfield  {title} {\enquote {\bibinfo {title} {{Universe's Worth of Electrons to Probe Long-Range Interactions of High-Energy Astrophysical Neutrinos}},}\ }\href {\doibase 10.1103/PhysRevLett.122.061103} {\bibfield  {journal} {\bibinfo  {journal} {Phys. Rev. Lett.}\ }\textbf {\bibinfo {volume} {122}},\ \bibinfo {pages} {061103} (\bibinfo {year} {2019})},\ \Eprint {http://arxiv.org/abs/1808.02042} {arXiv:1808.02042 [astro-ph.HE]} \BibitemShut {NoStop}%
\bibitem [{\citenamefont {Gonzalez-Garcia}\ \emph {et~al.}(2016)\citenamefont {Gonzalez-Garcia}, \citenamefont {Maltoni}, \citenamefont {Martinez-Soler},\ and\ \citenamefont {Song}}]{Gonzalez-Garcia:2016gpq}%
  \BibitemOpen
  \bibfield  {author} {\bibinfo {author} {\bibfnamefont {M.~C.}\ \bibnamefont {Gonzalez-Garcia}}, \bibinfo {author} {\bibfnamefont {Michele}\ \bibnamefont {Maltoni}}, \bibinfo {author} {\bibfnamefont {Ivan}\ \bibnamefont {Martinez-Soler}}, \ and\ \bibinfo {author} {\bibfnamefont {Ningqiang}\ \bibnamefont {Song}},\ }\bibfield  {title} {\enquote {\bibinfo {title} {{Non-standard neutrino interactions in the Earth and the flavor of astrophysical neutrinos}},}\ }\href {\doibase 10.1016/j.astropartphys.2016.07.001} {\bibfield  {journal} {\bibinfo  {journal} {Astropart. Phys.}\ }\textbf {\bibinfo {volume} {84}},\ \bibinfo {pages} {15--22} (\bibinfo {year} {2016})},\ \Eprint {http://arxiv.org/abs/1605.08055} {arXiv:1605.08055 [hep-ph]} \BibitemShut {NoStop}%
\bibitem [{\citenamefont {Abbasi}\ \emph {et~al.}(2024{\natexlab{c}})\citenamefont {Abbasi} \emph {et~al.}}]{IceCube:2024kel}%
  \BibitemOpen
  \bibfield  {author} {\bibinfo {author} {\bibfnamefont {R.}~\bibnamefont {Abbasi}} \emph {et~al.} (\bibinfo {collaboration} {IceCube}),\ }\bibfield  {title} {\enquote {\bibinfo {title} {{A search for an eV-scale sterile neutrino using improved high-energy $\nu_\mu$ event reconstruction in IceCube}},}\ }\href {\doibase 10.1103/PhysRevLett.133.201804} {\bibfield  {journal} {\bibinfo  {journal} {Phys. Rev. Lett.}\ }\textbf {\bibinfo {volume} {133}},\ \bibinfo {pages} {201804} (\bibinfo {year} {2024}{\natexlab{c}})},\ \Eprint {http://arxiv.org/abs/2405.08070} {arXiv:2405.08070 [hep-ex]} \BibitemShut {NoStop}%
\bibitem [{\citenamefont {Abbasi}\ \emph {et~al.}(2024{\natexlab{d}})\citenamefont {Abbasi} \emph {et~al.}}]{IceCube:2024uzv}%
  \BibitemOpen
  \bibfield  {author} {\bibinfo {author} {\bibfnamefont {R.}~\bibnamefont {Abbasi}} \emph {et~al.} (\bibinfo {collaboration} {IceCube}),\ }\bibfield  {title} {\enquote {\bibinfo {title} {{Methods and stability tests associated with the sterile neutrino search using improved high-energy $\nu_\mu$ event reconstruction in IceCube}},}\ }\href {\doibase 10.1103/PhysRevD.110.092009} {\bibfield  {journal} {\bibinfo  {journal} {Phys. Rev. D}\ }\textbf {\bibinfo {volume} {110}},\ \bibinfo {pages} {092009} (\bibinfo {year} {2024}{\natexlab{d}})},\ \Eprint {http://arxiv.org/abs/2405.08077} {arXiv:2405.08077 [hep-ex]} \BibitemShut {NoStop}%
\bibitem [{\citenamefont {Aartsen}\ \emph {et~al.}(2018{\natexlab{d}})\citenamefont {Aartsen} \emph {et~al.}}]{IceCube:2017qyp}%
  \BibitemOpen
  \bibfield  {author} {\bibinfo {author} {\bibfnamefont {M.~G.}\ \bibnamefont {Aartsen}} \emph {et~al.} (\bibinfo {collaboration} {IceCube}),\ }\bibfield  {title} {\enquote {\bibinfo {title} {{Neutrino Interferometry for High-Precision Tests of Lorentz Symmetry with IceCube}},}\ }\href {\doibase 10.1038/s41567-018-0172-2} {\bibfield  {journal} {\bibinfo  {journal} {Nature Phys.}\ }\textbf {\bibinfo {volume} {14}},\ \bibinfo {pages} {961--966} (\bibinfo {year} {2018}{\natexlab{d}})},\ \Eprint {http://arxiv.org/abs/1709.03434} {arXiv:1709.03434 [hep-ex]} \BibitemShut {NoStop}%
\bibitem [{\citenamefont {Coloma}\ \emph {et~al.}(2018)\citenamefont {Coloma}, \citenamefont {Lopez-Pavon}, \citenamefont {Martinez-Soler},\ and\ \citenamefont {Nunokawa}}]{Coloma:2018idr}%
  \BibitemOpen
  \bibfield  {author} {\bibinfo {author} {\bibfnamefont {Pilar}\ \bibnamefont {Coloma}}, \bibinfo {author} {\bibfnamefont {Jacobo}\ \bibnamefont {Lopez-Pavon}}, \bibinfo {author} {\bibfnamefont {Ivan}\ \bibnamefont {Martinez-Soler}}, \ and\ \bibinfo {author} {\bibfnamefont {Hiroshi}\ \bibnamefont {Nunokawa}},\ }\bibfield  {title} {\enquote {\bibinfo {title} {{Decoherence in Neutrino Propagation Through Matter, and Bounds from IceCube/DeepCore}},}\ }\href {\doibase 10.1140/epjc/s10052-018-6092-6} {\bibfield  {journal} {\bibinfo  {journal} {Eur. Phys. J. C}\ }\textbf {\bibinfo {volume} {78}},\ \bibinfo {pages} {614} (\bibinfo {year} {2018})},\ \Eprint {http://arxiv.org/abs/1803.04438} {arXiv:1803.04438 [hep-ph]} \BibitemShut {NoStop}%
\bibitem [{\citenamefont {Stuttard}\ and\ \citenamefont {Jensen}(2020)}]{Stuttard:2020qfv}%
  \BibitemOpen
  \bibfield  {author} {\bibinfo {author} {\bibfnamefont {Thomas}\ \bibnamefont {Stuttard}}\ and\ \bibinfo {author} {\bibfnamefont {Mikkel}\ \bibnamefont {Jensen}},\ }\bibfield  {title} {\enquote {\bibinfo {title} {{Neutrino decoherence from quantum gravitational stochastic perturbations}},}\ }\href {\doibase 10.1103/PhysRevD.102.115003} {\bibfield  {journal} {\bibinfo  {journal} {Phys. Rev. D}\ }\textbf {\bibinfo {volume} {102}},\ \bibinfo {pages} {115003} (\bibinfo {year} {2020})},\ \Eprint {http://arxiv.org/abs/2007.00068} {arXiv:2007.00068 [hep-ph]} \BibitemShut {NoStop}%
\bibitem [{\citenamefont {Jones}\ and\ \citenamefont {Seidel}(2024)}]{Jones:2024qfr}%
  \BibitemOpen
  \bibfield  {author} {\bibinfo {author} {\bibfnamefont {B.~J.~P.}\ \bibnamefont {Jones}}\ and\ \bibinfo {author} {\bibfnamefont {O.~H.}\ \bibnamefont {Seidel}},\ }\bibfield  {title} {\enquote {\bibinfo {title} {{Collapse of Neutrino Wave Functions under Penrose Gravitational Reduction}},}\ }\href {\doibase 10.1103/PhysRevD.110.016026} {\bibfield  {journal} {\bibinfo  {journal} {Phys. Rev. D}\ }\textbf {\bibinfo {volume} {110}},\ \bibinfo {pages} {016026} (\bibinfo {year} {2024})},\ \Eprint {http://arxiv.org/abs/2405.03954} {arXiv:2405.03954 [hep-ph]} \BibitemShut {NoStop}%
\bibitem [{\citenamefont {Abbasi}\ \emph {et~al.}(2024{\natexlab{e}})\citenamefont {Abbasi} \emph {et~al.}}]{ICECUBE:2024fej}%
  \BibitemOpen
  \bibfield  {author} {\bibinfo {author} {\bibfnamefont {R.}~\bibnamefont {Abbasi}} \emph {et~al.} (\bibinfo {collaboration} {ICECUBE}),\ }\bibfield  {title} {\enquote {\bibinfo {title} {{Search for decoherence from quantum gravity with atmospheric neutrinos}},}\ }\href {\doibase 10.1038/s41567-024-02436-w} {\bibfield  {journal} {\bibinfo  {journal} {Nature Phys.}\ }\textbf {\bibinfo {volume} {20}},\ \bibinfo {pages} {913--920} (\bibinfo {year} {2024}{\natexlab{e}})},\ \Eprint {http://arxiv.org/abs/2308.00105} {arXiv:2308.00105 [hep-ex]} \BibitemShut {NoStop}%
\bibitem [{\citenamefont {Salvado}\ \emph {et~al.}(2017)\citenamefont {Salvado}, \citenamefont {Mena}, \citenamefont {Palomares-Ruiz},\ and\ \citenamefont {Rius}}]{Salvado:2016uqu}%
  \BibitemOpen
  \bibfield  {author} {\bibinfo {author} {\bibfnamefont {Jordi}\ \bibnamefont {Salvado}}, \bibinfo {author} {\bibfnamefont {Olga}\ \bibnamefont {Mena}}, \bibinfo {author} {\bibfnamefont {Sergio}\ \bibnamefont {Palomares-Ruiz}}, \ and\ \bibinfo {author} {\bibfnamefont {Nuria}\ \bibnamefont {Rius}},\ }\bibfield  {title} {\enquote {\bibinfo {title} {{Non-standard interactions with high-energy atmospheric neutrinos at IceCube}},}\ }\href {\doibase 10.1007/JHEP01(2017)141} {\bibfield  {journal} {\bibinfo  {journal} {JHEP}\ }\textbf {\bibinfo {volume} {01}},\ \bibinfo {pages} {141} (\bibinfo {year} {2017})},\ \Eprint {http://arxiv.org/abs/1609.03450} {arXiv:1609.03450 [hep-ph]} \BibitemShut {NoStop}%
\bibitem [{\citenamefont {Abbasi}\ \emph {et~al.}(2022{\natexlab{h}})\citenamefont {Abbasi} \emph {et~al.}}]{IceCube:2022ubv}%
  \BibitemOpen
  \bibfield  {author} {\bibinfo {author} {\bibfnamefont {R.}~\bibnamefont {Abbasi}} \emph {et~al.} (\bibinfo {collaboration} {IceCube}),\ }\bibfield  {title} {\enquote {\bibinfo {title} {{Strong Constraints on Neutrino Nonstandard Interactions from TeV-Scale $\nu_u$ Disappearance at IceCube}},}\ }\href {\doibase 10.1103/PhysRevLett.129.011804} {\bibfield  {journal} {\bibinfo  {journal} {Phys. Rev. Lett.}\ }\textbf {\bibinfo {volume} {129}},\ \bibinfo {pages} {011804} (\bibinfo {year} {2022}{\natexlab{h}})},\ \Eprint {http://arxiv.org/abs/2201.03566} {arXiv:2201.03566 [hep-ex]} \BibitemShut {NoStop}%
\bibitem [{\citenamefont {Ya\~nez}\ and\ \citenamefont {Fedynitch}(2023)}]{Yanez:2023lsy}%
  \BibitemOpen
  \bibfield  {author} {\bibinfo {author} {\bibfnamefont {Juan~Pablo}\ \bibnamefont {Ya\~nez}}\ and\ \bibinfo {author} {\bibfnamefont {Anatoli}\ \bibnamefont {Fedynitch}},\ }\bibfield  {title} {\enquote {\bibinfo {title} {{Data-driven muon-calibrated neutrino flux}},}\ }\href {\doibase 10.1103/PhysRevD.107.123037} {\bibfield  {journal} {\bibinfo  {journal} {Phys. Rev. D}\ }\textbf {\bibinfo {volume} {107}},\ \bibinfo {pages} {123037} (\bibinfo {year} {2023})},\ \Eprint {http://arxiv.org/abs/2303.00022} {arXiv:2303.00022 [hep-ph]} \BibitemShut {NoStop}%
\bibitem [{\citenamefont {Block}\ and\ \citenamefont {Halzen}(2011)}]{Block:2011vz}%
  \BibitemOpen
  \bibfield  {author} {\bibinfo {author} {\bibfnamefont {Martin~M.}\ \bibnamefont {Block}}\ and\ \bibinfo {author} {\bibfnamefont {Francis}\ \bibnamefont {Halzen}},\ }\bibfield  {title} {\enquote {\bibinfo {title} {{Experimental Confirmation that the Proton is Asymptotically a Black Disk}},}\ }\href {\doibase 10.1103/PhysRevLett.107.212002} {\bibfield  {journal} {\bibinfo  {journal} {Phys. Rev. Lett.}\ }\textbf {\bibinfo {volume} {107}},\ \bibinfo {pages} {212002} (\bibinfo {year} {2011})},\ \Eprint {http://arxiv.org/abs/1109.2041} {arXiv:1109.2041 [hep-ph]} \BibitemShut {NoStop}%
\bibitem [{\citenamefont {Arg\"uelles}\ \emph {et~al.}(2015{\natexlab{b}})\citenamefont {Arg\"uelles}, \citenamefont {Halzen}, \citenamefont {Wille}, \citenamefont {Kroll},\ and\ \citenamefont {Reno}}]{Arguelles:2015wba}%
  \BibitemOpen
  \bibfield  {author} {\bibinfo {author} {\bibfnamefont {Carlos~A.}\ \bibnamefont {Arg\"uelles}}, \bibinfo {author} {\bibfnamefont {Francis}\ \bibnamefont {Halzen}}, \bibinfo {author} {\bibfnamefont {Logan}\ \bibnamefont {Wille}}, \bibinfo {author} {\bibfnamefont {Mike}\ \bibnamefont {Kroll}}, \ and\ \bibinfo {author} {\bibfnamefont {Mary~Hall}\ \bibnamefont {Reno}},\ }\bibfield  {title} {\enquote {\bibinfo {title} {{High-energy behavior of photon, neutrino, and proton cross sections}},}\ }\href {\doibase 10.1103/PhysRevD.92.074040} {\bibfield  {journal} {\bibinfo  {journal} {Phys. Rev. D}\ }\textbf {\bibinfo {volume} {92}},\ \bibinfo {pages} {074040} (\bibinfo {year} {2015}{\natexlab{b}})},\ \Eprint {http://arxiv.org/abs/1504.06639} {arXiv:1504.06639 [hep-ph]} \BibitemShut {NoStop}%
\bibitem [{\citenamefont {Aartsen}\ \emph {et~al.}(2017{\natexlab{g}})\citenamefont {Aartsen} \emph {et~al.}}]{IceCube:2017roe}%
  \BibitemOpen
  \bibfield  {author} {\bibinfo {author} {\bibfnamefont {M.~G.}\ \bibnamefont {Aartsen}} \emph {et~al.} (\bibinfo {collaboration} {IceCube}),\ }\bibfield  {title} {\enquote {\bibinfo {title} {{Measurement of the multi-TeV neutrino cross section with IceCube using Earth absorption}},}\ }\href {\doibase 10.1038/nature24459} {\bibfield  {journal} {\bibinfo  {journal} {Nature}\ }\textbf {\bibinfo {volume} {551}},\ \bibinfo {pages} {596--600} (\bibinfo {year} {2017}{\natexlab{g}})},\ \Eprint {http://arxiv.org/abs/1711.08119} {arXiv:1711.08119 [hep-ex]} \BibitemShut {NoStop}%
\bibitem [{\citenamefont {Bustamante}\ and\ \citenamefont {Connolly}(2019)}]{Bustamante:2017xuy}%
  \BibitemOpen
  \bibfield  {author} {\bibinfo {author} {\bibfnamefont {Mauricio}\ \bibnamefont {Bustamante}}\ and\ \bibinfo {author} {\bibfnamefont {Amy}\ \bibnamefont {Connolly}},\ }\bibfield  {title} {\enquote {\bibinfo {title} {{Extracting the Energy-Dependent Neutrino-Nucleon Cross Section above 10 TeV Using IceCube Showers}},}\ }\href {\doibase 10.1103/PhysRevLett.122.041101} {\bibfield  {journal} {\bibinfo  {journal} {Phys. Rev. Lett.}\ }\textbf {\bibinfo {volume} {122}},\ \bibinfo {pages} {041101} (\bibinfo {year} {2019})},\ \Eprint {http://arxiv.org/abs/1711.11043} {arXiv:1711.11043 [astro-ph.HE]} \BibitemShut {NoStop}%
\bibitem [{\citenamefont {Zhou}\ and\ \citenamefont {Beacom}(2020)}]{Zhou:2019frk}%
  \BibitemOpen
  \bibfield  {author} {\bibinfo {author} {\bibfnamefont {Bei}\ \bibnamefont {Zhou}}\ and\ \bibinfo {author} {\bibfnamefont {John~F.}\ \bibnamefont {Beacom}},\ }\bibfield  {title} {\enquote {\bibinfo {title} {{W-boson and trident production in TeV\textendash{}PeV neutrino observatories}},}\ }\href {\doibase 10.1103/PhysRevD.101.036010} {\bibfield  {journal} {\bibinfo  {journal} {Phys. Rev. D}\ }\textbf {\bibinfo {volume} {101}},\ \bibinfo {pages} {036010} (\bibinfo {year} {2020})},\ \Eprint {http://arxiv.org/abs/1910.10720} {arXiv:1910.10720 [hep-ph]} \BibitemShut {NoStop}%
\bibitem [{\citenamefont {Abbasi}\ \emph {et~al.}(2020)\citenamefont {Abbasi} \emph {et~al.}}]{IceCube:2020rnc}%
  \BibitemOpen
  \bibfield  {author} {\bibinfo {author} {\bibfnamefont {R.}~\bibnamefont {Abbasi}} \emph {et~al.} (\bibinfo {collaboration} {IceCube}),\ }\bibfield  {title} {\enquote {\bibinfo {title} {{Measurement of the high-energy all-flavor neutrino-nucleon cross section with IceCube}},}\ }\href {\doibase 10.1103/PhysRevD.104.022001} {\  (\bibinfo {year} {2020}),\ 10.1103/PhysRevD.104.022001},\ \Eprint {http://arxiv.org/abs/2011.03560} {arXiv:2011.03560 [hep-ex]} \BibitemShut {NoStop}%
\bibitem [{\citenamefont {Bertone}\ \emph {et~al.}(2019)\citenamefont {Bertone}, \citenamefont {Gauld},\ and\ \citenamefont {Rojo}}]{Bertone:2018dse}%
  \BibitemOpen
  \bibfield  {author} {\bibinfo {author} {\bibfnamefont {Valerio}\ \bibnamefont {Bertone}}, \bibinfo {author} {\bibfnamefont {Rhorry}\ \bibnamefont {Gauld}}, \ and\ \bibinfo {author} {\bibfnamefont {Juan}\ \bibnamefont {Rojo}},\ }\bibfield  {title} {\enquote {\bibinfo {title} {{Neutrino Telescopes as QCD Microscopes}},}\ }\href {\doibase 10.1007/JHEP01(2019)217} {\bibfield  {journal} {\bibinfo  {journal} {JHEP}\ }\textbf {\bibinfo {volume} {01}},\ \bibinfo {pages} {217} (\bibinfo {year} {2019})},\ \Eprint {http://arxiv.org/abs/1808.02034} {arXiv:1808.02034 [hep-ph]} \BibitemShut {NoStop}%
\bibitem [{\citenamefont {Esteban}\ \emph {et~al.}(2022)\citenamefont {Esteban}, \citenamefont {Prohira},\ and\ \citenamefont {Beacom}}]{Esteban:2022uuw}%
  \BibitemOpen
  \bibfield  {author} {\bibinfo {author} {\bibfnamefont {Ivan}\ \bibnamefont {Esteban}}, \bibinfo {author} {\bibfnamefont {Steven}\ \bibnamefont {Prohira}}, \ and\ \bibinfo {author} {\bibfnamefont {John~F.}\ \bibnamefont {Beacom}},\ }\bibfield  {title} {\enquote {\bibinfo {title} {{Detector requirements for model-independent measurements of ultrahigh energy neutrino cross sections}},}\ }\href {\doibase 10.1103/PhysRevD.106.023021} {\bibfield  {journal} {\bibinfo  {journal} {Phys. Rev. D}\ }\textbf {\bibinfo {volume} {106}},\ \bibinfo {pages} {023021} (\bibinfo {year} {2022})},\ \Eprint {http://arxiv.org/abs/2205.09763} {arXiv:2205.09763 [hep-ph]} \BibitemShut {NoStop}%
\bibitem [{\citenamefont {Garcia~Soto}\ \emph {et~al.}(2023)\citenamefont {Garcia~Soto}, \citenamefont {Garg}, \citenamefont {Reno},\ and\ \citenamefont {Arg\"uelles}}]{GarciaSoto:2022vlw}%
  \BibitemOpen
  \bibfield  {author} {\bibinfo {author} {\bibfnamefont {Alfonso}\ \bibnamefont {Garcia~Soto}}, \bibinfo {author} {\bibfnamefont {Diksha}\ \bibnamefont {Garg}}, \bibinfo {author} {\bibfnamefont {Mary~Hall}\ \bibnamefont {Reno}}, \ and\ \bibinfo {author} {\bibfnamefont {Carlos~A.}\ \bibnamefont {Arg\"uelles}},\ }\bibfield  {title} {\enquote {\bibinfo {title} {{Probing quantum gravity with elastic interactions of ultrahigh-energy neutrinos}},}\ }\href {\doibase 10.1103/PhysRevD.107.033009} {\bibfield  {journal} {\bibinfo  {journal} {Phys. Rev. D}\ }\textbf {\bibinfo {volume} {107}},\ \bibinfo {pages} {033009} (\bibinfo {year} {2023})},\ \Eprint {http://arxiv.org/abs/2209.06282} {arXiv:2209.06282 [hep-ph]} \BibitemShut {NoStop}%
\bibitem [{\citenamefont {Alvarez-Muniz}\ \emph {et~al.}(2002)\citenamefont {Alvarez-Muniz}, \citenamefont {Feng}, \citenamefont {Halzen}, \citenamefont {Han},\ and\ \citenamefont {Hooper}}]{Alvarez-Muniz:2002snq}%
  \BibitemOpen
  \bibfield  {author} {\bibinfo {author} {\bibfnamefont {Jaime}\ \bibnamefont {Alvarez-Muniz}}, \bibinfo {author} {\bibfnamefont {Jonathan~L.}\ \bibnamefont {Feng}}, \bibinfo {author} {\bibfnamefont {Francis}\ \bibnamefont {Halzen}}, \bibinfo {author} {\bibfnamefont {Tao}\ \bibnamefont {Han}}, \ and\ \bibinfo {author} {\bibfnamefont {Dan}\ \bibnamefont {Hooper}},\ }\bibfield  {title} {\enquote {\bibinfo {title} {{Detecting microscopic black holes with neutrino telescopes}},}\ }\href {\doibase 10.1103/PhysRevD.65.124015} {\bibfield  {journal} {\bibinfo  {journal} {Phys. Rev. D}\ }\textbf {\bibinfo {volume} {65}},\ \bibinfo {pages} {124015} (\bibinfo {year} {2002})},\ \Eprint {http://arxiv.org/abs/hep-ph/0202081} {arXiv:hep-ph/0202081} \BibitemShut {NoStop}%
\bibitem [{\citenamefont {Dey}\ \emph {et~al.}(2018)\citenamefont {Dey}, \citenamefont {Kar}, \citenamefont {Mitra}, \citenamefont {Spannowsky},\ and\ \citenamefont {Vincent}}]{Dey:2017ede}%
  \BibitemOpen
  \bibfield  {author} {\bibinfo {author} {\bibfnamefont {Ujjal~Kumar}\ \bibnamefont {Dey}}, \bibinfo {author} {\bibfnamefont {Deepak}\ \bibnamefont {Kar}}, \bibinfo {author} {\bibfnamefont {Manimala}\ \bibnamefont {Mitra}}, \bibinfo {author} {\bibfnamefont {Michael}\ \bibnamefont {Spannowsky}}, \ and\ \bibinfo {author} {\bibfnamefont {Aaron~C.}\ \bibnamefont {Vincent}},\ }\bibfield  {title} {\enquote {\bibinfo {title} {{Searching for Leptoquarks at IceCube and the LHC}},}\ }\href {\doibase 10.1103/PhysRevD.98.035014} {\bibfield  {journal} {\bibinfo  {journal} {Phys. Rev. D}\ }\textbf {\bibinfo {volume} {98}},\ \bibinfo {pages} {035014} (\bibinfo {year} {2018})},\ \Eprint {http://arxiv.org/abs/1709.02009} {arXiv:1709.02009 [hep-ph]} \BibitemShut {NoStop}%
\bibitem [{\citenamefont {Babu}\ \emph {et~al.}(2020)\citenamefont {Babu}, \citenamefont {Dev}, \citenamefont {Jana},\ and\ \citenamefont {Sui}}]{Babu:2019vff}%
  \BibitemOpen
  \bibfield  {author} {\bibinfo {author} {\bibfnamefont {K.~S.}\ \bibnamefont {Babu}}, \bibinfo {author} {\bibfnamefont {P.~S.}\ \bibnamefont {Dev}}, \bibinfo {author} {\bibfnamefont {Sudip}\ \bibnamefont {Jana}}, \ and\ \bibinfo {author} {\bibfnamefont {Yicong}\ \bibnamefont {Sui}},\ }\bibfield  {title} {\enquote {\bibinfo {title} {{Zee-Burst: A New Probe of Neutrino Nonstandard Interactions at IceCube}},}\ }\href {\doibase 10.1103/PhysRevLett.124.041805} {\bibfield  {journal} {\bibinfo  {journal} {Phys. Rev. Lett.}\ }\textbf {\bibinfo {volume} {124}},\ \bibinfo {pages} {041805} (\bibinfo {year} {2020})},\ \Eprint {http://arxiv.org/abs/1908.02779} {arXiv:1908.02779 [hep-ph]} \BibitemShut {NoStop}%
\bibitem [{\citenamefont {Greisen}(1966)}]{Greisen:1966jv}%
  \BibitemOpen
  \bibfield  {author} {\bibinfo {author} {\bibfnamefont {Kenneth}\ \bibnamefont {Greisen}},\ }\bibfield  {title} {\enquote {\bibinfo {title} {{End to the cosmic ray spectrum?}}}\ }\href {\doibase 10.1103/PhysRevLett.16.748} {\bibfield  {journal} {\bibinfo  {journal} {Phys. Rev. Lett.}\ }\textbf {\bibinfo {volume} {16}},\ \bibinfo {pages} {748--750} (\bibinfo {year} {1966})}\BibitemShut {NoStop}%
\bibitem [{\citenamefont {Zatsepin}\ and\ \citenamefont {Kuzmin}(1966)}]{Zatsepin:1966jv}%
  \BibitemOpen
  \bibfield  {author} {\bibinfo {author} {\bibfnamefont {G.~T.}\ \bibnamefont {Zatsepin}}\ and\ \bibinfo {author} {\bibfnamefont {V.~A.}\ \bibnamefont {Kuzmin}},\ }\bibfield  {title} {\enquote {\bibinfo {title} {{Upper limit of the spectrum of cosmic rays}},}\ }\href@noop {} {\bibfield  {journal} {\bibinfo  {journal} {JETP Lett.}\ }\textbf {\bibinfo {volume} {4}},\ \bibinfo {pages} {78--80} (\bibinfo {year} {1966})}\BibitemShut {NoStop}%
\bibitem [{\citenamefont {Aguilar}\ \emph {et~al.}(2021{\natexlab{b}})\citenamefont {Aguilar}, \citenamefont {Allison}, \citenamefont {Beatty}, \citenamefont {Bernhoff}, \citenamefont {Besson}, \citenamefont {Bingefors}, \citenamefont {Botner}, \citenamefont {Buitink}, \citenamefont {Carter}, \citenamefont {Clark}, \citenamefont {Connolly}, \citenamefont {Dasgupta}, \citenamefont {de~Kockere}, \citenamefont {de~Vries}, \citenamefont {Deaconu}, \citenamefont {DuVernois}, \citenamefont {Feigl}, \citenamefont {García-Fernández}, \citenamefont {Glaser}, \citenamefont {Hallgren}, \citenamefont {Hallmann}, \citenamefont {Hanson}, \citenamefont {Hendricks}, \citenamefont {Hokanson-Fasig}, \citenamefont {Hornhuber}, \citenamefont {Hughes}, \citenamefont {Karle}, \citenamefont {Kelley}, \citenamefont {Klein}, \citenamefont {Krebs}, \citenamefont {Lahmann}, \citenamefont {Magnuson}, \citenamefont {Meures}, \citenamefont {Meyers}, \citenamefont {Nelles}, \citenamefont {Novikov}, \citenamefont {Oberla}, \citenamefont
  {Oeyen}, \citenamefont {Pandya}, \citenamefont {Plaisier}, \citenamefont {Pyras}, \citenamefont {Ryckbosch}, \citenamefont {Scholten}, \citenamefont {Seckel}, \citenamefont {Smith}, \citenamefont {Southall}, \citenamefont {Torres}, \citenamefont {Toscano}, \citenamefont {Broeck}, \citenamefont {van Eijndhoven}, \citenamefont {Vieregg}, \citenamefont {Welling}, \citenamefont {Wissel}, \citenamefont {Young},\ and\ \citenamefont {Zink}}]{Aguilar_2021}%
  \BibitemOpen
  \bibfield  {author} {\bibinfo {author} {\bibfnamefont {J.A.}\ \bibnamefont {Aguilar}}, \bibinfo {author} {\bibfnamefont {P.}~\bibnamefont {Allison}}, \bibinfo {author} {\bibfnamefont {J.J.}\ \bibnamefont {Beatty}}, \bibinfo {author} {\bibfnamefont {H.}~\bibnamefont {Bernhoff}}, \bibinfo {author} {\bibfnamefont {D.}~\bibnamefont {Besson}}, \bibinfo {author} {\bibfnamefont {N.}~\bibnamefont {Bingefors}}, \bibinfo {author} {\bibfnamefont {O.}~\bibnamefont {Botner}}, \bibinfo {author} {\bibfnamefont {S.}~\bibnamefont {Buitink}}, \bibinfo {author} {\bibfnamefont {K.}~\bibnamefont {Carter}}, \bibinfo {author} {\bibfnamefont {B.A.}\ \bibnamefont {Clark}}, \bibinfo {author} {\bibfnamefont {A.}~\bibnamefont {Connolly}}, \bibinfo {author} {\bibfnamefont {P.}~\bibnamefont {Dasgupta}}, \bibinfo {author} {\bibfnamefont {S.}~\bibnamefont {de~Kockere}}, \bibinfo {author} {\bibfnamefont {K.D.}\ \bibnamefont {de~Vries}}, \bibinfo {author} {\bibfnamefont {C.}~\bibnamefont {Deaconu}}, \bibinfo {author} {\bibfnamefont {M.A.}\
  \bibnamefont {DuVernois}}, \bibinfo {author} {\bibfnamefont {N.}~\bibnamefont {Feigl}}, \bibinfo {author} {\bibfnamefont {D.}~\bibnamefont {García-Fernández}}, \bibinfo {author} {\bibfnamefont {C.}~\bibnamefont {Glaser}}, \bibinfo {author} {\bibfnamefont {A.}~\bibnamefont {Hallgren}}, \bibinfo {author} {\bibfnamefont {S.}~\bibnamefont {Hallmann}}, \bibinfo {author} {\bibfnamefont {J.C.}\ \bibnamefont {Hanson}}, \bibinfo {author} {\bibfnamefont {B.}~\bibnamefont {Hendricks}}, \bibinfo {author} {\bibfnamefont {B.}~\bibnamefont {Hokanson-Fasig}}, \bibinfo {author} {\bibfnamefont {C.}~\bibnamefont {Hornhuber}}, \bibinfo {author} {\bibfnamefont {K.}~\bibnamefont {Hughes}}, \bibinfo {author} {\bibfnamefont {A.}~\bibnamefont {Karle}}, \bibinfo {author} {\bibfnamefont {J.L.}\ \bibnamefont {Kelley}}, \bibinfo {author} {\bibfnamefont {S.R.}\ \bibnamefont {Klein}}, \bibinfo {author} {\bibfnamefont {R.}~\bibnamefont {Krebs}}, \bibinfo {author} {\bibfnamefont {R.}~\bibnamefont {Lahmann}}, \bibinfo {author}
  {\bibfnamefont {M.}~\bibnamefont {Magnuson}}, \bibinfo {author} {\bibfnamefont {T.}~\bibnamefont {Meures}}, \bibinfo {author} {\bibfnamefont {Z.S.}\ \bibnamefont {Meyers}}, \bibinfo {author} {\bibfnamefont {A.}~\bibnamefont {Nelles}}, \bibinfo {author} {\bibfnamefont {A.}~\bibnamefont {Novikov}}, \bibinfo {author} {\bibfnamefont {E.}~\bibnamefont {Oberla}}, \bibinfo {author} {\bibfnamefont {B.}~\bibnamefont {Oeyen}}, \bibinfo {author} {\bibfnamefont {H.}~\bibnamefont {Pandya}}, \bibinfo {author} {\bibfnamefont {I.}~\bibnamefont {Plaisier}}, \bibinfo {author} {\bibfnamefont {L.}~\bibnamefont {Pyras}}, \bibinfo {author} {\bibfnamefont {D.}~\bibnamefont {Ryckbosch}}, \bibinfo {author} {\bibfnamefont {O.}~\bibnamefont {Scholten}}, \bibinfo {author} {\bibfnamefont {D.}~\bibnamefont {Seckel}}, \bibinfo {author} {\bibfnamefont {D.}~\bibnamefont {Smith}}, \bibinfo {author} {\bibfnamefont {D.}~\bibnamefont {Southall}}, \bibinfo {author} {\bibfnamefont {J.}~\bibnamefont {Torres}}, \bibinfo {author} {\bibfnamefont
  {S.}~\bibnamefont {Toscano}}, \bibinfo {author} {\bibfnamefont {D.J. Van~Den}\ \bibnamefont {Broeck}}, \bibinfo {author} {\bibfnamefont {N.}~\bibnamefont {van Eijndhoven}}, \bibinfo {author} {\bibfnamefont {A.G.}\ \bibnamefont {Vieregg}}, \bibinfo {author} {\bibfnamefont {C.}~\bibnamefont {Welling}}, \bibinfo {author} {\bibfnamefont {S.}~\bibnamefont {Wissel}}, \bibinfo {author} {\bibfnamefont {R.}~\bibnamefont {Young}}, \ and\ \bibinfo {author} {\bibfnamefont {A.}~\bibnamefont {Zink}},\ }\bibfield  {title} {\enquote {\bibinfo {title} {Design and sensitivity of the radio neutrino observatory in greenland (rno-g)},}\ }\href {\doibase 10.1088/1748-0221/16/03/P03025} {\bibfield  {journal} {\bibinfo  {journal} {Journal of Instrumentation}\ }\textbf {\bibinfo {volume} {16}},\ \bibinfo {pages} {P03025} (\bibinfo {year} {2021}{\natexlab{b}})}\BibitemShut {NoStop}%
\bibitem [{\citenamefont {collaboration}\ \emph {et~al.}(2021)\citenamefont {collaboration}, \citenamefont {Abarr}, \citenamefont {Allison}, \citenamefont {Yebra}, \citenamefont {Alvarez-Muñiz}, \citenamefont {Beatty}, \citenamefont {Besson}, \citenamefont {Chen}, \citenamefont {Chen}, \citenamefont {Xie}, \citenamefont {Clem}, \citenamefont {Connolly}, \citenamefont {Cremonesi}, \citenamefont {Deaconu}, \citenamefont {Flaherty}, \citenamefont {Frikken}, \citenamefont {Gorham}, \citenamefont {Hast}, \citenamefont {Hornhuber}, \citenamefont {Huang}, \citenamefont {Hughes}, \citenamefont {Hynous}, \citenamefont {Ku}, \citenamefont {Kuo}, \citenamefont {Liu}, \citenamefont {Martin}, \citenamefont {Miki}, \citenamefont {Nam}, \citenamefont {Nichol}, \citenamefont {Nishimura}, \citenamefont {Novikov}, \citenamefont {Nozdrina}, \citenamefont {Oberla}, \citenamefont {Prohira}, \citenamefont {Prechelt}, \citenamefont {Rauch}, \citenamefont {Roberts}, \citenamefont {Romero-Wolf}, \citenamefont {Russell},
  \citenamefont {Seckel}, \citenamefont {Shiao}, \citenamefont {Smith}, \citenamefont {Southall}, \citenamefont {Varner}, \citenamefont {Vieregg}, \citenamefont {Wang}, \citenamefont {Wang}, \citenamefont {Wissel}, \citenamefont {Young}, \citenamefont {Zas},\ and\ \citenamefont {Zeolla}}]{Abarr_2021}%
  \BibitemOpen
  \bibfield  {author} {\bibinfo {author} {\bibfnamefont {The~PUEO}\ \bibnamefont {collaboration}}, \bibinfo {author} {\bibfnamefont {Q.}~\bibnamefont {Abarr}}, \bibinfo {author} {\bibfnamefont {P.}~\bibnamefont {Allison}}, \bibinfo {author} {\bibfnamefont {J.~Ammerman}\ \bibnamefont {Yebra}}, \bibinfo {author} {\bibfnamefont {J.}~\bibnamefont {Alvarez-Muñiz}}, \bibinfo {author} {\bibfnamefont {J.J.}\ \bibnamefont {Beatty}}, \bibinfo {author} {\bibfnamefont {D.Z.}\ \bibnamefont {Besson}}, \bibinfo {author} {\bibfnamefont {P.}~\bibnamefont {Chen}}, \bibinfo {author} {\bibfnamefont {Y.}~\bibnamefont {Chen}}, \bibinfo {author} {\bibfnamefont {C.}~\bibnamefont {Xie}}, \bibinfo {author} {\bibfnamefont {J.M.}\ \bibnamefont {Clem}}, \bibinfo {author} {\bibfnamefont {A.}~\bibnamefont {Connolly}}, \bibinfo {author} {\bibfnamefont {L.}~\bibnamefont {Cremonesi}}, \bibinfo {author} {\bibfnamefont {C.}~\bibnamefont {Deaconu}}, \bibinfo {author} {\bibfnamefont {J.}~\bibnamefont {Flaherty}}, \bibinfo {author} {\bibfnamefont
  {D.}~\bibnamefont {Frikken}}, \bibinfo {author} {\bibfnamefont {P.W.}\ \bibnamefont {Gorham}}, \bibinfo {author} {\bibfnamefont {C.}~\bibnamefont {Hast}}, \bibinfo {author} {\bibfnamefont {C.}~\bibnamefont {Hornhuber}}, \bibinfo {author} {\bibfnamefont {J.J.}\ \bibnamefont {Huang}}, \bibinfo {author} {\bibfnamefont {K.}~\bibnamefont {Hughes}}, \bibinfo {author} {\bibfnamefont {A.}~\bibnamefont {Hynous}}, \bibinfo {author} {\bibfnamefont {Y.}~\bibnamefont {Ku}}, \bibinfo {author} {\bibfnamefont {C.-Y.}\ \bibnamefont {Kuo}}, \bibinfo {author} {\bibfnamefont {T.C.}\ \bibnamefont {Liu}}, \bibinfo {author} {\bibfnamefont {Z.}~\bibnamefont {Martin}}, \bibinfo {author} {\bibfnamefont {C.}~\bibnamefont {Miki}}, \bibinfo {author} {\bibfnamefont {J.}~\bibnamefont {Nam}}, \bibinfo {author} {\bibfnamefont {R.J.}\ \bibnamefont {Nichol}}, \bibinfo {author} {\bibfnamefont {K.}~\bibnamefont {Nishimura}}, \bibinfo {author} {\bibfnamefont {A.}~\bibnamefont {Novikov}}, \bibinfo {author} {\bibfnamefont {A.}~\bibnamefont
  {Nozdrina}}, \bibinfo {author} {\bibfnamefont {E.}~\bibnamefont {Oberla}}, \bibinfo {author} {\bibfnamefont {S.}~\bibnamefont {Prohira}}, \bibinfo {author} {\bibfnamefont {R.}~\bibnamefont {Prechelt}}, \bibinfo {author} {\bibfnamefont {B.F.}\ \bibnamefont {Rauch}}, \bibinfo {author} {\bibfnamefont {J.M.}\ \bibnamefont {Roberts}}, \bibinfo {author} {\bibfnamefont {A.}~\bibnamefont {Romero-Wolf}}, \bibinfo {author} {\bibfnamefont {J.W.}\ \bibnamefont {Russell}}, \bibinfo {author} {\bibfnamefont {D.}~\bibnamefont {Seckel}}, \bibinfo {author} {\bibfnamefont {J.}~\bibnamefont {Shiao}}, \bibinfo {author} {\bibfnamefont {D.}~\bibnamefont {Smith}}, \bibinfo {author} {\bibfnamefont {D.}~\bibnamefont {Southall}}, \bibinfo {author} {\bibfnamefont {G.S.}\ \bibnamefont {Varner}}, \bibinfo {author} {\bibfnamefont {A.G.}\ \bibnamefont {Vieregg}}, \bibinfo {author} {\bibfnamefont {S.-H.}\ \bibnamefont {Wang}}, \bibinfo {author} {\bibfnamefont {Y.-H.}\ \bibnamefont {Wang}}, \bibinfo {author} {\bibfnamefont {S.A.}\
  \bibnamefont {Wissel}}, \bibinfo {author} {\bibfnamefont {R.}~\bibnamefont {Young}}, \bibinfo {author} {\bibfnamefont {E.}~\bibnamefont {Zas}}, \ and\ \bibinfo {author} {\bibfnamefont {A.}~\bibnamefont {Zeolla}},\ }\bibfield  {title} {\enquote {\bibinfo {title} {The payload for ultrahigh energy observations (pueo): a white paper},}\ }\href {\doibase 10.1088/1748-0221/16/08/P08035} {\bibfield  {journal} {\bibinfo  {journal} {Journal of Instrumentation}\ }\textbf {\bibinfo {volume} {16}},\ \bibinfo {pages} {P08035} (\bibinfo {year} {2021})}\BibitemShut {NoStop}%
\bibitem [{\citenamefont {Wissel}\ \emph {et~al.}(2020)\citenamefont {Wissel} \emph {et~al.}}]{Wissel:2020sec}%
  \BibitemOpen
  \bibfield  {author} {\bibinfo {author} {\bibfnamefont {Stephanie}\ \bibnamefont {Wissel}} \emph {et~al.},\ }\bibfield  {title} {\enquote {\bibinfo {title} {{Prospects for high-elevation radio detection of \ensuremath{>}100 PeV tau neutrinos}},}\ }\href {\doibase 10.1088/1475-7516/2020/11/065} {\bibfield  {journal} {\bibinfo  {journal} {JCAP}\ }\textbf {\bibinfo {volume} {11}},\ \bibinfo {pages} {065} (\bibinfo {year} {2020})},\ \Eprint {http://arxiv.org/abs/2004.12718} {arXiv:2004.12718 [astro-ph.IM]} \BibitemShut {NoStop}%
\bibitem [{\citenamefont {Beacom}\ \emph {et~al.}(2002)\citenamefont {Beacom}, \citenamefont {Crotty},\ and\ \citenamefont {Kolb}}]{Beacom:2001xn}%
  \BibitemOpen
  \bibfield  {author} {\bibinfo {author} {\bibfnamefont {John~F.}\ \bibnamefont {Beacom}}, \bibinfo {author} {\bibfnamefont {Patrick}\ \bibnamefont {Crotty}}, \ and\ \bibinfo {author} {\bibfnamefont {Edward~W.}\ \bibnamefont {Kolb}},\ }\bibfield  {title} {\enquote {\bibinfo {title} {{Enhanced Signal of Astrophysical Tau Neutrinos Propagating through Earth}},}\ }\href {\doibase 10.1103/PhysRevD.66.021302} {\bibfield  {journal} {\bibinfo  {journal} {Phys. Rev. D}\ }\textbf {\bibinfo {volume} {66}},\ \bibinfo {pages} {021302} (\bibinfo {year} {2002})},\ \Eprint {http://arxiv.org/abs/astro-ph/0111482} {arXiv:astro-ph/0111482} \BibitemShut {NoStop}%
\bibitem [{\citenamefont {Soto}\ \emph {et~al.}(2022)\citenamefont {Soto}, \citenamefont {Zhelnin}, \citenamefont {Safa},\ and\ \citenamefont {Arg\"uelles}}]{Soto:2021vdc}%
  \BibitemOpen
  \bibfield  {author} {\bibinfo {author} {\bibfnamefont {Alfonso~Garcia}\ \bibnamefont {Soto}}, \bibinfo {author} {\bibfnamefont {Pavel}\ \bibnamefont {Zhelnin}}, \bibinfo {author} {\bibfnamefont {Ibrahim}\ \bibnamefont {Safa}}, \ and\ \bibinfo {author} {\bibfnamefont {Carlos~A.}\ \bibnamefont {Arg\"uelles}},\ }\bibfield  {title} {\enquote {\bibinfo {title} {{Tau Appearance from High-Energy Neutrino Interactions}},}\ }\href {\doibase 10.1103/PhysRevLett.128.171101} {\bibfield  {journal} {\bibinfo  {journal} {Phys. Rev. Lett.}\ }\textbf {\bibinfo {volume} {128}},\ \bibinfo {pages} {171101} (\bibinfo {year} {2022})},\ \Eprint {http://arxiv.org/abs/2112.06937} {arXiv:2112.06937 [hep-ph]} \BibitemShut {NoStop}%
\bibitem [{\citenamefont {Aab}\ \emph {et~al.}(2019)\citenamefont {Aab} \emph {et~al.}}]{PierreAuger:2019ens}%
  \BibitemOpen
  \bibfield  {author} {\bibinfo {author} {\bibfnamefont {Alexander}\ \bibnamefont {Aab}} \emph {et~al.} (\bibinfo {collaboration} {Pierre Auger}),\ }\bibfield  {title} {\enquote {\bibinfo {title} {{Probing the origin of ultra-high-energy cosmic rays with neutrinos in the EeV energy range using the Pierre Auger Observatory}},}\ }\href {\doibase 10.1088/1475-7516/2019/10/022} {\bibfield  {journal} {\bibinfo  {journal} {JCAP}\ }\textbf {\bibinfo {volume} {10}},\ \bibinfo {pages} {022} (\bibinfo {year} {2019})},\ \Eprint {http://arxiv.org/abs/1906.07422} {arXiv:1906.07422 [astro-ph.HE]} \BibitemShut {NoStop}%
\bibitem [{\citenamefont {Abbasi}\ \emph {et~al.}(2021{\natexlab{b}})\citenamefont {Abbasi} \emph {et~al.}}]{Abbasi:2021ryj}%
  \BibitemOpen
  \bibfield  {author} {\bibinfo {author} {\bibfnamefont {R.}~\bibnamefont {Abbasi}} \emph {et~al.},\ }\bibfield  {title} {\enquote {\bibinfo {title} {{A Convolutional Neural Network based Cascade Reconstruction for the IceCube Neutrino Observatory}},}\ }\href {\doibase 10.1088/1748-0221/16/07/P07041} {\bibfield  {journal} {\bibinfo  {journal} {JINST}\ }\textbf {\bibinfo {volume} {16}},\ \bibinfo {pages} {P07041} (\bibinfo {year} {2021}{\natexlab{b}})},\ \Eprint {http://arxiv.org/abs/2101.11589} {arXiv:2101.11589 [hep-ex]} \BibitemShut {NoStop}%
\bibitem [{\citenamefont {Kharuk}\ \emph {et~al.}(2023)\citenamefont {Kharuk}, \citenamefont {Safronov}, \citenamefont {Matseiko},\ and\ \citenamefont {Leonov}}]{Kharuk:2023xnl}%
  \BibitemOpen
  \bibfield  {author} {\bibinfo {author} {\bibfnamefont {I.}~\bibnamefont {Kharuk}}, \bibinfo {author} {\bibfnamefont {G.}~\bibnamefont {Safronov}}, \bibinfo {author} {\bibfnamefont {A.}~\bibnamefont {Matseiko}}, \ and\ \bibinfo {author} {\bibfnamefont {A.}~\bibnamefont {Leonov}},\ }\bibfield  {title} {\enquote {\bibinfo {title} {{Machine learning in Baikal-GVD experiment}},}\ }\href {\doibase 10.22323/1.444.1077} {\bibfield  {journal} {\bibinfo  {journal} {PoS}\ }\textbf {\bibinfo {volume} {ICRC2023}},\ \bibinfo {pages} {1077} (\bibinfo {year} {2023})}\BibitemShut {NoStop}%
\bibitem [{\citenamefont {Reck}\ \emph {et~al.}(2021)\citenamefont {Reck}, \citenamefont {Guderian}, \citenamefont {Vermari\"en},\ and\ \citenamefont {Domi}}]{Reck:2021zqw}%
  \BibitemOpen
  \bibfield  {author} {\bibinfo {author} {\bibfnamefont {S.}~\bibnamefont {Reck}}, \bibinfo {author} {\bibfnamefont {D.}~\bibnamefont {Guderian}}, \bibinfo {author} {\bibfnamefont {G.}~\bibnamefont {Vermari\"en}}, \ and\ \bibinfo {author} {\bibfnamefont {A.}~\bibnamefont {Domi}} (\bibinfo {collaboration} {KM3NeT}),\ }\bibfield  {title} {\enquote {\bibinfo {title} {{Graph neural networks for reconstruction and classification in KM3NeT}},}\ }\href {\doibase 10.1088/1748-0221/16/10/C10011} {\bibfield  {journal} {\bibinfo  {journal} {JINST}\ }\textbf {\bibinfo {volume} {16}},\ \bibinfo {pages} {C10011} (\bibinfo {year} {2021})},\ \Eprint {http://arxiv.org/abs/2107.13375} {arXiv:2107.13375 [astro-ph.IM]} \BibitemShut {NoStop}%
\bibitem [{\citenamefont {Abbasi}\ \emph {et~al.}(2022{\natexlab{i}})\citenamefont {Abbasi} \emph {et~al.}}]{IceCube:2022njh}%
  \BibitemOpen
  \bibfield  {author} {\bibinfo {author} {\bibfnamefont {R.}~\bibnamefont {Abbasi}} \emph {et~al.} (\bibinfo {collaboration} {IceCube}),\ }\bibfield  {title} {\enquote {\bibinfo {title} {{Graph Neural Networks for low-energy event classification \& reconstruction in IceCube}},}\ }\href {\doibase 10.1088/1748-0221/17/11/P11003} {\bibfield  {journal} {\bibinfo  {journal} {JINST}\ }\textbf {\bibinfo {volume} {17}},\ \bibinfo {pages} {P11003} (\bibinfo {year} {2022}{\natexlab{i}})},\ \Eprint {http://arxiv.org/abs/2209.03042} {arXiv:2209.03042 [hep-ex]} \BibitemShut {NoStop}%
\bibitem [{\citenamefont {Yu}\ \emph {et~al.}(2023)\citenamefont {Yu}, \citenamefont {Lazar},\ and\ \citenamefont {Arg\"uelles}}]{Yu:2023ehc}%
  \BibitemOpen
  \bibfield  {author} {\bibinfo {author} {\bibfnamefont {Felix~J.}\ \bibnamefont {Yu}}, \bibinfo {author} {\bibfnamefont {Jeffrey}\ \bibnamefont {Lazar}}, \ and\ \bibinfo {author} {\bibfnamefont {Carlos~A.}\ \bibnamefont {Arg\"uelles}},\ }\bibfield  {title} {\enquote {\bibinfo {title} {{Trigger-level event reconstruction for neutrino telescopes using sparse submanifold convolutional neural networks}},}\ }\href {\doibase 10.1103/PhysRevD.108.063017} {\bibfield  {journal} {\bibinfo  {journal} {Phys. Rev. D}\ }\textbf {\bibinfo {volume} {108}},\ \bibinfo {pages} {063017} (\bibinfo {year} {2023})},\ \Eprint {http://arxiv.org/abs/2303.08812} {arXiv:2303.08812 [hep-ex]} \BibitemShut {NoStop}%
\bibitem [{\citenamefont {Capel}\ \emph {et~al.}(2023)\citenamefont {Capel}, \citenamefont {Spannfellner}, \citenamefont {Haack},\ and\ \citenamefont {Prottung}}]{Capel:2023ijl}%
  \BibitemOpen
  \bibfield  {author} {\bibinfo {author} {\bibfnamefont {Francesca}\ \bibnamefont {Capel}}, \bibinfo {author} {\bibfnamefont {Christian}\ \bibnamefont {Spannfellner}}, \bibinfo {author} {\bibfnamefont {Christian}\ \bibnamefont {Haack}}, \ and\ \bibinfo {author} {\bibfnamefont {Janik}\ \bibnamefont {Prottung}},\ }\bibfield  {title} {\enquote {\bibinfo {title} {{Evaluation of an FPGA-based fast machine-learning trigger for neutrino telescopes}},}\ }\href {\doibase 10.22323/1.444.1104} {\bibfield  {journal} {\bibinfo  {journal} {PoS}\ }\textbf {\bibinfo {volume} {ICRC2023}},\ \bibinfo {pages} {1104} (\bibinfo {year} {2023})}\BibitemShut {NoStop}%
\bibitem [{\citenamefont {Jin}\ \emph {et~al.}(2023)\citenamefont {Jin}, \citenamefont {Hu},\ and\ \citenamefont {Arg\"uelles}}]{Jin:2023xts}%
  \BibitemOpen
  \bibfield  {author} {\bibinfo {author} {\bibfnamefont {Miaochen}\ \bibnamefont {Jin}}, \bibinfo {author} {\bibfnamefont {Yushi}\ \bibnamefont {Hu}}, \ and\ \bibinfo {author} {\bibfnamefont {Carlos~A.}\ \bibnamefont {Arg\"uelles}},\ }\bibfield  {title} {\enquote {\bibinfo {title} {{Two Watts is All You Need: Enabling In-Detector Real-Time Machine Learning for Neutrino Telescopes Via Edge Computing}},}\ }\href {\doibase 10.1088/1475-7516/2024/06/026} {\bibfield  {journal} {\bibinfo  {journal} {JCAP}\ }\textbf {\bibinfo {volume} {06}},\ \bibinfo {pages} {026} (\bibinfo {year} {2023})},\ \Eprint {http://arxiv.org/abs/2311.04983} {arXiv:2311.04983 [hep-ex]} \BibitemShut {NoStop}%
\bibitem [{\citenamefont {Eller}(2023)}]{Eller:2023myr}%
  \BibitemOpen
  \bibfield  {author} {\bibinfo {author} {\bibfnamefont {Philipp}\ \bibnamefont {Eller}} (\bibinfo {collaboration} {IceCube}),\ }\bibfield  {title} {\enquote {\bibinfo {title} {{Public Kaggle Competition ''IceCube -- Neutrinos in Deep Ice''}},}\ }in\ \href@noop {} {\emph {\bibinfo {booktitle} {{38th International Cosmic Ray Conference}}}}\ (\bibinfo {year} {2023})\ \Eprint {http://arxiv.org/abs/2307.15289} {arXiv:2307.15289 [astro-ph.HE]} \BibitemShut {NoStop}%
\bibitem [{\citenamefont {Lazar}\ \emph {et~al.}(2023)\citenamefont {Lazar}, \citenamefont {Meighen-Berger}, \citenamefont {Haack}, \citenamefont {Kim}, \citenamefont {Giner},\ and\ \citenamefont {Arg\"uelles}}]{Lazar:2023rol}%
  \BibitemOpen
  \bibfield  {author} {\bibinfo {author} {\bibfnamefont {Jeffrey}\ \bibnamefont {Lazar}}, \bibinfo {author} {\bibfnamefont {Stephan}\ \bibnamefont {Meighen-Berger}}, \bibinfo {author} {\bibfnamefont {Christian}\ \bibnamefont {Haack}}, \bibinfo {author} {\bibfnamefont {David}\ \bibnamefont {Kim}}, \bibinfo {author} {\bibfnamefont {Santiago}\ \bibnamefont {Giner}}, \ and\ \bibinfo {author} {\bibfnamefont {Carlos~A.}\ \bibnamefont {Arg\"uelles}},\ }\bibfield  {title} {\enquote {\bibinfo {title} {{Prometheus: An Open-Source Neutrino Telescope Simulation}},}\ }\href {\doibase 10.1016/j.cpc.2024.109298} {\bibfield  {journal} {\bibinfo  {journal} {Comput. Phys. Commun.}\ }\textbf {\bibinfo {volume} {304}},\ \bibinfo {pages} {109298} (\bibinfo {year} {2023})},\ \Eprint {http://arxiv.org/abs/2304.14526} {arXiv:2304.14526 [hep-ex]} \BibitemShut {NoStop}%
\bibitem [{\citenamefont {Lazar}\ \emph {et~al.}(2024)\citenamefont {Lazar}, \citenamefont {Olavarrieta}, \citenamefont {Gatti}, \citenamefont {Arg\"uelles},\ and\ \citenamefont {Sanz}}]{Lazar:2024luq}%
  \BibitemOpen
  \bibfield  {author} {\bibinfo {author} {\bibfnamefont {Jeffrey}\ \bibnamefont {Lazar}}, \bibinfo {author} {\bibfnamefont {Santiago~Giner}\ \bibnamefont {Olavarrieta}}, \bibinfo {author} {\bibfnamefont {Giancarlo}\ \bibnamefont {Gatti}}, \bibinfo {author} {\bibfnamefont {Carlos~A.}\ \bibnamefont {Arg\"uelles}}, \ and\ \bibinfo {author} {\bibfnamefont {Mikel}\ \bibnamefont {Sanz}},\ }\bibfield  {title} {\enquote {\bibinfo {title} {{New Pathways in Neutrino Physics via Quantum-Encoded Data Analysis}},}\ }\href@noop {} {\  (\bibinfo {year} {2024})},\ \Eprint {http://arxiv.org/abs/2402.19306} {arXiv:2402.19306 [hep-ex]} \BibitemShut {NoStop}%
\bibitem [{VLV()}]{VLVnT}%
  \BibitemOpen
  \href@noop {} {\enquote {\bibinfo {title} {Very large volume neutrino telescopes},}\ }\bibinfo {howpublished} {\url{http://www.vlvnt.nl/}},\ \bibinfo {note} {2024}\BibitemShut {NoStop}%
\bibitem [{NuT()}]{NuTeL}%
  \BibitemOpen
  \href@noop {} {\enquote {\bibinfo {title} {Neutrino telescopes},}\ }\bibinfo {howpublished} {\url{https://agenda.infn.it/event/33107/}},\ \bibinfo {note} {2024}\BibitemShut {NoStop}%
\bibitem [{GNN()}]{GNN}%
  \BibitemOpen
  \href@noop {} {\enquote {\bibinfo {title} {The global neutrino network},}\ }\bibinfo {howpublished} {\url{https://www.globalneutrinonetwork.org/}},\ \bibinfo {note} {2024}\BibitemShut {NoStop}%
\end{thebibliography}%
